\documentclass[12pt]{article}
\usepackage{amssymb}
\usepackage{graphics}
\usepackage{psfrag}
\usepackage[mathscr]{eucal}

\DeclareFontFamily{U}{euc}{}
\DeclareFontShape{U}{euc}{m}{n}{<-6>eurm5<6-8>eurm7<8->eurm10}{}% 
\DeclareSymbolFont{AMSc}{U}{euc}{m}{n} 
\DeclareMathSymbol{\upsi}{\mathord}{AMSc}{"20}
\DeclareMathSymbol{\ulambda}{\mathord}{AMSc}{"15}
\DeclareMathSymbol{\usigma}{\mathord}{AMSc}{"1B}
\DeclareMathSymbol{\utau}{\mathord}{AMSc}{"1C}
\DeclareMathSymbol{\uups}{\mathord}{AMSc}{"1D}

\catcode`\@=11
\def\@citex[#1]#2{%
\if@filesw \immediate \write \@auxout {\string \citation {#2}}\fi
\@tempcntb\m@ne \let\@h@ld\relax \def\@citea{}%
\@cite{%
  \@for \@citeb:=#2\do {%
    \@ifundefined {b@\@citeb}%
      {\@h@ld\@citea\@tempcntb\m@ne{\bf ?}%
      \@warning {Citation `\@citeb ' on page \thepage \space undefined}}%
      {\@tempcnta\@tempcntb \advance\@tempcnta\@ne%
      \@tempcntb\number\csname b@\@citeb \endcsname \relax%
      \ifnum\@tempcnta=\@tempcntb %Number follows previous--hold on to it
        \ifx\@h@ld\relax%
          \edef \@h@ld{\@citea\csname b@\@citeb\endcsname}%
        \else%
          \edef\@h@ld{\ifmmode{-}\else--\fi\csname b@\@citeb\endcsname}%
        \fi%
      \else%   %  non-successor--dump what's held and do this one
        \@h@ld\@citea\csname b@\@citeb \endcsname%
        \let\@h@ld\relax%
      \fi}%
    \def\@citea{,\penalty\@highpenalty\,}%
  }\@h@ld
}{#1}}

\def\@citeb#1#2{{[#1]\if@tempswa , #2\fi}}
\def\@citeu#1#2{{$^{#1}$\if@tempswa , #2\fi }}
\def\@citep#1#2{{#1\if@tempswa , #2\fi}}

\def\bcites{         % cite with []'s
        \catcode`\@=11
        \let\@cite=\@citeb
        \catcode`\@=12
}

\def\upcites{         % cite with exponents
        \catcode`\@=11
        \let\@cite=\@citeu
        \catcode`\@=12
}

\def\plaincites{      % cite without brackets
        \catcode`\@=11
        \let\@cite=\@citep
        \catcode`\@=12
}

\newcount\hour
\newcount\minute
\newtoks\amorpm
\hour=\time\divide\hour by 60
\minute=\time{\multiply\hour by 60 \global\advance\minute by-\hour}
\edef\standardtime{{\ifnum\hour<12 \global\amorpm={am}%
        \else\global\amorpm={pm}\advance\hour by-12 \fi
        \ifnum\hour=0 \hour=12 \fi
        \number\hour:\ifnum\minute<10 0\fi\number\minute\the\amorpm}}
\edef\militarytime{\number\hour:\ifnum\minute<10 0\fi\number\minute}

\def\draftlabel#1{{\@bsphack\if@filesw {\let\thepage\relax
   \xdef\@gtempa{\write\@auxout{\string
      \newlabel{#1}{{\@currentlabel}{\thepage}}}}}\@gtempa
   \if@nobreak \ifvmode\nobreak\fi\fi\fi\@esphack}
        \gdef\@eqnlabel{#1}}
\def\@eqnlabel{}
\def\@vacuum{}
\def\marginnote#1{}
\def\draftmarginnote#1{\marginpar{\raggedright\scriptsize\tt#1}}
\def\draft{
        \pagestyle{plain}
        \overfullrule=2pt
        \oddsidemargin -.5truein
        \def\@oddhead{\sl \phantom{\today\quad\militarytime} \hfil
        \smash{\Large\sl DRAFT} \hfil \today\quad\militarytime}
        \let\@evenhead\@oddhead
        \let\label=\draftlabel
        \let\marginnote=\draftmarginnote
        \def\ps@empty{\let\@mkboth\@gobbletwo
        \def\@oddfoot{\hfil \smash{\Large\sl DRAFT} \hfil}
        \let\@evenfoot\@oddhead}
        \def\@eqnnum{(\theequation)\rlap{\kern\marginparsep\tt\@eqnlabel}%
        \global\let\@eqnlabel\@vacuum}  }

\def\section{\@startsection {section}{1}{\z@}{3.ex plus 1ex minus
 .2ex}{2.ex plus .2ex}{\large\bf}}
\def\subsection{\@startsection{subsection}{2}{\z@}{2.75ex plus 1ex minus
 .2ex}{1.5ex plus .2ex}{\bf}}        

\def\appendix{{\newpage\section*{Appendix}}\let\appendix\section%
        {\setcounter{section}{0}
        \gdef\thesection{\Alph{section}}}\section}

\def\abstract{\if@twocolumn
\section*{Abstract}
\else %\small
\begin{center}
{\bf Abstract\vspace{-.5em}\vspace{0pt}}
\end{center}
\quotation
\fi}

\catcode`\@=12

\newcommand{\beq}{\begin{equation}}
\newcommand{\eeq}{\end{equation}}
\newcommand{\beqa}{\begin{eqnarray}}
\newcommand{\eeqa}{\end{eqnarray}}
\newcommand{\dd}{{\rm d}}

\newcommand{\Z}{{\bf Z}}

\newcommand{\R}{{\bf R}}

\newcommand{\C}{{\bf C}}
\newcommand{\CC}{{\mathbb C}}
\newcommand{\PP}{{\mathbb P}}
\newcommand{\e}{\,{\rm e}}
\newcommand{\CP}{{\CC\PP}}

\newcommand{\be}{\begin{equation}}
\newcommand{\ee}{\end{equation}}
\newcommand{\bea}{\begin{eqnarray}}
\newcommand{\eea}{\end{eqnarray}}

\def\to{\rightarrow}

\def\lae{\mathrel{\mathop{\smash{\lower .5 ex \hbox{$\stackrel<\sim$}}}}}
\def\lae{\mathrel{\mathop{\smash{\lower .5 ex \hbox{$\stackrel>\sim$}}}}}

\def\Tr{{\rm Tr}}
\def\l:{\mathopen{:}\,}
\def\r:{\,\mathclose{:}}

\catcode`\@=11
\def\theequation{\arabic{equation}}
\catcode`\@=12

\bcites

\catcode`\@=11
\def\theequation{\thesection.\arabic{equation}}
\@addtoreset{equation}{section}
\@addtoreset{footnote}{section}
\@addtoreset{footnote}{subsection}
\catcode`\@=12

\newcommand{\bepsilon}{\overline{\epsilon}}

\newcommand{\bphi}{\overline{\phi}}
\newcommand{\bpsi}{\overline{\psi}}

\newcommand{\bi}{\overline{\imath}}
\newcommand{\bj}{\overline{\jmath}}
\newcommand{\bz}{\overline{z}}

\newcommand{\blambda}{\overline{\lambda}}
\newcommand{\bsigma}{\overline{\sigma}}

\newcommand{\wh}{\widehat}
\newcommand{\wt}{\widetilde}

\newcommand{\nn}{\nonumber}
\newcommand{\cL}{{\mathcal L}}
\newcommand{\Dirac}{\,\!\not{\!\!D}}
\newcommand{\dirac}{\!\not{\!\partial}}
\newcommand{\nbirac}{\!\not{\!\nabla}}
\newcommand{\leftnbirac}{\!\not{\!\overleftarrow{\nabla}}}
\newcommand{\ttt}{\mathfrak{t}}
\newcommand{\im}{{\rm i}}

\newcommand{\wtL}{\widetilde{L}}
\newcommand{\wtG}{\widetilde{G}}
\newcommand{\wtJ}{\widetilde{J}}
\newcommand{\tepsilon}{\widetilde{\epsilon}}
\newcommand{\btepsilon}{\overline{\widetilde{\epsilon}}}
\newcommand{\half}{{1\over 2}}
\newcommand{\ch}{{\rm ch}}
\newcommand{\longto}{\longrightarrow}
\newcommand{\whn}{\wh{n}}
\newcommand{\bvarepsilon}{\overline{\varepsilon}}
\newcommand{\sDelta}{\mathit{\Delta}}

\begin{document}

\begin{titlepage}

\begin{center}

\today\hfill
hep-th/yymmnnn\\

\vskip 2.5 cm
{\large \bf Exact Results
In Two-Dimensional (2,2) Supersymmetric\\[0.1cm]
 Gauge Theories
With Boundary}
\vskip 1 cm

{Kentaro Hori ~ and ~ Mauricio Romo}\\
\vskip 0.5cm
{\it Kavli IPMU, The University of Tokyo, Kashiwa, Japan}

\end{center}

\vskip 0.5 cm
\begin{abstract}
We compute the partition function on the hemisphere of a class
of two-dimensional (2,2) supersymmetric field theories including gauged linear
sigma models. The result provides a general exact formula for the
central charge of the D-brane placed at the boundary.
It takes the form of Mellin-Barnes integral and the question of
its convergence leads to the grade restriction rule
concerning branes near the phase boundaries.
We find expressions in various phases including
the large volume formula in which a characteristic class called
the Gamma class shows up.
The two sphere partition function factorizes
into two hemispheres glued by inverse to the annulus.
The result can also be written in a form familiar in mirror symmetry,
and suggests a way to find explicit mirror correspondence between branes.
\end{abstract}

\end{titlepage}

\newpage

%{\footnotesize
\tableofcontents
%}

\newpage
\section{Introduction}

Localization has been a powerful tool to obtain exact results
in supersymmetric quantum field theories since the very beginning
\cite{Windex}. Recently, the idea was applied to a part
of superconformal symmetry which exists for a class of spacetime
even if the theory is not conformally invariant.
After the pioneering work by Pestun \cite{Pestun}, several important results are
obtained in three, four and five dimensions, and
qualitatively new information of the respective theory is obtained.
Along this line, the partition function on the two sphere of
$(2,2)$ supersymmetric gauge theories was computed by Benini et al 
\cite{Beninietal} and Doroud et al \cite{Doroudetal}.
Interestingly, it was observed in some examples \cite{Romoetal}
that it computes the K\"ahler potential for the family of superconformal
fixed points of the theory.

These developments motivate
 us to study the partition function on the hemisphere
of two-dimensional $(2,2)$ supersymmetric gauge theories.
The result will surely depend on the choice of boundary condition,
or the D-brane, at the boundary of the hemisphere.
There is a chance that it will tell us something non-trivial
about D-branes or even about the theory itself.
In this paper, we formulate $(2,2)$ supersymmetric gauge theories
on the hemisphere, compute the partition function based on localization,
and study some of its properties. Our main target is the
class of theories called the gauged linear sigma models 
with the supersymmetry that admits the type of branes called
the B-branes at the boundary.

One obvious question is: what does it compute?
We find in general that it depends holomorphically on
twisted chiral parameters and has no dependence on chiral parameters.
This suggests that it is the central charge of the D-brane.
Indeed, after the computation, we find that it agrees with the central charge
whenever the computation can be completed on both sides.
In particular, the match can be made in various phases
of the theories.
In the Landau-Ginzburg orbifold point, the result agrees with the
formula for the central charge proposed in \cite{WalcherLG}.
In the geometric phase, in which the theory reduces to
the non-linear sigma model with a K\"ahler target space $X$,
the large volume limit of our result is
\beq
Z_{D^2}({\mathcal B})~=~\int_X\wh{\Gamma}_X\e^{B+{\im\over 2\pi}\omega}
\ch({\mathcal B})\,+\,\cdots
\eeq
where $\wh{\Gamma}_X$ is a characteristic class of $X$ called the
Gamma class \cite{Hosono,Iritani09,Katzarkov,Iritani11},
$B$ and $\omega$ are the B-field and the K\"ahler class of $X$
respectively.
This is the expected behaviour for the central charge of the brane.
$+\cdots$ is the worldsheet instanton corrections, and we provide a precise
form of such corrections, to all orders in the instanton number.
In a class of theories, the exact expression for the central charge
is known by mirror symmetry and/or by detailed instanton calculus.
Our results matches with that in such cases.
In other cases where the expression is not known, our result can
be regarded as a prediction for the central charge.

The conjecture of \cite{Romoetal} 
and our observation suggests that the two sphere partition function can be
factorized into two hemispheres glued by the inverse to the
annulus. We show that this is indeed the case in the geometric
phase in which the formula for  the annulus is known.
We find though that the formula given in \cite{Beninietal,Doroudetal}
should be corrected by a shift of theta angle. This is extremely
subtle becuase it is just a matter of sign in the sum over
different topological sectors, and it is not always non-trivial.

The most interesting aspect of our study is that the formula for
the partition function is written as an integral of some meromorphic form
and the choice of contour is related to the choice of boundary condition for
the vector multiplet of the gauge theory. There is no apriori rule
to decide the boundary condition, and the convergence of the integral
can give us some hint to find it. 
Deep in phases one can usually find the contour so that the integral
converges for an arbitrary brane. However, near the phase boundary,
a convergent
contour can be found only for a very restricted class of branes.
This reproduces the {\it grade restriction rule} found in
\cite{HHP} in the Abelian and Calabi-Yau cases, and generalizes it
or provides a way of generalization in non-Abelian and non-Calabi-Yau cases.
This is of importance to the study of analytic behaviour of
the partition function, especially across the phase boundary.
The integral is of the form called Mellin-Barnes integral.
The present work shows that the issue of convergence of such
integrals encodes a rich physics content.

Using the most famous formula for the gamma function,
we can convert our result into the formula for the central charge
found in \cite{HV} during the derivation of the mirror symmetry.
This provides a proof of the conjecture that
the hemisphere computes the central charge, 
in the gauge theories we study.
The precise corrspondence between the original B-brane
and the mirror A-brane was out of reach in the method of \cite{HV,HIV}.
Our results can now be used to find the correspondence,
at least at the level of the Ramond-Ramond charge.

Our work may be of interest from another point of view.
Supersymmetric field theories on higher dimensional
spacetime with boundary can be an interesting subject of study
in its own right and also in relation to the dymnamics
of branes in superstring theory and M theory.
We believe that some experience in two dimensions will be of some
help in such investigations.

While our work was in progress but no sentence was written,
we were informed by two groups of people, one by
Daigo Honda and Takuya Okuda, and another by
Sotaro Sugishita and Seiji Terashima, that they are working on a possibly
related subject and that they were getting to be ready or were ready
to publish their papers.
We asked them to wait for us to write up our
results, and they kindly agreed to do so.
We would like to thank them for their generosity.

\section{Supersymmetry}

In this section, we write down the supersymmetry transformation
rule of component fields in various supermultiplets on
the sphere and the hemisphere.

\subsection*{Notation}

We follow the convention originated from Wess-Bagger
\cite{WessBagger} concerning the notation for the variational
parameters as well as the component fields
$(\phi,\psi,F)$ and $(v,\sigma,\lambda,D)$ of the chiral and vector
multiplets.
Dimensional reduction \cite{Wphases} and a $\sqrt{2}$ rescaling
yields a notation in two-dimensional
Minkowski space \cite{cmibook}. The relation to \cite{WessBagger}
and \cite{Wphases} (superscripts ``WB'' and ``W'' respectively) is
\beqa
&\epsilon_{\pm}=\sqrt{2}\epsilon^{\rm W}_{\pm},\nn\\
&\lambda_{\pm}=\sqrt{2}\lambda^{\rm W}_{\pm},\qquad
\sigma=\sigma_1+\im \sigma_2=
\sqrt{2}\sigma^{\rm W}=v^{\rm WB}_1-\im v^{\rm WB}_2,\nn\\
&v_0=v^{\rm W}_0=v^{\rm WB}_0,\quad v_1=v^{\rm W}_1=v^{\rm WB}_3.\nn
\eeqa
The same relation holds for the hermitian conjugates.
Other components, $\phi,\psi_{\pm}$, their hermitian conjugates, and $D$,
are trivially related.
Notation in the Euclidean space is obtained by
Wick rotation, $x^0\to -\im x^2$, $v_0\to \im v_2$.
We also write
\beq
D=\im D_E,\qquad
F=\im f\stackrel{\rm or}{=}\im F_E
\quad\overline{F}=\im \overline{f}\stackrel{\rm or}{=}\im\overline{F}_E.
\eeq
In Euclidean signature, $D_E$ is real
and $(f,\overline{f})$ (or $(F_E,\overline{F}_E)$) is the complex
conjugate pair,
\beq
D_E^{\dag}=D_E,\qquad
\overline{f}=f^{\dag}\quad
\mbox{(or $\overline{F}_E=F_E^{\dag}$)}.
\eeq
The fermionic parameters and fields are to be regarded as sections of
the spinor bundle $S=S_-\oplus S_+$ on the Euclidean space $\R^2$,
with metric $\dd s^2=|\dd z|^2$, ($g_{z\bz}=1/2$), where
$z=x^1+\im x^2$. See Appendix~\ref{app:spinors} for conventions and facts on
spinors on two-dimensional manifolds.
For example,
\beq
\psi=\psi_-\sqrt{\dd z}+\psi_+\sqrt{\dd \bz},\qquad
\bpsi=\bpsi_+\sqrt{\dd z}+\bpsi_-\sqrt{\dd \bz},
\eeq
and similarly for the variational parametes $\epsilon,\bepsilon$.
For the fermions in vector multiplets it is more useful
to write
\beq
\lambda=-\im\lambda_-\sqrt{\dd z}+\im\blambda_+\sqrt{\dd \bz},\qquad
\blambda=\im \blambda_-\sqrt{\dd z}-\im \lambda_+\sqrt{\dd \bz}.
\eeq

\subsection{Superconformal Transformations}

We first write down the $(2,2)$ superconformal transformations
of various supermultiplet fields (taken from \cite{Beninietal,Doroudetal}).
They are obtained by replacing the pair $(\epsilon,\bepsilon)$ of 
constant spinors on the Euclidean space $\R^2$ by a pair of
(local) conformal Killing spinors on a curved surface with a spin structure
$(\Sigma,g)$, via the Weyl covariantization procedure \cite{Doroudetal}.

\subsubsection*{\rm \underline{Chiral multiplet} (vector R-charge $R$):}
\beqa
&\delta\phi=\langle\epsilon,\psi\rangle,
\quad
\delta\bphi=-\langle\bepsilon,\bpsi\rangle,\nn\\
&\delta \psi=\im\dirac\phi\bepsilon+\im{R\over 2}\phi\nbirac\bepsilon
+\im f\epsilon,\quad
\delta \bpsi=-\im\dirac\bphi\epsilon-\im {R\over 2}\bphi\nbirac\epsilon
+\im\overline{f}\bepsilon,\nn\\
&\delta f=\langle \bepsilon,\nbirac\psi\rangle
-{R\over 2}\langle\nbirac\bepsilon,\psi\rangle,\quad
\delta\overline{f}=\langle\epsilon,\nbirac\bpsi\rangle
-{R\over 2}\langle\nbirac\epsilon,\bpsi\rangle
\label{scchiral}
\eeqa

\subsubsection*{\rm \underline{Twisted chiral multiplet} (axial R-charge $R$):}

The transformation is obtained from the one for the chiral multiplet
by swapping $\epsilon_+$ and $\bepsilon_+$ while
keeping $\epsilon_-$ and $\bepsilon_-$ intact.
In other words, replacing $(\epsilon,\bepsilon)$
by $(\tepsilon,\btepsilon)$ defined by
\beq
\tepsilon:=P_-\epsilon+P_+\bepsilon,
\qquad
\btepsilon:=P_-\bepsilon+P_+\epsilon.
\label{deftepsilon}
\eeq
This has full information but let us anyway write down the transformations
\beqa
&\delta u=\langle\tepsilon,\chi\rangle,
\quad
\delta\overline{u}=-\langle\btepsilon,\overline{\chi}\rangle,\nn\\
&\delta \chi=\im\dirac u\btepsilon+\im{R\over 2} u\nbirac\btepsilon
+\im g\tepsilon,\quad
\delta \overline{\chi}
=-\im\dirac\overline{u}\tepsilon-\im {R\over 2}\overline{u}
\nbirac\tepsilon
+\im\overline{g}\btepsilon,\nn\\
&\delta g=\langle \btepsilon,\nbirac\chi\rangle
-{R\over 2}\langle\nbirac\btepsilon,\chi\rangle,\quad
\delta\overline{g}=\langle\tepsilon,\nbirac\overline{\chi}\rangle
-{R\over 2}\langle\nbirac\tepsilon,\overline{\chi}\rangle
\label{sctwistedchiral}
\eeqa

\subsubsection*{\rm \underline{Vector multiplet}:}

\beqa
&\delta v_{\mu}=\half\langle\tepsilon,\gamma_{\mu}\gamma_3\blambda\rangle
-\half\langle\btepsilon,\gamma_{\mu}\gamma_3\lambda\rangle,\nn\\
&\delta\sigma=\langle\tepsilon,\lambda\rangle,\quad
\delta\bsigma=-\langle\btepsilon,\blambda\rangle,\nn\\
&\delta\lambda=\im \Dirac\sigma\btepsilon+\im\sigma\nbirac\btepsilon
+\im\left(D_E+\im{v_{12}\over\sqrt{g}}\right)\tepsilon
-\half[\sigma,\bsigma]\gamma_3\tepsilon,\nn\\
&\delta\blambda=-\im \Dirac\bsigma\tepsilon-\im\bsigma\nbirac\tepsilon
+\im\left(D_E-\im {v_{12}\over\sqrt{g}}\right)\btepsilon
-\half[\sigma,\bsigma]\gamma_3\btepsilon,\nn\\
&\delta D_E=\half\left(
\langle\tepsilon,\Dirac\blambda-\im\gamma_3[\bsigma,\lambda]\rangle
-\langle\nbirac\tepsilon,\blambda\rangle
+\langle\btepsilon,\Dirac\lambda-\im\gamma_3[\sigma,\blambda]\rangle
-\langle\nbirac\btepsilon,\lambda\rangle
\right).
\label{scvector}
\eeqa
We also note
\beq
\delta\left(\im {v_{12}\over\sqrt{g}}\right)
=\half\left(
-\langle\tepsilon,\Dirac\blambda\rangle
+\langle\nbirac\tepsilon,\blambda\rangle
+\langle\btepsilon,\Dirac\lambda\rangle
-\langle\nbirac\btepsilon,\lambda\rangle
\right).
\eeq
We see that $(\sigma,\lambda,D_E+\im {v_{12}\over\sqrt{g}})$
transform as fields in a twisted chiral multiplet of axial R-charge $2$,
i.e. as $(u,\chi,g)$ in (\ref{sctwistedchiral}) with $R=2$,
up to commutator terms.

\subsubsection*{\rm \underline{Charged chiral multiplet}:}

\beqa
&\delta\phi=\langle\epsilon,\psi\rangle,
\quad
\delta\bpsi=-\langle\bepsilon,\bpsi\rangle,\nn\\
&\delta \psi=\im\left(\Dirac\phi+{R\over 2}\phi\nbirac
-\im\sigma_1\phi\gamma_3+\sigma_2\phi
\right)\bepsilon
+\im f\epsilon,\nn\\
&\delta \bpsi=-\im\left(\Dirac\bphi+{R\over 2}\bphi\nbirac
+\im\bphi\sigma_1\gamma_3+\bphi\sigma_2
\right)\epsilon
+\im\overline{f}\bepsilon,\nn\\
&\delta f=\langle \bepsilon,\Dirac\psi
+{R\over 2}\leftnbirac\psi-\im\gamma_3\sigma_1\psi-\sigma_2\psi
-\overline{\widetilde{\lambda}}\phi\rangle,
\nn\\
&\delta\overline{f}=\langle
\Dirac\bpsi
+{R\over 2}\bpsi\nbirac+\im\gamma_3\bpsi\sigma_1-\bpsi\sigma_2
-\bphi\widetilde{\lambda},\epsilon\rangle.
\label{scchargedchiral}
\eeqa
Here $\widetilde{\lambda}=\lambda_-\sqrt{\dd z}+\lambda_+\sqrt{\dd \bz}$ and
$\overline{\widetilde{\lambda}}=\blambda_-\sqrt{\dd z}
+\blambda_+\sqrt{\dd \bz}$.

\subsection*{Commutation Relations}

The above superconformal transformations form a closed algebra
together with conformal and R-symmetry transformations,
up to gauge transformations.
Under conformal transformations, generated by vector fields
$X=X^{\mu}\partial_{\mu}$ such that 
$\partial_{\bz}X^z=\partial_zX^{\bz}=0$,
all the fields transform like primary fields (in the sense of \cite{BPZ})
\beq
\delta^{\rm conf}_X{\mathcal O}
=X^{\mu}D_{\mu}{\mathcal O}+(\Delta\nabla_zX^z
+\wt{\Delta}\nabla_{\bz}X^{\bz}){\mathcal O},
\eeq
except the gauge potential which transforms as
\beq
\delta_X^{\rm conf}v_{\mu}=X^{\nu}v_{\nu\mu}.
\eeq
$(\Delta,\wt{\Delta})$ are the ``conformal weights'' of ${\mathcal O}$
listed below together with the R-charges
$$
\begin{array}{c|cccccccc}
{\mathcal O}&\phi&\psi_-&\psi_+&f&\sigma&\lambda_-&\blambda_+&D_E\\
\hline
(\Delta,\wt{\Delta})&({R\over 4},{R\over 4})&\!({R\over 4}+\half,{R\over 4})\!&
({R\over 4},{R\over 4}+\half)&\!\!({R\over 4}+\half,{R\over 4}+\half)\!\!&
(\half,\half)&(1,\half)&(\half,1)&(1,1)\\
(F_V,F_A)&(R,0)&\!(R-1,1)\!&\!(R-1,-1)\!&(R-2,0)&(0,2)&(1,1)&(-1,1)&(0,0)
\end{array}
$$
Under $(\phi,\psi_{\pm},f)\to (\bphi,\bpsi_{\pm},\overline{f})$
and $(\sigma,\lambda_{\pm})\to (\bsigma,\blambda_{\pm})$,
conformal weights do not change but the R-charges changes their signs.
Those for twisted chiral multiplet fields
are obtained from the ones for chiral multiplet fields
by the exchange of $F_V$ and $F_A$ (no change in $(\Delta,\wt{\Delta})$).
It is useful to introduce right and left handed R-symmetries
$F_R$ and $F_L$ defined by
$F_V=F_R+F_L$ and $F_A=-F_R+F_L$.
We write te corresponding symmetry transformations by
$\delta^{\rm right}$ and $\delta^{\rm left}$.

Let $\delta_1$ and $\delta_2$ be the superconformal transformations
with conformal Killing spinors $(\epsilon_1,\bepsilon_1)$
and $(\epsilon_2,\bepsilon_2)$. For all fields ${\mathcal O}$,
we find
\beq
[\delta_2,\delta_1]{\mathcal O}=
\delta^{\rm conf}_{X}{\mathcal O}
+\delta^{\rm right}_{\Theta}{\mathcal O}
+\delta^{\rm left}_{\wt{\Theta}}{\mathcal O}
+\delta^{\rm gauge}_{\im\Lambda}{\mathcal O}.
\label{commutator}
\eeq
where
\beqa
X^{\mu}&=&\im\langle \epsilon_{[1},\gamma^{\mu}\bepsilon_{2]}\rangle,
\\[0.25cm]
\Theta&=&{\im\over 2}\left(\langle\nbirac\epsilon_{[1},P_-\bepsilon_{2]}\rangle
+\langle P_-\epsilon_{[1},\nbirac\bepsilon_{2]}\rangle\right),
\\
\wt{\Theta}&=&{\im\over 2}\left(\langle\nbirac\epsilon_{[1},
P_+\bepsilon_{2]}\rangle
+\langle P_+\epsilon_{[1},\nbirac\bepsilon_{2]}\rangle\right),
\\[0.25cm]
\im\Lambda&=&\langle\epsilon_{[1},(\gamma_3\sigma_1+\im\sigma_2)
\bepsilon_{2]}\rangle.
\eeqa
Here
$\langle \epsilon_{[1},\gamma^{\mu}\bepsilon_{2]}\rangle:=
\langle \epsilon_{1},\gamma^{\mu}\bepsilon_{2}\rangle
-\langle \epsilon_{2},\gamma^{\mu}\bepsilon_{1}\rangle$, etc.
It is straightforward to find
\beq
X^z=2\im \epsilon^-_{[1}\bepsilon^-_{2]},\qquad
X^{\bz}=-2\im \epsilon^+_{[1}\bepsilon^+_{2]},
\eeq
\beq
\Theta=\im\epsilon^-_{[1}\partial_z\bepsilon^-_{2]}
-\im\partial_z\epsilon^-_{[1}\bepsilon^-_{2]},\qquad
\wt{\Theta}=-\im\epsilon^+_{[1}\partial_{\bz}\bepsilon^+_{2]}
+\im\partial_{\bz}\epsilon^+_{[1}\bepsilon^+_{2]}.
\eeq
We see that $X^z$ and $\Theta$ are holomorphic and
$X^{\bz}$ and $\wt{\Theta}$ are antiholomorphic, as they should be.
The commutation relation (\ref{commutator}) and other obvious ones
form the $(2,2)$ superconformal algebra. 
Let us spell out a correspondence to the more standard notation.
We write the superconformal
transformation for the pair $(\epsilon,\bepsilon)$ of conformal Killing
spinors by $\delta=\delta^-_{\epsilon}+\delta^+_{\bepsilon}$.
As a local basis of conformal Killing spinors, we use
\beq
{\bf s}_r=z^{r+\half}\sqrt{\partial\over\partial z},\qquad
\wt{\bf s}_r=\bz^{r+\half}\sqrt{\partial\over\partial \bz}.
\eeq
Then, the correspondence is
\beqa
&&L_n=\delta^{\rm conf}_{z^{n+1}{\partial\over\partial z}},\qquad
\wt{L}_n=\delta^{\rm conf}_{\bz^{n+1}{\partial\over\partial \bz}},\\
&&G^{\pm}_r=\e^{-{\pi\im\over 4}}\delta^{\pm}_{{\bf s}_r},\qquad
\wt{G}^{\pm}_r=\e^{\pi\im\over 4}\delta^{\pm}_{\wt{\bf s}_r},\\
&&J_n=\delta^{\rm right}_{z^n},\qquad \wt{J}_n=\delta^{\rm left}_{\bz^n}.
\eeqa
They indeed obey the standard commutation relations,
such as those in \cite{LVW},
in which the central terms are set equal to zero.

\subsection{Supersymmetry On The Sphere And The Hemisphere}

\newcommand{\hatL}{\wh{L}}

We shall formulate $(2,2)$ supersymmetric field theories on 
the sphere and the hemisphere in such a way that a part of the
$(2,2)$ superconformal symmetry is preserved.
We consider the Riemann sphere with coordinates $z$ and $w$, $zw=1$,
and the southern or northern hemisphere defined by $|z|\leq 1$ or
$|w|\leq 1$ respectively.
%The sphere is covered by the complex $z$-plane and $w$-plane related by $zw=1$.
%It is divided at the equator $|z|=1$ into two --- the southern
%and northern hemispheres, $D^2_0$ and $D^2_{\infty}$,
%consisting of points with $|z|\leq 1$ and $|w|\leq 1$ respectively.
See Appendix~\ref{app:spinors} for more details on the facts on
the sphere and the hemispheres.
Since we consider theories which are not necessarily conformally invariant,
out of the conformal generators we can at most keep isometry generators.
On the round sphere,  the isometry group is $O(3)$ generated by
\beq
\ell_3=-z{\partial\over\partial z}+\bz{\partial\over\partial \bz},\quad
\ell_+=z^2{\partial\over\partial z}+{\partial\over\partial \bz},\quad
\ell_-=-{\partial\over\partial z}-\bz^2{\partial\over\partial \bz}.
\eeq
Note also that only ${\bf s}_{\pm\half}$ and $\wt{\bf s}_{\pm\half}$
are globally defined conformal Killing spinors.
On the hemisphere, only $\ell_3$ can be the isometry generator, and
only the linear combinations
${\bf s}_{\half}\pm\wt{\bf s}_{-\half}$ and
${\bf s}_{-\half}\pm\wt{\bf s}_{\half}$ satisfy the boundary condition
at $|z|=1$ for some spin structure.
Thus, what can be included are the isometry generators,
\beqa
\hatL_3&=&\delta^{\rm conf}_{\ell_3}=-L_0+\wtL_0,\nn\\
\hatL_+&=&\delta^{\rm conf}_{\ell_+}=L_1+ \wtL_{-1},\nn\\
\hatL_-&=&\delta^{\rm conf}_{\ell_-}=-L_{-1}-\wtL_{1},
\eeqa
the supercharges
\beq
G^+_{\pm\half},\quad G^-_{\pm\half},\quad
\wtG^+_{\pm\half},\quad \wtG^-_{\pm\half},
\eeq
and the R-symmetry generators,
\beqa
F_V&=&\delta^{\rm vector}_1=J_0+\wtJ_0,\nn\\
F_A&=&\delta^{\rm axial}_1=-J_0+\wtJ_0.
\eeqa
We would like to find subsets of these, closed under the
commutation relation, which include all the isometries
of the respective geometry and a maximum number of supercharges.

On the round two-sphere, there are two possibilities
\beqa
\mbox{(A-type)}&&\hatL_3,\,\,\hatL_{\pm},\,\,F_V,\nn\\
&&Q^{A\pm}_{(+)}=\delta^\pm_{{\bf s}_{\pm\half}}
+\delta^\pm_{\wt{\bf s}_{\mp\half}}
=\e^{\pi\im\over 4}(G^\pm_{\pm\half}-\im \wtG^\pm_{\mp\half}),\nn\\
&&Q^{A\pm}_{(-)}=\delta^\pm_{{\bf s}_{\mp\half}}
-\delta^\pm_{\wt{\bf s}_{\pm\half}}
=\e^{\pi\im\over 4}(G^\pm_{\mp\half}+\im \wtG^\pm_{\pm\half}),
\label{QAdef}\\
\mbox{(B-type)}&&\hatL_3,\,\,\hatL_{\pm},\,\,F_A,\nn\\
&&Q^{B\pm}_{(+)}=\delta^\mp_{{\bf s}_{\pm\half}}
+\delta^\pm_{\wt{\bf s}_{\mp\half}}
=\e^{\pi\im\over 4}(G^\mp_{\pm\half}-\im \wtG^\pm_{\mp\half}),\nn\\
&&Q^{B\pm}_{(-)}=\delta^\mp_{{\bf s}_{\mp\half}}
-\delta^\pm_{\wt{\bf s}_{\pm\half}}
=\e^{\pi\im\over 4}(G^\mp_{\mp\half}+\im \wtG^\pm_{\pm\half}),
\label{QBdef}
\eeqa
and their axial and vector R-rotations, (A$^\beta$-type)
or (B$^{\alpha}$-type), which are obtained by the replacement
$G^{\pm}_r\to\e^{\mp\im\beta}G^{\pm}_r$,
$\wtG^{\pm}_r\to\e^{\pm\im\beta}\wtG^{\pm}_r$, or
$G^{\pm}_r\to\e^{\pm\im\alpha}G^{\pm}_r$,
$\wtG^{\pm}_r\to\e^{\pm\im\alpha}\wtG^{\pm}_r$, respectively.
The generators from each set form
a closed algebra which is isomorphic to $\mathfrak{osp}(2|2)$.
At the south or the north pole, parts of the 
A-type ({\it resp}. B-type) supercharges define twisted chiral 
({\it resp}. chiral) operators. For example, let us look at the A-type 
supercharges. At the south pole $z=0$, operators annihilated by
$Q^{A+}_{(+)}\sim \e^{-{\pi\im\over 4}}\wtG^+_{-\half}$ and 
$Q^{A-}_{(+)}\sim \e^{\pi\im\over 4}G^-_{-\half}$
are twisted chiral
while those annihilated by
$Q^{A-}_{(-)}\sim -\e^{-{\pi\im\over 4}}\wtG^-_{-\half}$ and
$Q^{A+}_{(-)}\sim \e^{\pi \im\over 4}G^+_{-\half}$
are twisted antichiral, and things are the opposite at the north pole
$z=\infty$.
This is the motivation for the name ``A'' and ``B''.
In the litarature \cite{Beninietal,Doroudetal}, partition function of
gauged linear sigma model preserving A-type supersymmetry is studied, and 
the result depends on the twisted chiral parameters but not on the chiral
parameters--- the K\"ahler parameters but not the complex structure parameters
when there is a non-linear sigma model interpretation.

On the hemisphere, there are four possibilities
\beqa
\mbox{(A$_{(+)}$-type)}&&\hatL_3,
\,\,Q^{A\pm}_{(+)},\,\,F_V,\\
\mbox{(A$_{(-)}$-type)}&&\hatL_3,
\,\,Q^{A\pm}_{(-)},\,\,F_V,\\
\mbox{(B$_{(+)}$-type)}&&\hatL_3,
\,\,Q^{B\pm}_{(+)},\,\,F_A,\\
\mbox{(B$_{(-)}$-type)}&&\hatL_3,
\,\,Q^{B\pm}_{(-)},\,\,F_A,
\eeqa
and their axial and vector R-rotations, (A$^\beta_{(\pm)}$-type)
or (B$^{\alpha}_{(\pm)}$-type).
The generators from each set form a closed algebra:
$(\e^{\pi\im\over 4}Q,\e^{\pi\im\over 4}\overline{Q},F)$ $=$
$(Q^{A\pm}_{(\pm)}, Q^{A\mp}_{(\pm)},\pm F_V)$ or
$(Q^{B\pm}_{(\pm)},Q^{B\mp}_{(\pm)},\pm F_A)$
obeys
\beqa
&Q^2=\overline{Q}^2=0,\nn\\
&\{Q,\overline{Q}\}=-2\hatL_3+F,\nn\\
&[\hatL_3,Q]={1\over 2}Q,\quad 
[\hatL_3,\overline{Q}]=-{1\over 2}\overline{Q},\nn\\
&[F,Q]=Q,\quad
[F,\overline{Q}]=-\overline{Q}.
\label{susyalgebra}
\eeqa

A boundary condition must be specified at the boundary $|z|=1$.
There are basically two types of subalgebra of the $(2,2)$
supersymmetry with half the amount of supercharges
that can be preserved at the boundary \cite{OOY,HIV} --- A-type and B-type
--- and the boundary conditions preserving these are called 
A-branes and B-branes.
In a superconformal field theory, the preserved generators are \cite{OOY}
\beqa
\mbox{A-branes:}&&
L_n-\wtL_n,\quad
G^+_r\pm\im \wtG_{-r}^-,\quad G^-_r\pm\im \wtG_{-r}^+,\quad J_n-\wtJ_{-n},\\
\mbox{B-branes:}&&
L_n-\wtL_n,\quad
G^-_r\pm\im \wtG_{-r}^-,\quad G^+_r\pm\im \wtG_{-r}^+,\quad J_n+\wtJ_{-n},
\eeqa
We see that the boundary conditions preserving
A$_{(\pm)}$ are B-branes while 
those preserving B$_{(\pm)}$ are A-branes.

The sign in the parenthesis, $(\pm)$, corresponds to
the choice of spin structure at the boundary ---
$(\pm)_0$ for the southern hemisphere $D^2_0$ and 
$(\mp)_{\infty}$ for the northern hemisphere $D^2_{\infty}$. 
See Appendix~\ref{app:spinors}. 
We shall denote the partition function on the southern 
hemisphere $D^2_0$ preserving the A$_{(\pm)}$-type and B$_{(\pm)}$-type
supersymmetry by
$Z^{\rm A}_{D^2_0{}_{(\pm)}}$ and $Z^{\rm A}_{D^2_0{}_{(\pm)}}$ respectively,
while 
 the partition function on the northern 
hemisphere $D^2_{\infty}$ preserving the A$_{(\pm)}$-type and B$_{(\pm)}$-type
supersymmetry by
$Z^{\rm A}_{D^2_0{}_{(\mp)}}$ and $Z^{\rm A}_{D^2_0{}_{(\mp)}}$ respectively.
Note that the label $(\pm)$ in the partition function
is correlated with the spin structure.
Since there is really no difference between
the southern and northern hemispheres, we have the equality
\beq
Z_{D^2_0{}_{(\pm)}}^{\rm A}=Z_{D^2_{\infty}{}_{(\pm)}}^{\rm A},\qquad
Z_{D^2_0{}_{(\pm)}}^{\rm B}=Z_{D^2_{\infty}{}_{(\pm)}}^{\rm B},
\label{southnorth}
\eeq
When there is no room of confusion between A and B,
we shall drop the superscript.

\subsection{Some Useful Formulae}

For convenience in later sections, we collect some useful properties of
the variational parameters for the supersymmetry transformations of each type.

We first write down the action of the Dirac operator.
Let us first look at the B-type supersymmetry. The parameters of the four
supercharges are $(\epsilon,\bepsilon)=({\bf s}_{\half},\wt{\bf s}_{-\half})$,
$({\bf s}_{-\half},-\wt{\bf s}_{\half})$,
$(-\wt{\bf s}_{-\half},{\bf s}_{\half})$ and
$(\wt{\bf s}_{\half},{\bf s}_{-\half})$, times a constant 
anticommuting variational parameter. Using (\ref{diracons}), we see that
each satisfies $\nbirac\epsilon=-\bepsilon/r$
and $\nbirac\bepsilon=\epsilon/r$. 
These can also be written as
$\nbirac\tepsilon=\gamma_3\tepsilon/r$,
$\nbirac\btepsilon=-\gamma_3\btepsilon/r$ using
$(\tepsilon,\btepsilon)$ introduced in (\ref{deftepsilon}).
The same applies to the A-type if we replace $(\epsilon,\bepsilon)$
by $(\tepsilon,\btepsilon)$.
To summarize,
\beqa
\mbox{(A-type)}&&\nbirac\tepsilon=-\btepsilon/r,\quad
\nbirac\btepsilon=\tepsilon/r,\quad\mbox{or equivalently}\nn\\
&&\nbirac\epsilon=\gamma_3\epsilon/r,\quad 
\nbirac\bepsilon=-\gamma_3\bepsilon/r,\label{diraconepsilonsA}\\
\mbox{(B-type)}&&\nbirac\epsilon=-\bepsilon/r,\quad
\nbirac\bepsilon=\epsilon/r,\quad\mbox{or equivalently}\nn\\
&&\nbirac\tepsilon=\gamma_3\tepsilon/r,\quad
\nbirac\btepsilon=-\gamma_3\btepsilon/r.
\label{diraconepsilonsB}
\eeqa

We next write down the action of
$\gamma^{\whn}=g_{\mu\nu}\wh{n}^{\mu}\gamma^{\nu}$
on the variational parameters
at the boundary $|z|=1$, where $\wh{n}$ is the outward unit normal
to the southern hemisphere $D^2_0$.
Using (\ref{ngammaons}), we find
\beqa
\mbox{(A$_{(\pm)}$-type)}&&
\gamma^{\whn}\,\epsilon=\mp \epsilon,\quad
\gamma^{\whn}\,\bepsilon=\mp\bepsilon,\label{ngonepsilonsA}\\
\mbox{(B$_{(\pm)}$-type)}&&
\gamma^{\whn}\,\epsilon=\mp\bepsilon,\quad
\gamma^{\whn}\,\bepsilon=\mp\epsilon.
\label{ngonepsilonsB}
\eeqa

\section{Formulation}

In this section, we formulate a class of theories on the hemisphere 
in such a way that some of the supersymmetry studied in the previous section
are preserved.
We shall first find a bulk action with appropriate boundary interaction
so that the total is automatically supersymmetric, and then
discuss the boundary conditions. 
The main target is the gauged linear sigma models
with A-type supersymmetry (B-branes at the boundary),
but we start with Landau-Ginzburg models with B-type supersymmetry
(A-branes at the boundary) as a warm up.
In view of (\ref{southnorth}) it is enough to consider the southern
hemisphere, so we set $D^2=D^2_0$.

\subsection{Bulk Action}

\subsubsection{Warm Up: 
Landau-Ginzburg Model (B-Type Supersymmetry)}
\label{subsec:LGaction}

We consider the Landau-Ginzburg model of $n$ chiral multiplets
$(\phi^i,\psi^i,f^i)$, $i=1,\ldots, n$, with superpotential
$W(\phi)=W(\phi^1,\ldots,\phi^n)$.

Before starting, we comment on a useful fact concerning 
B-type supersymmetry transformation of chiral multiplets.
Using (\ref{diraconepsilonsB}) in (\ref{scchiral}), 
we find that if $(\phi,\psi,f)$ is a chiral multiplet of vector
R-charge $R$, then $(\phi,\psi,f_!)$ with
\beq
f_!=f+ {R\over 2r}\phi,\quad\,\,
\overline{f}_!=\overline{f}+ {R\over 2r}\bphi,
\label{fshift}
\eeq
transforms under the B-type supersymmetry as a chiral multiplet
of vanishing vector R-charge. 
This remark applies equally well to A-type supersymmetry transformation of
twisted chiral multiplets, as will be used
in Section~\ref{subsec:LSMaction}.

\subsection*{Kinetic term}

First, let us find the kinetic term of a single chiral multiplet
$(\phi,\psi,f)$.
Let $\delta_1$ and $\delta_2$ be the B$_{(+)}$-type supersymmetry
with parameters $(\epsilon_1,\bepsilon_1)$ and $(\epsilon_2,\bepsilon_2)$.
We compute $\delta_2\delta_1$ of some combination of fields
and see if something like a kinetic term appears. 
After some try and error, we find
\beq
\im\delta_2\delta_1\left(\overline{f}_!\phi+\bphi f_!
\mp {1\over r}\bphi\phi\right)
=\nabla_{\mu}J^{\mu}-2c_-{\mathcal L}_{\rm kin},
\label{exactLG}
\eeq
where
\beq
{\mathcal L}_{\rm kin}=\partial^{\mu}\bphi\partial_{\mu}\phi
+{\im\over 2}\langle\nbirac\bpsi,\psi\rangle
+{\im\over 2}\langle\bpsi,\nbirac\psi\rangle
+\overline{f}_!f_!,
\label{LLGkin}
\eeq
and
\beqa
J^{\mu}&=&c_-\partial^{\mu}(\bphi\phi)
-\langle\epsilon_1,\gamma^{\mu}\bepsilon_2\rangle\overline{f}_!\phi
-\langle\bepsilon_1,\gamma^{\mu}\epsilon_2\rangle\bphi f_!\nn\\
&&+c_+\left(\partial^{\mu}\bphi\phi-\bphi\partial^{\mu}\phi
+\langle\bpsi,\gamma^{\mu}\psi\rangle\right)
+c_{3-}{\im\over\sqrt{g}}\epsilon^{\mu\nu}\partial_{\nu}(\bphi\phi),
\label{JLG}
\eeqa
in which $c_{\pm}=\half(\langle \epsilon_1,\epsilon_2\rangle
\pm\langle \bepsilon_1,\bepsilon_2\rangle)$,
$c_{3\pm}=\half(\langle \epsilon_1,\gamma_3\epsilon_2\rangle
\pm\langle \bepsilon_1,\gamma_3\bepsilon_2\rangle)$.
In deriving the above we used (\ref{diraconepsilonsB})
and some of its consequences, such as
the fact that $c_-$ and $c_{3+}$ are constants.
It is also useful to note that $c_+=c_{3-}=0$ at the equator $|z|=1$.
In particular, at $|z|=1$ we have
\beqa
\wh{n}\cdot \! J&=&c_-\wh{n}^{\mu}\partial_{\mu}(\bphi\phi)
-\langle\epsilon_1,\gamma^{\whn}\, \bepsilon_2\rangle\overline{f}_!\phi
-\langle\bepsilon_1,\gamma^{\whn}\,\epsilon_2\rangle\bphi f_!\nn\\
&=&c_-\left(\wh{n}^{\mu}\partial_{\mu}(\bphi\phi)
\pm (\overline{f}\phi- \bphi f)\right)
\eeqa
for B$_{(\pm)}$-type supersymmetry, where we used (\ref{ngonepsilonsB})
as well as (\ref{fshift}).
Evaluating $c_-$ and integrating over the hemisphere, we find
\beqa
\lefteqn{
\int_{D^2}{\mathcal L}_{\rm kin}\sqrt{g}\dd^2 x
\,-\,\half\int_{\partial D^2}\Bigl[\,\wh{n}^{\mu}\partial_{\mu}(\bphi\phi)
\pm (\overline{f}\phi- \bphi f)\,\Bigr]\dd\tau}\nn\\
&&~~~~~~~~~~~~~~~~~~~~~
=\pm{\im\over 2r}\int_{D^2}Q^{B-}_{(\pm)}Q^{B+}_{(\pm)}\left(
\overline{f}_!\phi+\bphi f_!\mp {1\over r}\bphi\phi\right)\sqrt{g}\dd^2x.
\label{LGkin}
\eeqa
Here we used a periodic coordinate
$\tau\equiv\tau+2\pi r$ of the boundary
$\partial D^2$, defined by $z=\e^{\im\tau/r}$ for $|z|=1$.
Using the algebra (\ref{susyalgebra}),
invariance of $D^2$ under the rotation $\ell_3$, 
and the fact that $\overline{f}_!\phi+\bphi f_!-{1\over r}\bphi\phi$
has vanishing axial R-charge, we find that the right hand side is
 $Q^{B-}_{(\pm)}$-exact as well as $Q^{B+}_{(\pm)}$-exact, and
in particular, invariant under both.
Thus, we can take the left hand side of (\ref{LGkin})
as the action we wanted. It is the usual type of kinetic term plus
a particular boundary term.

With a little more hard work, we can generalize the above construction
to the case of $n$ variables with a K\"ahler potential $K$ and the
K\"ahler metric $g_{i\bj}=\partial_i\partial_{\bj}K$.
We shall use the notation $K_i=\partial_iK$ etc.
We have the relation of the form (\ref{exactLG}), in which we
make the replacement
$\bphi\phi\to K$,
$\overline{f}_!\phi\to \overline{f}^{\bi}_!K_{\bi}
-{\im \over 2}K_{\bi\bj}\langle\bpsi^{\bi},\bpsi^{\bj}\rangle$,
$\bphi f_!\to K_i f^i_!
+{\im \over 2}K_{ij}\langle\psi^i,\psi^j\rangle$,
$\partial^{\mu}\bphi\phi\to \partial^{\mu}\bphi^{\bi}K_{\bi}$
and $\bphi\partial^{\mu}\phi\to K_i\partial^{\mu}\phi^i$,
in the expressions for the left hand side and for $J^{\mu}$,
and
\beqa
{\mathcal L}_{\rm kin}&=&g_{i\bj}\partial^{\mu}\bphi^{\bj}
\partial_{\mu}\phi^i
+{\im\over 2}g_{i\bj}\langle\Dirac\bpsi^{\bj},\psi^i\rangle
+{\im\over 2}g_{i\bj}\langle\bpsi^{\bj},\Dirac\psi^i\rangle
+{1\over 4}R_{i\bj k\overline{l}}
\langle \psi^i,\psi^k\rangle\langle\bpsi^{\bj},\bpsi^{\overline{l}}\rangle
\nn\\
&&+g_{i\bj}\left(\overline{f}^{\bj}_!
-{\im\over 2}\Gamma^{\bj}_{\overline{k}\overline{l}}
\langle\bpsi^{\overline{k}},\bpsi^{\overline{l}}\rangle\right)
\left(f^i_!+{\im\over 2}\Gamma^i_{kl}\langle\psi^k,\psi^l\rangle\right).
\label{LLGkinnvar}
\eeqa
As the action, we may take
\beq
\int_{D^2}{\mathcal L}_{\rm kin}\sqrt{g}\dd^2 x
-\half\int_{\partial D^2}\left[\,\wh{n}^{\mu}\partial_{\mu}K
\pm \left(\overline{f}_!^{\bi}K_{\bi}-K_i f_!^i
-{\im\over 2}K_{\bi\bj}\langle\bpsi^{\bi},\bpsi^{\bj}\rangle
-{\im\over 2}K_{ij}\langle\psi^i,\psi^j\rangle
\right)\,\right]\dd\tau.
\label{LGkinnvar}
\eeq
It is not only supersymmetric but also $Q$-exact as long as
the K\"ahler potential is globally defined.

\subsection*{Superpotential term}

We next turn to the superpotential term. Let us put
\beq
{\mathcal L}_W={\im\over 2r}(W+\overline{W})
-{\im\over 2}f_!^i\partial_iW
-{\im\over 2}\overline{f}_!^{\bi}\partial_{\bi}\overline{W}
+{1\over 4}\langle\psi^i,\psi^j\rangle\partial_i\partial_jW
-{1\over 4}\langle\bpsi^{\bi},\bpsi^{\bj}\rangle
\partial_{\bi}\partial_{\bj}\overline{W}.
\label{defLW}
\eeq
Under B-type supersummetry, it transforms as
$\delta{\mathcal L}_W=\nabla_{\mu}J^{\mu}$ where
\beq
J^{\mu}={\im\over 2}\langle\gamma^{\mu}\bepsilon,\psi^i\rangle\partial_iW
+{\im\over 2}\langle\gamma^{\mu}\epsilon,\bpsi^{\bi}\rangle
\partial_{\bi}\overline{W}.
\label{defJmu}
\eeq
Note that
\beqa
\wh{n}\cdot J&=&
{\im \over 2}\langle\gamma^{\whn}\,\bepsilon,\psi^i\rangle\partial_iW
+{\im\over 2}\langle\gamma^{\whn}\epsilon,\bpsi^{\bi}\rangle\partial_{\bi}
\overline{W}\nn\\
&=&\mp{\im\over 2}\left(\langle\epsilon,\psi^i\rangle\partial_iW
+\langle\bepsilon,\bpsi^{\bi}\rangle\partial_{\bi}\overline{W}\right)
=\mp{\im\over 2}\delta(W-\overline{W}),
\label{luckyyou}
\eeqa
where we used (\ref{ngonepsilonsB}).
We therefore find that
\beq
\int_{D^2}{\mathcal L}_W\sqrt{g}\dd^2x\,
\pm\,\int_{\partial D^2}{\im\over 2}\Bigl(\,W-\overline{W}\,\Bigr)
\dd\tau
\label{LGW}
\eeq
is invariant under B$_{(\pm)}$-type supersymmetry.
Again, for this we do not need to use any boundary condition.

When the superpotential is quasi-homogeneous,
the system on the flat space has the vector
$U(1)$ R-symmetry under the assignment of the R-charges so that
$W(\lambda^R\phi)=\lambda^2W(\phi)$, or equivalently,
$\sum_iR_i\phi^i\partial_iW=2W$.
Then, the expression (\ref{defLW}) simplifies as
\beq
{\mathcal L}_W=
-{\im\over 2}f^i\partial_iW
-{\im\over 2}\overline{f}^{\bi}\partial_{\bi}\overline{W}
+{1\over 4}\langle\psi^i,\psi^j\rangle\partial_i\partial_jW
-{1\over 4}\langle\bpsi^{\bi},\bpsi^{\bj}\rangle
\partial_{\bi}\partial_{\bj}\overline{W}.
\label{fake}
\eeq
This itself is invariant under the vector $U(1)$ R-rotation.
However, the last term of the bulk kinetic term (\ref{LLGkin})
as well as the boundary term in (\ref{LGW})
violate this symmetry. Thus, the systems on the sphere and the
hemisphere do
not inherite the vector $U(1)$ R-symmetry.

\subsection*{B$^{\alpha}$-type supersymmetry}

The above actions (\ref{LGkin}) and (\ref{LGW}) can be made invariant under
B$^{\alpha}_{(\pm)}$-type supersymmetry provided we make the folowing changes:

(i) The shift
(\ref{fshift}) is modified into
$f_!=f+\e^{2\im\alpha}{R\over 2r}\phi$ and
$\overline{f}_!=\overline{f}+\e^{-2\im\alpha} {R\over 2r}\bphi$.

(ii) $\overline{f}\phi-\bphi f$ in the boundary term of (\ref{LGkin})
is changed to $\e^{2\im\alpha}\overline{f}\phi-\e^{-2\im\alpha}\bphi f$.

(iii) $W+\overline{W}$ in the expression  (\ref{defLW}) for
${\mathcal L}_W$ is changed to 
$\e^{2\im\alpha}W+\e^{-2\im \alpha}\overline{W}$.

(iv) $W-\overline{W}$ in the boundary term of (\ref{LGW})
is changed to $\e^{2\im\alpha}W-\e^{-2\im \alpha}\overline{W}$.

\noindent
When $W$ is quasihomogeneous, this change is done simply by operating the
vector R-symmetry transformation $\e^{i\alpha F_V}$
on all field variables.

\subsubsection{Gauge Theory (A-Type Supersymmetry)}\label{subsec:LSMaction}

We consider a gauge theory with gauge group $G$ (a complact Lie group)
and a matter representation $V$ (a unitary representation of $G$).
We write $(\phi,\psi,f)$ for the chiral multiplet valued in $V$,
and $(\sigma,v_{\mu},\lambda, D_E)$ for the vector multiplet fields.
We denote the superpotential by $W(\phi)$ and the twisted superpotential
by $\wt{W}(\sigma)$. 
Since the A-type supersymmetry includes the vector $U(1)$ R-symmetry,
$W(\phi)$ must be quasi-homogeneous and we need to assign the vector
R-charges so that
\beq
W(\lambda^R\phi)=\lambda^2 W(\phi).
\label{qho}
\eeq
We assume that $R$ commutes with the gauge symmetry.
The twisted superpotential $\wt{W}(\sigma)$ is arbitrary at the moment,
although we shall later study in detail the
gauged linear sigma models in which it takes a special form
\beq
\wt{W}=-{1\over 2\pi}t(\sigma),
\label{FITheta}
\eeq
where $t=\zeta-\im\theta$ is the complex combination of
Fayet-Iliopoulos and Theta parameters.

\subsection*{Gauge kinetic term}

Recall that 
$(\sigma,\lambda,D_E+\im{v_{12}\over\sqrt{g}})$ transforms like a twisted
chiral multiplet with axial R charge 2.
Twisted chiral multiplets transform under A-type supersymmetry
in the same way as chiral multiplets do under B-type supersymmetry.
Therefore, the construction of the action of
the Landau-Ginzburg models with B-type supersymmetry can give us a guide
to construct gauge kinetic term and the twisted superpotential term.
In view of the fact
that the axial R-charge of $\sigma$ is fixed to be $2$,
it is convenient to introduce, following (\ref{fshift}),
\beq
\mathcal{E}_!:=\left(D_E+\im{v_{12}\over\sqrt{g}}\right)
+{1\over r}\sigma,\qquad
\overline{\mathcal{E}}_!:=
\left(D_E-\im{v_{12}\over\sqrt{g}}\right)+{1\over r}\bsigma.
\eeq
Using (\ref{LGkin}) as a guide, we obtain the gauge kinetic term
\beqa
\lefteqn{\int_{D^2}\cL^{\rm gauge}_{\rm kin}\sqrt{g}\,\dd^2x
\,-{1\over 4e^2}\oint_{\partial D^2}
\Tr\left[\,n^{\mu}\partial_{\mu}(\bsigma\sigma)
\pm 2\im\left(D_E\sigma_2-{v_{12}\over\sqrt{g}}\sigma_1\right)\right]
\dd\tau
}\nn\\
&&\qquad\qquad\qquad\quad\,\,
=\,\,\pm{\im\over 4e^2r}\int_{D^2}Q^{A-}_{(\pm)}Q^{A+}_{(\pm)}\Tr\left[\,
\overline{\mathcal{E}}_!\sigma+\bsigma\mathcal{E}_!
\mp{1\over r}\bsigma\sigma\,\right]\sqrt{g}\,\dd^2x,\qquad
\label{LSMgkin}
\eeqa
where
\beqa
\cL^{\rm gauge}_{\rm kin}
\!&=&\!
{1\over 2e^2}\Tr\Biggl[\,D^{\mu}\bsigma D_{\mu}\sigma
+{1\over 4}[\sigma,\bsigma]^2
+\left(D_E+{1\over r}\sigma_1\right)^2
+\left({v_{12}\over\sqrt{g}}+{1\over r}\sigma_2\right)^2
\nn\\
&&~~~~~~~~+{\im\over 2}\langle\blambda,\Dirac\lambda\rangle
+{\im\over 2}\langle\Dirac\blambda,\lambda\rangle
+{1\over 2}\langle\lambda,\gamma_3[\bsigma,\lambda]\rangle
+{1\over 2}\langle\blambda,\gamma_3[\sigma,\blambda]\rangle\,\Biggr].
~~~
\label{Lgaugekin}
\eeqa
Here ``${1\over e^2}\Tr(XY)$'' is an invariant inner product of the
adjoint representation. ``$e^2$'' is a collective notation
for the gauge coupling constant for each gauge group factor.
The term (\ref{LSMgkin}) is not only A$_{(\pm)}$-type
supersymmetric but also $Q^{A+}_{(\pm)}$ and
$Q^{A-}_{(\pm)}$ exact.

\subsection*{Twisted superpotential term}

Copying (\ref{LGW}), we obtain the
twisted superpotential term having  A$_{(\pm)}$-type supersymmetry:
\beq
\int_{D^2}\cL_{\widetilde{W}}\sqrt{g}\,\dd^2x
\,\pm\,\oint_{\partial D^2}\left(
{\im\over 2}\widetilde{W}-{\im\over 2}\overline{\widetilde{W}}\right)
\dd \tau
\label{LSMtW}
\eeq
where
\beqa
\cL_{\widetilde{W}}
&=&{\im\over 2r}\left(\widetilde{W}
+\overline{\widetilde{W}}\right)
-{\im\over 2}\left(\mathcal{E}_!^a\partial_{a}\widetilde{W}
+\overline{\mathcal{E}}_!^a\partial_{\bar a}\overline{\widetilde{W}}\right)
+{1\over 4}\langle\lambda^a,\lambda^b\rangle\partial_a\partial_b\widetilde{W}
-{1\over 4}\langle\blambda^a,\blambda^b\rangle
\partial_{\bar a}\partial_{\bar b}\overline{\widetilde{W}}.\nn\\
\eeqa
In the particular case (\ref{FITheta}), it reads
\beq
\int_{D^2}\left({\im\over 2\pi}\zeta(D_E)\sqrt{g}\dd^2x
-{\im \over 2\pi}\theta(F_v)\right)
\pm\int_{\partial D^2}{\rm Im}\left({1\over 2\pi}t(\sigma)\right)
\dd\tau,
\label{LSMFItheta}
\eeq
where $F_v=\dd v+{\im\over 2}[v,v]$
is the curvature of the gauge potential $v$. We see that
$\theta$ is indeed a theta parameter.

\subsection*{Matter kinetic term}

Finding the A-type supersymmetric kinetic term for the chiral multiplet
with a possibly non-trivial vector R-charge is a whole new story. 
However, just as we have done in
finding (\ref{LGkin}), we compute $\delta_2\delta_1$ of some combination of
fields and see whether the result looks like a kinetic term.
After some try and error, we arrive at the following result:
\beqa
\lefteqn{\int_{D^2}\cL^{\rm matter}_{\rm kin}\sqrt{g}\,\dd^2x
\,\pm\,\oint_{\partial D^2}
\left[{\im\over 2}\langle\bpsi,\psi\rangle
-\bphi\sigma_2\phi
\right]\dd \tau
}\nn\\
&&
=\,\,\pm{1\over 2r}\int_{D^2}
Q^{A-}_{(\pm)}Q^{A+}_{(\pm)}\left[
\langle\bpsi,\gamma_3\psi\rangle
+{\im\over r}\bphi\phi+2\bphi\left(-\im{R\over 2r}+\sigma_1\right)\phi
\right]\sqrt{g}\,\dd^2x,
\label{LSMmkin}
\eeqa
where
\beqa
\cL^{\rm matter}_{\rm kin}&=&
D^{\mu}\bphi D_{\mu}\phi+\bphi\left[
{2R-R^2\over 4r^2}-\im D_E-\im{R\over r}\sigma_1+(\sigma_1^2+\sigma_2^2)\right]
\phi
+\overline{f}f
\nn\\
&&+{\im\over 2}\langle\bpsi,\Dirac\psi\rangle
+{\im\over 2}\langle\Dirac\bpsi,\psi\rangle
+\left\langle\bpsi,
\left[\left(-\im{R\over 2r}+\sigma_1\right)\gamma_3-\im\sigma_2\right]\psi
\right\rangle\nn\\
&&-\im\langle\bpsi,\overline{\tilde{\lambda}}\rangle\phi
-\im\bphi\langle\tilde{\lambda},\psi\rangle.
\label{Lmatterkin}
\eeqa
We take the left hand side of (\ref{LSMmkin}) as the matter kinetic term.
This is not only A$_{(\pm)}$-type supersymmetric but also $Q^{A+}_{(\pm)}$ and
$Q^{A-}_{(\pm)}$ exact. 

\subsection*{Matter superpotential}

Finally, let us discuss the superpotential term. Let us put 
\beq
\cL_W=-{\im\over 2}f^i\partial_iW
-{\im\over 2}\overline{f}^{\bi}\partial_{\bi}\overline{W}
+{1\over 4}\langle\psi^i,\psi^j\rangle\partial_i\partial_jW
-{1\over 4}\langle\bpsi^{\bi},\bpsi^{\bj}\rangle
\partial_{\bi}\partial_{\bj}\overline{W}.
\label{LLSMW}
\eeq
Under the condition (\ref{qho}) or equivalently
$\sum_iR_i\phi^i\partial_iW=2W$, any superconformal transformation
(\ref{scchiral}) of ${\mathcal L}_W$ can be written as
$\delta{\mathcal L}_W=\nabla_{\mu}J^{\mu}$
where $J^{\mu}$ is the same as (\ref{defJmu}).
Note that
\beqa
\wh{n}\cdot J&=&
{\im \over 2}\langle\gamma^{\whn}\,\bepsilon,\psi^i\rangle\partial_iW
+{\im\over 2}\langle\gamma^{\whn}\epsilon,\bpsi^{\bi}\rangle\partial_{\bi}
\overline{W}\nn\\
&=&\mp{\im\over 2}\left(\langle\bepsilon,\psi^i\rangle\partial_iW
+\langle\epsilon,\bpsi^{\bi}\rangle\partial_{\bi}\overline{W}\right)
\eeqa
for the A$_{(\pm)}$-type supersymmetry, where (\ref{ngonepsilonsA}) is used.
Unlike in (\ref{luckyyou}), it cannot be written as a supersymmetry
variation of some combination of bulk fields.
Thus, we can only say
\beq
\delta\int_{D^2}\cL_W\sqrt{g}\,\dd^2x
\,\,=\,
\mp\,{\im\over 2}\oint_{\partial D^2}\left[\langle\bepsilon,
\psi^i\rangle\partial_iW
+\langle\epsilon,
\bpsi^{\bi}\rangle\partial_{\bi}\overline{W}\right]
\dd\tau
\label{Warner0}
\eeq
The right hand side is the so called Warner term \cite{Warner}.
It can only be cancelled by the supersymmetry transformation
of a boundary interaction on a Chan-Paton factor of rank greater than one,
which we turn to next.

\newcommand{\bupsi}{\overline{\upsi}}
\newcommand{\tupsi}{\wt{\upsi}}
\newcommand{\bulambda}{\overline{\ulambda}}
\newcommand{\tulambda}{\wt{\ulambda}}
\newcommand{\busigma}{\overline{\usigma}}

\subsection{Chan-Paton Factor}\label{subsec:CP}

We introduce a class of boundary interactions in the
gauge theory which are important by themselves but also
can be used in cancellation of the Warner term. 

First, let us introduce some notations that are suited to the boundary.
The most relevant ones are the fermions
\beq
\upsi\,:=\,{1\over\sqrt{r}}\left[\,z^{\half}\psi_-^{\{z\}}
\,\pm\, \bz^{\half}\psi_+^{\{z\}}\,\right],\qquad
\bupsi\,:=\,{1\over\sqrt{r}}\left[\,z^{\half}\bpsi_-^{\{z\}}
\,\pm\, \bz^{\half}\bpsi_+^{\{z\}}\,\right].
\label{upsidef}
\eeq
The superscript $\{z\}$ is there to emphasize that the field components
are in the $z$-frame, $\sqrt{\dd z}, \sqrt{\dd\bz}$,
 as in (\ref{spinorexpression}).
$\upsi$ and $\bupsi$ can be regarded as the boundary value of $\psi$
and $\bpsi$ with respect to the natural frame at the boundary $\partial D^2$,
$\sqrt{r\dd z/z}\equiv\pm\sqrt{r\dd \bz/\bz}$, where
the sign $\pm$ corresponds to the spin structure $(\pm)_0$
which is correlated with the supersymmetry type A$_{(\pm)}$.
Note that $\upsi$ and $\bupsi$ are antiperiodic along $\partial D^2$.
Let $\varepsilon_0$ and $\overline{\varepsilon}_0$
be the constant and anticommuting variational parameters
for $Q^{A-}_{(\pm)}$ and $Q^{A+}_{(\pm)}$ respectively. By definition,
the supersymmetry parameters $\epsilon$ and $\bepsilon$ are given by
\beq
\epsilon=\varepsilon_0
\left({\bf s}_{\mp\half}\pm\wt{\bf s}_{\pm\half}
\right),\quad\,\,
\bepsilon=\overline{\varepsilon}_0
\left({\bf s}_{\pm\half}\pm\wt{\bf s}_{\mp\half}\right).
\eeq
We now introduce a non-contant and antiperiodic
variational parameters along $\partial D^2$:
\beq
\varepsilon(\tau)=\sqrt{r}\varepsilon_0\e^{\mp\im{\tau\over 2r}},\quad\,\,
\overline{\varepsilon}(\tau)=\sqrt{r}\overline{\varepsilon}_0
\e^{\pm\im{\tau\over 2r}}.
\label{defvarep}
\eeq
The supersymmetry transformation of 
the boundary values of the fields can now be expressed in a simple way,
\beqa
\delta \phi\,=\,\varepsilon\,\upsi,&&
\delta\bphi\,=\,-\overline{\varepsilon}\,\bupsi,\nn\\
\delta\upsi\,=\,2\overline{\varepsilon}\left[\,
D_{\tau}\phi\pm\im{R\over 2r}\phi\mp\sigma_1\phi\,
\right],&&
\delta\bupsi\,=\,2\varepsilon\left[\,
-D_{\tau}\bphi\pm \im\bphi{R\over 2r}\mp\bphi\sigma_1\,
\right].
\eeqa
Also, the Warner term can be written as
\beq
\delta\int_{D^2}\cL_W\sqrt{g}\,\dd^2x
\,\,=\,
\mp\,{\im\over 2}\oint_{\partial D^2}\left[\,
\overline{\varepsilon}\,\upsi^i\partial_iW
+\varepsilon\,\bupsi^{\bi}\partial_{\bi}\overline{W}\,\right]
\dd\tau
\label{Warner}
\eeq

A boundary interaction is specified for a choice of
a homogeneous and gauge invariant matrix factorization of the superpotential
\cite{KapLi,BHLS,HW,HHP}. The latter consists of the following data:
a $\Z_2$ graded hermitian Chan-Paton vector space $M$,
a polynomial function $Q(\phi)$ of $\phi\in V$ 
with values in ${\rm End}^{\it od}(M)$ obeying 
\beq
Q(\phi)^2=\mp\,\im\, W(\phi)\cdot {\rm id}_M,
\label{mfW}
\eeq
even and unitary actions
of the vector R-symmetry and the gauge symmetry
on $M$, $\lambda\mapsto \lambda^{{\bf r}_*}$ and $g\mapsto \rho(g)$,
which commute with each other and satisfy
\beqa
&&\lambda^{{\bf r}_*}Q(\lambda^R\phi)\lambda^{-{\bf r}_*}=\lambda Q(\phi),
\label{homogQ}\\
&&\rho(g)^{-1}Q(g\phi)\rho(g)=Q(\phi).
\label{GinvQ}
\eeqa
Given such a data, we can write down the boundary interaction

\beq
{\mathcal A}_{\tau}=
\rho\left(\im v_{\tau}\mp\sigma_1\right)
-\half\,\upsi^i\partial_iQ
+\half\,\bupsi^{\bi}\partial_{\bi}Q^{\dag}
+{1\over 2}\,\{Q,Q^{\dag}\}
\,\mp{\im\over 2r}\,{\bf r}_*,
\label{calA}
\eeq

\medskip
\noindent
which is to be placed in the Chan-Paton factor,

\beq
{\rm tr}^{}_M 
\!\left[P\exp\left(-\oint_{\partial D^2}{\mathcal A}_{\tau}\dd\tau\right)
\right].
\label{CPfactor}
\eeq

\medskip

In (\ref{calA}), the fermionic and anti-periodic fields
$\upsi$ and $\bupsi$ come with $\partial Q$ and $\partial Q^{\dag}$
which appear to be bosonic and periodic. This might look strange.
However, we should note that (\ref{CPfactor}) needs to
be understood as the {\it graded} Chan-Paton factor
where $Q$ and $Q^{\dag}$ are regarded as fermionic 
and anti-periodic in a specific sense. See Appenidx~\ref{app:gradedCP}
for detail. Then, (\ref{CPfactor}) makes a perfect sense.

Let us study the supersymmetry transformation of the Chan-Paton factor.
We first note that the combination
$(\im v_{\tau}\mp\sigma_1)$ is invariant under
the A$_{(\pm)}$-type supersymmetry. Thus, we have
\beqa
\delta{\mathcal A}_\tau&=&
-\half 2\overline{\varepsilon}\left(
D_{\tau}\phi^i\pm{\im\over 2r}(R\phi)^i\mp(\sigma_1\phi)^i\right)\partial_iQ
\nn\\
&&+\half 2\varepsilon\left(
-D_{\tau}\bphi^{\bi}
\pm{\im\over 2r}(\bphi R)^{\bi}
\mp(\bphi\sigma_1)^{\bi}\right)\partial_{\bi}Q^{\dag}\nn\\
&&+\half\Bigl\{\varepsilon \upsi^i\partial_iQ,Q^{\dag}\Bigr\}
+\half\Bigl\{Q,-\overline{\varepsilon}\bupsi^{\bi}\partial_{\bi}Q^{\dag}
\Bigr\}.
\label{threelines}
\eeqa
Using the infinitesimal forms of (\ref{homogQ}) and (\ref{GinvQ}),
the first two lines are written as
\beqa
&&-\overline{\varepsilon}\left(
D_{\tau}Q\pm{\im\over 2r}Q\mp{\im\over 2r}[{\bf r}_*,Q]
\mp[\rho(\sigma_1),Q]\right)\nn\\
&&+\varepsilon\left(
-D_{\tau}Q^{\dag}\pm{\im\over 2r}Q^{\dag}
\pm{\im\over 2r}[{\bf r}_*,Q^{\dag}]
\pm[\rho(\sigma_1),Q^{\dag}]\right).\nn
\eeqa
If we use 
${\dd\over\dd\tau}\varepsilon=\mp{\im\over 2r}\varepsilon$ and
${\dd\over\dd\tau}\overline{\varepsilon}=\pm{\im\over 2r}
\overline{\varepsilon}$ that follows from the definition,
it simplifies as
\beq
-D_{\tau}(\overline{\varepsilon}Q+\varepsilon Q^{\dag})
-\left[\mp\rho(\sigma_1)\mp{\im\over 2r}{\bf r}_*\,,\,
\overline{\varepsilon}Q+\varepsilon Q^{\dag}\right].
\eeq
If we write
${\mathcal D}_\tau(-)={\dd\over \dd \tau}(-)+[{\mathcal A}_\tau,(-)]$,
it can be written as
\beq
-{\mathcal D}_{\tau}(\overline{\varepsilon}Q+\varepsilon Q^{\dag})
+\left[-\half\,\upsi^i\partial_iQ
+\half\,\bupsi^{\bi}\partial_{\bi}Q^{\dag}
+{1\over 2}\,\{Q,Q^{\dag}\}
\,,\,
\overline{\varepsilon}Q+\varepsilon Q^{\dag}\right].
\eeq
By the fermionic nature of $Q$ and $Q^{\dag}$,
a part of it cancels with the third line of (\ref{threelines}) and
another part can be simplified as
\beq
[\upsi^i\partial_iQ,Q]=\psi^i\partial_iQQ-Q\upsi^i\partial_iQ
=\upsi^i(\partial_iQQ+Q\partial_iQ)=\upsi^i\partial_i(Q^2).
\eeq
Collecting all, we have
\beqa
\delta{\mathcal A}_{\tau}
&=&-{\mathcal D}_{\tau}\left(\overline{\varepsilon} Q
+\varepsilon Q^{\dag}\right)\nn\\[0.2cm]
&&
-\half\,\overline{\varepsilon}\,\upsi^i\partial_iQ^2
+\half\,\varepsilon\,\bupsi^{\bi}\partial_{\bi}(Q^{\dag})^2
+\half\overline{\varepsilon}\Bigl[\,Q^{\dag},Q^2\,\Bigr]
+\half\varepsilon\Bigl[\,Q,Q^{\dag 2}\,\Bigr].
\label{deltacalA}
\eeqa
Finally, if we use the matrix factorization property (\ref{mfW}), 
the last two commutator terms vanish and the
two preceding terms become
$\pm{\im\over 2}
\left(\overline{\varepsilon}\upsi^i\partial_iW
+\varepsilon\bupsi^{\bi}\partial_{\bi}\overline{W}\right){\rm id}_M$,
which is equal to the Warner term (\ref{Warner}) except
that the sign is opposite.
Note that the term of the form ${\mathcal D}_\tau(-)$ can be ignored
if we consider the variation of the Chan-Paton factor
(\ref{CPfactor}).
Thus, the combination
\beq
\exp\left(-\int_{D^2}\cL_W\sqrt{g}\,\dd^2x\right)
{\rm tr}^{}_M 
\!\left[P\exp\left(-\oint_{\partial D^2}{\mathcal A}_{\tau}\dd\tau\right)
\right]\eeq
is invariant under the A$_{(\pm)}$-type supersymmetry.

\subsection{Boundary Condition}\label{subsec:BC}

Let us now discuss the boundary conditions of the field variables.
Since we have constructed the action which is automatically supersymmetric,
the main requirement is the supersymmetry of the boundary conditions
themselves as well as compatibility with the Euler-Lagrange equations.
We shall consult the analysis of \cite{HHP} which studied
the boundary conditions
for A-branes in Landau-Ginzburg models and B-branes in gauge theories 
with the type of boundary interactions discussed above,
in a half of the flat Minkowski space with a timelike boundary.

As in the discussion on boundary interactions,
it is convenient to use the spinor components
with respect to the natural frames near the boundary;
$\sqrt{r\dd z/z}$ for $S_-$
and $\pm\sqrt{r\dd\bz/\bz}$ for $S_+$
which are identified at the boundary in the spin structure $(\pm)_0$.
We denote them in upright symbols as
\beqa
&\upsi_-:=\sqrt{z\over r}\psi_-^{\{z\}},\quad
\bupsi_-:=\sqrt{z\over r}\bpsi_-^{\{z\}},\quad
\upsi_+:=\pm\sqrt{\bz\over r}\psi_+^{\{z\}},\quad
\bupsi_+:=\pm\sqrt{\bz\over r}\bpsi_+^{\{z\}},\nn\\
&\ulambda_-:=\sqrt{z\over r}\lambda_-^{\{z\}},\quad
\bulambda_-:=\sqrt{z\over r}\blambda_-^{\{z\}},\quad
\ulambda_+:=\pm\sqrt{\bz\over r}\lambda_+^{\{z\}},\quad
\bulambda_+:=\pm\sqrt{\bz\over r}\blambda_+^{\{z\}}.
\label{fcomp}
\eeqa
We shall use the real coordinates near the boundary, $\rho$ and
$\tau$, which are related to the complex coordinate by
$z=\exp((\rho+\im\tau)/r)$.

\subsection*{A-branes in the Landau-Ginzburg model}

For concreteness, we consider the Landau-Ginzburg model of $n$ variables
with a purely quadratic K\"ahler potential.
To study A-branes, it is convenient to use real components
$x^I$ and $f_0^I$ ($I=1,\ldots, 2n$) of the scalars
$\phi^i=x^{2i-1}+\im x^{2i}$ and $f^i=\mp\im (f_0^{2i-1}+\im f_0^{2i})$.
We also use
linear combinations $\upsi^I$ and $\tupsi^I$ of the fermions,
$\upsi_+^i-\upsi_-^i=\upsi^{2i-1}+\im\upsi^{2i}$,
$\bupsi_+^{\bi}-\bupsi_-^{\bi}=\upsi^{2i-1}-\im\upsi^{2i}$,
$\upsi_+^i+\upsi_-^i=\tupsi^{2i-1}+\im\tupsi^{2i}$ and
$\bupsi_+^{\bi}+\bupsi_-^{\bi}=\tupsi^{2i-1}-\im\tupsi^{2i}$.
We denote by ${\mathcal J}^I_{\,\,J}$ the complex structure of $\R^{2n}$,
with non-zero entries 
${\mathcal J}^{2i}_{\,\,2i-1}=-{\mathcal J}^{2i-1}_{\,\,2i}=1$,
and by $g_{IJ}$ the flat K\"ahler metric.
It is also convenient to use $\varepsilon_1$ and $\varepsilon_2$
defined by
$\varepsilon=\im\varepsilon_1-\varepsilon_2$,
$\bvarepsilon=-\im\varepsilon_1-\varepsilon_2$, and
\beq
N^I:=\partial_{\rho}x^I+\im f_0^I.
\eeq
Note that there is no reality for the fermionic fields and parameters
in Euclidean signature, and also that $N^I$ are complex valued. 
The B$_{(\pm)}$-type supersymmetry transformation at the boundary reads, 
\beqa
\delta x^I&=&\im\varepsilon_1\upsi^I
+\im\varepsilon_2{\mathcal J}^I_{\,\,J}\tupsi^J,\nn\\
\delta\upsi^I&=&-2\im \varepsilon_1\dot{x}^I
+2\varepsilon_2{\mathcal J}^I_{\,\,J}N^J,\nn\\
\delta\tupsi^I&=&-2\varepsilon_1N^I
+2\im\varepsilon_2{\mathcal J}^I_{\,\,J}\dot{x}^J,\nn\\
\delta N^I&=&\varepsilon_1\left(
-\dot{\tupsi}^I\pm{1\over 2r}{\mathcal J}^I_{\,\,J}\upsi^J\right)
+\varepsilon_2\left(-{\mathcal J}^I_{\,\,J}\dot{\upsi}^J
\mp{1\over 2r}\tupsi^I\right),\label{AbraneSUSY}
\eeqa
where $\dot{\mathcal O}={\dd\over\dd\tau}{\mathcal O}$.
Except the $1/r$ terms, this is exactly the same as the Wick rotated
version of the expression for A-type supersymmetry in the flat Minkowski
space \cite{HHP}.
An invariant set of boundary conditions is found for a
{\it totally real} submanifold of $\C^n=(\R^{2n},{\mathcal J})$, that is,
a middle dimensnional submanifold $L\subset \R^{2n}$ such that
the tangent space ${\rm T}_xL$ at each point $x\in L$ is transversal to
its ${\mathcal J}_x$-image. The conditions are, at each
point of the boundary,
\beq
x\in L,\quad\,\,
\upsi\in {\rm T}_{x}L\otimes \C,
\quad\,\,
\tupsi,\,\,N
\in {\mathcal J}_{x}{\rm T}_{x}L\otimes \C.
\label{Abc}
\eeq

The next constraint is compatibility with the Euler-Lagrange equations. 
Here we make a discrimination between the kinetic term and
the superpotential term. We consider the superpotential term
as a perturbation and take into account the Euler-Lagrange equations
only from the kinetic term.
This approach is suitable in the localization computation where we take the 
limit of large K\"ahler metric, $g_{IJ}\to \infty$.
As analyzed in \cite{HHP}, the compatibility requires that
at each point $x\in L$ the tangent space ${\rm T}_xL$ is orthogonal to
its ${\mathcal J}_x$-image, or equivalently,
$L$ is a {\it Lagrangian} submanifold with respect to
the symplectic structure $\omega_{IJ}={\mathcal J}^K_{\,\,\,I}g_{KJ}$.
Moreover, if we stick to the boundary term as in
(\ref{LGkin}) or (\ref{LGkinnvar}), only a {\it linear} Lagrangian subspace
is allowed. We can have a more general Lagrangian submanifold
by adding a boundary term which is itself $Q$-exact. Alternatively,
we can have an arbitrary Lagrangian submanifold by simply dropping
the boundary term of (\ref{LGkin}). In that approach, however, $Q$-exactness
of the kinetic term is lost.

Although we consider the superpotential term as a perturbation,
there is one constraint from its presence.
It is that the boundary potential
in the superpotential term (\ref{LGW}) must be bounded below.
This requires that $\mp{\rm Im}(W)$ is bounded below at every infinity of $L$.
(For the B$^{\alpha}_{(\pm)}$-type supersymmetry,
 $\mp{\rm Im}(\e^{2\im\alpha}W)$ must be bounded below.)

\subsection*{B-branes in the gauge theory}

\newcommand{\normal}{n}

Let us discuss the boundary conditions for B-branes in the gauge theory.
Our main interests are gauge linear sigma models where in a generic locus
of the FI-parameter space, the gauge group is mostly broken and
we have the theory on the Higgs branch at low energies.
In such a theory, the main part of the information on the brane is expected
to be carried by the Chan-Paton data $(M,Q,\rho,{\bf r}_*)$.
This is in contrast to the A-branes discussed above where the main part
of the information is carried by the choice of a Lagrangian submanifold $L$.
Nevertheless, we need to select boundary conditions for all bulk fields
in order to complete the formulation of the theory.

Before starting, we write down the essential part of the supersymmetry
transformation of the fields at the the boundary. For the chiral multiplet,
\beqa
&&\delta\phi=\varepsilon(\upsi_-+\upsi_+),\quad\,\,
\delta\bphi=-\bvarepsilon(\bupsi_-+\bupsi_+),\nn\\
&&\delta(\upsi_-+\upsi_-)=2\bvarepsilon D_{\tau}'\phi,\quad\,\,
\delta(\bupsi_-+\bupsi_-)=-2\varepsilon D_{\tau}'\bphi,\nn\\
&&\delta(\upsi_--\upsi_+)=2\im\bvarepsilon
\left[D_{\rho}\phi\mp \sigma_2\phi\right]\mp2\im\varepsilon f,\nn\\
&&\delta(\bupsi_--\bupsi_+)=-2\im\bvarepsilon
\left[D_{\rho}\bphi\mp \bphi\sigma_2\right]\mp2\im\bvarepsilon \overline{f},
\nn\\
&&\delta f=\bvarepsilon\left[\pm D_{\rho}(\upsi_-+\upsi_+)
-\sigma_2(\upsi_-+\upsi_+)
-(\bulambda_-+\bulambda_+)\phi\pm\im D'_{\tau}(\upsi_--\upsi_+)\right],\nn\\
&&\delta \overline{f}=\varepsilon\left[\pm D_{\rho}(\bupsi_-+\bupsi_+)
-(\bupsi_-+\bupsi_+)\sigma_2
-\bphi(\bulambda_-+\bulambda_+)\pm\im D'_{\tau}(\bupsi_--\bupsi_+)\right].
\label{chSUSYb}
\eeqa
For the vector multiplet,
\beqa
\delta \sigma^a&=&\im\varepsilon_1\ulambda^a
+\im\varepsilon_2{\mathcal J}^a_{\,\,b}\tulambda^b,\nn\\
\delta\ulambda^a&=&-2\im \varepsilon_1D'_{\tau}\sigma^a
+2\varepsilon_2{\mathcal J}^a_{\,\,b}N^b,\nn\\
\delta\tulambda^a&=&-2\varepsilon_1N^a
+2\im\varepsilon_2{\mathcal J}^a_{\,\,b}D'_{\tau}\sigma^b,\nn\\
\delta N^a&=&\varepsilon_1\left(
-D'_{\tau}\tulambda^a\pm{1\over 2r}{\mathcal J}^a_{\,\,b}\ulambda^b\right)
+\varepsilon_2\left(-{\mathcal J}^a_{\,\,b}D'_{\tau}\ulambda^b
\mp{1\over 2r}\tulambda^b\right).\label{vctSUSYb}
\eeqa
In the above expressions, $D'_{\tau}$ is defined to be
$D'_{\tau}\varphi
=D_{\tau}\varphi\mp(\sigma_1-{\im \over 2r}R)\varphi$
and $D'_{\tau}\overline{\varphi}
=D_{\tau}\overline{\varphi}\pm\overline{\varphi}
(\sigma_1-{\im \over 2r}R)$ for the components of the chiral multiplet
of R-charge $R$ and 
\beq
D'_{\tau}\upsilon=D_{\tau}\upsilon\mp[\sigma_1,\upsilon]
\label{Dprime}
\eeq
for the components of the vector multiplet.
For other notation and for more detail, see Appendix~\ref{app:SUSY}.

Let us first discuss the boundary conditions for the chiral multiplet.
In order for the boundary interaction (\ref{calA}) to be non-trivial,
we would like the boundary values of $\phi$ 
as well as the boundary values $\upsi$ and $\bupsi$
of $\upsi_++\upsi_-$ and $\bupsi_++\bupsi_-$
 to be as free as possible.
This leaves us with no choice on the boundary conditions:
\beqa
&&D_{\rho}\phi\mp\sigma_2\phi=0,\quad\,\,
D_{\rho}\bphi\mp\bphi\sigma_2=0\nn\\
&&\upsi_+-\upsi_-=0,\quad\,\,\bupsi_+-\bupsi_-=0,\nn\\
&&D_{\rho}(\upsi_++\upsi_-)\mp\sigma_2(\upsi_++\upsi_-)
\mp(\blambda_++\blambda_-)\phi=0,\nn\\
&&D_{\rho}(\bupsi_++\bupsi_-)\mp(\bupsi_++\bupsi_-)\sigma_2
\mp\bphi(\lambda_++\lambda_-)=0,\nn\\
&&f=0,\quad\,\,\overline{f}=0.
\label{BCchiral0}
\eeqa
This set of boundary conditions is closed under the supersymmetry
--- the supersymmetry
transformation of the left hand sides all vanish if we use the boundary
conditions. This is so for any configuration
of the vector multiplet fields. See (\ref{chSUSYb})-(\ref{vctSUSYb})
and Appendix~\ref{app:SUSY}.
The above boundary conditions are also compatible with the
Euler-Lagrange equations coming from the kinetic term (\ref{LSMmkin})
which includes a particular boundary interaction.
If we have the superpotential $W$ and the matrix factorization $Q$,
the Euler-Lagrange equation changes. However, as long as we can treat
these F-terms as perturbation, we can still use (\ref{BCchiral0})
as the boundary condition.
This approach is particularly suited to the localization computation
in which we take $1/e^2$ and the K\"ahler potential for $\phi$
to be infinitely large.

For the vector multiplet, the boundary condition is analogous to the 
A-brane boundary conditions for the chiral multiplet in the Landau-Ginzburg
model. Indeed, the supersymmetry transformation (\ref{vctSUSYb})
is of the same form as
(\ref{AbraneSUSY}) except that the $\tau$-derivative
is replaced by the $D'_{\tau}$-derivative given in (\ref{Dprime}).
As in the discussion there,
we need to choose a Lagrangian submanifold $L$ of the space
$\mathfrak{g}_{\C}$ of the values of $\sigma=\sigma_1+\im\sigma_2$
which is equipped with a flat K\"ahler metric.
Because of the commutator terms $[\sigma_1,\upsilon]$ in $D'_{\tau}\upsilon$
for $\upsilon=\sigma_2,\tulambda$ and $\ulambda$,
we also have additional conditions
\beqa
&&[\sigma_1,\sigma_2]=0\quad\mbox{on $L$},\label{sigma12}\\
&&[\sigma_1,{\rm T}_{\sigma}L]\subset {\rm T}_{\sigma}L\quad
\forall\sigma\in L.
\label{sigma1TL}
\eeqa
Because we are mainly interested in gauged linear sigma models,
we do not want to break the gauge symmetry at the boundary
by the choice of boundary conditions on the vector multiplet fields.
That is, 
we do not want to have any constraint on the boundary values of
the gauge transformations.
This requires that $L$ is invariant under the adjoint $G$ action,
\beq
GL=L.
\label{Ginvariance}
\eeq
Finally, we would like to require that the boundary potential is bounded below.
However, the precise meaning of the boundary potential is not
so clear because the vector multiplet is interacting with the chiral
multiplet and also with itself. 
In \cite{HHP}, we studied the effective boundary potential on
the Coulomb branch in Abelian gauged linear sigma models and obtained
a general set of D-branes by choosing $L$ to be the real locus
$\im\mathfrak{g}\subset \mathfrak{g}_{\C}$
where $\sigma_1$ is free and $\sigma_2$ is zero, 
or its small deformations.
For a general compact Lie group $G$, the real locus $L=\im\mathfrak{g}$
obviously satisfies the conditions
(\ref{sigma12}), (\ref{sigma1TL}) and (\ref{Ginvariance}).
This motivates us to take the Lagrangian to be the real locus
\beq
L=\im \mathfrak{g}\subset \mathfrak{g}_{\C},
\label{realL}
\eeq
or its deformations satisfying
(\ref{sigma12}), (\ref{sigma1TL}) and (\ref{Ginvariance}).

Let us determine the supersymmetric boundary conditions
corresponding to the real locus $L=\im\mathfrak{g}$.
A set of boundary conditions containing $\sigma_2=0$ is
\beqa
&&v_{\rho}=0,\quad\,\,\partial_{\rho}v_{\tau}=0\nn\\
&&\sigma_2=0,\quad\,\, \partial_{\rho}\sigma_1=0,\nn\\
&&\ulambda_++\ulambda_-=0,\quad\,\, \bulambda_++\bulambda_-=0,\nn\\
&&\partial_{\rho}(\ulambda_+-\ulambda_-)=0,\quad\,\,
\partial_{\rho}(\bulambda_+-\bulambda_-)=0,\nn\\
&&\partial_{\rho}D_E=0.
\label{BCvector}
\eeqa
These are obtained from the corresponding boundary conditions
in Minkowski space \cite{HHP}.
Because of the Wick rotation which changed the reality of the fields,
a part of the conditions in \cite{HHP} need to be split into
the real and imaginary parts. 
As a consequence, these boundary conditions are
not closed under the supersymmetry.
The supersymmetry transformation of (\ref{BCvector}) generates an infinite
series of new conditions, consisting of even number of normal derivatives of
each, $\partial_\rho^{2k}v_\rho=0$,
 $\ldots$ , 
$\partial_{\rho}^{2k+1}D_E=0$, $k=1,2,3,\ldots$
This might look problematic, 
but we will find in Section~\ref{subsec:mode}
a reasonable space of fields on the hemisphere
which satisfies all these boundary conditions.\footnote{The same problem
existsed in (\ref{Abc}) where the last condition requires that
the real and imaginary parts of $N$, $\partial_{\rho}x$ and $f_0$,
should independently belong to ${\mathcal J}_{x(p)}{\rm T}_{x(p)}L$.
This is stronger compared to the condition in Minkowski space
where $\im f_0$ were real and only the sum $N=\partial_{\rho}x+\im f_0$
needs to be in that real subspace. As we shall see,
the same solution applies when $L$ is a linear Lagrangian subspace.}
%We shall also consider the same type of
%boundary conditions on the flat annulus in Section~\ref{sec:annulus}
%where we again find a nice space of fields.
By the condition $v_{\rho}=0$, the gauge symmetry is broken to
those $g:D^2\to G$ satisfying the Neumann
boundary condition $\partial_{\rho}g=0$, but the boundary values of $g$
are unconstrained. 
The boundary conditions (\ref{BCchiral0})-(\ref{BCvector})
are invariant under this residual gauge symmetry.
If we also require $\partial_{\rho}^{2k+1}g=0$ for $k=0,1,2,\ldots$,
then the extended boundary conditions
are also gauge invariant.

The choice of real locus (\ref{realL}) has some simplifying features.
First, under (\ref{BCvector}) the condition for the chiral multiplet
becomes the purely Neumann boundary condition,
\beqa
&&\partial_{\rho}\phi=0,\quad\,\,
\partial_{\rho}\bphi=0\nn\\
&&\upsi_+-\upsi_-=0,\quad\,\,\bupsi_+-\bupsi_-=0,\nn\\
&&\partial_{\rho}(\upsi_++\upsi_-)=0,\quad\,\,
\partial_{\rho}(\bupsi_++\bupsi_-)=0,\nn\\
&&f=0,\quad\,\,\overline{f}=0.
\label{BCchiral}
\eeqa
This will facilitates the analysis considerably.
Second, under these conditions, the boundary terms in the gauge kinetic term
(\ref{LSMgkin}) and the matter kinetic term (\ref{LSMmkin})
both vanish. Thus, we may simply take
${\mathcal L}_{\rm kin}^{\rm gauge}+{\mathcal L}_{\rm kin}^{\rm matter}$
as the total kinetic terms.
If we had taken another Lagrangian submanifold, even if it is a small
deformation of (\ref{realL}), the computation becomes suddenly very
hard.

In the direct computation of the partition function,
we shall take the real locus (\ref{realL}), that is,
the boundary condition (\ref{BCchiral})-(\ref{BCvector})
(plus the infinite series).
However, as we shall see in Section~\ref{subsec:offreal},
there is a simple trick to find the result for the deformations of
(\ref{realL}), once the result for (\ref{realL}) is found.

\subsection{Remarks On R-Symmetry}

Here we make some remarks on the vector $U(1)$ R-symmetry of the gauge theory
preserving A$_{(\pm)}$-type supersymmetry.

\subsection*{Charge integrality}

An important class of theories are those in which {\it A-twist}
is possible. It requires not only the existence
of a vector $U(1)$ R-symmetry but also its {\it charge integrality}:
The R-charges of gauge invariant operators must be
integers and they reduce modulo 2 to the statistics of the operators.
The quasihomogeneity (\ref{qho}) of the superpotential $W(\phi)$
only assures the existence of the symmetry.
The condition for the charge integrality is
\beq
\e^{\pi i R}=J\in G,
\label{defJ}
\eeq
that is, the linear transformation $\e^{\pi i R}:V\to V$ is the
same as the action of an element $J$ of $G$.
The charge integrality is extended to the boundary sector 
as the following condition on
the brane data $(M,Q,\rho,{\bf r}_*)$:
\beq
\e^{\pi i {\bf r}_*}\rho(J)=\left\{
\begin{array}{ll}
+1&\mbox{on $M^{\rm ev}$}\\
-1&\mbox{on $M^{\rm od}$}.
\end{array}\right.
\label{intr*}
\eeq
%In Section~\ref{subsec:what}, we shall discuss the interpretation
%of the partition function in theories where A-twist is possible.

\subsection*{Gauge shift of R-charges}

If the gauge group $G$ has a center $Z_G$ with non-zero
Lie algebra $\mathfrak{z}^{}_G$, we may shift the R-charges as
\beq
R\to R+\sDelta
\label{shiftR}
\eeq
for any element $\im \sDelta$ of $\mathfrak{z}^{}_G$.
Indeed, if $R$ satisfies (\ref{qho}) and commutes with $G$,
so does $R+\sDelta$. We shall call this ``gauge shift'' of the R-charges.
Note that it necessarily changes
the matrix factorization data $(M,Q,\rho,{\bf r}_*)$ as
\beq
{\bf r}_*\to {\bf r}_*-\rho(\sDelta),
\label{shiftr*}
\eeq
in order for the condition (\ref{homogQ}) to remain satisfied.

When the charge integrality is assumed, and if the element $J$ in (\ref{defJ})
belongs to the identity component of the center $Z_G$,
then, by the above shift with $\sDelta$ given by
$J=\e^{-\pi\im \sDelta}$, we may assume that all the bulk R-charges
$R_i$ are even integers and that all the boundary R-charges $r_j$
(eigenvalues of ${\bf r}_*$) are integers which reduce modulo 2 to the
$\Z_2$-grading of $M$. We shall refer to such a choice as
the ``R$^o$-frame'' and denote the R-charges with the
superscript ``$o$'':
\beq
R^o_i\in 2\Z,\qquad
r^o_j\in\left\{
\begin{array}{ll}
2\Z&\mbox{on $M^{\rm ev}$}\\
2\Z+1&\mbox{on $M^{\rm od}$}.
\end{array}\right.
\label{Roframe}
\eeq

Dressing by gauge transformation should not change any physics,
and therefore, the gauge shift of the R-charges is expected to be
an unphysical operation. However, that is far from obvious
if we look into the $R$ (and ${\bf r}_*$) dependence of the action which
we have constructed. If we look more closely, however, we find that
it might be possible to undo the shift by suitable change of variables.
We assume the boundary conditions, (\ref{BCchiral}) and (\ref{BCvector}),
so that we can avoid complication coming from the boundary terms
of the gauge and matter kinetic terms.
The $R$ dependence appears in the matter kinetic term (\ref{Lmatterkin}).
The shift (\ref{shiftR}) can be absorbed under the change of variables
\beq
\sigma_1\,\to\,\sigma_1+\im{\sDelta\over 2r},\qquad
D_E\,\to\, D_E-\im{\sDelta\over 2r^2}.
\label{changevar}
\eeq
Note that this violates the original reality of the fields.
The shift of variables (\ref{changevar}) 
does not change the gauge kinetic term (\ref{Lgaugekin}), nor
the matter superpotential term plus
the boundary interaction (\ref{calA}), provided we also
do the gauge shift of ${\bf r}_*$ (\ref{shiftr*}).
However, this does change the twisted superpotential term.
Therefore, the gauge shift of the R-charge is not unphysical in general.
However, if $\wt{W}(\sigma)$ is linear in $\sigma$ as in (\ref{FITheta}),
then the change is simply a constant shift of the action:
\beq
\Delta S=\left\{\begin{array}{ll}
\displaystyle \half t(\sDelta)&\mbox{for $(+)_0$}\\[0.3cm]
\displaystyle \half \overline{t}(\sDelta)&\mbox{for $(-)_0$}.
\end{array}\right.
\label{shiftS}
\eeq
%This is precisely of the form which we shall discuss momentarily
%as ``intrinsic ambiguity'' in Section~\ref{subsec:intrinsic}.
Whether the reality violating change of variables (\ref{changevar})
is allowed is a subtle question. That would be OK as long as
it does not change physical observables. 
We shall examine the effect on the partition function when we compute it.

\subsubsection*{Range of R-charges}

Under the original reality of the field variables, 
each term of the real part of the bulk Lagrangian is
non-negative except possibly the term $\bphi{2R-R^2\over 4r^2}\phi$ 
in the matter kinetic Lagrangian (\ref{Lmatterkin}).
This motivates us to require $2R-R^2\geq 0$, that is,
the R-charge of each component $\phi_i$ must be in the range
\beq
0\leq R_i\leq 2.
\label{Rbound0}
\eeq
In any known models of interest, we can find R-charges in this range.
Indeed, since the R-charge of the superpotential $W(\phi)$ is $2$,
as long as the fields $\phi_i$ entering into $W(\phi)$ are concerned,
if we choose all $R_i$ to be non-negative they must also
satisify the upper bound $R_i\leq 2$.

\section{Parameter Dependence}
\label{sec:parameter}

The kinetic terms with appropriate boundary interaction
which we constrcuted in the previous section are $Q$-exact where $Q$
is one or both of the two preserved supercharges.
See (\ref{LGkin}), (\ref{LSMgkin}) and (\ref{LSMmkin}). This means
that the partition function does not change if we multiply any positive
number in front of these terms. For example, the result should not depend 
on the gauge coupling constant $e$.
This fact is very important and will be used in a crucial way
in the computation (Section~\ref{sec:computation}).
In this section, we study how the partition function
depends on other coupling constansts
--- chiral parameters that enter into the
superpotential and the matrix factorization and
twisted chiral parameters that enter into the twisted superpotential.
We will again find some kind of $Q$-exactness and
show that it depends holomorphically ({\it resp}. anti-holomorphically)
on the twisted chiral parameters and does not depend on
the chiral parameters if the system preserves
the A$_{(+)}$-type ({\it resp}. A$_{(-)}$-type) supersymmetry.

\subsection{Holomorphy}\label{subsec:holomorphy}

\newcommand{\balpha}{\overline{\alpha}}

Let us consider the Landau-Ginzburg model preserving the 
B$_{(\pm)}$-type supersymmetry. For $\delta_1$ and $\delta_2$ as in 
Section~\ref{subsec:LGaction}, we have 
\beq
\delta_2\delta_1\overline{W}=
\im\langle\bepsilon_1,\gamma^{\mu}\epsilon_2\rangle\partial_{\mu}\overline{W}
+\langle\bepsilon_1,\bepsilon_2\rangle\left(
-\im\overline{f}^{\bi}_{!}\partial_{\bi}\overline{W}
-\half\langle\bpsi^{\bi},\bpsi^{\bj}\rangle
\partial_{\bi}\partial_{\bj}\overline{W}\right).
\eeq
In what follows in this subsection,
we take off the anticommuting variational parameters 
from $\delta_i,\epsilon_i,\bepsilon_i$ but denote
the result by the same symbols.
If we use (\ref{diraconepsilonsB}) and the Fierz identity (\ref{Fierzetc}),
where we should be careful that $\epsilon_i$'s are now bosonic,
we find
\beq
\nabla_{\mu}{\langle\bepsilon_1,\gamma^{\mu}\epsilon_2\rangle\over
\langle \bepsilon_1,\bepsilon_2\rangle}=-{1\over r}.
\eeq
Using this we find
\beq
{1\over \langle\bepsilon_1,\bepsilon_2\rangle}\delta_2\delta_1\overline{W}=
\nabla_{\mu}\left(\im{\langle\bepsilon_1,\gamma^{\mu}\epsilon_2\rangle
\over \langle \bepsilon_1,\bepsilon_2\rangle}\overline{W}\right)
+{\im\over r} \overline{W}-\im\overline{f}^{\bi}_{!}\partial_{\bi}\overline{W}
-\half\langle\bpsi^{\bi},\bpsi^{\bj}\rangle
\partial_{\bi}\partial_{\bj}\overline{W}.
\eeq
Integrating over $D^2=D^2_0$ and using (\ref{ngonepsilonsB}), we have
\beqa
\int_{D^2}{1\over 2\langle\bepsilon_1,\bepsilon_2\rangle}
\delta_2\delta_1\overline{W}\sqrt{g}\dd^2x&=&
\mp\int_{\partial D^2}{\im\over 2}\overline{W}\,\dd\tau\nn\\
&&
\!\!\!\!\!\!\!\!\!\!\!\!\!\!\!\!\!\!\!\!\!
+\int_{D^2}\left({\im\over 2r}\overline{W}
-{\im\over 2}\overline{f}^{\bi}_{!}\partial_{\bi}\overline{W}
-{1\over 4}\langle\bpsi^{\bi},\bpsi^{\bj}\rangle
\partial_{\bi}\partial_{\bj}\overline{W}\right)\sqrt{g}\dd^2x.
\label{eqneqn1}
\eeqa
Similarly, we have
\beqa
\int_{D^2}{-1\over 2\langle\epsilon_1,\epsilon_2\rangle}
\delta_2\delta_1 W\sqrt{g}\dd^2x&=&
\pm\int_{\partial D^2}{\im\over 2}W\,\dd\tau\nn\\
&&
\!\!\!\!\!\!\!\!\!\!\!\!\!\!\!\!\!\!\!\!\!
+\int_{D^2}\left({\im\over 2r}W
-{\im\over 2}f^i_{!}\partial_{i}W
+{1\over 4}\langle\psi^{i},\psi^{j}\rangle
\partial_{i}\partial_{j}W\right)\sqrt{g}\dd^2x.
\label{eqneqn2}
\eeqa
The right hand sides of (\ref{eqneqn1}) and (\ref{eqneqn2})
are precisely the $\overline{W}$ and $W$ parts of the superpotential term
(\ref{LGW}). So, it appears that the entire superpotential term is
$Q$-exact for both B$_{(+)}$ and B$_{(-)}$-type supersymmetry.
However, note that
\beqa
(\delta_1,\delta_2)=(Q^{B+}_{(+)},Q^{B-}_{(+)}):&&
\langle\bepsilon_1,\bepsilon_2\rangle={2r\over 1+|z|^2},\quad
\langle\epsilon_1,\epsilon_2\rangle={-2r|z|^2\over 1+|z|^2},\nn\\
(\delta_1,\delta_2)=(Q^{B+}_{(-)},Q^{B-}_{(-)}):&&
\langle\bepsilon_1,\bepsilon_2\rangle={-2r|z|^2\over 1+|z|^2},\quad
\langle\epsilon_1,\epsilon_2\rangle={2r\over 1+|z|^2}.\nn
\eeqa
We see that
division by $\langle\bepsilon_1,\bepsilon_2\rangle$
({\it resp}. $\langle\epsilon_1,\epsilon_2\rangle$)
is possible on $D^2_0$ only for B$_{(+)}$-type
({\it resp} B$_{(-)}$-type) supersymmetry. 
Therefore, if the system preserves the B$_{(+)}$-type supersymmetry,
the $\overline{W}$-part of the superpotential term (\ref{LGW})
is $Q$-exact while the $W$-part is not.
Hence the partition function does not depend on the anti-chiral parameters
but it can depend on the chiral parameters. 
In other words, it depends holomorphically on the chiral parameters.
If the B$_{(-)}$-type supersymmetry is preserved,
the partition function depends anti-holomorphically on the chiral parameters.

By the A-B exchange, that is,
by the replacement $(\epsilon,\bepsilon)\to (\tepsilon,\btepsilon)$,
we have also shown that the partition function 
of the system preserving A$_{(+)}$-type ({\it resp}. A$_{(-)}$-type)
supersymmetry depends holomorphically ({\it resp}. anti-holomorphically)
on the twisted chiral parameters.

\subsection{No Dependence}\label{subsec:nodep}

Let us next study the dependence of the chiral parameters in the systems
preserving A$_{(\pm)}$-type supersymmetry.
We note that deformation of the superpotential $W$ 
and/or the matrix factorization $Q$ is constrained by
\beq
Q\sDelta Q+\sDelta Q=\mp \im \sDelta W,
\label{condsDelta}
\eeq
so that the condition (\ref{mfW}) remains satisfied.
In particular, any deformation of $W$ should be accompanied by
some deformation of $Q$, while deformation of $Q$ for a fixed $W$
must satisfy $\{Q,\sDelta Q\}=0$.

Let $\epsilon'$ and $\bepsilon'$ be the variational
parameters for $Q^{A-}_{(\pm)}$ and $Q^{A+}_{(\pm)}$ in which the
anticommuting parameters are stripped off.
(I.e., $\epsilon'={\bf s}_{\mp\half}\pm \wt{\bf s}_{\pm\half}$
and $\bepsilon'={\bf s}_{\pm\half}\pm \wt{\bf s}_{\mp\half}$.
See (\ref{QAdef}).) Then we have
\beqa
\delta\left(\langle \gamma_3\bepsilon',\psi^i\rangle\partial_iW\right)
\!\!&=&\!\!
\nabla_{\mu}\left(\im\langle\gamma_3\bepsilon',\gamma^{\mu}\bepsilon\rangle
W\right)
+\langle\gamma_3\bepsilon',\epsilon\rangle\left(
\im f^i\partial_iW-\half\langle\psi^i,\psi^j\rangle\partial_i\partial_jW\right),
~~
\label{varnd1}\\
\delta\left(\langle \gamma_3\epsilon',\bpsi^{\bi}\rangle\partial_{\bi}
\overline{W}\right)
\!\!&=&\!\!
\nabla_{\mu}\left(-\im\langle\gamma_3\epsilon',\gamma^{\mu}\epsilon\rangle
W\right)
+\langle\gamma_3\epsilon',\bepsilon\rangle\left(
\im\overline{f}^{\bi}\partial_{\bi}\overline{W}
+\half\langle\bpsi^{\bi},\bpsi^{\bj}\rangle\partial_{\bi}\partial_{\bj}
\overline{W}\right),~~\label{varnd2}
\eeqa
where we used
$\nabla_{\mu}\langle\gamma_3\bepsilon',\gamma^{\mu}\bepsilon\rangle=0$ etc,
that follows from (\ref{diraconepsilonsA}).
Note that the big parentheses on the right hand sides
are parts of the superpotential term (\ref{LLSMW}) and that the
coefficient in front,
$\langle\gamma_3\bepsilon',\epsilon'\rangle$, is a constant
(which is $\pm 2r$). This means that
the $W$-part of the superpotential term ${\mathcal L}_W$
given by (\ref{LLSMW}) is $Q^{A-}_{(\pm)}$-exact
while the $\overline{W}$-part is $Q^{A+}_{(\pm)}$-exact.
However, this fact does not mean that the superpotential term is
supersymmetric since inside the parenthesis of $\delta(\,?\,)$
on the left hand sides have non-zero R-charges. (This is another way to
see that the supersymmetry variation 
of ${\mathcal L}_W$ is the Warner term (\ref{Warner0}).)
But it can used to study the effect of deformation of $W$.

Deformation of the matrix factorization
results in the following change in the Chan-Paton factor,
\beq
\sDelta \,{\rm tr}^{}_M 
\!\left[P\e^{-\oint_{\partial D^2}
{\mathcal A}}
\right]
=-\oint_{\partial D^2}
{\rm tr}^{}_M\Bigl[
\left(P\e^{-\int_{\tau}^{\tau+2\pi r}\!{\mathcal A}}\right)
\sDelta {\mathcal A}_{\tau}(\tau)\,\Bigr]\dd\tau,
\label{defoCP}
\eeq
where
\beq
\sDelta{\mathcal A}_{\tau}=
-\half\,\upsi^i\partial_i\sDelta Q
+\half\,\bupsi^{\bi}\partial_{\bi}\sDelta Q^{\dag}
+{1\over 2}\,\{\sDelta Q,Q^{\dag}\}
+{1\over 2}\,\{ Q,\sDelta Q^{\dag}\}.
\eeq
The supersymmetry transformation of expressions of the form
${\rm tr}^{}_M\left[\left(P\e^{-\int_{\tau}^{\tau+2\pi r}\!{\mathcal A}}\right)
{\mathcal B}(\tau)\right]$ is, due to
 (\ref{deltacalA}),
\beq
\delta\,{\rm tr}^{}_M\Bigl[\left(P\e^{-\int_{\tau}^{\tau+2\pi r}
\!{\mathcal A}}\right)
{\mathcal B}(\tau)\,\Bigr]
={\rm tr}^{}_M\Bigl[\left(P\e^{-\int_{\tau}^{\tau+2\pi r}\!{\mathcal A}}
\right)
\delta'{\mathcal B}(\tau)\,\Bigr]+\cdots
\eeq
where
\beq
\delta'{\mathcal B}(\tau)\,:=\,\delta{\mathcal B}(\tau)
-[\,\bvarepsilon Q+\varepsilon Q^{\dag},\,{\mathcal B}\,](\tau),
\eeq
and $+\cdots$ is the term that cancels the Warner term from
the bulk.
Let us note
\beqa
\delta'(\sDelta Q)
&=&\varepsilon\Bigl(\upsi^i\partial_i\sDelta Q-\{\sDelta Q,Q^{\dag}\}\Bigr)
-\bvarepsilon\{Q,\sDelta Q\},\\
\delta'(\sDelta Q^{\dag})
&=&-\bvarepsilon\Bigl(\bupsi^{\bi}\partial_{\bi}\sDelta Q^{\dag}
+\{Q,\sDelta Q^{\dag}\}\Bigr)
-\varepsilon\{Q^{\dag},\sDelta Q^{\dag}\}.
\eeqa
We see that the $\sDelta Q$-part of the change (\ref{defoCP}) of
the Chan-Paton factor is $Q^{A-}_{(\pm)}$-exact while the
 $\sDelta Q^{\dag}$-part is $Q^{A+}_{(\pm)}$-exact.

Let us now consider 
the deformation preserving the supersymmetry (\ref{condsDelta}).
The effects $\sDelta {\mathcal L}_W$ and (\ref{defoCP})
consist of terms which are exact under either $Q^{A+}_{(\pm)}$
or $Q^{A-}_{(\pm)}$. Therefore, the partition function does not change
under the deformation.
It is also reassuring to note that the total variation
is exact under the sum $Q_{\rm tot}=Q^{A+}_{(\pm)}+Q^{A-}_{(\pm)}$,
\beqa
\lefteqn{
\sDelta\left\{
\e^{-\int_{D^2}\cL_W\sqrt{g}\,\dd^2x}\,
{\rm tr}^{}_M 
\!\left[P\e^{-\oint_{\partial D^2}{\mathcal A}}
\right]\right\}}\nn\\
&&
=Q_{\rm tot}\left\{\e^{-\int_{D^2}\cL_W\sqrt{g}\,\dd^2x}\,
{\rm tr}^{}_M 
\!\left[P\e^{-\oint_{\partial D^2}{\mathcal A}}
\right]\int_{D^2}{\mathcal C}\sqrt{g}\dd^2x
\right.\nn\\
&&~~~~~~~~~~~~+\left.
\e^{-\int_{D^2}\cL_W\sqrt{g}\,\dd^2x}
\oint_{\partial D^2}
{\rm tr}^{}_M 
\!\Bigl[\left(P\e^{-\int_{\tau}^{\tau+2\pi r}
{\mathcal A}}\right){\mathcal B}(\tau)\,
\Bigr]\dd\tau
\right\},
\eeqa
where
\beqa
{\mathcal C}&:=&
\pm{1\over 4r}\langle\gamma_3\bepsilon',\psi^i\rangle\partial_i\sDelta W
\,\pm\,
{1\over 4r}\langle\gamma_3\epsilon',\bpsi^{\bi}\rangle\partial_{\bi}
\sDelta\overline{W},
\nn\\
{\mathcal B}&:=&{1\over 2\sqrt{r}}\e^{\pm \im\tau/2r}\sDelta Q
\,+\,{1\over 2\sqrt{r}}\e^{\mp \im\tau/2r}\sDelta Q^{\dag}.
\nn
\eeqa
To summarize, the partition function does not change under
deformation of $(W,Q)$. That is, it is independent of
the chiral parameters.

Analogous statement for systems preserving B$_{(\pm)}$-type
supersymmetry would be that the partition function is independent
of the twisted chiral parameters. For example, 
in the non-linear sigma model with a K\"ahler manifold $X$ as
the target space, the partition function does not change under
deformation of the K\"ahler class $\omega$
of $X$ and the necessary deformation
of the A-brane data; a Lagrangian submanifold $L$ of $X$
and a flat bundle $E$ on $L$.
This seems to be difficult to prove.
Even if we were able to
show that the deformation of $(\omega, L,E)$
changes the action by $Q$-exact terms,
that would not be sufficient. The path-integral
measure is usually constructed using the target space metric and therefore
is expected to change if the K\"ahler class $\omega$ is deformed.
Also the deformation of $L$ results in the change of the boundary condition
whose effect needs to be analyzed.
This is in sharp contrast with what we did above: The path integral
meaure and the boundary conditions
are defined with no reference to the data of $(W,Q)$ and hence 
$Q$-exactness of the change in the action under the deformation 
was sufficient to prove the invariance of the result.
A related statement in the Landau-Ginzburg model is that
the partition function does not depend on
the deformation of the Lagrangian submanifold $L$ on which
$\mp{\rm Im}(W)$ is bounded from below.
As discussed in \cite{HHP}, $Q$-exact terms at the boundary
generate Hamiltonian deformations of $L$ which are general deformations as
Lagrangian submanifold when $L$ has a trivial topology.  
However, since the boundary condition necessarily changes, it is again
difficult to prove that the result does not change.
In this paper, we shall simply assume or postulate the invariance
under such deformations. 
Based on this postulate and explicit computation in simple cases,
we shall find in Section~\ref{subsec:Abranes}
a reasonable proposal on the general expression for the partition function.

\subsection{What Does It Compute?}\label{subsec:what}

\newcommand{\braGS}{{}^{}_{{}_{\rm RR}}\!\langle 0}
\newcommand{\GSket}{0\rangle_{{}_{\rm RR}}}

\footnote{What
is said in this subsection holds when A and B are swapped provided
 `chiral' and `twisted chiral' are swapped at the same time.}
It was conjectured in \cite{Romoetal} that
the partition function $Z_{S^2}$ on the round two-sphere
with A-type supersymmetry computes $\e^{-K}$ where
$K$ is the K\"ahler potential
of the space of twisted chiral parameters,
when there is a spacetime physics interpretation. More generally,
if the theory is A-twistable, the conjecture is
\beq
Z_{S^2}\,=\,
\braGS|\GSket.
\eeq
$|\GSket$ is the canonical ground state defined via
the infinitely long half-cigar in which the curved region is A-twisted
\cite{CV}. In other words $Z_{S^2}$ is equal to the partition function
of the infintely long cigar in which the two curved regions are A and
anti-A twisted.
The latter is known as a component of the $tt^*$ metric which is known
to satisfy special differential equations \cite{CV}.
Although there is some attempt \cite{GomisLee}, 
the real understanding of the realtion between $Z_{S^2}$ and
$\braGS|\GSket$ is still missing.

Now we would like to ask what does the hemisphere partition function compute?
The combination of A-type supersymmetry and B-branes,
holomorphic dependence on the twisted chiral parameters,
no dependence of the chiral parameters
all points to one possibility:
the overlaps of supersymmetric ground states and the D-brane boundary
states in the Ramond-Ramond sector, as studied in \cite{HIV}. 
If the latter are defined as the partition function of
the infinitely long half-cigar, in which the curved region is A-twisted
and B-branes are placed at the boundary, then, they are
independent on the chiral parameters but depend holomorphically
on twisted chiral parameters in a partucular way, so that
Picard-Fuchs type equations hold \cite{HIV}. When nothing is inserted at the
tip of the cigar, they are the overlaps of the state
$|\GSket$ and the boundary states,
called {\it the central charges} of the D-branes.
So, we would like to ask: {\it Does the partition function on the round
hemisphere computes the D-brane central charge?}
\beq
Z_{D_0^2{}_{(+)}\!}(\mathfrak{B})\,
\stackrel{?}{=}\,
\braGS|\mathfrak{B}\rangle_{{}_{\rm RR}},
\qquad
Z_{D_{\infty}^2{}_{(-)}\!}(\mathfrak{B})\,
\stackrel{?}{=}\,
{}_{{}_{\rm RR}}\!\langle \mathfrak{B}|\GSket.
\eeq
We shall compute the partition functions in a large classes of examples
and will observe that this is indeed the case whenever the D-brane central
charge is known.

%\subsection{Intrinsic Ambiguity}
%\label{subsec:intrinsic}

%\beq
%Z_{S^2}\to \e^{f(t)+\overline{g(t)}}Z_{S^2}
%\eeq
%\beq
%Z_{D^2\!{}_{(+)}\!}(\mathfrak{B})\to \e^{f(t)}
%Z_{D^2\!{}_{(+)}\!}(\mathfrak{B}),\qquad
%Z_{D^2\!{}_{(-)}\!}(\mathfrak{B})\to \e^{\overline{g(t)}}
%Z_{D^2\!{}_{(-)}\!}(\mathfrak{B}),
%\eeq

\section{Computation}\label{sec:computation}

We now compute the partition function.
In the gauge theory, we perform the direct computation by
choosing the simplest boundary condition for the vector multiplet
in which the Lagrangian submanifold is the real locus (\ref{realL}).
We also compute the partition function for
A-branes in Landau-Ginzburg model, where we discuss the
choice of integration measure.
Using that discussion and
employing the holomorphy discussed in Section~\ref{subsec:holomorphy},
we find the expression for the gauge theory partition function
for more general choice of Lagrangian submanifold.

\newcommand{\mc}{{\rm g}}

\subsection{Supersymmetric Configuration}

Since the kinetic terms (\ref{LSMgkin})
and (\ref{LSMmkin}) for the vector and the chiral multiplets
are $Q$-exact, the result of the path-integral does not depend on the
gauge coupling constant $e$ and the constant $1/\mc^2$ which we may put
in front of the matter kinetic terms. If we take the limit $e\to 0$ and
$\mc\to 0$, the path-integral localizes at the configurations in which
the real parts of these kinetic terms are minimized.

Let us find the condition for the minimization.
Recall that the boundary condition (\ref{BCchiral}) and (\ref{BCvector})
annihilates the boundary terms of  (\ref{LSMgkin})
and (\ref{LSMmkin}).
Recall also that, as long as the R-charges are in the range (\ref{Rbound0}),
$0\leq R_i\leq 2$,
the real part of the bulk kinetic Lagrangian,
 (\ref{Lgaugekin}) and (\ref{Lmatterkin}), is the sum of non-negative terms.
Therefore, it is minimized when each term vanishes.
This condition reads
\beqa
&&D_{\mu}\sigma\,=\,[\sigma,\bsigma]
\,=\,D_E+{1\over r}\sigma_1\,=\,{v_{12}\over \sqrt{g}}+{1\over r}\sigma_2
\,=\,0,
\label{minimivector}\\
&&
D_{\mu}\phi\,=\,(2R-R^2)\phi\,=\,\sigma_1\phi\,=\,\sigma_2\phi
\,=\,f\,=\,0.
\label{minimichiral}
\eeqa
Since $\sigma_2$ must vanish at the boundary by the boundary condition
(\ref{BCvector}), being required to be covariantly constant,
it must vanish everywhere.
This then implies $v_{12}=0$, that is, the gauge field is flat.
Since the hemisphere is contractible, we may set $v_{\mu}=0$ everywhere.
Then, $\sigma_1$ and $\phi$ are literally constants.
Note that a component $\phi_i$ must vanish unless its R-charge $R_i\in [0,2]$
is either $0$ or $2$.

Almost the same condition follows from the supersymmetry.
Vanishing of the supersymmetry transformation of
the gaugino, $\delta\lambda=0$ and $\delta\blambda=0$, 
 requires precisely the same condition as
(\ref{minimivector}).
Vanishing for the matter fermion,
 $\delta\psi=0$ and $\delta\bpsi=0$, requires
\beq
D_{\rho}\phi-\left(x_3{R\over 2r}\mp\sigma_2\right)\phi
\,=\,D_\tau\phi-\im\left(\mp {R\over 2r}-x_3\sigma_2\right)\phi
\,=\,\sigma_1\phi\,=\,f\,=\,0,
\label{SUSYchiral}
\eeq
where $x_3={|z|^2-1\over |z|^2+1}$. If we put $\sigma_2=v_{\mu}=0$,
the first and the second conditions read
$\partial_\rho\phi_i-x_3{R_i\over 2r}\phi_i=0$ and 
$\partial_\tau\phi_i\pm\im {R_i\over 2r}\phi_i=0$. 
The first has a non-zero and regular solution only when $R_i=0$
while the second has a non-zero and single valued solution
only when $R_i$ is an even integer.
That is, $\phi_i$ is required to vanish unless $R_i=0$.
When $R_i=0$, $\phi_i$ must be a constant.
Thus, this is stronger than the minimization condition in that
$\phi_i$ is required to vanish when $R_i=2$.

In what follows, we shall assume that all the R-charges are
in the range
\beq
0<R_i<2,
\label{narrowrange}
\eeq
or can be made into this range by using the gauge shift (\ref{shiftR})
if necessary. This is certainly the case in all known examples of
interest. Then, the supersymmetry requires all fields including $\phi$
 to vanish except that $\sigma_1=-rD_E$ must have a constant value
$\usigma_1\in\im\mathfrak{g}$.
The moduli space of supersymmetric configurations is
the space of $\usigma_1$ modulo the contant gauge transformations,
that is, the quotient of $\im\mathfrak{g}$ by the adjoint action of 
$G$. Or equivalently,
\beq
\im\mathfrak{g}/G\,\,\cong\,\,\im\ttt/W_G,
\label{moduliisom}
\eeq
where $\ttt$ is the Lie algebra of a maximal torus $T_G$ of $G$
and $W_G$ is the Weyl group of $G$.

Let us evaluate the action at the supersymmetric background.
Since the hemisphere has area $2\pi r^2$ and the boundary has length
$2\pi r$, the twisted superpotential term is
\beqa
S_0&=&2\pi r^2{\im\over 2r}\left(\wt{W}(\usigma_1)
+\overline{\wt{W}}(\usigma_1)\right)
\pm2\pi r{\im\over 2}\left(\wt{W}(\usigma_1)-\overline{\wt{W}}(\usigma_1)\right)
\nn\\
&=&\left\{\begin{array}{ll}
2\pi \im\, r \wt{W}(\usigma_1)&\mbox{~for $(+)_0$}\\
2\pi \im\, r\overline{\wt{W}}(\usigma_1)&
\mbox{~for $(-)_0$}
\end{array}\right.\label{classS0}\\
&\stackrel{(\ref{FITheta})}{=}&
\left\{\begin{array}{ll}
\im \,r t(\usigma_1)&\mbox{~for $(+)_0$}\\
\im \,r\overline{t}(\usigma_1)&\mbox{~for $(-)_0$}
\end{array}\right.
\eeqa
The superpotential term vanishes, but the 
boundary interaction remains, 
${\mathcal A}_{\tau}=\mp\im \rho(\usigma_1)\mp{\im\over 2r}{\bf r}_*$.
It gives the Chan-Paton factor
\beq
{\rm tr}_M^{}\e^{\pm \pi\im {\bf r}_*}\e^{\pm 2\pi\im r \rho(\usigma_1)}
\label{classCP}
\eeq

For the gauge fixing, we take the standard Lorentz gauge.
The gauge fixing term is given by
\beq
{\mathcal L}_{\rm gauge\,\, fixing}=
{1\over 2e^2}\Tr\left[(\nabla^{\mu}v_{\mu})^2+
\overline{c}\nabla^{\mu}D_{\mu}c\right],
\eeq

In the $e\to 0$ and $\mc\to 0$ limit,
all the terms that is cubic or higher in the fluctuation fields become
irrelevant,
and we are left with the classical action computed above, plus
the terms that are quadratic in the fluctuation fields.
The quadratic terms are given by the sum of the following
\beqa
S_1&=&\int_{D^2}\Biggl\{
\bphi\left[\Delta+{2R-R^2\over 4r^2}+\im {1-R\over r}\usigma_1
+\usigma_1^2\right]\phi\,+\overline{f}f\nn\\
&&~~~~~~~~~~~~
+\left\langle\bpsi,\left[
\im\nbirac+\left(-\im{R\over 2r}+\usigma_1\right)\gamma_3\right]\psi\right\rangle
\Biggr\}\sqrt{g}\dd^2 x,\label{defS1}\\
S_2&=&\int_{D^2}\Tr\Biggl[\sigma_1'\Delta\sigma_1'
+\sigma_2\left(\Delta+\usigma_1^2+{1\over r^2}\right)\sigma_2
+2g^{z\bz}v_{\bz}\left(\Delta+\usigma_1^2\right)v_z+(D_E')^2
\nn\\
&&~~~~~~~~~~~~
+2\im g^{z\bz}(\partial_z v_{\bz}+\partial_{\bz}v_z)\usigma_1\sigma_1'
-2\im g^{z\bz}(\partial_z v_{\bz}-\partial_{\bz}v_z){1\over r}\sigma_2\nn\\
&&~~~~~~~~~~~~
+\im \langle\blambda,\nbirac\lambda\rangle
+\half\langle\lambda,\usigma_1\gamma_3\lambda\rangle
+\half\langle\blambda,\usigma_1\gamma_3\blambda\rangle
\Biggr]\sqrt{g}\dd^2x,\label{defS2}\\
S_3&=&\int_{D^2}\Tr\bigl[\,\overline{c}\,\Delta c\,\bigr] \,\sqrt{g}\dd^2x.
\label{defS3}
\eeqa
In the above expressions, we have absorbed the factor of $e$ and $g$
by a field redefinition. $\sigma_1'$ is the non-zero modes of
$\sigma_1$ and $D_E':=D_E+\sigma_1/r$. 
$\Delta$ is the Laplace operator $\Delta=dd^{\dag}+d^{\dag}d$ 
on functions and one-forms. For a $\mathfrak{g}$-valued field ${\mathcal O}$,
we denoted $[\usigma_1,{\mathcal O}]$ simply by $\usigma_1{\mathcal O}$.

\subsection{Mode Expansion}\label{subsec:mode}

We shall regard the fields on the hemisphere as
the restriction of the fields on the whole sphere.
All the field components on the sphere can be considered as
differentiable sections of the line bundle ${\mathcal O}(n)$ over
$\CP^1$ for some $n$.
Indeed, scalars, spinors, and vectors are sections of
${\mathcal O}(0)$, $S_{\pm}={\mathcal O}(\pm 1)$,
and ${\mathcal O}(\pm 2)$ respectively.
We may also regard the two-sphere as the coset space
$SU(2)/U(1)$, where $U(1)$ is the diagonal subgroup
consisting of elements of the form
$h_u={\rm diag}(u,u^{-1})$ with $|u|=1$.
In this description, the sections of ${\mathcal O}(n)$ are functions on
$SU(2)$ obeying the condition $F(gh_u)=u^{-n}F(g)$.
If $g^{(j)}_{m,m'}$ denote matrix elements of the spin
$j$ representation\footnote{It is the $2j$-th symmetic
tensor power of the doublet. The orthonormal basis
$\{|m\rangle\}_{m=-j}^j$ is the natural one in that realization
 so that we have the reality (\ref{realityWigner}).}
of $SU(2)$, we find that $g^{(j)}_{m,-{n\over 2}}$ satisfies this condition.
Of course, we need $j-{n\over 2}$ to be an integer.
In fact, such matrix elements span the space of global sections 
of ${\mathcal O}(n)$ as orthogonal basis \cite{Pontrjagin},
\beq
\int_{S^2}\left(g^{(j)}_{m,-{n\over 2}}\right)^*g^{(j')}_{m',-{n\over 2}}
\sqrt{g}\dd^2 x
\,=\,4\pi r^2\,{\delta_{j,j'}\delta_{m,m'}\over
2j+1}.
\label{orthogonality}
\eeq
We also notice the reality
\beq
(g^{(j)}_{m,m'})^*=(-1)^{2j-m-m'}g^{(j)}_{-m,-m'}.
\label{realityWigner}
\eeq
The Laplace and the Dirac operator act on these elements as
\beqa
&&\Delta g^{(j)}_{m,-{n\over 2}}={j(j+1)\over r^2}g^{(j)}_{m,-{n\over 2}},
\quad \mbox{for $n=0,\pm 2$,}\\
&&\nbirac g^{(j)}_{m,\half}=-{j+\half\over r}g^{(j)}_{m,-\half},\qquad
\nbirac g^{(j)}_{m,-\half}={j+\half\over r}g^{(j)}_{m,\half}.
\eeqa

\newcommand{\oa}{\overline{a}}
\newcommand{\ob}{\overline{b}}

Let us find the relation to the coordinate $z$ 
and the frames $\sqrt{\dd z}$, etc, which we have been using.
If we write an element of $SU(2)$ as
\beq
g=\left(\begin{array}{cc}
a&-\ob\\
b&\oa
\end{array}\right),\qquad
|a|^2+|b|^2=1,
\eeq
then
\beq
z=\oa/\ob.
\eeq
Also, we may identify
\beq
\sqrt{\dd z}={1\over \sqrt{2r}\cdot \ob},\qquad
\sqrt{\dd\bz}={1\over\sqrt{2r}\cdot b},
\eeq
as well as $\dd z=1/(2r\ob^2)$ and $\dd\bz=1/(2rb^2)$.
The mode expansion of the fields takes the form
\beqa
\phi&=&\sum_{j,m}\phi_{j,m}\,g^{(j)}_{-m,0}\mu_j,\nn\\
\psi_-^{\{z\}}&=&\sqrt{2r}\sum_{j,m}\psi_{j,m}\,
\ob\, g^{(j)}_{-m,\half}\mu_j,\nn\\
\psi_+^{\{z\}}&=&\sqrt{2r}\sum_{j,m}\wt{\psi}_{j,m}\,
b\,g^{(j)}_{-m,-\half}\mu_j,\nn\\
v_z&=&2r\sum_{j,m}v_{j,m}\,
\ob^2g^{(j)}_{-m,1}\mu_j,\nn\\
v_{\bz}&=&2r\sum_{j,m}\wt{v}_{j,m}\,
b^2g^{(j)}_{-m,-1}\mu_j,\nn
\eeqa
where $\mu_j=\sqrt{(2j+1)/(2\pi r^2)}$,
and similarly for $\bphi$, $\bpsi_{\pm}^{\{z\}}$, $f,\overline{f}$
as well as other
components of the vector multiplet fields and the ghosts.
Note that $\sigma_1'$ and the ghosts do not include the $j=0$ mode.
We would like to find which of the terms to be kept in order for
the fields to satisfy the boundary conditions discussed
in Section~\ref{subsec:BC}.

For this we need some information on the matrix elements
$g^{(j)}_{m,m'}$, and relevant ones can be found in standard textbooks
on angular momentum such as \cite{Rose}. The elements for
$g=R(\alpha,\beta,\gamma)
=\e^{-\im\alpha\hatL_3}\e^{-\im\beta\hatL_2}\e^{-\im\gamma \hatL_3}$
 is written as
\beq
g^{(j)}_{m,m'}=\e^{-\im m\alpha}d^j_{m,m'}(\beta)\e^{-\im m'\gamma}.
\eeq
This $g$ has
\beq
a=\e^{-\im {\alpha\over 2}}\cos{\beta\over 2}\e^{-\im{\gamma\over 2}},\quad
b=\e^{\im{\alpha\over 2}}\sin{\beta\over 2}\e^{-\im{\gamma\over 2}},\quad
\eeq
and hence
\beq
z=\e^{\rho+\im\tau\over r}=\e^{\im \alpha}\cot{\beta\over 2}.
\eeq
We see that $\beta=0$, ${\pi\over 2}$ and $\pi$ correspond to
the north pole $z=\infty$, the equator $|z|=1$ and the south pole $z=0$
respectively. The hemisphere $D^2_0$ is in the region
${\pi\over 2}\leq\beta\leq\pi$.
The functions $d^j_{m,m'}(\beta)$ satisfy some identities \cite{Rose}.
The most important for us is
\beq
d^j_{m,m'}(\pi-\beta)=(-1)^{j+m}d^j_{m,-m'}(\beta).
\label{imprel}
\eeq
Note that $\beta\to \pi-\beta$ precisely correspopnds to
$\rho\to-\rho$ 
and is nothing but the reflection with respect to the equator.

We see from (\ref{imprel}) that the function $g^{(j)}_{-m,0}$ is even or odd
under the reflection $\rho\to -\rho$,
depending on $j-m$ is even or odd. In particular
they satify Neumann or Dirichlet boundary condition at the boundary,
\beq
\begin{array}{ll}
\partial_{\rho}g^{(j)}_{-m,0}\Bigr|_{\partial D^2}=0
&\mbox{if $j-m$\, is even},\\[0.2cm]
g^{(j)}_{-m,0}\Bigr|_{\partial D^2}=0
&\mbox{if $j-m$\, is odd}.
\end{array}
\label{relmodes1}
\eeq
Out of the $(2j+1)$ spin $j$ scalar modes,
$(j+1)$ of them satisfy the Neumann boundary
condition while the remaining $j$ of them satisfy the Dirichlet boundary
condition.
To see the boundary conditions for the spinors and the vectors,
it is best to look at the componets in the natural frames at the boundary.
The spinor modes in the frames
$\sqrt{r\dd z/z}$ and $\sqrt{r\dd\bz/\bz}$ are
\beqa
\varphi^{S_-}_{j,m}&=&\sqrt{2}\mu_j(\oa\ob)^{\half}g^{(j)}_{-m,\half}
=\mu_j\e^{\im m\alpha}(\sin\beta)^{\half}\,d^j_{-m,\half}(\beta),\nn\\
\varphi^{S_+}_{j,m}&=&\sqrt{2}\mu_j(ab)^{\half}g^{(j)}_{-m,-\half}
=\mu_j\e^{\im m\alpha}(\sin\beta)^{\half}\,d^j_{-m,-\half}(\beta),\nn
\eeqa
and the vector modes in the frames $r\dd z/z$ and $r\dd\bz/\bz$ are
\beqa
\varphi^{V_-}_{j,m}&=&2\mu_j\oa\ob\, g^{(j)}_{-m,1}
=\mu_j\e^{\im m\alpha}\sin\beta\,d^j_{-m,1}(\beta),\nn\\
\varphi^{V_+}_{j,m}&=&2\mu_jab\,g^{(j)}_{-m,-1}
=\mu_j\e^{\im m\alpha}\sin\beta\,d^j_{-m,-1}(\beta).\nn
\eeqa
We see from (\ref{imprel}) that the reflection
$\rho\to-\rho$ does
$\varphi_{j,m}^{\bullet_-}\to (-1)^{j-m}\varphi_{j,m}^{\bullet_+}$ 
for both spinor and vector modes.
In particular, they satisfy
\beq
\begin{array}{l}
\left(\varphi^{\bullet_-}_{j,m}
-(-1)^{j-m}\varphi^{\bullet_+}_{j,m}\right)\Bigr|_{\partial D^2}=0,
\\[0.3cm]
\left(\partial_{\rho}\varphi^{\bullet_-}_{j,m}+
(-1)^{j-m}\partial_{\rho}\varphi^{\bullet_+}_{j,m}\right)
\Bigr|_{\partial D^2}=0.
\end{array}
\label{relmodes2}
\eeq

In view of (\ref{relmodes1})-(\ref{relmodes2}), the
boundary conditions (\ref{BCchiral}) and (\ref{BCvector})
requires the following constraints on the modes. For the
chiral multiplet,
\beqa
&&\mbox{$\phi_{j,m}$, $\bphi_{j,m}$: $j-m$ even},\nn\\
&&\psi_{j,m}=\pm(-1)^{j-m}\wt{\psi}_{j,m},\quad\,\,
\bpsi_{j,m}=\pm(-1)^{j-m}\overline{\wt{\psi}}_{j,m},\nn\\
&&\mbox{$f_{j,m}$, $\overline{f}_{j,m}$: $j-m$ odd.}
\label{cmodesconstr}
\eeqa
For the vector multiplet,
\beqa
&&v_{j,m}=-(-1)^{j-m}\wt{v}_{j,m},\nn\\
&&\mbox{$(\sigma'_1)_{j,m}$, $(D_E)_{j,m}$: $j-m$ even},\nn\\
&&\mbox{$(\sigma_2)_{j,m}$: $j-m$ odd},\nn\\
&&\lambda_{j,m}=\mp(-1)^{j-m}\wt{\lambda}_{j,m},\quad\,\,
\blambda_{j,m}=\mp(-1)^{j-m}\overline{\wt{\lambda}}_{j,m}.
\label{vmodesconstr}
\eeqa
Recall that we also had infinitely many conditons: even number of
$\rho$-derivatives of (\ref{BCvector}).
In fact, the relations (\ref{relmodes1}) and (\ref{relmodes2}) hold also when 
$g^{(j)}_{-m,0}$ and $\varphi^{\bullet_{\pm}}_{j,m}$ are replaced by
$\partial_{\rho}^{2k}g^{(j)}_{-m,0}$ and
$\partial_{\rho}^{2k}\varphi^{\bullet_{\pm}}_{j,m}$.
Therefore, the vector multiplet fields with the mode expansion
obeying (\ref{vmodesconstr}) satisfy also these infinitely many
boundary conditions.
For the ghosts, we have
\beq
\mbox{$c_{j,m}$, $\overline{c}_{j,m}$: $j-m$ even.}
\eeq

Note that the reality of fields, $\bphi=\phi^{\dag}$,
$\overline{f}=f^{\dag}$, $v_{\mu}=v_{\mu}^{\dag}$ and
${\mathcal O}={\mathcal O}^{\dag}$ for
${\mathcal O}=\sigma'_1$, $\sigma_2$, $D_E'$,
yields via (\ref{realityWigner}) the following constraints:
\beqa
&&\bphi_{j,m}=(-1)^m\phi^{\dag}_{j,-m},\quad\,\,
\overline{f}_{j,n}=(-1)^mf^{\dag}_{j,-m},\\
&&\wt{v}_{j,m}=(-1)^{m-1}v_{j,-m}^\dag,\quad\,\,
{\mathcal O}_{j,m}=(-1)^m{\mathcal O}_{j,-m}^\dag.
\eeqa

Let us write down the kinetic terms.
To simplify the computation, we do the following trick.
Given the fields on the hemisphere $D^2=D^2_0$ we define the fields
on the other hemisphere $D^2_{\infty}$ in such a way that the action
on $D^2_{\infty}$ is equal to the one on $D^2_0$.
This is done as follows. First let us denote by $x\mapsto x'$ the
reflection at the equator, given by $z\mapsto \bz^{-1}$, or equivalently
$(\tau,\rho)\to (\tau, -\rho)$, or
$(\alpha,\beta)\mapsto (\alpha,\pi-\beta)$.
For a scalar ${\mathcal O}_N$ or ${\mathcal O}_D$
 obeying the
Neumann or Dirichlet boundary condition,
we define the extension by ${\mathcal O}_N(x')={\mathcal O}_N(x)$
or ${\mathcal O}_D(x')=-{\mathcal O}_D(x)$. For the spinors,
the extension is defined by $\upsi_{\pm}(x')=\upsi_{\mp}(x)$,
$\ulambda_{\pm}(x')=-\ulambda_{\mp}(x)$ (and similarly for the ``bared''
fields). For the vectors, we define it by
$v_z(x')=-(\bz^2v_{\bz})(x)$ and $v_{\bz}(x')=-(z^2v_z)(x)$.
Then, it is easy to see that the action on $D^2_{\infty}$ is the same as
the original action on $D^2_0$.
We can also see that the fields on $D^2_{\infty}$ defined this way
is equal to the na\"ive extension of the above mode expansions, from
$D^2_0$ to $D^2_{\infty}$.
Thus, we find
\beq
\int_{D^2}{\mathcal L}\sqrt{g}\dd^2x={1\over 2}\int_{S^2}{\mathcal L}
\Bigr|{}_{\rm \!\!\!naive\,\,\,\,\,\atop extension}
\sqrt{g}\dd^2x.
\eeq
Once the action is expressed as an integral on the whole sphere,
we can use the orthogonality (\ref{orthogonality}) for the evaluation.

Let us express the quadratic part of the action, 
(\ref{defS1}), (\ref{defS2}) and (\ref{defS3}), in terms of
the mode variables. For computation involving
$g^{z\bz}\partial_{\bz}v_z$ and $g^{z\bz}\partial_zv_{\bz}$,
it is useful to note
\beq
g^{z\bz}\partial_{\bz}\left(\ob^2g^{(j)}_{-m,1}\right)
=-{\sqrt{j(j+1)}\over 2r^2}g^{(j)}_{-m,0},\quad\,\,
g^{z\bz}\partial_{z}\left(b^2g^{(j)}_{-m,-1}\right)
={\sqrt{j(j+1)}\over 2r^2}
g^{(j)}_{-m,0}.
\eeq
The expressions are\footnote{The
bosonic variables of the vector multiplet are rescaled as
follows. For the scalars 
${\mathcal O}=\sigma_1',\sigma_2, D_E'$, we do
${\mathcal O}_{j,m}\to{1\over \sqrt{2}}{\mathcal O}_{j,m}$ for $m\geq 1$
but keep ${\mathcal O}_{j,0}$ intact.
For the vector, we do $v_{j,m}\to{1\over \sqrt{8}}v_{j,m}$ for $m\geq 1$
and $v_{j,0}\to{1\over \sqrt{2}}v_{j,0}$.}
\beqa
S_1
&=&\sum_{j-m\,\,{\rm even}}
\phi_{j,m}^{\dag}\left[{j(j+1)\over r^2}+{2R-R^2\over 4r^2}
+\im {1-R\over r}\usigma_1+\usigma_1^2\right]\phi_{j,m}
+\!\!\sum_{j-m\,\,{\rm odd}}f_{j,m}^{\dag}f_{j,m}\nn\\
&&
+\,2\,\im\sum_{j,m}(-1)^{m+\half}\bpsi_{j,-m}\left[
{j+\half\over r}\mp\im(-1)^{j-m}\left(\usigma_1-\im{R\over 2r}\right)\right]
\psi_{j,m},
\label{S1}\\[0.2cm]
S_2&=&
\sum_{j\geq 1,\,m\geq 0\atop j-m\,\,{\rm even}}
(\sigma'_1)_{j,m}^{\dag}{j(j+1)\over r^2}(\sigma'_1)_{j,m}
+\sum_{j\geq 1,\,m\geq 0\atop j-m\,\,{\rm odd}}
(\sigma_2)_{j,m}^{\dag}\left[{j(j+1)\over r^2}
+\usigma_1^2+{1\over r^2}
\right](\sigma_2)_{j,m}
\nn\\
&&
+\sum_{j\geq 1\,m\geq 0}v_{j,m}^{\dag}
\left[{j(j+1)\over r^2}+\usigma_1^2\right]v_{j,m}
+\sum_{m\geq 0\atop j-m\,\,{\rm even}}(D_E')_{j,m}^{\dag}(D_E')_{j,m}
\nn\\
&&-\im\sum_{j\geq 1,\,m\geq 0\atop j-m\,\,{\rm even}}
\left(v_{j,m}^{\dag}{\sqrt{j(j+1)}\over r}\usigma_1(\sigma'_1)_{j,m}
+v_{j,m}{\sqrt{j(j+1)}\over r}\usigma_1(\sigma'_1)^{\dag}_{j,m}\right)\nn\\
&&-\im\sum_{j\geq 1,\,m\geq 0\atop j-m\,\,{\rm odd}}
\left(v_{j,m}^{\dag}{\sqrt{j(j+1)}\over r}{1\over r}(\sigma_2)_{j,m}
-v_{j,m}{\sqrt{j(j+1)}\over r}{1\over r}(\sigma_2)^{\dag}_{j,m}\right)\nn\\
&&
+\,2\,\im\sum_{j,m}(-1)^{m+\half}\blambda_{j,-m}\left[
{j+\half\over r}\pm\im(-1)^{j-m}\usigma_1\right]
\lambda_{j,m},
\label{S2}\\[0.2cm]
S_3&=&\sum_{j\geq 1\atop j-m\,\,{\rm even}}
(-1)^m\overline{c}_{j,-m}{j(j+1)\over r^2}c_{j,m}.
\eeqa

\subsection{Determinants}

We are now ready to compute the fluctuation determinants.
We choose a maximal torus $T_G$ of $G$ so that that the supersymmetric 
background $\sigma_1=\usigma_1$ belongs to its Lie algebra $\ttt$
times $\im$.
We choose a Weyl chamber in $\im\ttt^*$ and write $\alpha>0$ if
$\alpha$ is a positive root with respect to that.
We write $d_G$ and $l_G$ for the dimension and the rank of $G$,
and put $d_V:=\dim_{\C}V$.

Let us first consider a single chiral multiplet that has charge $+1$
under a single $U(1)$ gauge group and vector R-charge $R$.
The big parenthesis of the first line of (\ref{S1}) factorizes as
$$
\left({j\over r}+\im\left(\usigma_1-\im {R\over 2r}\right)\right)
\left({j+1\over r}-\im\left(\usigma_1-\im {R\over 2r}\right)\right)
$$
Thus the determinant is
\beqa
{{\rm det}_F\over {\rm det}_B^{\half}}&=&
{\displaystyle
\prod_{j=\half}^{\infty}
\left({j+\half\over r}
+\im\left(\usigma_1-\im {R\over 2r}\right)\right)^{j+\half}
\left({j+\half\over r}
-\im\left(\usigma_1-\im {R\over 2r}\right)\right)^{j+\half}
\over
\displaystyle
\prod_{j=1}^{\infty}
\left({j\over r}+\im\left(\usigma_1-\im {R\over 2r}\right)\right)^{j+1}
\left({j+1\over r}-\im\left(\usigma_1-\im {R\over 2r}\right)\right)^{j+1}
}\nn\\
&=&{1\over
\displaystyle
\prod_{j=0}^{\infty}
\left({j\over r}+\im\left(\usigma_1-\im {R\over 2r}\right)\right)}
\label{detchiral}
\eeqa
For a chiral multiplet of weight $Q$ under $T_G$,
the result is obtained from the above by the replacement
$\usigma_1\to Q(\usigma_1)$.

Next we consider the vector multiplet.
The fermionic determinant is straightforward,
\beq
{\rm det}_F\,=\,\prod_{j=\half}^{\infty}
\left[\left({j+\half\over r}\right)^{(2j+1)l_G}
\prod_{\alpha>0}\left(\left({j+\half\over r}\right)^2+\alpha(\usigma_1)^2
\right)^{2j+1}\right].
\eeq
The bosonic sector is complicated. We first notice that
it splits into $j-m$ even part involving $v$ and $\sigma'_1$
and $j-m$ odd part involving $v$ and $\sigma_2$. We also notice that the $m=0$
modes are real or pure imaginary while the $m\geq 1$ modes are complex.
After some computation, we find
\beq
{\rm det}_B^{\half}
\,=\,\prod_{j=1}^{\infty}\left[
\left({j(j+1)\over r^2}\right)^{(j+1)d_G+jl_G}
\prod_{\alpha>0}\left(\left({j(j+1)\over r^2}+\alpha(\usigma_1)^2\right)^2
+{\alpha(\usigma_1)^2\over r^2}\right)^j\,\right].
\eeq
The ratio is
\beq
{{\rm det}_F\over{\rm det}_B^{\half}}\,\,=\,\,{\displaystyle
\prod_{j=1}^{\infty}\left[
\left({j\over r}\right)^{l_G}
\prod_{\alpha>0}
\left({j\over r}+\im\alpha(\usigma_1)\right)
\left({j\over r}-\im\alpha(\usigma_1)\right)\right]\over
\displaystyle\prod_{j=1}^{\infty}\left({j(j+1)\over r^2}\right)^{(j+1)d_G}}.
\label{detvector}
\eeq

Finally, the ghost determinant is
\beq
{\rm det}_{gh}\,=\,
\prod_{j=1}^{\infty}\left({j(j+1)\over r^2}\right)^{(j+1)d_G}.
\eeq
We notice that it cancels againt the denominator of (\ref{detvector}).

We find two problems in the above result.
One is that it is a product of infinite factors and the other is
that each factor is dimensionful.
The former will be dealt with by regularization and renormalization.
The latter is simply because we were not careful in defining the measure,
 even formally.
If $\varphi$ is a field of canonical demension $d_{\varphi}$
and if there is a coupling constant factor $1/g_0^2$ in front of the
kinetic term, we should define the measure by
\beq
{\mathcal D}\varphi=
\sqrt{
\det\left({\Lambda_0^{D-2d_{\varphi}}(\varphi_n,\varphi_m)\over g_0^2}\right)}
\prod_n\dd a_n,
\label{defmeasure}
\eeq
for some mode expansion $\varphi(x)=\sum_n\varphi(x)a_n$,
where $D$ is the spacetime dimension, $(\varphi_n,\varphi_m)$ is the 
inner product of the modes defined by the spacetime integration,
and $\Lambda_0$ is a parameter of mass dimension which is usually
taken to be the ultra-violet cut-off. In the present context, we should take
$D=2$, $g_0=1$ (as $e$ and $\mc$ are absorbed into fields),
and we had chosen the modes so that $(\varphi_n,\varphi_m)=\delta_{n,m}$.
The net effect is to multiply $\Lambda^k_0$
to each factor of length dimension $k$. For example, we should do
the replacement
\beq
\left({j\over r}+\im\left(\usigma_1-\im {R\over 2r}\right)\right)\,
\longrightarrow\,
{1\over \Lambda_0}
\left({j\over r}+\im\left(\usigma_1-\im {R\over 2r}\right)\right),
\eeq
in the denominator factor of (\ref{detchiral}).

Let us now discuss the regularization. We take the
na\"ive cut off\footnote{This discussion is important and leads to the
identification of the 2d central charge (\ref{2dc}) below.
This had been done on the two-sphere
by Sungjay Lee as presented in conferences \cite{LeeTalks}
and by the other authors of \cite{Doroudetal} \cite{privatecomm},
and also in 4d by Pestun.}
 where we introduce an upper bound $N$ of the product
over $j$. We will eventually take the $N\to\infty$ limit after a suitable
renomalization of coupling constants.
Since $j/r$ corresponds to the energy scale, we may interpret
$\Lambda_0:=N/r$ as the ultra-violet cut-off which we take to be the
same $\Lambda_0$ in (\ref{defmeasure}).
Using the formula for the gamma function
\beq
\Gamma(z)=\lim_{N\to\infty}{N!\,(N+1)^z\over
\prod_{j=0}^N(j+z)},
\eeq
together with Stirling's formula
$N!\sim \sqrt{2\pi}N^{N+\half}\e^{-N}$, we find
\beq
\prod_{j=0}^{r\Lambda_0}{1\over \Lambda_0}\left({j\over r}+a\right)
\,=\,{1\over (r\Lambda_0)^{r\Lambda_0+1}}
\prod_{j=0}^{r\Lambda_0}(j+ra)
\,=\,\sqrt{2\pi}(r\Lambda_0)^{-\half+ra}\e^{-r\Lambda_0}{1\over\Gamma(ra)}.
\eeq

Now the determinants make sense.
The factor from the chiral multiplet is
\beqa
\lefteqn{Z_{\rm chiral}\,\,=\,\,(2\pi)^{-{d_V\over 2}}\times}\nn\\
&&
\exp\!\left(\,d_Vr\Lambda_0+\sum_i\left[{1-R_i\over 2}
-\im rQ_i(\usigma_1)\right]\log(r\Lambda_0)\right)
\prod_i\Gamma\!\left(\im r Q_i(\usigma_1)+{R_i\over 2}\right).
\eeqa
The factor from the vector multiplet and the ghost is
\beqa
Z_{\rm vector+ghost}
&=&{\displaystyle (2\pi)^{d_G\over 2}
\exp\left(-d_Gr\Lambda_0-{d_G\over 2}\log(r\Lambda_0)\right)
\over\displaystyle
\prod_{\alpha>0}\Gamma\left(\im r\alpha(\usigma_1)\right)
\Gamma\left(-\im r\alpha(\usigma_1)\right)r^2\alpha(\usigma_1)^2}
\nn\\
&=&(2\pi)^{d_G\over 2}
\exp\left(-d_Gr\Lambda_0-{d_G\over 2}\log(r\Lambda_0)\right)
\prod_{\alpha>0}{\displaystyle \sinh(\pi r\alpha(\usigma_1))
\over \pi r\alpha(\usigma_1)}
\eeqa
where we used $\Gamma(1+z)=z\Gamma(z)$ and
$\Gamma(z)\Gamma(1-z)=\pi/\sin(\pi z)$.

\subsection{The Result}

Following (\ref{defmeasure}), the zero mode measure is
\beq
\left({r\Lambda_0\over e}\right)^{d_G}\dd^{d_G}\usigma_1.
\label{zeromodemeasure}
\eeq
We use the following formula that holds for an adjoint invariant function
$F(\usigma_1)$,
\beq
{1\over {\rm vol}(G)}\int\limits_{\,\,\,\,\im \mbox{$\mathfrak{g}$}\!\!\!\!}
\dd^{d_G}\usigma_1\,F(\usigma_1)
\,\,=\,\,{1\over |W_G|}\int\limits_{\,\,\im \mbox{$\ttt$}\!\!\!}
\dd^{l_G}\usigma_1
\prod_{\alpha>0}\alpha(\usigma_1)^2\,\cdot\,F(\usigma_1).
\eeq
Collecting everything, for the brane data $\mathfrak{B}=(M,Q,\rho,{\bf r}_*)$
we find
\beqa
Z_{D^2\!{}_{(+)}\!}(\mathfrak{B})&=&C
\left({\Lambda_0\over e}\right)^{d_G}\exp\left((d_V-d_G)r\Lambda_0
+{\wh{c}\over 2}\log(r\Lambda_0)\right)\nn\\
&&\!\!\!\!\!\!\!\!\!\!\!\!
\times\int\limits_{\,\,\im \mbox{$\ttt$}\!\!\!}
r^{l_G}\dd^{l_G}\usigma_1
\prod_{\alpha>0}r\alpha(\usigma_1)\sinh(\pi r\alpha(\usigma_1))
\prod_{i}\Gamma\left(\im rQ_i(\usigma_1)+{R_i\over 2}\right)
\label{1stexpr}\\
&&~~~~\times
\exp\left(2\pi \im r\wt{W}(\usigma_1)
-\im r\sum_iQ_i(\usigma_1)\log(r\Lambda_0)\right)
{\rm tr}_M^{}\left(\e^{\pi i {\bf r}_*}\e^{2\pi r\rho(\usigma_1)}\right),
\nn
\eeqa
where $C$ is a numercal factor and
\beq
\wh{c}\,\,:=\,\,\sum_i(1-R_i)-d_G.
\label{whcdef}
\eeq
$Z_{D^2\!{}_{(-)}\!}(\mathfrak{B})$ is the same as (\ref{1stexpr}) except that
we need to replace $\wt{W}$ by $\overline{\wt{W}}$ and invert the exponents
of the Chan-Paton factor. (See (\ref{classS0}) and (\ref{classCP}).)

Before removing the cut-off $\Lambda_0$ we need to do a renormalization.
We consider the following cut-off dependent local counter terms:
\beqa
\mbox{dilaton}&=&{\wh{c}\over 2}\log(\Lambda_0/\Lambda),\\
\mbox{boundary potential}&=&{1\over 2\pi}(d_V-d_G)\Lambda_0,\\
\Delta\wt{W}(\sigma)&=&{1\over 2\pi}{\rm tr}^{}_V(\sigma)
\log(\Lambda_0/\Lambda).
\eeqa
Here $\Lambda$ is a finite energy scale.
There is also an overall multiplicative divergence $\Lambda_0^{d_G}$
which we decide to absorb by a multiplicative change of measure,
say, by replacing $e$ in (\ref{zeromodemeasure}) by $\Lambda_0$.
Then, we have cut-off independent expressions:

\beqa
Z_{D^2\!{}_{(+)}\!}(\mathfrak{B})\!\!&=&\!\!C(r\Lambda)^{\wh{c}/2}\!
\int\limits_{\,\,\im \mbox{$\ttt$}\!\!\!}
r^{l_G}\dd^{l_G}\usigma_1
\prod_{\alpha>0}r\alpha(\usigma_1)\sinh(\pi r\alpha(\usigma_1))
\prod_{i}\Gamma\!\left(\im r Q_i(\usigma_1)+{R_i\over 2}\right)
\label{theresult1}\\[-0.2cm]
&&~~~~~~~\times
\exp\left(2\pi \im r \wt{W}(\usigma_1)
-\im r \sum_iQ_i(\usigma_1)\log(r\Lambda)\right)
{\rm tr}_M^{}\left(\e^{\pi i {\bf r}_*}\e^{2\pi r \rho(\usigma_1)}\right),
\nn\\[0.5cm]
Z_{D^2\!{}_{(-)}\!}(\mathfrak{B})\!\!&=&\!\!C(r\Lambda)^{\wh{c}/2}\!
\int\limits_{\,\,\im \mbox{$\ttt$}\!\!\!}
r^{l_G}\dd^{l_G}\usigma_1
\prod_{\alpha>0}r\alpha(\usigma_1)\sinh(\pi r\alpha(\usigma_1))
\prod_{i}\Gamma\!\left(\im r Q_i(\usigma_1)+{R_i\over 2}\right)
\label{theresult2}\\[-0.2cm]
&&~~~~~~~\times
\exp\left(2\pi \im r \overline{\wt{W}}(\usigma_1)
-\im r \sum_iQ_i(\usigma_1)\log(r\Lambda)\right)
{\rm tr}_M^{}\left(\e^{-\pi i {\bf r}_*}\e^{-2\pi r \rho(\usigma_1)}\right).
\nn
\eeqa

\medskip

\subsection{A-Branes}\label{subsec:Abranes}

Let us compute the partition function of the Landau-Ginzburg model
preserving B$_{(\pm)}$-type supersymmetry.
We consider the model of $\C^n$ valued variable $\phi=(\phi^1,\ldots,\phi^n)$
with a superpotential $W$ and a flat K\"ahler metric
$\mc^2\sum_i|\dd\phi^i|^2$.
We start with the case where the brane $L_{\pm}$
is a linear Lagrangian subspace of $\R^{2n}$ such
that $\mp{\rm Im}(W)$ is bounded from below on $L_{\pm}$.
In this case, the boundary term in the action (\ref{LGkin}) or
(\ref{LGkinnvar}) vanishes and the usual kinetic term itself is $Q$-exact,
so that the usual localization is valid.
In the limit $\mc\to \infty$, the path-integral localizes on the
supersymmetric locus,
\beq
\partial_{\mu}\phi=0, \quad\,\, f_!=0.
\eeq
The classical Lagrangian is
\beq
S_0=\left\{\begin{array}{ll}
2\pi\im r W(\phi)&\mbox{for $(+)_0$},\\
2\pi\im r\overline{W(\phi)}&\mbox{for $(-)_0$}.
\end{array}\right.
\eeq
The fluctuation determinant is independent of the location $\phi$.
The scalars tangent ({\it resp}. normal) to the brane obey Neumann
({\it resp}. Dirichlet) boundary condition and the fermions obey the
the corresponding boundary conditon. We also need to omit the bosonic
zero modes. Employing the mode expansions obatined in the gauge theory,
we find
\beq
{{\rm det}_F\over {\rm det}_B^{\half}}
=\left[{\displaystyle \prod_{j=\half}^{\infty}{1\over \Lambda_0}
\left({j+\half\over r}\right)^{2j+1}
\over \displaystyle
\prod_{j=1}^{\infty}{1\over\Lambda_0^2}
\left({j(j+1)\over r^2}\right)^{{j+1\over 2}+{j\over 2}}
}\right]^n
={1\over (r\Lambda_0)^{n/2}},
\eeq
where $\Lambda_0$ is an ultra-violet cut off.
The measure for the scalar zero modes is, following (\ref{defmeasure}),
\beq
(r\Lambda_0)^n\mc^n\dd{\rm vol}^{}_{L_{\pm}},
\eeq
where $\dd {\rm vol}_{L_{\pm}}$ is the volume element of $L_{\pm}$
associated to the metric induced from the metric 
$\sum_i|\dd\phi^i|^2$ of $\C^n$.
The result has a cut-off dependence which can be renormalized by
the dilaton shift ${n\over 2}\log(\Lambda_0/\Lambda)$.
We shall also absorb the divergence as $\mc\to\infty$
by a multiplicative change of the measure.
Collecting all the elements, we find that the partition function is given by
\beq
Z_{D^2\!{}_{(\pm)}\!}(L_{\pm})
=(r\Lambda)^{n/2}\int_{L_{\pm}}\dd{\rm vol}^{}_{L_{\pm}}\left\{
\begin{array}{l}
\e^{-2\pi\im rW(\phi)}\\
\e^{-2\pi\im r\overline{W(\phi)}}
\end{array}\right.
\label{1stexprLG}
\eeq

Let us next consider deforming $L_{\pm}$ from a Lagrangian subspace
to a more general Lagrangian submanifold, while maintaining the condition
that $\mp{\rm Im}(W)$ is bounded from below on $L_{\pm}$. As discussed in
Section~\ref{subsec:nodep}, we require that the result does not change under
such a deformation. But the expression (\ref{1stexprLG})
does change if we deform $L_{\pm}$ and thus cannot be the correct answer.
We propose that we should replace the volume element
by holomorphic or antiholomorphic volume
form,
\beq
\dd {\rm vol}^{}_{L_+}\to
\dd^n\phi=\dd \phi^1\wedge\cdots\wedge\dd \phi^n,\quad\,\,
\dd {\rm vol}^{}_{L_-}
\to\dd^n\bphi=\dd \bphi^1\wedge\cdots\wedge\dd \bphi^n,\
\eeq
so that
\beqa
Z_{D^2\!{}_{(+)}\!}(L_+)
&=&(r\Lambda)^{n/2}\!\int_{L_+}\dd^n\phi\,\e^{-2\pi\im rW(\phi)},
\label{LGresult1}\\
Z_{D^2\!{}_{(-)}\!}(L_-)
&=&(r\Lambda)^{n/2}\!\int_{L_-}\dd^n\bphi\,\e^{-2\pi\im r\overline{W(\phi)}}.
\label{LGresult2}
\eeqa
Indeed, it meets the requirement of invariance under deformation of
$L_{\pm}$ and at the same time, when $L_{\pm}$ is linear, it
reduces to the result (\ref{1stexprLG}) up to a phase.

Let us discuss the issue of convergence of the integral
(\ref{LGresult1})-(\ref{LGresult2}).
Thanks to the asymptotic condition that
$\mp{\rm Im}(W)$ is bounded
from below on $L_{\pm}$, the exponential factor does not grow at infinity.
If $\mp{\rm Im}(W)$ grows fast enough at infinity,
the integral would be absolutely convergent.
Even if it does not, as long as the real part ${\rm Re}(W)$ changes
fast enough, the integral converges due to rapid oscillation
of the exponential factor.
If the infinity of $L_{\pm}$ consists of cones of linear subspaces,
the ``fast enough'' condition is met provided $|\partial W(\phi)|$
grows faster than a power of $\phi$. See
\cite{HPT} for a recent explanation on the conditional convergence
of the integral of this type.

Let us look at the $r$ dependence.
When $W(\phi)$ is quasi-homogeneous, $W(\lambda^R\phi)=\lambda^2W(\phi)$,
we can absorb the $r$ in the integrand
by a change of variables, $\phi\to r^{-R/2}\phi$, and we find that
 the $r$ dependence is just
an overall factor
\beq
Z_{D^2\!{}_{(\pm)}\!}(L_{\pm})
~\sim~ r^{n/2-{\rm tr}(R/2)}\quad\mbox{for all $r$.}
\label{power}
\eeq
This power behaviour is a characteristic feature of the
partition function of a conformally invariant
field theory \cite{Polyakov}, where the power must be identified
with one sixth of the central charge in the case of a hemisphere.
Indeed,
\beq
6\left({n\over 2}-{\rm tr}\left({R\over 2}\right)\right)
=3\sum_i(1-R_i),
\eeq
is the central charge of the superponformal field theory
to which the Landau-Ginzburg model is believed to flow
\cite{Martinec,VafaWarner}.
As the extreme opposite, let us consider the case where 
the superpotential $W(\phi)$ is a Morse function, having isolated
and non-degenerate critical points only.
The theory has supersymmetric ground states with mass gaps whose
wavefunctions are supported at the critical points.
To each critical point $p$,
one can associate a pair of Lagrangian submanifolds $L_{p,\pm}$
passing through $p$, called {\it Lefschetz thimbles}, 
whose $W$-values are straight semi-lines
emanating from $W(p)$ in the direction where
${\rm Im}(W)$ goes to negative/positive infinity \cite{HIV}.
For such Lagrangians,
one can employ the saddle point approximation for large values of $r$,
which finds
\beq
|Z_{D^2\!{}_{(\pm)}\!}(L_{p,\pm})|~\sim~ \e^{-2\pi r(\mp {\rm Im}W(p))}\quad
\mbox{as\, $ r\longrightarrow\infty$.}
\label{exponential}
\eeq
It is exponentially decaying or growing as a function of $r$,
depending on whether $\mp{\rm Im}(W)$ is
 positive or negative at $p$.
We shall consider this exponential behaviour as a signal of
the vacuum with a mass gap.

If we consider the system preserving B$^{\alpha}_{(\pm)}$-type supersymmetry, 
all we need to do is to replace $W$ by $\e^{2\im\alpha}W$.
In particular, the power behaviour (\ref{power}) for the case of
quasi-homogeneous $W$ is independent of the parameter $\alpha$, while
the exponential behaviour (\ref{exponential}) for Morse $W$
is changed so that
what matters is the sign of $\mp{\rm Im}(\e^{2\im\alpha}W(p))$.

The proposal can be extended to a more general Landau-Ginzburg
model and the non-linear sigma model preserving B$_{(\pm)}$-type
supersymmetry. Recall that an axial $U(1)$ R-symmetry is necessary for
B-type supersymmetry and the existence requires the target space $X$
have a trivial first Chern class, $c_1(X)=0$. In many cases, this also
means that there exists a holomorphic volume form $\Omega$.
The proposal in such a case is
\beq
Z_{D^2\!{}_{(+)}\!}(L_+)=
(r\Lambda)^{n/2}\int_{L_+}\Omega\,\e^{-2\pi\im r W},\quad\,\,
Z_{D^2\!{}_{(-)}\!}(L_-)
=(r\Lambda)^{n/2}\int_{L_-}\overline{\Omega}\,\e^{-2\pi\im r \overline{W}}.
\label{genLGresult}
\eeq
When $X$ is non-compact, the holomorphic volume form $\Omega$ is
not unique and therefore we must make a choice. 
In the non-linear sigma model on a compact Calabi-Yau manifold $X$,
the holomorphic volume form is unique up to constant multiplication.

The result (\ref{1stexprLG}) for linear Lagrangians as well as
the proposal
(\ref{LGresult1})-(\ref{LGresult2}) or (\ref{genLGresult}) for the general case
are indeed the same as the formula for the central charge of the 
A-branes in the Landau-Ginzburg model or the non-linear sigma model
\cite{OOY,HIV}.

\subsection{Deformation From The Real Locus}\label{subsec:offreal}

The above discussion allows us to propose a formula for the partition function 
in which the boundary condition for the vector multiplet is deformed from
the one (\ref{BCvector}) associated to the real locus
$L=\im\mathfrak{g}$ to a more general Lagrangian submanifold
$L$ of $\mathfrak{g}_{\C}$ satisfying the conditions
(\ref{sigma12}), (\ref{sigma1TL}) and (\ref{Ginvariance}).

First, we claim that such an $L$ is the adjoint $G$-orbit of
a Lagrangian submanifold $\gamma$ of $\ttt_{\C}$,
\beq
~~~L=G\gamma,\quad\,\,\, \gamma\subset\ttt_{\C}.
\eeq
To prove this claim,
let us take a point $\sigma=\sigma_1+\im \sigma_2$ of $L$.
By the condition $[\sigma_1,\sigma_2]=0$ of (\ref{sigma12}),
there is an element $g\in G$ which sends both
$\sigma_1$ and $\sigma_2$ to $\ttt$. 
By the condition (\ref{Ginvariance}) that $L$ is $G$-invariant,
we have $g(\sigma)\in L\cap \ttt_{\C}=:\gamma$.
Thus, we have seen $L/G\cong \gamma/W_G$. In particular,
the dimension of $\gamma$ is equal to the dimension of
$L/G$ which is equal to $\dim L-\dim (G/G_{\sigma})$
where $G_{\sigma}$ is the isotropy subgroup of $G$ at $\sigma\in \gamma$.
Generically, the dimension of $G_{\sigma}$ is equal to the rank $l_G$ of $G$.
Thus, $\dim \gamma=\dim L-(\dim G- l_G)=l_G$. Thus, $\gamma$
is a middle dimensional submanifold of $\ttt_{\C}$.
Let us now show that $\gamma$ is a Lagrangian submanifold of $\ttt_{\C}$.
Since $\ttt_{\C}\subset \mathfrak{g}_{\C}$  is a complex submanifold and 
$L\subset \mathfrak{g}_{\C}$ is a Lagrangian submanifold, 
at any point $\sigma\in\gamma$,
${\mathcal J}_{\sigma}{\rm T}_{\sigma}\gamma$ is a subspace of
${\rm T}_{\sigma}\ttt_{\C}$ which is orthogonal to ${\rm T}_{\sigma}\gamma$.
Since $\gamma$ is middle dimensional,
${\mathcal J}_{\sigma}{\rm T}_{\sigma}\gamma$ is the orthocomplement of 
${\rm T}_{\sigma}\gamma$ in ${\rm T}_{\sigma}\ttt_{\C}$.
That is, $\gamma$ is a Lagrangian submanifold of $\ttt_{\C}$.
Finally, we show that
the condition (\ref{sigma1TL}) is satisfied for such an $L$.
By the homogeneity, we may assume $\sigma\in\gamma$. The tangent
space ${\rm T}_{\sigma}L$ of $L=G\gamma$ is the direct sum of $T_{\sigma}\gamma$
and the orbit directions $\mathfrak{g}(\sigma)$.
The former component $T_{\sigma}\gamma$ commutes with $\sigma_1$.
The latter component $\mathfrak{g}(\sigma)$ is invariant under
commutator with $\sigma_1$, since
$[\sigma_1,[X,\sigma]]=[[\sigma_1,X],\sigma]$ where we used
$[\sigma_1,\sigma]=0$. This completes the proof of the claim.

The localization procedure works in the same way as the real locus
and we have an integral over $\gamma=L\cap \ttt_{\C}$ 
of the classical exponential factor times
the fluctuation determinant with respect to some measure.
However, the direct computation of the fluctuation determinant is
very complicated for a general choice of $L$. At this point, 
we employ the holomorphy discussed in Section~\ref{subsec:holomorphy}
and also take the lesson from the previous section concerning
the measure. 
The integrand can be regarded as the effective partition function
on the Coulomb branch in which the vector multiplet of the group $T_G$
is fixed to be a supersymmetric background satisfying (\ref{minimivector}).
In the background, the scalar $\sigma$ takes a constant value
\beq
\usigma=\usigma_1+\im\usigma_2,
\eeq
and the gauge field has the boundary holonomy
\beq
\oint_{\partial D^2}v=-2\pi r\usigma_2.
\label{boundaryholonomy}
\eeq
In this picture, $\usigma$ is a twisted chiral parameter and
the integrand of
$Z_{D^2\!{}_{(+)}\!}(\mathfrak{B})$ 
({\it resp}. $Z_{D^2\!{}_{(-)}\!}(\mathfrak{B})$)
must depend holomorphically ({\it resp}. antiholomorphically) on it.
It is uniquely determined by the
values at the real locus $\im\ttt$, given as the integrand of
(\ref{theresult1}) or (\ref{theresult2}).
The (anti)holomorphic extension of the Chan-Paton factor,
from $\e^{\pm 2\pi r\rho(\usigma_1)}$
to $\e^{\pm 2\pi r\rho(\usigma_1\pm\im \usigma_2)}$, can be understood
from (\ref{boundaryholonomy}), since it
originates from the factor
$\e^{-\oint_{\partial D^2}\rho(\im v_{\tau}\mp\sigma_1)\dd\tau}$
in (\ref{CPfactor}).
The measure $\dd^{l_G}\usigma_1$ is extended uniquely to 
the holomorphic or anti-holomorphic volume form of $\ttt_{\C}$,
denoted by $\dd^{l_G}\usigma$ or $\dd^{l_G}\busigma$.
In this way, we arrive at the following expressions

\beqa
Z_{D^2\!{}_{(+)}\!}(\mathfrak{B})\!&=&\!C(r\Lambda)^{\wh{c}/2}
\int\limits_{\gamma_+}
r^{l_G}\dd^{l_G}\usigma
\prod_{\alpha>0}r\alpha(\usigma)\sinh(\pi r\alpha(\usigma))
\prod_{i}\Gamma\left(\im r Q_i(\usigma)+{R_i\over 2}\right)
\label{thegenresult1}\\[-0.2cm]
&&~~~~~~~~~~~\times
\exp\left(2\pi \im r \wt{W}(\usigma)
-\im r \sum_iQ_i(\usigma)\log(r\Lambda)\right)
{\rm tr}_M^{}\left(\e^{\pi i {\bf r}_*}\e^{2\pi r \rho(\usigma)}\right),
\nn\\[0.5cm]
Z_{D^2\!{}_{(-)}\!}(\mathfrak{B})\!&=&\!C(r\Lambda)^{\wh{c}/2}
\int\limits_{\gamma_-}
r^{l_G}\dd^{l_G}\busigma
\prod_{\alpha>0}r\alpha(\busigma)\sinh(\pi r\alpha(\busigma))
\prod_{i}\Gamma\left(\im r Q_i(\busigma)+{R_i\over 2}\right)
\label{thegenresult2}\\[-0.2cm]
&&~~~~~~~~~~~\times
\exp\left(2\pi \im r \overline{\wt{W}(\usigma)}
-\im r \sum_iQ_i(\busigma)\log(r\Lambda)\right)
{\rm tr}_M^{}\left(\e^{-\pi i {\bf r}_*}\e^{-2\pi r \rho(\busigma)}\right).
\nn
\eeqa

\medskip
\noindent
In general, the Lagrangian submanifold $\gamma_+\subset \ttt_{\C}$
 for the A$_{(+)}$-type supersymmetry can be different from the one
$\gamma_-\subset \ttt_{\C}$ for the A$_{(-)}$-type supersymmetry.
As in the Landau-Ginzburg model, $\gamma_+$ and $\gamma_-$
must be chosen so that the effective boundary potential
is bounded from below. A concrete proposal for the right
choice will be given in Section~\ref{sec:contour}.

\subsection{The Case Of Linear Sigma Model}\label{subsec:thecaseof}

Let us now restrict our attention to the gauged linear sigma models
where the twisted superpotential is linear,
$\wt{W}(\sigma)={1\over 2\pi}t(\sigma)$.

In this case, the two expressions (\ref{thegenresult1})
and (\ref{thegenresult2}) are related by complex conjugation,
\beq
\left(Z_{D^2\!{}_{(+)}\!}(\mathfrak{B})\right)^*
=Z_{D^2\!{}_{(-)}\!}(\mathfrak{B}),
\label{ccreln}
\eeq
for the real locus $\gamma_+=\im\ttt=\gamma_-$ and the relation continues to
hold as long as $\gamma_+$ and $\gamma_-$ are mapped to each other
by the inversion $\usigma\mapsto-\usigma$. 
In what follows, we shall assume this relation between
$\gamma_+$ and $\gamma_-$
and will only mention $Z_{D^2\!{}_{(+)}\!}$ untill a special need of
$Z_{D^2\!{}_{(-)}\!}$ arizes in Section~\ref{sec:factorize}. Hence we shall 
drop the subscript $+$ from the expressions.

When $\wt{W}(\sigma)$ is linear,
we may absorb the radius $r$ into the integration variable as
\beq
\usigma'=r\usigma,
\eeq
so that the integral (\ref{thegenresult1})
can be written as
\beqa
Z_{D^2}(\mathfrak{B})\!&=&\!C(r\Lambda)^{\wh{c}/2}
\int\limits_{\gamma}
\dd^{l_G}\usigma'
\prod_{\alpha>0}\alpha(\usigma')\sinh(\pi \alpha(\usigma'))
\prod_{i}\Gamma\left(\im Q_i(\usigma')+{R_i\over 2}\right)
\nn\\
&&~~~~~~~~~~~~~~~~~~~~\times
\exp\left(\im t_{\rm R}(\usigma')\right)
{\rm tr}_M^{}\left(\e^{\pi i {\bf r}_*}\e^{2\pi \rho(\usigma')}\right).
\label{result}
\eeqa
Here we introduce the renormalized FI parameter
\beq
t_{\rm R}=t-{\rm tr}^{}_V\log(r\Lambda).
\eeq

As promised, we examine the effect of
the gauge shift of the R-charges (\ref{shiftR})-(\ref{shiftr*}),
which reads $R_i\to R_i+Q_i(\sDelta)$ and
${\bf r}_*\to {\bf r}_*-\rho(\sDelta)$, for a generator $\sDelta$
of the center of $G$.
Let us shift the integration variables as
\beq
\usigma'\to\usigma'+{\im\over 2}\sDelta
\label{shiftofv}
\eeq
Note that $\alpha(\sDelta)=0$ for any root $\alpha$ since $\sDelta$ is central.
Also, the exponent $\wh{c}$ may change but it
is absorbed by the shift of the part
$-\im{\rm tr}^{}_V(\usigma')\log(r\Lambda)$ of $\im t_{\rm R}(\usigma')$.
The net effect is the overall multiplication
\beq
Z\to \e^{-\half t(\sDelta)}Z,
\label{effectgshift}
\eeq
 plus the shift of integration contour,
$\gamma \to \gamma+{\im\over 2}\sDelta$.
Thus, as long as this shift does not cross any pole from the gamma functions,
the change is only by the multiplication by $\e^{-\half t(\sDelta)}$.
For the choice of real Lagrangian $\gamma=\im \ttt$, this is the case 
as long as the shift does not move the R-charges out of the range
$0<R_i<2$. The shift of variables (\ref{shiftofv}) has two-fold
interpretation. One is the reality violating change of variables, as in
(\ref{changevar}), in which the result $Z\to \e^{-\half t(\sDelta)}Z$
was anticipated in  (\ref{shiftS}).
The other is a change of boundary conditions which
moves the Lagrangian submanifold 
in the $\sigma_2$-direction,
$\gamma \to \gamma+{\im\over 2}\sDelta$.

Let us further specialize to the {\it Calabi-Yau case}:
\beq
G\subset SL(V).
\label{CY}
\eeq
Then, the trace ${\rm tr}^{}_V(\sigma)$ 
vanishes and the FI parameter is not renormalized,
\beq
t_{\rm R}=t.
\eeq
Accordingly, the number $\wh{c}$
does not change under the gauge shift of the R-charges.
The dependence on the size $r$ of the hemisphere is only in the
factor $(r\Lambda)^{\wh{c}/2}$.
As remarked in the Landau-Ginzburg model,
this is precisely the form of conformal anomaly \cite{Polyakov}
in a conformal field theory of central charge
\beq
c\,=\,3\,\wh{c}.
\label{2dc}
\eeq
With $\wh{c}$ given by (\ref{whcdef}), this
is indeed the central charge of the infra-red fixed point
of the gauge theory obtained by identifying the conformal algebra in the
$\overline{Q}_+$ chiral ring as in \cite{SilWi} or by a short-cut
argument \cite{HoTo}.

If the charge integrality (\ref{defJ})-(\ref{intr*}) holds,
the brane factor ${\rm tr}_M(\cdots)$ in (\ref{result})
can be written as ${\rm Str}_M\rho(J^{-1}\e^{2\pi\usigma'})$.
If the gauge group $G$ is a finite group, where
the theory is a Landau-Ginzbutg orbifold, 
we do not have the $\usigma$ integral and the result is simply
\beq
Z_{D^2}(B)=C(r\Lambda)^{\wh{c}/2}\cdot
{\rm Str}^{}_M\rho(J^{-1}).
\label{LGOres}
\eeq
Up to the prefactor, this indeed agrees with the central charge for the
B-brane $B=(M,Q,\rho,{\bf r}_*)$ of the Landau-Ginzburg orbifold
proposed in \cite{WalcherLG}.
If, instead, $J$ is in the identity component of the center,
we may gauge shift the R-charges of bulk fields 
from $0<R_i<2$ to the R$^o$-frame (\ref{Roframe}),
where $R^o_i=0$ or $2$ by continuity:
\beqa
Z_{D^2}(\mathfrak{B})\!&=&\!C(r\Lambda)^{\wh{c}/2}
\int\limits_{\gamma}
\dd^{l_G}\usigma'
\prod_{\alpha>0}\alpha(\usigma')\sinh(\pi \alpha(\usigma'))
\prod_{i}\Gamma\left(\im Q_i(\usigma')+{R^o_i\over 2}\right)
\nn\\
&&~~~~~~~~~~~~~~~~~~~~\times
\exp\left(\im t_{\rm R}(\usigma')\right)
{\rm Str}_M^{}\e^{2\pi \rho(\usigma')}.
\label{resultint}
\eeqa
If the contour before the gauge shift was the real locus $\im\ttt$,
then $\gamma$ is such that $Q_i(\usigma')$ for $R_i^o=0$ has a small
negative imaginary part.

In what follows, we shall concentrate
on the study of the hemisphere partition function (\ref{result}) of
the linear sigma model.
We shall often specialize to the Calabi-Yau case (\ref{CY})
or to the case with charge integrality
(\ref{defJ})-(\ref{intr*}).

\section{The Contour}
\label{sec:contour}

\newcommand{\Arg}{{\rm Arg}}

We are left with one and the most important problem:
decide which Lagrangian submanifold $L\subset\mathfrak{g}_{\C}$ to take
for the boundary condition on the vector multiplet, or equivalently
(see Section~\ref{subsec:offreal}),
which Lagrangian submanifold $\gamma\subset\ttt_{\C}$
to take as the contour of the integration (\ref{result}).
Let us copy the integral for convenience,
\beqa
Z_{D^2}(\mathfrak{B})\!&=&\!(r\Lambda)^{\wh{c}/2}
\int\limits_{\gamma}
\dd^{l_G}\usigma'
\prod_{\alpha>0}\alpha(\usigma')\sinh(\pi \alpha(\usigma'))
\prod_{i}\Gamma\left(\im Q_i(\usigma')+{R_i\over 2}\right)
\nn\\
&&~~~~~~~~~~~~~~~~~~~~\times
\exp\left(\im t_{\rm R}(\usigma')\right)
\sum_j\e^{\pi\im r_j}\e^{2\pi q_j(\usigma')}.
\label{resultagain}
\eeqa
Here, we wrote the brane factor as a sum,
$$
{\rm tr}_M^{}\left(\e^{\pi \im {\bf r}_*}\e^{2\pi\rho(\usigma')}\right)
=\sum_j\e^{\pi\im r_j}\e^{2\pi\rho(\usigma')},
$$
where $r_j$ and $q_j$ are the R-charge and
the $T$-weight of the basis element of the Chan-Paton vector space $M$.

\subsection{A Proposal}

To attack this problem,
we would like to have some idea on the integrand of (\ref{resultagain}).
In particular, we would like to know the location of singularity as well as
the growth or decay rate at infinity.

The gamma function has simple poles
at non-positive integers,
\beq
\Gamma(z)~\sim ~ {(-1)^n\over n!}{1\over z+n},\qquad z\sim -n.
\label{GammaPoles}
\eeq
Therefore, the integrand of (\ref{resultagain}) has poles at infinitely many
hyperplanes
\beq
~~~~~~~~~~~~~~~~~~
Q_i(\usigma')~=~\im\left(n_i+{R_i\over 2}\right),\qquad n_i=0,1,2,3,\ldots\, .
\label{poles}
\eeq
These are where $Q_i(\usigma')$ are on the positive imaginary axis 
if we choose $R_i>0$.
In particular, the real locus $\gamma=\im\ttt$ does not hit the singularity.
Of course we anticipated this since the scalar $\phi_i$ has no zero mode
when $\usigma=\usigma_1\in\im \ttt$
as long as we put the R-charges in the range (\ref{narrowrange}).
In fact, the poles (\ref{poles}) must be associated with the
zero modes of the scalars $\phi_i$, in the presence of the boundary term
$-Q_i(\usigma_2)|\phi_i|^2$ in (\ref{LSMmkin}) which
takes negative values for a positive $Q_i(\usigma_2)$.

The gamma function has the asymptotic behaviour (Stirling's formula),
\beq
\Gamma(z)~\sim~
\sqrt{2\pi}\e^{-z}z^{z-\half}\left(1+O(1/z)\right),
\label{Stirling}
\eeq
as $|z|\to\infty$ with $\Arg(z)\in (-\pi,\pi)$. We also know that
\beq
\sinh(z)~\sim~\pm\half\e^{\pm z}
\label{sinh}
\eeq
as ${\rm Re}(z)\to\pm\infty$.
These allow us to find the asymptotic behaviour of the integrand
in a generic direction in the $\usigma'$-space:
The term of Chan-Paton weight $q$ of the integrand (\ref{resultagain})
behaves as
\beqa
{\rm integrand}_q&=&
{\rm const}\cdot\prod_{\alpha>0}\alpha(\usigma)\sinh(\pi\alpha(\usigma))
\prod_i\Gamma\!\left(\im Q_i(\usigma')+{R_i\over 2}\right)\,
\e^{\im t_R(\usigma')+2\pi q(\usigma')}
\nn\\
&\sim&{\rm const}\cdot
\prod_{\alpha>0}\alpha(\usigma')\prod_iQ_i(\usigma')^{R_i-\half}
\cdot \exp\left(-2\pi \im r\wt{W}_{\!{\it eff},q}(\usigma)\right)
\eeqa
with
\beqa
2\pi\wt{W}_{\!{\it eff},q}(\usigma)
&=&\sum_{\alpha>0}\pm\pi\im \alpha(\usigma)
-\sum_iQ_i(\usigma)
\left(\log\left({Q_i(\usigma)\over-\im \Lambda}\right)-1\right)\nn\\
&&-t(\usigma)+2\pi\im q(\usigma).
\label{Weff}
\eeqa
The above is valid when $|{\rm Re}(\alpha(\usigma'))|\gg 1$ for all $\alpha$,
$|Q_i(\usigma')|\gg 1$ for all $i$, and $Q_i(\usigma')$
are not on the positive imaginary axis.
The sign $\pm\pi \im \alpha(\usigma)$ is chosen
when $\pm{\rm Re}(\alpha(\usigma'))$ is positive.
The imaginary part of the logarithm
is defined to have values in the open interval $(-\pi,\pi)$.

The function $\wt{W}_{\!{\it eff},q}(\usigma)$
is equal to the effective twisted superpotential on the Coulomb branch.
We see the well-known $-\sigma(\log\sigma-1)$ from the
1-loop integral of the matter multiplet.
The term $\pm\pi\im \alpha(\usigma)$ may be less familiar, but it comes from
the 1-loop integral of the W-boson multiplet. See 
\cite{2dduality} for the explanation based on \cite{WVerlinde}.
In the bulk, or in the closed string sector,
the shift of $\wt{W}_{\!\it eff}(\sigma)$ by
$2\pi \im w(\sigma)$ does not matter for any weight $w$ of the maximal
torus $T$ 
since it is just a $2\pi$ shift of the theta angle.
In the presence of boundary,
on the other hand, the shift does matter, since the $2\pi$ shift of the
theta angle amounts to the shift of Chan-Paton weight.
The last term $2\pi\im q(\usigma)$ is nothing but the
contribution from the classical Chan-Paton factor.
Also, the precise choice of the sign $\pm\pi\im\alpha(\usigma)$
and the imaginary part of the logarithm, which is irrelevant in the bulk,
does matter here.

The imaginary part,
$E_{{\it eff},q}(\usigma)=-{\rm Im}(\wt{W}_{\!{\it eff},q}(\usigma))$,
may be interpreted as the effective boundary potential.
For the evaluation, we use the formula
\beq
\Arg(iz)={\rm sgn}\left({\rm Re}(z)\right)
\left({\pi\over 2}+\arctan\left[{{\rm Im}(z)\over |{\rm Re}(z)|}\right]
\right),
\label{Argid}
\eeq
which holds if we assume that $\Arg(-)$ and
$\arctan(-)$
take values in the intervals $(-\pi,\pi)$
and $(-{\pi\over 2},{\pi\over 2})$ respectively ---
we assume this in what follows too. We find
\beqa
E_{{\it eff},q}(\usigma)\!\!\!&=&\!\!\!
-\half\sum_{\alpha>0}|\alpha(\usigma_1)|\nn\\
&&\!\!\!\!\!+{1\over 2\pi}\sum_i\left\{
Q_i(\usigma_2)\left(\log\Bigl|{Q_i(\usigma)\over\Lambda}\Bigr|-1\right)
+|Q_i(\usigma_1)|\left({\pi\over 2}+\arctan\left[{Q_i(\usigma_2)\over
|Q_i(\usigma_1)|}\right]\right)\right\}\nn\\
&&\!\!\!\!\!
+{1\over 2\pi}\zeta(\usigma_2)-\left({\theta\over 2\pi}+q\right)(\usigma_1).
\label{Eeff}
\eeqa
The second line is nothing but the 
effective boundary energy of the matter system,
which was obtained in \cite{HHP}
by computing the energy density of the ground state
of the canonically quantized matter system on an interval or a half line
with the same boundary condition as (\ref{BCchiral0}),
in which we set $(\sigma_1,\sigma_2)$ to be a constant $(\usigma_1,\usigma_2)$.
See Section 6, Eqn (6.79) of \cite{HHP}.
The first line is regarded as
the boundary energy of the W-boson multiplet.
It would be interesting to check it directly by a computation like \cite{HHP}.
The third line is already there in the classical action as the 
classical boundary potential, see (\ref{LSMFItheta}) and (\ref{calA}).

This $E_{\it eff}(\usigma)$ shows the asymptotic growth or decay
of the integrand. In order to isolate the dependence on
the size $r$ of the hemisphere, 
it is more convenient to use the $\usigma'$ variables. We have
\beq
\Bigl|\,\mbox{integrand}_q\,\Bigr|~\sim~
P(\usigma')\cdot\exp\Bigl(-A_q(\usigma')\Bigr),
\label{behav}
\eeq
where $P(\usigma')$ is a power of $\usigma'$ and
\beqa
A_q(\usigma')&=&-\,\sum_{\alpha>0}\pi|\alpha(\usigma_1')|\nn\\
&&+\sum_i\left\{
Q_i(\usigma_2')\Bigl(\log|Q_i(\usigma')|-1\Bigr)
+|Q_i(\usigma_1')|\left({\pi\over 2}+\arctan\left[{Q_i(\usigma'_2)\over
|Q_i(\usigma'_1)|}\right]\right)\right\}\nn\\
&&+\,\zeta_R(\usigma'_2)-(\theta+2\pi q)(\usigma_1').
\label{Aq}
\eeqa
The $r$ dependence is only in the renormalized FI parameter,
\beq
\zeta_R=\zeta-\sum_iQ_i\log(r\Lambda).
\label{renFI}
\eeq
Notice that $A_q(\usigma')$ is essentially piecewise linear
at infinity in the $\usigma'$ space. 
One can also see from (\ref{Weff}) that the oscilation 
part, i.e. the imaginary part of the exponent, is also essentially 
piecewise linear in $\usigma'$.
For the absolute convergence of the integral,
we need to choose the Lagrangian $\gamma\subset\ttt_{\C}$ so that
$A_q(\usigma')$ grows at infinity of $\gamma$.
One may also allow $A_q(\usigma')$ to approach a constant at infinity,
hoping for the conditional convergence due to rapid oscillation.
However, the linear growth of the imaginary part makes the case very subtle
(see for example \cite{HPT}).
This motivates us to make the following proposal:

\fbox{
\begin{minipage}{14.35cm}

\medskip
{\it The Lagrangian submanifold $\gamma$ is a
deformation of the real locus $\im \ttt$, avoiding the poles
(\ref{poles}), so that for any Chan-Paton weight $q$ of the brane,
$A_q(\usigma')$ in (\ref{Aq}) grows to infinity in every asymptotic
direction of $\gamma$.}

\medskip
\end{minipage}
}

\noindent
We shall refer to the asymptotic region in which $A_q(\usigma')$
grows to infinity as the {\it admissible region}.
Thus, $\gamma$ is obtained
from the real locus $\im\ttt$ by ``bending'' the infinity, if necessary,
so that every asymptotic direction
is in the admissible region, and we require that the poles (\ref{poles})
are not hit in the bending process. We shall also
refer to the Lagrangian $\gamma$ satisfying this condition as
{\it admissible}.

The main question is whether there exists an admissible Lagrangian submanifold
$\gamma$, and if so, whether it is unique up to deformation.
In the rest of this section, we shall examine this question in several
examples, and at the same time identify the deformation class of admissible
Lagrangian submanifolds, when that is possible. 
In particular, we will find that, at some special loci in
the parameter space, called windows between phase boundaries,
it is not always possible to find an admissible
Lagrangian submanifold for an arbitrary brane $\mathfrak{B}$.
The factor $\e^{2\pi q(\usigma')}$ is exponentially growing in a certain
direction of the $\usigma'$ space, and the other factors cannot rule this
divergence for any choice of $\gamma$, if the parameter is
on a window.
This means that there is a severe constraint,
depending on the window, on the possible range of
Chan-Paton weight $q$ of the brane $\mathfrak{B}$.

The problem of identifying a Lagrangian submanifold $\gamma$
for the boundary condition on the vector multiplet was studied in
\cite{HHP} for the Abelian and Calabi-Yau cases,
and essentially the same condition on $\gamma$ was obtained:
The condition of adimissible asymptotic direction of $\gamma$
matches because,
as we have just seen, the effective boundary potential in that case
is precisely equal to $E_{{\it eff},q}(\usigma)$.
Also, there is a singularity along the entire positive imaginary axis
of $Q_i(\usigma)$ due to the zero mode of $\phi_i$ localized near the
boundary. Avoiding that is the counterpart of avoiding the poles
(\ref{poles}) at discrete points along
the same axis, which are associated with the zero mode of $\phi_i$
on the hemi-sphere.
And in \cite{HHP}, a constraint 
on the possible range of Chan-Paton weight $q$ on windows
was obtained and was named the {\it grade restriction rule} for
D-brane transport across windows.

In Abelain and Calabi-Yau cases, we will
indeed reproduce the grade restriction rule. This is of course of no
surprize regarding what we have just said,
but the convergence of the integral
(\ref{resultagain}) provides a somewhat sharper constraint.
We shall also discuss the non Calabi-Yau and/or non-Abelian theories
as well.

The integrals of the type (\ref{resultagain}) are known as the (multiple)
Mellin-Barnes integrals and have been a subject of mathematical study,
from the old time to more recent days, especially after the discovery
\cite{Candelasetal} of the importance of mirror symmetry.
The present discussion shows that
the issue of convergence, or the problem of identifying convergent domains,
of such integrals  encodes a rich physical content.

\subsection{$U(1)$ Theories}\label{subsec:U1}

We first consider the theories with gauge group $G=U(1)$.
As the basic class of examples, we consider
the theory with matter fields $P$, $X_1,\ldots, X_N$
of charge $-d$, $1,\ldots, 1$, and with superpotential
$W=Pf(X_1,\ldots, X_N)$ where $f$ is a homogeneous polynomial
of degree $d$. We assume that $f$ is generic so that the 
projective hypersurface $X_f=(f=0)\subset \CP^{N-1}$ is smooth.
The R-charge assignment is unique up to the gauge shift,
$2-d\epsilon$ to $P$ and $\epsilon$ to
$X_1,\ldots, X_N$. The bound (\ref{narrowrange})
is ensured by $0<\epsilon<2/d$. The R$^o$-frame is obtained by
the limit $\epsilon\searrow 0$.
Before attacking the problem to identify the admissible contour $\gamma$,
we recall some basic facts on the bulk theory \cite{Wphases}.

The nature of the classical theory depends very much on the sign of the
FI parameter $\zeta$, which enters into the D-term equation,
\beq
\sum_{i=1}^N|x_i|^2-d|p|^2=\zeta.
\eeq
When $\zeta$ is positive, this requires some $x_i$ to have a non-zero
value which breaks the $U(1)$ gauge group completely.
When $\zeta$ is negative, this requires $p$ to have a non-zero
value which breaks the $U(1)$ gauge group 
to the cyclic subgroup $\Z_d$ or order $d$.
When $\zeta$ is zero, there is a locus $x=p=0$ in which the
$U(1)$ is unbroken. The classical low energy theory is
 the non-linear sigma model with the target $X_f$ for $\zeta\gg 0$
and the $\Z_d$ orbifold of the Landau-Ginzburg model
 with superpotential $W=f(X_1,\ldots,X_N)$ for $\zeta\ll 0$.
The theory is said to be in
the geometric phase and the Landau-Ginzburg orbifold phase respectively,
 when $\zeta\gg 0$ and $\zeta\ll 0$.

The nature of the quantum theory depends very much on the sign of
$(N-d)$, since the FI parameter runs as $\zeta_R=\zeta-(N-d)\log(r\Lambda)$.
When $d<N$ ({\it resp}. $d>N$), it runs from positive to negative
({\it resp}. negtaive to positive) as the distance scale $r$ is increased.
There are also massive vacua on the Coulomb branch where $\sigma$ is non-zero:
They are found by solving
\beq
t_{\it eff}(\sigma):=-2\pi\partial_\sigma\wt{W}_{\!{\it eff}}\equiv 0\quad\,
\mbox{mod $2\pi\im \Z$},
\label{teff}
\eeq
where $\wt{W}_{\!{\it eff}}$ is the effective twisted superpotential, which
is equal to (\ref{Weff}) but omitting $q$ since it does not matter for
this purpose.
This equation takes the form
$\sigma^{N-d}=(-d)^d\e^{-\zeta}(-\im\Lambda)^{N-d}:=
(-1)^d(-\im \wt{\Lambda})^{N-d}$ and hence has $|N-d|$ solutions.
When $d=N$ (Calabi-Yau case), the FI parameter does not run,
and we have a family of theories parametrized by
$t=\zeta-\im \theta\in \C/2\pi\im \Z$.
In this case (\ref{teff}) is a constant,
$t_{\it eff}\equiv t-N\log(-N)$. The theory is singular at
$t_{\it eff}\equiv 0$ due to the presence of non-compact Coulomb branch.
This point, $\zeta=N\log N$ and $\theta\equiv \pi N$,
is the quantum remnant of the phase boundary which `separates'
the $\zeta\gg 0$ geometric phase and the $\zeta\ll 0$ Landau-Ginzburg
orbifold phase, although it does not really separate the two regimes
since we can go around a complex codimension one locus.

\subsubsection{Calabi-Yau Case}\label{subsub:CY}

Let us first consider the Calabi-Yau case, $d=N$.
In this case, the expression for $A_q$ is very simple
\beq
A_q(\usigma')=\zeta_{\it eff}\usigma'_2
+\Bigl(N\pi-{\rm sgn}(\usigma'_1)(\theta+2\pi q)\Bigr)|\usigma'_1|,
\eeq
where $\zeta_{\it eff}=\zeta-N\log N$. When $\zeta_{\it eff}>0$
({\it resp}. $\zeta_{\it eff}<0$), 
the entire region of the $\usigma'$-plane above ({\it resp}. below)
the broken line $A_q(\usigma')=0$ is admissible.
\begin{figure}[thb]
\psfrag{pos}{\footnotesize $\zeta_{\it eff}>0$}
\psfrag{neg}{\footnotesize $\zeta_{\it eff}<0$}
\centerline{\includegraphics{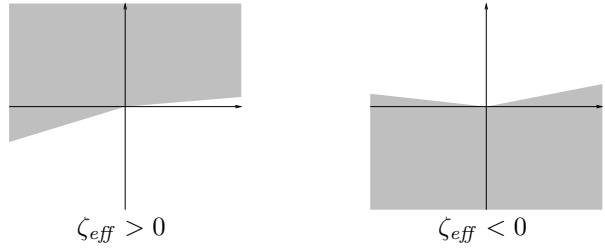}}
\caption{Admissible regions (Calabi-Yau case)}
\label{fig:wedgeCY}
\end{figure}
Figure~\ref{fig:wedgeCY}
depicts the $\usigma'$-planes for two values of
$(\zeta,\theta+2\pi q)$, where
the admissible regions are shaded. 
For any finite $q$, a sector of positive angle including the
positive ({\it resp}. negative) imaginary axis is inside the admissible region.
Therefore, 
\begin{figure}[htb]
\psfrag{pos}{\footnotesize $\zeta_{\it eff}>0$}
\psfrag{neg}{\footnotesize $\zeta_{\it eff}<0$}
\psfrag{gamma}{$\gamma$}
\centerline{\includegraphics{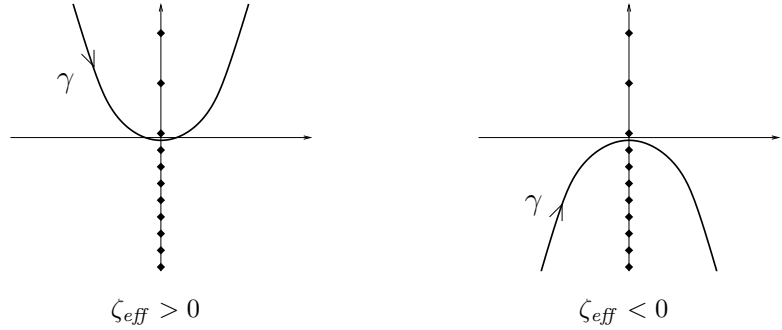}}
\caption{Admissible contours (Calabi-Yau case)}
\label{fig:contour1}
\end{figure}
we can take $\gamma$ to be the curve
obtained by bending the real line $\R$
toward the positive  ({\it resp}. negative) imaginary direction,
like a graph of a function which grows or decays faster than
a linear function.
See Figure~\ref{fig:contour1}-Left
({\it resp}. -Right),
where the poles (\ref{poles}) are also drawn together as dots.
Then, the integral (\ref{resultagain}) is convergent
for any brane whose Chan-Paton charge $q$ ranges over an {\it arbitary}
but finite set of integers.

When $\zeta_{\it eff}=0$, the situation is very different.
If $\theta+2\pi q\geq N\pi$, the entire right half plane is not admissible.
There is no way to move the right end of the real line
to the left half plane without hitting the poles.
Therefore, an admissible contour does not exist.
Similarly for the case $\theta+2\pi q\leq -N\pi$
where the entire left half plane is not admissible.
If $-N\pi<\theta+2\pi q<N\pi$, the entire directions is admissible
except infinitesimally small sectors including the imaginary axis. 
Therefore, the real line $\gamma=\R$ itself is admissible, as well as any of
its deformation that keeps a non-zero angle against the imaginary axis.
Thus, we have a strong constraint on the brane
$\mathfrak{B}=(M,Q,\rho,{\bf r}_*)$:

\fbox{
\begin{minipage}{14.5cm}

\vspace{0.1cm}
~{\it At $\zeta_{\it eff}=0$,
all the Chan-Paton charges $q$ of $\mathfrak{B}$ must be in the range}
\beqa
&\displaystyle -{N\over 2}~<~{\theta\over 2\pi}+q~<~{N\over 2}.
\label{GRR1}\\
&&\nn
\eeqa

\end{minipage}
}

\noindent
The allowed charges form a set of $N$ consecutive integers
provided $\theta$ avoids $N\pi+2\pi \Z$, which are singular values for
$\theta$ at $\zeta_{\it eff}=0$.
This set does not change if $\theta$ moves inside
an open interval, or a window, of length $2\pi$ of regular values.
If a brane $\mathfrak{B}$ obeys the condition (\ref{GRR1})
for $\theta$ in such an interval, we shall call it
{\it grade restricted with respect to the window}.

This is strange. If $\zeta_{\it eff}$ is positive or negative,
an arbitrary brane has an admissible Lagrangian submanifold $\gamma$
for the boundary condition on the vector multiplet.
At $\zeta_{\it eff}=0$, that is possible 
only for grade restricted branes 
which form a tiny subset of the set of all branes 
in the linear sigma model.

\begin{figure}[htb]
\psfrag{q}{\Large{\bf ?}}
\psfrag{pos}{\footnotesize $\zeta_{\it eff}>0$}
\psfrag{neg}{\footnotesize $\zeta_{\it eff}<0$}
\psfrag{lpos}{\footnotesize $\zeta_{\it eff}\gg 0$}
\psfrag{lneg}{\footnotesize $\zeta_{\it eff}\ll 0$}
\psfrag{zero}{\footnotesize $\zeta_{\it eff}=0$}
\centerline{\includegraphics{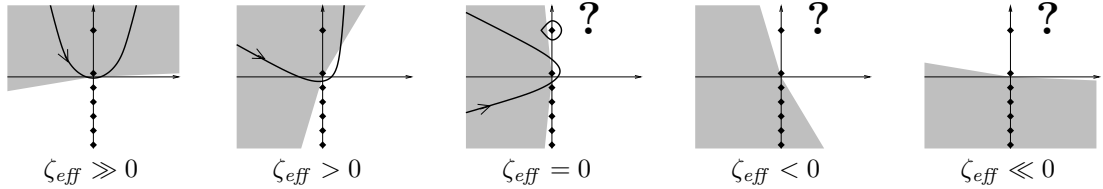}}
\caption{Not Grade Restricted Case}
\label{fig:vpos}
\end{figure}

To illustrate the problem,
let us see what happens to the partition function for
 a brane $\mathfrak{B}=(M,Q,{\bf r}_*,\rho)$ if we move
the parameter $t=\zeta-\im \theta$ along a path
from one phase to another, say
from the geometric phase $\zeta\gg 0$ to 
the Landau-Ginzburg orbifold phase $\zeta\ll 0$.
The path must avoid the singularity
$t\equiv N\log N+N\pi\im$ and hence must go through one of the
windows at $\zeta=N\log N$.
First, let us consider the case where the brane $\mathfrak{B}$ 
is grade restricted with respect to that window.
\begin{figure}[htb]
\psfrag{pos}{\footnotesize $\zeta_{\it eff}>0$}
\psfrag{neg}{\footnotesize $\zeta_{\it eff}<0$}
\psfrag{lpos}{\footnotesize $\zeta_{\it eff}\gg 0$}
\psfrag{lneg}{\footnotesize $\zeta_{\it eff}\ll 0$}
\psfrag{zero}{\footnotesize $\zeta_{\it eff}=0$}
\centerline{\includegraphics{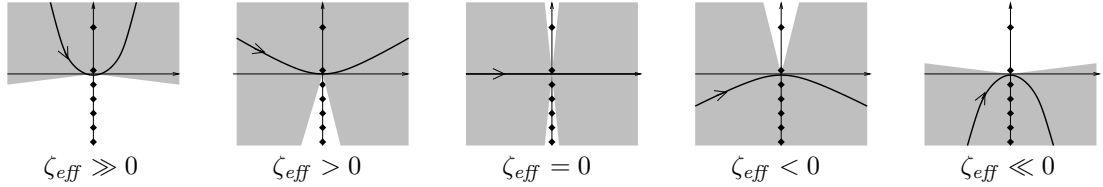}}
\caption{Grade Restricted Case}
\label{fig:vgrr}
\end{figure}
The move of the admissible region for any Chan-Paton charge $q$ of the brane
is shown in Fig.~\ref{fig:vgrr}.
We see that one can find a continuous family of
admissible contours as depicted in the same Figure.
Therefore, the partition function for the brane $\mathfrak{B}$
in the phase $\zeta\gg 0$
is related to the one for the same brane $\mathfrak{B}$
in the phase $\zeta\ll 0$ 
by anlytic continuation along the path.
Let us next consider the case where the brane $\mathfrak{B}$
is not grade restricted with respect to the window.
Then, it includes a Chan-Paton charge $q$ which is outside, say above,
 the set (\ref{GRR1}).
The move of the admissible region for such a charge
is depicted in Figure~\ref{fig:vpos}.
We see that, before $\zeta_{\it eff}$ approaches $0$,
an admissible contour is forced to hit the singularity
along the positive imaginary axis. 
The integral must pick these infinitely many 
poles when $\zeta_{\it eff}$ goes negative, but the convergence of
the infinite sum is not obvious at all.
So, we do not know what happens to the partition function
if we try to see it this way.
The same problem arizes if there is a charge $q$ below the set (\ref{GRR1}).

In a sense, only grade resticted branes can cross the window safely.
This is how the grade restriction rule was stated in \cite{HHP}.
What does this mean?
Is there a real phase boundary between the geometric phase and
the Landau-Ginzburg orbifold phase across which some of the branes
cannot cross? That would be strange since the points with
$\zeta_{\it eff}=0$, $\theta\not\equiv \pi N$
have no special status compared to other points in the parameter space.
The answer to this problem, given in \cite{HHP}, is that
in either phase, there is a huge equivalence relation
among the branes, and each equivalence class has a
grade restricted representative.
This point, which we shall call the
``classical grade restriction rule'',
will be revisited in the next section, where we show that the
partition function takes the same value on branes in the same
equivalence class. This will give a solution to
the above problem of analytic continuation of the partition function
for a brane which is not grade restricted with respect to the window.

In this paper, we shall call the constraint (\ref{GRR1}) itself
the grade restriction rule.

\subsubsection{Non Calabi-Yau Case}

In the non-Calabi-Yau case, $d\ne N$, the function $A_q$ can be written as
\beqa
A_q(\usigma')&=&
(N-d)\left(\log\left|{\usigma'\over r\wt{\Lambda}}\right|-1\,\right)
\usigma'_2\label{AqFano}\\
&&\!\!\!\!\!\!+\left({\pi\over 2}(N+d)
+(N-d)\arctan\left[{\usigma_2'\over |\usigma'_1|}\right]
-{\rm sgn}(\usigma'_1)(\theta+2\pi q)\right)|\usigma'_1|.\nn
\eeqa
We see that the coefficient of $\usigma'_2$ changes its sign at the cricle
$|\usigma'|=r\wt{\Lambda}\e=r\wt{\Lambda}\times 2.1718...$ ---
it is postive outside the circle and negative inside when $d<N$
and the other way around when $d>N$.
Also, the coefficient of $|\usigma'_1|$
is positive on the real axis if $|\theta+2\pi q|<{\pi\over 2}(N+d)$.
\begin{figure}[htb]
\psfrag{pos}{\footnotesize $|\theta+2\pi q|<\pi d$}
\psfrag{neg}{\footnotesize $\pi d<|\theta+2\pi q|<\pi{N+d\over 2}$}
\psfrag{abc}{\footnotesize $\!\!\pi{N+d\over 2}<|\theta+2\pi q|<\pi N$}
\psfrag{rL}{\footnotesize $r\wt{\Lambda}\e$}
\centerline{\includegraphics{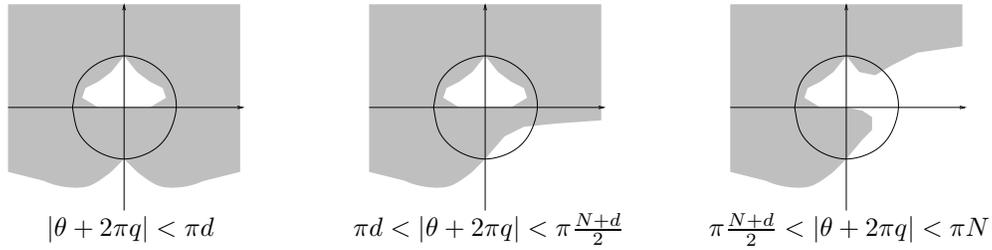}}
\caption{Regions with positive $A_q$ (the case $d<N$).}
\label{fig:Fano}
\end{figure}
Fig~\ref{fig:Fano} shows the $\usigma'$ planes for three values
of $(\zeta,\theta+2\pi q)$ for the case $d<N$.
We shade the region with positive $A_q$ and draw the circle at
$|\usigma'|=r\wt{\Lambda}\e$. 
We assume $r\Lambda\gg 1$ so that (\ref{behav}) is 
a good approximation at the scale $\usigma'\sim r\wt{\Lambda}$. 
For any value of $(\zeta,\theta+2\pi q)$,
the function $A_q$ grows at least linearly in any ray direction 
on the upper half plane. 
\begin{figure}[htb]
\psfrag{pos}{\footnotesize $~~d<N$}
\psfrag{neg}{\footnotesize $~\,d>N$}
\psfrag{gamma}{$\gamma$}
\centerline{\includegraphics{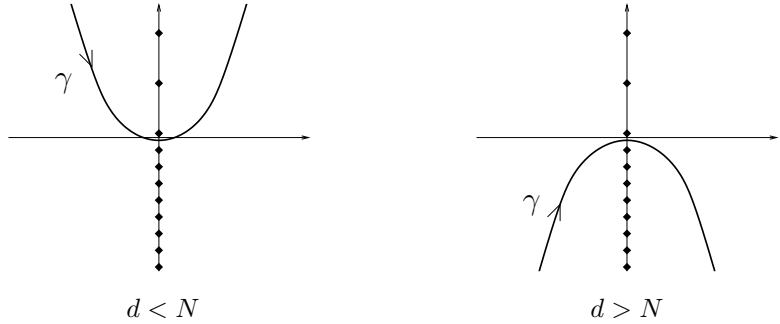}}
\caption{Admissible contours (Non Calabi-Yau cases)}
\label{fig:contournonCY}
\end{figure}
Therefore, for any brane, the contour $\gamma$ can be taken to be a curve,
 as in Fig.~\ref{fig:contournonCY}-Left,
which comes in from and goes out to the region where
${\rm Im}(\usigma')$ is positive infinity. 
This is as in the $\zeta_{\it eff}>0$ phase of the Calabi-Yau case, which
may be understood by the fact that the effective FI parameter
$\zeta_{\it eff}(\usigma)=(N-d)\log|\usigma/\wt{\Lambda}|$
goes to positive infinity as $|\usigma'|\to \infty$.
For the case $d>N$, the picture for the $A_q>0$ region is upside down
compared to Fig.~\ref{fig:Fano},
with $d$ and $N$ exchanged in the subtitles.
Also, the effective FI-parameter
$\zeta_{\it eff}(\usigma)$ goes to negative infinity as $|\usigma'|\to \infty$.
Thus, the contour $\gamma$ for any brane can be taken as in 
Fig.~\ref{fig:contournonCY}-Right,
coming in from and going out to the region where
${\rm Im}(\usigma')$ is negative infinity.

\subsubsection{More General Theories}

What is said on this particular class of examples applies more generally.
Let us consider the $U(1)$ theory with fields $X_i$ of 
R- and gauge charge $(R_i,Q_i)$ and with some superpotential $W$.
We assume that each $Q_i$ is non-zero.
We put $N_{\pm}:=\sum_{\pm Q_i>0}|Q_i|$.
In the Calabi-Yau case $N_+=N_-$, we have a family of theories parameterized
by $t$, with the singularity at $t\equiv -\sum_iQ_i\log Q_i$.
The contour $\gamma$ can be chosen as in Fig.~\ref{fig:contour1} if
$\zeta_{\it eff}=\zeta-\sum_iQ_i\log |Q_i|$ is non-zero. 
At $\zeta_{\it eff}=0$,
the brane must obey the grade restrcition rule (\ref{GRR1}),
with $N$ replaced by $N_+=N_-$.
In the case $N_+>N_-$ ({\it resp}. $N_+<N_-$),
the FI parameter runs from positive to negative ({\it resp}. negtaive to
positive) and there are $|N_+-N_-|$ massive vacua on the Coulomb branch.
The contour can be chosen as in Fig.~\ref{fig:contournonCY}-Left
({\it resp}. -Right).

\subsection{Higher Rank Abelian Theories}\label{subsec:higher}

\newcommand{\Rs}{\utau}
\newcommand{\Is}{\uups}
\newcommand{\rmP}{{\rm P}}
\newcommand{\HP}{\mathfrak{H}}

In the rest of this paper, except when we discuss $U(1)$ theories,
we write the real and imaginary parts of 
$\usigma$ as
\beq
\Rs={\rm Re}(\usigma),\quad
\Is={\rm Im}(\usigma),
\label{newn}
\eeq
instead of $\usigma_1$ and $\usigma_2$,
in order to avoid confusion between the index of coordinates on $\im\ttt$
and the $(1,2)$ for the (real, imaginary) part.

In this subsection, we consider theories with Abelian gauge group $G$.
For simplicity we take it to be a connected group so that $G=T$.
We write  $k:=d_G=l_G$.
The FI parameter $\zeta$ takes values in $\im \ttt^*$.
We consider a theory
with a matter chiral multiplet $\phi=(\phi_i)_{i\in I}$ of R-charge
$(R_i)_{i\in I}$ and gauge charge
$(Q_i)_{i\in I}$. We assume some superpotential $W$ of R-charge $2$.
We shall only consider the Calabi-Yau case, $\sum_iQ_i=0$,
 so that we decide not to distinguish $\usigma'$
from $\usigma$.

As in the $U(1)$ theory, the space of FI parameters is divided into phases.
For $\zeta$ in a phase, any solution to the D-term equation
\beq
\sum_{i\in I}Q_i|\phi_i|^2=\zeta
\label{Deqn}
\eeq
breaks the gauge group to a finite subgroup.
An interface between two phases, called a phase boundary,
is a positive linear span of $(k-1)$ independent charges
from $\{Q_i\}_{i\in I}$. For $\zeta$ in such a phase boundary,
there is a solution to (\ref{Deqn}) which breaks the gauge group
to a subgroup of rank one whose Lie algebra is the common kernel
of the $(k-1)$ charges.
The rank of the possibly unbroken subgroup will be higher for
intersection of phase boundaries.
The quantum theory is parametrized by the FI-theta parameter
$t=\zeta-\im\theta\in \ttt^*_{\C}/2\pi\im \rmP$ where
$\rmP\subset\im \ttt^*$ is the weight lattice of $T$.
The theory is singular at a hypersurface in which
$t_{\it eff}:=-2\pi \dd\wt{W}_{\!\it eff}\equiv 0$, i.e.
\beq
t+\sum_{i\in I}Q_i\log Q_i(\sigma)\equiv 0\qquad
\mbox{mod $2\pi\im \rmP$,}
\eeq
has a set of solutions, i.e. a non-compact Coulomb branch.
There can also be additional singularity from mixed Coulomb-Higgs branches.
The $\zeta$ images of the singular hypersurfaces
asymptote to the phase boundaries, and the Coulomb branch
approaches the one for the unbroken gauge group at
each of them.

\begin{figure}[htb]
\psfrag{i}{\footnotesize Phase I}
\psfrag{ii}{\footnotesize Phase II}
\psfrag{iii}{\footnotesize Phase III}
\psfrag{iv}{\footnotesize Phase IV}
\psfrag{12}{\footnotesize $X_{1,2}$}
\psfrag{345}{\footnotesize $X_{3,4,5}$}
\psfrag{6}{\footnotesize $X_6$}
\psfrag{p}{\footnotesize $P$}
\psfrag{z}{\footnotesize $\im\ttt^*$}
\centerline{\includegraphics{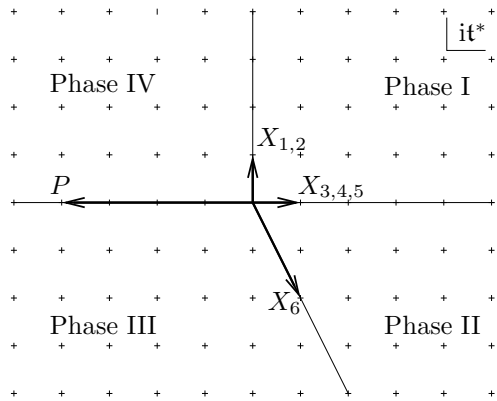}}
\caption{A two parameter model}
\label{fig:twopara}
\end{figure}
As an illustration, let us consider
the $U(1)\times U(1)$ linear sigma model
familiar to physicists \cite{MP}: Fields are $X_1,\ldots, X_6$ and $P$ whose
charges are as in Fig.~\ref{fig:twopara},
and with superpotential $W=Pf(X)$ where $f(X)$ is a polynomial of 
$X_1,\ldots, X_6$ of bidegree $(0,4)$.
The theory has four phases,
I, II, III and IV, which are respectively the geometric, orbifold,
Landau-Ginzbirg orbifold and hybrid phases.
The quantum theory is singular at the two curves
\beqa
C_1:&&\e^{-t^1}=4^{-4}(1-2u),\quad\e^{-t^2}={u^2\over (1-2u)^2},\\
C_2:&&\e^{-t^2}=2^{-2}.
\eeqa
The curve $C_1$ 
is associated to the pure Coulomb branch with $\sigma_2/\sigma_1=u$.
(Here, $\sigma_1$ and $\sigma_2$ are not the real and the
imaginary parts of $\sigma$, but the first and the second $U(1)$
components of $\sigma$. This is why we introduce the new notation 
(\ref{newn}) for the real and imaginary parts.) 
The limit points $u=0,\half,\infty$ correspond to
the I-IV, II-III, III-IV boundaries with the right unbroken gauge groups,
$(\sigma_1,\sigma_2)\in \C(1,0), \C(2,1), \C(0,1)$.
The curve $C_2$ is associated to a mixed Higgs-Coulomb branch
in which the second $U(1)$ is unbroken. It corresponds to the
I-II and III-IV boundaries.
For more detail of the relevant aspects of the theory, see \cite{MP,HHP}.

Assuming the Calabi-Yau condition,
the function (\ref{Aq}) can be written as
\beq 
A_q(\usigma)=\zeta_{\it eff}(\Is)-\theta_{{\it eff},q}(\Rs),
\label{AqCY}
\eeq
where $\zeta_{\it eff}$ and $\theta_{{\it eff},q}$ are defined by
$t_{{\it eff},q}:=-2\pi\dd\wt{W}_{\!{\it eff},q}$ using (\ref{Weff}).
Although we hide from the notation to avoid clatter,
$\zeta_{\it eff}$ and $\theta_{{\it eff},q}$ depend on
$\sigma$. In fact they depend only on
the direction $\wh{\sigma}=\sigma/|\!|\sigma|\!|$.
If $t$ is on the singular hypersurface, they `vanish', i.e.,
$\zeta_{\it eff}=0$ and $\theta_{\it eff}\equiv 0$
(mod $2\pi\im \rmP$), for $\wh{\sigma}$ in the Coulomb branch direction.
If $\zeta$ is deep inside a phase,
$\zeta_{\it eff}(\Is)$ is dominated by the classical part $\zeta(\Is)$
for any direction $\wh{\usigma}$,
and nearly the entire half space
$$
\Bigl\{\,\usigma=\Rs+\im\Is\,\Bigl|\,\zeta(\Is)>0\,\Bigr\}
\subset \ttt_{\C}
$$
is admissible. When $\zeta$ approaches
a phase boundary, the quantum correction becomes comparable to
the classical part and a careful analysis will be needed.

\begin{figure}[htb]
\psfrag{a}{\footnotesize A}
\psfrag{b}{\footnotesize B}
\psfrag{c}{\footnotesize C}
\psfrag{d}{\footnotesize D}
\psfrag{e}{\footnotesize E}
\psfrag{f}{\footnotesize F}
\psfrag{12}{\footnotesize $H_{1,2}$}
\psfrag{345}{\footnotesize $\!\!\!\!\!\!\!\!H_P=H_{3,4,5}$}
\psfrag{p}{}
\psfrag{6}{\footnotesize $H_{6}$}
\psfrag{z}{\footnotesize $\im\ttt$}
\centerline{\includegraphics{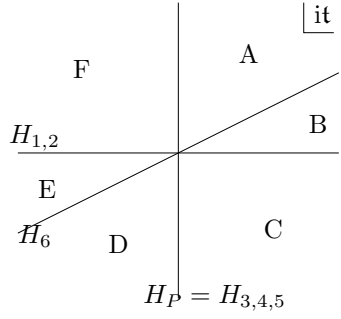}}
\caption{The chamber decomposition of $\im\ttt$
in the model of Fig.~\ref{fig:twopara}.}
\label{fig:ttt}
\end{figure}

Since the contour $\gamma$ is defined to be a deformation of the real locus,
$\Is=0$, it may be regarded as a graph of a map,
$\Rs\in\im\ttt\mapsto \Is(\Rs)\in\im \ttt$.
It must obey the condition that $A_q(\Rs+\im\Is(\Rs))$
grows to infinity as $|\Rs|\to\infty$ in any direction. 
Also, the deformation should avoid the poles at (\ref{poles}), that is,
 $Q_i(\Rs)=0$ and $Q_i(\Is)=n_i+{R_i\over 2}$ with $n_i=0,1,2,\ldots$ .
This condition is satisfied if the contour $\gamma$ avoids the wedge,
$Q_i(\Rs)=0$ and $Q_i(\Is)>0$, i.e., if we choose the map $\Is=\Is(\Rs)$
to avoid positive values of $Q_i(\Is)$ over the
hyperplane $Q_i(\Rs)=0$.
Let us introduce some notations.
Let $H_i\subset \im\ttt$ be the hyperplane annihilated by $Q_i$, i.e,
$H_i={\rm Ker}\ Q_i$, and let $D^{\pm}_i\subset \im \ttt$ be the half space
with positive values of $\pm Q_i$.
The hyperplanes $\{H_i\}$ define a chamber decomposition of $\im\ttt$.
See Fig.~\ref{fig:ttt} for an example.

As the first step, we look for a piecewise linear map
$\Rs\mapsto\Is(\Rs)$ satisfying the conditions,
which is linear on each chamber.
The wedge condition to avoid poles is
a condition on the values at the walls of the chambers,
$\Is(H_i)\subset \overline{D}_i^-$.

\begin{figure}[htb]
\psfrag{i}{\footnotesize $C_{\rm I}$}
\psfrag{ii}{\footnotesize $C_{\rm II}$}
\psfrag{iii}{\footnotesize $C_{\rm III}$}
\psfrag{iii'}{\footnotesize $C_{\rm III'}$}
\psfrag{iv}{\footnotesize $C_{\rm IV}$}
\psfrag{z}{\footnotesize $\Is$}
\centerline{\includegraphics{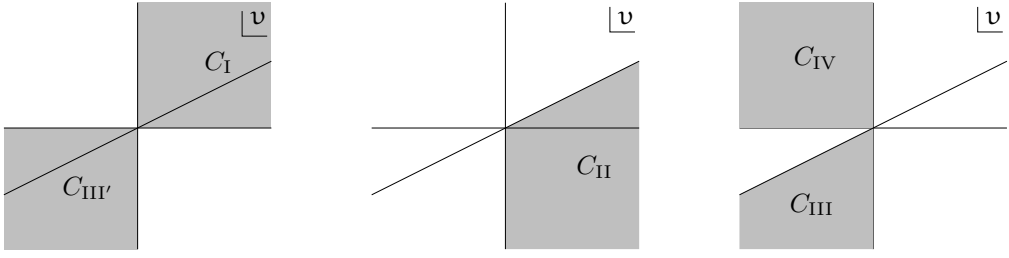}}
\caption{The image cones}
\label{fig:cones}
\end{figure}

If $\zeta$ is deep inside a phase, the growth condition
is satisfied if the image is deep inside the $\zeta$-positive
half space $D^+_{\zeta}$ and if it is of full rank on each chamber.
In particular, the image is a cone of full dimension inside the
half space $D^+_{\zeta}$. We shall call it the image cone of the map.
Let us show examples of such maps in the two parameter model:
\beqa
\zeta\in\mbox{Phase I}:&(\Is_1,\Is_2)=(|\Rs_1|,|\Rs_2|),\label{ph1}\\
\zeta\in\mbox{Phase II}:&(\Is_1,\Is_1-2\Is_2)=(|\Rs_1|,|\Rs_1-2\Rs_2|),
\label{ph2}\\
\zeta\in\mbox{Phase III}:&(\Is_1,\Is_1-2\Is_2)=(-|\Rs_1|,|\Rs_1-2\Rs_2|),
\label{ph3}\\
\zeta\in\mbox{Phase IV}:&(\Is_1,\Is_2)=(-|\Rs_1|,|\Rs_2|),
\label{ph4}
\eeqa
The choice may not be unique. For example, if $\zeta$ is in
the subset $\zeta^1<0$, $\zeta^2<0$ of Phase III, we may also take
$(\Is_1,\Is_2)=(-|\Rs_1|,-|\Rs_2|)$.
However, the two can be continuously connected to each other by a 
homotopy which stays inside the admissible region.
That is, they are in the same deformation class.
In Fig.~\ref{fig:cones}, we show the image cones of these maps.
$C_{\rm I},\ldots, C_{\rm IV}$ are the image cones of
(\ref{ph1}), ..., (\ref{ph4}), and $C_{\rm III'}$
is the one for the other map on a part of Phase III.
%We see that the image cones is always a chamber or a union of chambers.
%At this moment, we do not know if that must be the case.
We may try to generalize the examples (\ref{ph1})-(\ref{ph4}).
Suppose that $\zeta$ is a positive linear span of a set $\{Q_j\}_{j\in J}$
of $k$ charges, which must be linearly independent if $\zeta$ is
deep inside a phase.
Then, define $\Is(\Rs)$ by
\beq
Q_j(\Is(\Rs))=|Q_j(\Rs)|,\quad \,\,\forall j\in J.
\label{daijoubu}
\eeq
It certainly satisfies the growth condition, but the question is the wedge
condition to avoid poles. The latter is always satisfied when $k=2$
and also in many other examples with higher $k$.
However, it is easy to find counter examples with $k=3$.

The graph of such a piecewise linear map is not always Lagrangian.
For example, the maps (\ref{ph1}) and (\ref{ph4})
already define (piecewise) Lagrangian submanifolds
with respect to
$\omega=\dd\Rs_1\wedge \dd\Is_1+\dd\Rs_2\wedge \dd\Is_2$,
but the maps (\ref{ph2}) and (\ref{ph2}) do not.
In fact we may modify the maps as
\beq
(\Is_1,\Is_1-2\Is_2)=(\pm f(\Rs)|\Rs_1|,g(\Rs)|\Rs_1-2\Rs_2|)
\label{asin}
\eeq
for some positive valued functions $f(\Rs)$ and $g(\Rs)$.
It is straightforward though technically involved to find such functions
so that the graph is a Lagrangian.
Such a modification is also useful even if the graph is already a Lagrangian.
For a piecewise linear map,
no matter how deep inside $\zeta$ is, if we consider a very large Chan-Paton
charge $q$, the growth condition can be violated.
However, the graph can be `bent' by multiplying positive functions
to the maps, as in (\ref{asin}). For example, in Phase I, the map (\ref{ph1})
can be modified to $(\Is_1,\Is_2)=(|\Rs_1|^{1+\epsilon},|\Rs_2|^{1+\epsilon})$
for some positive $\epsilon$, say $1$. Then, the growth condition is satisfied
for any brane $\mathfrak{B}$ with an arbitrary set of Chan-Paton charges.

At this moment, we do not have a general proof of the existence and 
uniqueness of the deformation class of a map $\Rs\mapsto \Is(\Rs)$
satisfying the conditions. 
We leave it as a problem for a future work.

Let us now consider the region where $\zeta$ is not deep inside a phase.
Since the analysis is very complicated in general,
in this paper, we focus on the region near an 
``asymptotic phase boundary'', that is,
deep in the interior of the boundary between two phases.
Take a phase boundary spanned by
$(k-1)$ charges $\{Q_i\}_{i\in I_b}$.
We denote by $T^u$ the unbroken subgroup at the boundary
and take its integral generator $e^u\in \im \ttt$.
We write $\xi(e^u)=\xi^u$ for $\xi\in\ttt^*_{\C}$.
Since $e^u$ is the common kernel of $\{Q_i\}_{i\in I_b}$ we have
$Q^u_i=0$ for $i\in I_b$.
If we choose some element $e_u\in\im\ttt^*$ such that $e_u(e^u)=1$
we can write $t=\sum_{i\in I_b}Q_it^i+e_ut^u$.
We are looking at the regime where $\zeta^i\gg 0$ for all $i\in I_b$.
For any fixed $(t^i)_{i\in I_b}$ in that regime, we have an array
of singular `points' in the $t^u$-plane, separated by $2\pi\im$.
(Each `point' is in general a collection of a number of points
which are very close to each other.) In any limit with
$\zeta^i\to+\infty$, the singular `points' approach
the points
\beq
t^u=-\sum_{i\in I}Q_i^u\log Q_i^u+2\pi\im n,\quad n\in \Z.
\eeq
The line $\zeta^u=-\sum_{i\in I}Q_i^u\log |Q_i^u|$
is the asymptotic phase boundary, and the open intervals of length $2\pi$
between the adjacent singular points shall be called
the windows between the phases in the asymptotic regime.
In the two parameter model, the asymptotic singular points of
the four phase boundaries are
\beqa
\mbox{I-II}:&& t^2\equiv 2\log 2,\quad [1],\\
\mbox{II-III}:&& 2t^1+t^2\equiv 9\log 4,\quad [1],\\
\mbox{III-IV}:&& t^2\equiv 2\log 2.\quad [2],\\
\mbox{IV-I}:&& t^1\equiv 4\log 4,\quad [2].
\eeqa
The number in the bracket shows the number of points in 
the collection.

Let us examine the image of the map $\Rs\mapsto\Is(\Rs)$ over the line 
$\Rs\in \R e^u$ of the unbroken gauge group $T^u$.
This line is equal to the intersection of
the hyperplanes $H_i$ for $i\in I_b$. By the wedge
condition, we need
$Q_i(\Is(\Rs))\leq 0$ for $i\in I_b$. On the other hand, 
by the growth condition,  none of 
$Q_i(\Is(\Rs))$ with $i\in I_b$ cannot go large negative
since we are looking at the regime $\zeta^i\gg 0$ for all $i\in I_b$.
Therefore, $Q_i(\Is(\Rs))$ with $i\in I_b$ are frozen to be small on the line 
$\Rs\in \R\e^u$. 
This is the incarnation of the Higgs mechanism in which
$Q_i(\sigma)=0$ is enforced by the non-vanishing value of $\phi_i$
at a solution to the D-term equation with $\zeta^i\gg 0$.
As a consequence, on this line, $\Rs=\Rs_ue^u$,
the function $A_q(\Rs,\Is(\Rs))$  is dominated by
the one for the theory with the gauge group $T^u$ only, that is,
\beq
A_q=\left(\zeta^u+\sum_{i\in I}Q^u_i\log |Q^u_i|\right)\Is_u(\Rs)
+\left(\sum_{Q^u_i>0}Q^u_i\pi-{\rm sgn}(\Rs_u)
(\theta^u+2\pi q^u)\right)|\Rs_u|.
\eeq
If $\zeta^u$ is exactly on the asymptotic phase boundary,
where the first term vanishes, we obtain a constraint
\beq
-\half\sum_{Q^u_i>0}Q^u_i~<~{\theta^u\over 2\pi}+q^u~<~
\half \sum_{Q^u_i>0}Q^u_i.
\eeq
This is the grade restriction rule.
It is a constraint on the Chan-Paton charges with respect only to
 the unbroken gauge group $T^u$ at the phase boundary.
(It is called the {\it band restriction rule} in \cite{HHP}.)
The set of charges satisfying this condition depends only on the window,
as in the $U(1)$ theories.
If $\zeta^u$ is above ({\it resp}. below) the asymptotic phase boundary,
we may choose $\Is_u(\Rs)$ on the line $\Rs=\Rs_ue^u$
to be a function that goes to postive ({\it resp}.
negative) infinity faster than $|\Rs_u|$ ({\it resp}. $-|\Rs_u|$).
Then, at least along this line, the growth condition is satisfied
for any charge $q^u$.
In fact, this behaviour is consistent with the choice of contour $\gamma$
deep inside either of the two phases, provided the latter
is constructed based on the map (\ref{daijoubu}).
In the phase above the boundary, as the set $\{Q_j\}_{j\in J}$
we take $\{Q_i\}_{i\in I_b}\cup\{Q_{j_+}\}$ where $Q_{j_+}$
is one of the charges such that $Q_{j_+}^u>0$. Then, up to a positive
rescaling, we may assume $e_u=Q_{j_+}$. This means
that $\Is_u(\Rs)=|\Rs_u|$ on the line (before the further bending). 
In the phase below the boundary, we take $Q_{j_-}$ with $Q_{j_-}^u<0$
instead of $Q_{j_+}$, and we have
$\Is_u(\Rs)=-|\Rs_u|$ on the line (before the further bending).

\subsection{Non-Abelian Examples}

The linear sigma model with non-Abelian gauge groups
is a surprisingly rich subject of study.
One interesting feature is that there can be phases in which
a continuous subgroup of the gauge group is totally unbroken.
The low energy physics of such a strongly coupled system
is usually hard to understand. 
Exact results obtained in this paper may provide some clue towards
better understanding.
In this subsection, we describe some examples with geometric phase
where we can find admissible contours, as well as an example where
a simple grade restriction rule can be obtained.
Full exploration is beyond the scope of the present paper and will
be left for future works. 

The models treated are all Calabi-Yau and hence we write $\usigma$ for
$\usigma'$.
Also reminded is the notation (\ref{newn}) for the
real and imaginary parts of $\usigma$.

\subsubsection{R\o dland Model}

The first example is the R\o dland model \cite{Rodland,HoTo}.
It is a $U(2)$ gauge theory with seven fundamental
doublets, $X_1,\ldots, X_7$
and seven fields $P^1,\ldots, P^7$ in the $\det^{-1}$ representation.
The superpotential is of the form
$W=\sum_{i,j,k=1}^7A^{ij}_kP^k[X_iX_j]$ where
$[X_iX_j]$ are the baryons $X^1_iX^2_j-X^2_iX^1_j$ and $A^{ij}_k$ are generic
complex coefficients which are antisymmetric in the upper indices. 
The R-charge assignment is unique up to te gauge shift,
$2-2\epsilon$ for $P^i$'s and $\epsilon$ for $X_i$'s, with
$0<\epsilon<1$.
There is one FI and one theta parameters,
$\zeta\in \R$, $\theta\in\R/2\pi\im\Z$.
$\zeta\gg 0$ is the usual geometric phase where
the gauge group is completely broken and the low energy theory
is the non-linear sigma model on the complete intersection of 
seven hypersurfaces,
$A_k^{ij}[x_ix_j]=0$, $k=1,\ldots, 7$, in the Grassmannian $G(2,7)$.
$\zeta\ll 0$ is the phase in which the $SU(2)$ subgroup is totally unbroken.
Obtaining and applying some understanding of $SU(2)$ gauge theories,
it is found \cite{HoTo}
that the low energy theory is the non-linear sigma model after all,
whose target space is the Pfaffian locus of $p\in \CP^6$
where the $7\times 7$ antisymmetric matrix $(A^{ij}(p))=(\sum_kA^{ij}_kp^k)$
is of rank $4$. $\zeta\gg 0$ is called the Grassmannian phase while
$\zeta\ll 0$ is called the Pfaffian phase.
The quantum theory is parametrized by $t=\zeta-\im\theta$ and there are
three singular points in the middle, at
$\e^t=(1+\omega)^7, (1+\omega^2)^7, (1+\omega^3)^7$ with
$\omega=\e^{2\pi\im /7}$.\footnote{Compared to \cite{HoTo}, there is a sign
difference. This is the effect of the W-boson integral.
The formulae in \cite{HoTo} should be corrected by the replacement
 $\e^{-t}\to\e^{-t}(-1)^{k+1}$ for $U(k)$ gauge theory.
The relationship between the theta angle and the B-field mentioned in
\cite{HoTo}
is totally explainable by Morrison-Plesser mechanism \cite{MP}.
The same formula as \cite{HoTo} is copied in
v1 of \cite{2dduality}. That is a careless mistake.}

Let us write down the function $A_q(\usigma)$,
\beqa
A_q(\usigma)
&=&\zeta(\Is_1+\Is_2)-\theta(\Rs_1+\Rs_2)-2\pi q^1\Rs_1-2\pi q^2\Rs_2
-\pi|\Rs_1-\Rs_2|\nn\\
&&+7(-\Is_1-\Is_2)\log|\usigma_1+\usigma_2|
+7|\Rs_1+\Rs_2|\left({\pi\over 2}+\arctan\left[{-\Is_1-\Is_2\over|\Rs_1+\Rs_2|
}\right]\right)\nn\\
&&+7\Is_1\log|\usigma_1|
+7|\Rs_1|\left({\pi\over 2}+\arctan\left[{\Is_1\over|\Rs_1|
}\right]\right)\nn\\
&&+7\Is_2\log|\usigma_2|
+7|\Rs_2|\left({\pi\over 2}+\arctan\left[{\Is_2\over|\Rs_2|
}\right]\right).
\eeqa
It can also be written as (\ref{AqCY}), i.e.,
$A_q(\usigma)=\zeta_{\it eff}(\Is)-\theta_{{\it eff},q}(\Rs)$,
where $t_{\it eff}=-2\pi\dd \wt{W}_{\!{\it eff}, q}$
is the effective FI-theta parameter
which depends on the direction $\wh{\usigma}=\usigma/|\!|\usigma|\!|$.
When $\zeta\gg 0$ or $\zeta\ll 0$, the term
$\zeta_{\it eff}(\Is)$ is dominated by $\zeta(\Is_1+\Is_2)$.
Therefore the admissible region is the region with
$(\Is_1+\Is_2)\gg 0$ or $(\Is_1+\Is_2)\ll 0$.

As in the higher rank Abelian theories, we would like to think of
$\gamma$ as the graph of a map $\Rs\mapsto \Is=\Is(\Rs)$.
The wedge condition to avoid poles (\ref{poles}) is
\beqa
\Rs_1+\Rs_2=0&\Longrightarrow&\Is_1+\Is_2\geq 0,\label{wedgeRodland}\\
\Rs_1=0&\Longrightarrow&\Is_1\leq 0,\\
\Rs_2=0&\Longrightarrow&\Is_2\leq 0.
\eeqa

In the Grassmannian phase $\zeta\gg 0$, an admissible contour is easy to find.
For example, we can take
\beq
\Is_1=(\Rs_1)^2,
\quad\,\,
\Is_2=(\Rs_2)^2.
\label{contourG2N}
\eeq
It may be replaced by $\Is_1=|\Rs_1|^{\alpha}$, $\Is_2=|\Rs_2|^{\alpha}$
for any $\alpha>1$.
For such a choice, the growth condition is satisfied for any $q$.
Therefore, this can be used for any brane $\mathfrak{B}$
with an arbitrary set of Chan-Paton representations.

In the Pfaffian phase $\zeta\ll 0$, on the other hand, it is hard to find any
admissible contour of the above type.
The growth condition $(\Is_1+\Is_1)\ll 0$ is in conflict with the
wedge condition (\ref{wedgeRodland}).
We may need to select the allowed set of Chan-Paton representations
over the entire phase.
We plan to explore this problem in the future works.

The contour choice of the type (\ref{contourG2N}) works in
the usual geometric phase in a Calabi-Yau model.
For example,  take a $U(k)$ gauge theory with $N$ fundamentals $X_1,\ldots, X_N$
and a number of powers of $\det^{-1}$ representations, $P^1,\ldots, P^S$,
and a gauge invariant superpotential
$W=P^1f_1(B)+\cdots+P^Sf_S(B)$ with $f_i(B)$'s being polynomials of the baryons
$B_{i_1\cdots i_k}=[X_{i_1}\cdots X_{i_k}]$.
$\zeta\gg 0$ is a geometric phase where the gauge group is completely broken
and the low energy theory is the non-linear sigma model on the complete
intersection of hypersurfaces
$f_1=\cdots=f_S=0$ of the Grassmannian $G(k,N)$.
In this phase, the contour
\beq
\Is_a=(\Rs_a)^2,\quad\,\, a=1,\ldots, k,
\eeq
is admissible for any brane $\mathfrak{B}$.

\subsubsection{A Model With A Simple Grade Restriction Rule}

The next example is the $U(2)$ gauge theory with four fundamentals,
$X_1,\ldots, X_4$, and four antifundamentals, $Y^1,\ldots,Y^4$.
Choice of superpotential and R-charge assignment are
not relevant for the matters we would like to discuss.
$\zeta\gg 0$ and $\zeta\ll 0$ are both phases where
the gauge group is completely broken and
$X_i$'s and $Y^i$'s span the Grassmannian $G(2,4)$ respectively.
The quantum theory is parametrized by $t=\zeta-\im\theta$
and there is a single singularity in the middle,
\beq
t\equiv \pi\im\quad\,\,\mbox{mod $2\pi\im \Z$}.
\eeq
This $\pi$ shift of the theta angle comes from the single pair of
the W-bosons.

The function $A_q(\usigma)$ is astonishingly simple,
\beqa
A_q(\usigma)&=&\zeta(\Is_1+\Is_2)-\theta(\Rs_1+\Rs_2)-2\pi(q^1\Rs_1+q^2\Rs_2)
\nn\\
&&-\pi|\Rs_1-\Rs_2|+4\pi|\Rs_1|+4\pi|\Rs_2|.
\eeqa
The wedge condition to avoid poles is
\beq
\Rs_1=0~\Rightarrow ~\Is_1=0,\quad\,\,
\Rs_2=0~\Rightarrow ~\Is_2=0.
\eeq

\begin{figure}[htb]
\psfrag{q1}{\footnotesize $q^1$}
\psfrag{q2}{\footnotesize $q^2$}
\centerline{\includegraphics{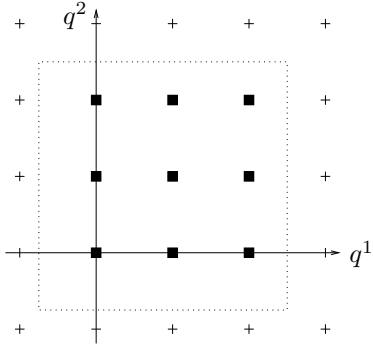}}
\caption{The grade restriction rule for the window
$-3\pi<\theta<-\pi$.}
\label{fig:G24}
\end{figure}

In the $\zeta\gg 0$ phase, as an admissible contour, we can take
\beq
\Is_1=(\Rs_1)^2,\quad\,\,\Is_2=(\Rs_2)^2.
\eeq
In the $\zeta\ll 0$ phase, as an admissible contour, we can take
\beq
\Is_1=-(\Rs_1)^2,\quad\,\,\Is_2=-(\Rs_2)^2.
\eeq
At the phase boundary, $\zeta=0$, the $\Is$ dependence disappears
and the choice of graph $\Is=\Is(\Rs)$ does not matter.
The growth condition simply requires that
$$
A_q(\usigma)~=~-\theta(\Rs_1+\Rs_2)-2\pi(q^1\Rs_1+q^2\Rs_2)
-\pi|\Rs_1-\Rs_2|+4\pi|\Rs_1|+4\pi|\Rs_2|
$$
goes to positive infinity as $|\Rs|\to\infty$ in any direction.
After some elementary exercise, we find that this condition
is equivalent to
\beq
-{3\over 2}\,<\,{\theta\over 2\pi}+q^1\,<\,{3\over 2},\quad\,\,
-{3\over 2}\,<\,{\theta\over 2\pi}+q^2\,<\,{3\over 2}.
\eeq
This is the grade restriction rule.
As long as $\theta\not\equiv \pi$ (mod $2\pi\Z$),
this defines a set of nine weights in a square of size $3$
on the diagonal. 
This set does not change as long as $\theta$ moves in a window,
i.e. and open interval of length $2\pi$ sandwitched between
singular ponts.
See Fig.~\ref{fig:G24} for example.

We decided to look at this example because this is one of the first examples
in a work by Donnovan-Segal \cite{DoSe} which studies aspects of
(classical) grade restriction rule in a class of
non-Abelian linear sigma models.
There it is found that as the relevant ``window category'' one can take
the one generated by Kapranov's exceptional collection of $G(2,4)$, which are
the vector bundles associated with the following representations
of $U(2)$:
\beq
\C,~~\C^2,~~{\rm Sym}^2\C^2,~~\det,~~\C^2\otimes \det,~~{\det}^{\otimes 2},
\eeq
where $\C$ is the trivial representation. We see that the weights of
these representations fits precisely to the one in Fig~\ref{fig:G24}.

\newcommand{\MFarrow}[2]{\begin{picture}(50,20)(0,20)
  \put(22,33){{\small $#1$}}
    \put(5,27){\vector(1,0){40}}
  \put(45,23){\vector(-1,0){40}}
  %\put(22,10){{\small $#2$}}                                                   
  \put(22,13){{\small $#2$}}
  \end{picture}}
\newcommand{\MFarrowshort}[2]{\begin{picture}(32,20)(0,20)
  \put(12,31){{\small $#1$}}
    \put(5,27){\vector(1,0){22}}
  \put(27,23){\vector(-1,0){22}}
  %\put(22,10){{\small $#2$}}                                                   
  \put(14,14){{\small $#2$}}
  \end{picture}}
\newcommand{\MFarrowt}[2]{\begin{picture}(40,20)(0,20)
  \put(16,31){{\small $#1$}}
    \put(5,27){\vector(1,0){30}}
  \put(35,23){\vector(-1,0){30}}
  %\put(22,10){{\small $#2$}}                                                   
  \put(17,14){{\small $#2$}}
  \end{picture}}
\newcommand{\sMFarrow}[2]{\begin{picture}(50,20)(0,20)
  \put(22,29){{\small $#1$}}
  \put(5,27){\vector(1,0){40}}
  \put(45,24){\vector(-1,0){40}}
  %\put(22,10){{\small $#2$}}                                                   
  \put(22,17){{\small $#2$}}
  \end{picture}}
\newcommand{\ssMFarrow}[2]{\begin{picture}(50,20)(0,20)
  \put(22,31){{\small $#1$}}
  \put(5,27){\vector(1,0){40}}
  \put(45,23){\vector(-1,0){40}}
  \put(22,12){{\small $#2$}}
  %\put(22,17){{\small $#2$}}                                                   
  \end{picture}}
\newcommand{\dMFarrow}[2]{\begin{picture}(50,10)(0,20)
  \put(11,22){{\small $#1$}}
  \put(45,32){\vector(-2,-1){40}}
  \put(5,8){\vector(2,1){40}}
  \put(34,15){{\small $#2$}}
  \end{picture}}
\newcommand{\MFmorph}[2]{\begin{picture}(50,30)(0,20)
  \put(10,10){{\small $#1$}}
  \put(5,15){\vector(2,1){40}}
  \put(45,15){\vector(-2,1){40}}
  \put(10,35){{\small $#2$}}
  \end{picture}}
\newcommand{\MFmorphA}[2]{\begin{picture}(50,25)(0,15)
  \put(0,14){{\small $#1$}}
  \put(45,35){\vector(-3,-2){40}}
  \put(5,35){\vector(3,-2){40}}
  \put(42,14){{\small $#2$}}
  \end{picture}}
\newcommand{\MFlsm}{\mathfrak{MF}_W(\C^N,T)}
\newcommand{\MFlsmT}{\mathfrak{MF}_W({\cal T}^w_{{\rm I},{\rm II}})}
\newcommand{\MFT}{\mathfrak{MF}_W({\cal T}^w)}

%Studying the behaviour of the submanifold for various values of
%Chan-Paton weight $q$, we will obtain a good understanding of the
%properties of the branes under variation of parameters like $(\zeta,\theta)$
%or $r$. In particular, we find Calabi-Yau/Landau-Ginzburg
%(or McKay) type correspondence --- the rule of brane transfer 
%in the Calabi-Yau cases and the rule of brane map under the bulk
%renormalization group flow for non-Calabi-Yau cases. This reproduces the
%grade restriction rule for the Abelian and Calabi-Yau cases and
%generalizes it to non-Abelian and/or non-Calabi-Yau cases.

\section{Low Energy Behaviour}

In this section, we check the partition function against
the expected low energy physics of the theory.
In the Calabi-Yau case, we fulfill the promise to
show that, deep inside a phase, the partition function takes the same value
for branes that descend to the same brane in the
low energy theory. This in particular shows that
the analytic property
of the partition function is consistent with the rule of D-brane transport
along a path in the parameter space.
In the non Calabi-Yau case, we look at the behaviour 
of the partition function in the large $r$ limit.
Some consistency check can be made, and moreover, the study leads us to
 find the rule of D-brane map under the bulk renormalization group flow.
The key is to look at the partition function of
a particular class of branes,  called ``empty branes''.

For concreteness, we consider in detail the particular $U(1)$ theory
introduced in Section~\ref{subsec:U1}.
Let us write down the formula for
the partirion function for a brane $\mathfrak{B}$ in this theory,
\beq
Z_{D^2}(\mathfrak{B})=(r\Lambda)^{\wh{c}/2}\int\limits_{\gamma}\dd\usigma'\,
\Gamma\left(-d\,\im\usigma'+1-{d\epsilon\over 2}\right)
\Gamma\left(\im\usigma'+{\epsilon\over 2}\right)^N
\e^{\im t_R(\usigma')}f_\mathfrak{B}(\usigma'),
\label{integralU1}
\eeq
where $f_\mathfrak{B}(\usigma')
={\rm tr}^{}_M(\e^{\pi\im {\bf r}_*}\e^{2\pi\rho(\usigma')})$
and
\beqa
\wh{c}&=&N-2-(N-d)\epsilon,\label{chatU1}\\
t_R&=&t-(N-d)\log(r\Lambda).
\eeqa
The integrand has poles at
\beq
\usigma'=\left\{\begin{array}{lll}
\im \left(n_x+{\epsilon\over 2}\right)&n_x=0,1,2,\ldots&
\mbox{(order $N$),}\\
\im \left(-{n_p+1\over d}+{\epsilon\over 2}\right)&n_p=0,1,2,\ldots&
\mbox{(simple)}.
\end{array}\right.
\label{poles1}
\eeq

\subsection{Tachyon Condensation}\label{subsub:classical}

Let us first describe how the 
branes in the linear sigma model reduce to branes
in the classical low energy theory,
deep in either the geometric phase $\zeta_c\gg 0$
or the Landau-Ginzburg orbifold phase $\zeta_c\ll 0$.
To emphasize that the analysis is purely classical,
we denoted the FI parameter by $\zeta_c$.
An important r\^ole is played again by the D-term equation,
\beq
\sum_{i=1}^N|x_i|^2-d|p|^2=\zeta_c.
\label{Dtermeqn}
\eeq
We shall only give an outline since all the detail can be found in \cite{HHP}.
Only the Calabi-Yau case was discussed in \cite{HHP}, but 
the classical discussion applies equally well without such a restriction.

The descent of branes can be decomposed into two steps: (i) impose the
D-term equation (\ref{Dtermeqn}) strictly but keep all
the chiral multiplets, and (ii) integrate out the heavy chiral multiplets.
Step (i) suffices for the present purposes.
Step (ii) will be described in the next section.

The main ingredient in the descent is the brane-antibrane annihilation
by tachyon condensation. Recall that the matrix factorization $Q$
enters into the boundary potential $\{Q,Q^{\dag}\}$.
I.e., it plays the r\^ole of a profile of the open string tachyon.
If the D-term equation (\ref{Dtermeqn}) is strictly imposed,
it is possible that $\{Q,Q^{\dag}\}$ is everywhere positive definite.
In such a case the brane can be regarded as {\it empty} in the 
low energy limit by the complete brane-antibrane annihilation.
Since the space of solutions to (\ref{Dtermeqn}) depends
very much on the sign of $\zeta_c$, 
which branes are empty and which branes are not
depends also on the sign of $\zeta_c$.

Let us introduce two basic examples:
\beqa
\mathfrak{B}_1:&&\C(0,0)\MFarrowt{f}{p}\C(1-d\epsilon,d)\\[0.2cm]
\mathfrak{B}_2:&&\C(0,0)\MFarrowt{x}{\!pf'}E
\MFarrowt{x}{\!pf'}\wedge^{\! 2}\!E
\MFarrowt{x}{\!pf'}\cdots
\MFarrowt{x}{\!pf'}\wedge^{\!N}\!E
\eeqa
where $E=\C(1-\epsilon,1)^{\oplus N}$. 
Here we used the notation of \cite{HHP}
except that the component $\mathscr{W}(q)_j$ 
of R-charge $j$ and gauge charge $q$ is here denoted by
$\C(j,q)$. Let us explain what the data $(M_i,Q_i,\rho_i,{\bf r}_{*i})$ is
for $\mathfrak{B}_i$, $i=1,2$.
The vector space $M_i$ is the direct sum of
the spaces appearing, i.e.,
$M_1=\C(0,0)\oplus \C(1-d\epsilon,d)$
and $M_2=\wedge E$, ${\bf r}_{*i}$ and $\rho_i$ are specified by the
numbers $(j,q)$ of each component $\C(j,q)$,
and the matrix factorization is given by
\beqa
Q_1&=&\left(\begin{array}{cc}
0&p\\
f(x)&0
\end{array}\right),\\
Q_2&=&\sum_{i=1}^N\left(\,x_i\overline{\eta}_i
+{1\over d}\,p\,
\partial_if(x)\eta_i\,\right),
\eeqa
where $\eta_i$ and $\overline{\eta}_i$ are generators of the Clifford algebra,
$\{\eta_i,\overline{\eta}_j\}=\delta_{i,j}$, $\{\eta_i,\eta_j\}
=\{\overline{\eta}_i,\overline{\eta}_j\}=0$,
that is used to construct $\wedge E$. 
We may also consider the shifts, $\mathfrak{B}_1(j,q)$ and
$\mathfrak{B}_2(j,q)$,
where $\mathfrak{B}\mapsto \mathfrak{B}(j,q)$ for $(j,q)\in\Z^{\oplus 2}$
is the uniform shift of the R-charges by $j$ and the gauge charges by $q$.
The boundary potentials are
\beqa
\{Q_1,Q_1^{\dag}\}&=&\left(|p|^2+|f(x)|^2\right){\rm id}^{}_{M_1},\\
\{Q_2,Q_2^{\dag}\}&=&\sum_{i=1}^N
\left(|x_i|^2+{1\over N^2}|p\partial_if(x)|^2\right){\rm id}^{}_{M_2}.
\eeqa
In the $\zeta_c\gg 0$ phase, $\mathfrak{B}_2$ and all of its shifts are
empty at low energies. This is because $\sum_i|x_i|^2\geq \zeta_c$ 
by the D-term equation (\ref{Dtermeqn}) and hence the boundary potential
$\{Q_2,Q_2^{\dag}\}$ is positive definite everywhere
with a strictly positive lower bound.
Likewise, in the $\zeta_c\ll 0$ phase, $\mathfrak{B}_1$ and all of its shifts
are empty at low energies since $\{Q_1,Q_1^{\dag}\}$ is positive
definite everywhere on the D-term locus (\ref{Dtermeqn})
where $|p|^2\geq |\zeta_c/N|$.
On the other hand, they are non-empty in the opposite phases,
since the boundary potentials fail to be positive definite at some locus:
$\{Q_1,Q_1^{\dag}\}$ vanishes
at $p=f(x)=0$ which is allowed in the $\zeta_c\gg 0$ phase,
while $\{Q_2,Q_2^{\dag}\}$ vanishes at $x=0$ (assuming $d>1$)
which is allowed in the $\zeta_c\ll 0$ phase.
After the step (ii), see \cite{HHP} or the next section, we find that
$\mathfrak{B}_1$ in the $\zeta_c\gg 0$ phase
is (a shift of) the structure sheaf ${\mathcal O}_{X_f}$, that is,
the single D-brane wrapped on the entire target space $X_f$
and supporting the trivial line bundle. 
When $f$ is a Fermat polynomial,
$\mathfrak{B}_2$ in the $\zeta_c\to-\infty$ limit is one of the 
${\bf L}={\bf 0}$ Recknagel-Schomerus branes \cite{ReSc}.

Brane-antibrane annihilation
implies that the descent map of branes in the linear
sigma model to branes in the low energy theory is not one to one
but many to one, as is always the case in renormalization group flow.
In fact it is huge to one since any number of copies of empty branes
should be regarded as ``nothing'' in the low energy theory.
It would be convenient if we have a subset, or a slice, in the set of branes 
in the linear sigma model such that the map is one to one when restricted
to that subset.
In fact, such subsets exist!
In the $\zeta_c\gg 0$ phase, let us consider a set of branes whose Chan-Paton
charges are within a zone ${\bf w}$ of length $N$, i.e., a set of
$N$ consecutive integers, say ${\bf w}=\{1,\ldots, N\}$
or ${\bf w}=\{17,\ldots, 16+N\}$. Then, the map of branes in that subset
to branes in the low energy theory at $\zeta_c\gg 0$ is one to one.
Likewise, in the $\zeta_c\ll 0$ phase,
let us consider a set of branes whose Chan-Paton
charges are within a zone ${\bf w}$ of length $d$, say
${\bf w}=\{0,\ldots, d-1\}$. Then, the map of branes in that subset
to branes in the low energy theory at $\zeta_c\ll 0$ is one to one.
We shall call this the {\it classical grade restriction rule}.

We put the adjective ``classical'' in order to distinguish it from
the (quantum) grade restriction rule which we have discussed
in the previous section, for branes of the theory sitting at or going through
a window between different phases, in the Calabi-Yau case.
However, these are certainly related.
Note that (\ref{GRR1}) defines a zone of length $N$,
and we shall call it the zone of the window.
Carrying over the terminology,
a brane in the subset determined by a zone ${\bf w}$
is said to be {\it grade restricted} with respect to ${\bf w}$.

The classical grade restriction rule means that
any brane can be replaced by a unique grade restricted brane
by a brane-antibrane creation and annihilation process.
We can show this by employing
the empty branes introduced above, i.e, $\mathfrak{B}_2$ and its shifts
in the $\zeta_c\gg 0$ phase and $\mathfrak{B}_1$ and its shifts in the
$\zeta_c\ll 0$ phase.
It goes as follows.
Let us take any brane $\mathfrak{B}=(M,Q,\rho,{\bf r}_*)$.
If $\mathfrak{B}$ has a Chan-Paton charge outside the zone ${\bf w}$,
then, we can bind an empty brane to $\mathfrak{B}$
at the vector of that charge so that
the resulting brane has charges closer to ${\bf w}$.
We repeat this process.
Note that $\mathfrak{B}_2$ has the smallest charge $0$
and the largest charge $N$,
while $\mathfrak{B}_1$ has the smallest charge $0$ and the largest charge $d$.
This guarantees that this binding process can eventually
put all the charges inside the zone ${\bf w}$ of length $N$
for $\zeta_c\gg 0$
and $d$ for $\zeta_c\ll 0$.

\newcommand{\LSM}{\mbox{\large $\mathfrak{D}$}}

To summarize the discussion, let us introduce some notations.
We denote the set of all linear sigma model branes
by $\LSM$, the set of branes in the 
classical low energy theory in the phase
$\zeta_c\gg 0$ ({\it resp}. $\zeta_c\ll 0$)
by $D_+$ ({\it resp}. $D_-$) and the
set of grade restricted branes with respect to a zone ${\bf w}$ by
${\mathcal T}_{\bf w}$.
Then, we have maps of branes:\\[-0.2cm]
\begin{figure}[h]
\psfrag{a}{${\mathcal T}_{{\bf w}_-}$}
\psfrag{b}{$\LSM$}
\psfrag{c}{$D_-$}
\psfrag{equiv}{$\cong$}
\psfrag{subset}{$\subset$}
\psfrag{ap}{${\mathcal T}_{{\bf w}_+}$}
\psfrag{cp}{$D_+$}
\psfrag{p}{$\pi_-$}
\psfrag{pp}{$\pi_+$}
\centerline{\includegraphics{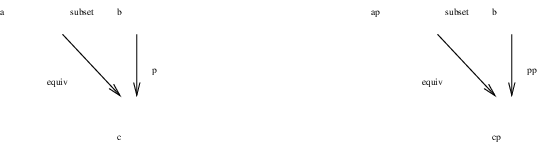}}
\end{figure}\\
The vertical arrow $\pi_{\pm}:\LSM\to D_{\pm}$ is the huge-to-one
descent map to the low energy theory.
The restriction to the subset ${\mathcal T}_{{\bf w}_{\pm}}\subset\LSM$
associated to a zone ${\bf w}_{\pm}$ is one to one, if
the length of ${\bf w}_-$ is $d$ and the length of ${\bf w}_+$ is $N$.
The above diagrams of ``sets'' and ``maps'' may be regarded as
diagrams of categories and functors.
In that case ``the one to one map'' should be regarded as an equivalence of
categories.
There is a recent development concerning such equivalences of categories,
 motivated by the classical
grade restriction rule \cite{Segal,DHL,BFK,DoSe}.

Finally let us compute the factor
$f_\mathfrak{B}(\usigma')$ in the integrand of (\ref{integralU1})
for the branes $\mathfrak{B}_1$ and $\mathfrak{B}_2$:
\beqa
f_{\mathfrak{B}_1}(\usigma')&=&1-\e^{-\pi\im d\epsilon}\e^{2\pi d\usigma'},\\
f_{\mathfrak{B}_2}(\usigma')&=&1-N\e^{-\pi\im \epsilon}\e^{2\pi\usigma'}
+{N\choose 2}\e^{-2\pi\im \epsilon}\e^{4\pi\usigma'}-\cdots
+(-1)^N\e^{-N\pi\im \epsilon}\e^{2N\pi\usigma'}\nn\\
&=&(1-\e^{-\pi\im\epsilon}\e^{2\pi\usigma'})^N.
\eeqa
Notice that $f_{\mathfrak{B}_2}(\usigma')$ 
cancels the poles of $\Gamma(\im\usigma'+{\epsilon\over 2})^N$
on the positive imaginary axis but cannot cancel all the poles of
$\Gamma(-d\im\usigma'+1-{d\epsilon\over 2})$
on the negative imaginary axis.
On the other hand,
$f_{\mathfrak{B}_1}(\usigma')$ 
cancels the poles of $\Gamma(-d\im\usigma'+1-{d\epsilon\over 2})$
on the negative imaginary axis but cannot cancel the higher order
poles of $\Gamma(\im\usigma'+{\epsilon\over 2})^N$ on the positive imaginary
axis.
This is a reflection of the fact that $\mathfrak{B}_2$ is empty in the
$\zeta_c\gg 0$ phase but not in the $\zeta_c\ll 0$ phase,
while  $\mathfrak{B}_1$ is empty in the
$\zeta_c\ll 0$ phase but not in the $\zeta_c\gg 0$ phase.
The real significance of this observation in the quantum theory
will be discussed below.

We now look at the partition function and compare it with the
expectation of the low energy behaviour of the theory and of
the branes, including the above descent map of branes.
We separate the discussion into the three cases, $d=N$,
$d<N$ and $d>N$.

\subsection{\underline{$d=N$}: Family of conformal field theories}

In the Calabi-Yau case, $d=N$,
the family of theories parametrized by $t\in \C/2\pi\im \Z$
is expected to flow to a family of superconformal field theories
with $c/3=N-2$. Note that the last number is equal to
the exponent $\wh{c}$ in (\ref{chatU1}) as already remaked (\ref{2dc}).
In the two extreme regimes, $\zeta\gg 0$ and $\zeta\ll 0$,
the degrees of freedom other than those in the classical
low energy theory are infinitely heavy.
Therefore, the classical analysis of the previous subsection is expected
to hold, with $\zeta\sim \zeta_c$.

As an examination,
let us look at the partition function
for the branes $\mathfrak{B}_1$ and $\mathfrak{B_2}$. 
Recall that $f_{\mathfrak{B}_2}(\usigma')$
cancels the poles of the gamma function factor
on the positive imaginary axis while $f_{\mathfrak{B}_1}(\usigma')$
cancels the poles on the negative imaginary axis.
In view of the contour choice in Fig.~\ref{fig:contour1},
we see that they indeed have vanishing partition function
in the phase where they are said to be empty, 
\beq
\begin{array}{ll}
Z_{D^2}(\mathfrak{B}_2)=0&\mbox{for $\,\zeta_{\it eff}>0$,}\\[0.2cm]
Z_{D^2}(\mathfrak{B}_1)=0&\mbox{for $\,\zeta_{\it eff}<0$.}
\end{array}
\label{vanishZ}
\eeq
The same holds for $\mathfrak{B}_2(j,q)$ and
$\mathfrak{B}_1(j,q)$ as the shift $\mathfrak{B}\mapsto \mathfrak{B}(j,q)$
changes the brane factor $f_\mathfrak{B}(\usigma')$ simply 
by multiplication of the entire function $(-1)^j\e^{2\pi q\usigma'}$
which does not affect the cancellation of poles.
Recall also that the brane factors for $\mathfrak{B}_1$ and $\mathfrak{B_2}$
fail to cancel the poles on the opposite sides of
the imaginary axis. Thus, the partition functions do not have to vanish
in the opposite phases. In the next section, we will compute
them and see that they are equal to the (expected) parition function
of the low energy images, i.e. the structure sheaf ${\mathcal O}_{X_f}$
of $X_f$ for $\mathfrak{B}_1$ and the Recknagel-Schomerus brane 
of the Landau-Ginzburg orbifold for
$\mathfrak{B}_2$.

The vanishing (\ref{vanishZ}) means that, in each phase,
 branes related by binding the empty branes have exactly
the same partition function.
In particular, a given brane and its grade restricted replacement
have the same partition function.
Therefore, the partition function takes the same value
on the branes which descend to the same brane in the classical
low energy theory.

Now let us come back to the problem in Section~\ref{subsub:CY}
concerning analytic continuation of
the partition function $Z_{D^2}(\mathfrak{B})$
along a path from one phase to another, say from 
$\zeta\gg 0$ to $\zeta\ll 0$.
There was a problem if the brane $\mathfrak{B}$ is not grade
restricted with respect to the window through which the path goes.
We now know what to do: while in the $\zeta\gg 0$ phase,
we replace $\mathfrak{B}$ by a grade restricted brane $\mathfrak{B}'$
by binding the empty branes $\mathfrak{B}_2(j,q)$.
We have just learned that $Z_{D^2}(\mathfrak{B})$ is exactly equal to
$Z_{D^2}(\mathfrak{B}')$ in the $\zeta\gg 0$ phase.
Now that $\mathfrak{B}'$ is grade restricted with respect to the window,
the partition function can be analytically continued to
$\zeta\ll 0$ through that window.
We therefore conclude that
the partition function of the brane $\mathfrak{B}$ at $\zeta\gg 0$
is analytically continued along the path to the
partiction function of the brane $\mathfrak{B}'$ at $\zeta\ll 0$.

This matches the rule of D-brane transport \cite{HHP}. 
In the Calabi-Yau case $d=N$,
the lengths of the zones for the classical grade restriction 
are the same between the two phases.
Hence we can take a common grade restricted
subset ${\mathcal T}_{\bf w}$ to make a bridge
between low energy branes in one phase and the low energy
branes in the other.\\[-0.2cm]
\begin{figure}[h]
\psfrag{a}{${\mathcal T}_{{\bf w}}$}
\psfrag{b}{$\LSM$}
\psfrag{c}{$D_-$}
\psfrag{equiv}{$\cong$}
\psfrag{cup}{$\cup$}
\psfrag{cp}{$D_+$}
\psfrag{p}{$\pi_-$}
\psfrag{pp}{$\pi_+$}
\centerline{\includegraphics{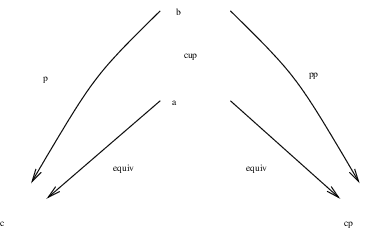}}
\end{figure}\\
If we take ${\bf w}$ to be the zone of a window, this gives the
rule of D-brane transport through that window. Once again,
we have seen that the analytic continuation of the partition function
matches with this rule.

We may also consider a closed loop in the parameter space, starting from
one phase, going to the other phase through a window, and then coming
back through a different window.
If we analytically continue the partition function of a brane
$\mathfrak{B}$ along such a path, it comes back as the partition function of
another brane $\mathfrak{B}''$.
\\[-0.2cm]
\begin{figure}[h]
\psfrag{a2}{${\mathcal T}_{{\bf w}'}\!\ne {\mathcal T}_{{\bf w}}$}
\psfrag{b}{$\LSM$}
\psfrag{c}{$D_-$}
\psfrag{equiv}{$\cong$}
\psfrag{cup}{$\cup$}
\psfrag{cp}{$D_+$}
\psfrag{p}{$\pi_-$}
\psfrag{pp}{$\pi_+$}
\centerline{\includegraphics{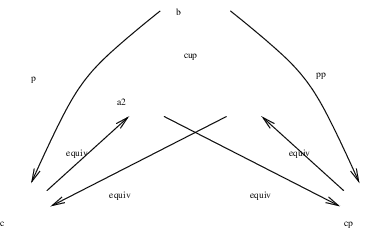}}
\end{figure}\\
The tranform $\mathfrak{B}\mapsto \mathfrak{B}''$ is what is known as
the D-brane
 monodromy. As in the above discussion, this is done by the brane replacement
via binding empty branes at appropriate phases.
If the loop goes around one singular point, it is to bind a brane which becomes
massless at the singular point, in accord with the picture found
by Strominger \cite{Strominger}.

At the level of categories,
the D-brane transport along a path from one phase to the other
gives an equivalence of the categories,
$D_+\stackrel{\cong}{\longrightarrow} D_-$. 
The equivalences for various windows are
the same as the equivalences first found by Orlov \cite{Orlov}.
The D-brane monodromy for a closed loop gives an autoequivalence
of the category, say, $D_+\stackrel{\cong}{\longrightarrow} D_+$.
Construction of such autoequivalences had been given in \cite{SeTh,Horja}
and is called Seidel-Thomas twist.

\subsection{\underline{$d<N$}: Flow from the non-linear sigma model}

\newcommand{\boldsigma}{\mbox{\boldmath$\sigma$}}

When $d<N$, the FI parameter is larger at higher energies and
the theory describes the asymptotically free non-linear sigma model on
the Fano manifold $X_f$, with $c/3=N-2$ in the ultra-violet limit.
At low energies, the theory reduces to
the Landau-Ginzburg orbifold $W=f(X_1,\ldots, X_N)/\Z_d$ or one of
the $(N-d)$ massive vacua.
The Landau-Ginzburg orbifold is expected to flow to a superconformal
field theory with
\beq
{c\over 3}=N\left(1-{2\over d}\right).
\label{LGOc}
\eeq
The massive vacua are at
\beq
\boldsigma_k
=-\im\wt{\Lambda}\exp\left(\im{\theta+\pi d+2\pi k\over N-d}\right),
\quad\,k\in \Z/(N-d)\Z,
\label{massivevacua}
\eeq
($\wt{\Lambda}^{N-d}:=\Lambda^{N-d}d^d\e^{-\zeta}$, 
see Section~\ref{subsec:U1})
with the value
$2\pi\wt{W}_{\!{\it eff}}=(N-d)\boldsigma_k$
for the twisted superpotential.

We would like to ask which branes correspond to
the superconformal field theory and which branes correspond to
the massive vacua at low energies.
Can we see that by looking at the behaviour of the partition function
in the large size limit $r\to\infty$? 
Taking the lesson from Section~\ref{subsec:Abranes}, we may try to see if
it has a power or exponential behaviour.
We suppose that the poles on the negative imaginary axis
are relevant for the Landau-Ginzburg orbifold. 
For $r\Lambda\gg 1$, the pole $\usigma'=\im(-1/d+\epsilon/2)$
closest to the origin yields the dominant contribution, which 
is of the order of
\beq
(r\Lambda)^{\wh{c}/2}\e^{\im t_R\im (-1/d+\epsilon/2)}
~\sim~ (r\Lambda)^{N(1-2/d)/2}.
\label{negpol}
\eeq
This is indeed the expected power behaviour for the conformal field theory
of central charge (\ref{LGOc}).
Therefore, if the partition function is dominated by 
the pole contribution (\ref{negpol}), we may say that the brane
corresponds to a brane in the superconformal field theory, or 
more precisely, has such a component.

\begin{figure}[htb]
\psfrag{pos}{\footnotesize $|\theta+2\pi q|<\pi d$}
\psfrag{neg}{\footnotesize $\pi d<|\theta+2\pi q|<\pi{N+d\over 2}$}
\psfrag{abc}{\footnotesize $\!\!\pi{N+d\over 2}<|\theta+2\pi q|<\pi N$}
\psfrag{rL}{\footnotesize $r\wt{\Lambda}\e$}
\centerline{\includegraphics{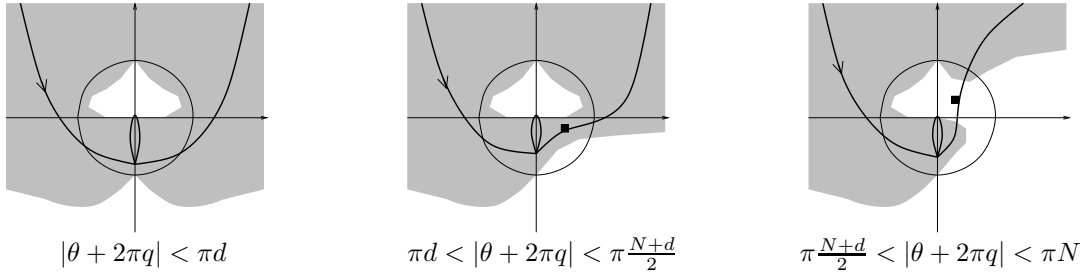}}
\caption{Deformed contours}
\label{fig:contour2}
\end{figure}

For which values of $q$ does
the integral have the residue (\ref{negpol}) as the dominant contribution?
We recall that the contour is decided to to be as in
Fig.~\ref{fig:contournonCY}-Left. We deform it
so that it picks this and some other poles 
on the negative imaginary axis as in Fig.~\ref{fig:contour2}.
(The meaning of the dots will be explained later.)
The part along the negative imaginary axis will have (\ref{negpol}).
The question is what the other part of $\gamma$ gives.
If $|\theta+2\pi q|<{\pi\over 2}(N+d)$, we can choose $\gamma$
so that it goes through the region in which
the integrand is exponentially small,
$\e^{-C'r|\usigma'|}$, as $r\to \infty$ where $C'$ is positive and with
a strictly positive lower bound along the way. Therefore
the integral from that part vanishes in the $r\to\infty$ limit,
and is dominated by (\ref{negpol}).
If $|\theta+2\pi q|>{\pi\over 2}(N+d)$, on the other hand,
it is unavoidable that $\gamma$ goes through a region where the
integrand is exponentially large. Therefore, 
the integral on the other part is generically
exponentially {\it growing} as $r\to\infty$.

For more detailed evaluation, let us see if the integrand has a critical
point. We assume $\theta\not\equiv \pi d, \pi N$ so that the massive vacua
(\ref{massivevacua}) are not on the imaginary axis.
For large values of $\usigma'$, we may omit the power factor and
only look at the exponent, which is
$2\pi\im r\wt{W}_{\!{\it eff},q}(\usigma)$, now with the $q$ dependence. 
The equation $\partial_{\usigma}\wt{W}_{\!{\it eff},q}(\usigma)=0$
reads $|\usigma|=\wt{\Lambda}$ and
\beq
(N-d){\rm Arg}(\im\usigma)=\theta+2\pi q-{\rm sgn}(\usigma_1)\pi d.
\eeq
(This is equivalent to the vanishing of the coefficient of
$|\usigma_1'|$ in (\ref{AqFano}). See (\ref{Argid}).)
When $|\theta+2\pi q|<\pi d$ and $|\theta+2\pi q|>\pi N$, there is no solution.
When $\pi d<|\theta+2\pi q|<\pi N$, there is a unique solution which 
is equal to $\boldsigma_k$ of (\ref{massivevacua}) with
\beq
k=\left\{\begin{array}{cl}
q-d&~~~(\pi d<\theta+2\pi q<\pi N)\\[0.2cm]
q&~~~(-\pi N<\theta+2\pi q<-\pi d).
\end{array}\right.
\label{kcorr}
\eeq 
For these values of $q$, a part of the integral can be evaluated by
the saddle point approximation at the critical point
and has the exponential behaviour as $r\to\infty$.
The dots in Fig.~\ref{fig:contour2}-Middle and -Right
are the critical points. Indeed the contour $\gamma$
comes close to this point.
If $|\theta+2\pi q|<\pi d$, except the sum of poles on the
negative imaginary axis, the integral vanishes more rapidly as $r\to\infty$
than any of these exponentials.

Let us see the behaviour of the partition function  at large $r$
for the branes
$\mathfrak{B}_1$ and $\mathfrak{B}_2$ and their shifts.
Looking at the contour $\gamma$, we immediately see that it vanishes for
$\mathfrak{B}_2$ and all of its shifts,
\beq
Z_{D^2}(\mathfrak{B}_2(j,q))=0.
\eeq
This is exact vanishing, for any value of $r$.
For $\mathfrak{B}_1(j,q)$, let us consider the deformed contour
as in Fig.~\ref{fig:contour2}. We know that the poles on the negative
imaginary axis is cancelled by the
brane factor $f_{\mathfrak{B}_1}(\usigma')$,
and hence the contribution comes entirely
from the other part. Recall that $\mathfrak{B}_1(j,q)$ has two components,
$\C(j,q)$ and $\C(j+1-d\epsilon, q+d)$. By (\ref{kcorr}),
the contribution of the former ({\it resp}. latter) component
has the exponential behaviour $\sim \e^{-(N-k)\im r\boldsigma_q}$
for $-\pi N<\theta+2\pi q<-\pi d$ 
({\it resp}. $\pi d<\theta+2\pi(q+d)<\pi N$).
Therefore, for such a $q$,
\beq
Z_{D^2}(\mathfrak{B}_1(j,q))~\sim~\exp\Bigl(-(N-d)\im r
\boldsigma_q\Bigr),
\quad\,\,\, r\to\infty.
\label{expoFano}
\eeq
It is non-zero and therefore the brane $\mathfrak{B}_1(j,q)$ cannot
be empty in the full quantum theory,
even though it is so when reduced to the Landau-Ginzburg orbifold.

\begin{figure}[htb]
\psfrag{W}{\footnotesize $\wt{W}_{\!{\it eff}}$}
\psfrag{2}{\footnotesize ${\mathcal O}(2)$}
\psfrag{1}{\footnotesize ${\mathcal O}(1)$}
\psfrag{0}{\footnotesize ${\mathcal O}(0)$}
\psfrag{-1}{\footnotesize ${\mathcal O}(-1)$}
\psfrag{-2}{\footnotesize $\!{\mathcal O}(-2)$}
\psfrag{-3}{\footnotesize $\!{\mathcal O}(-3)$}
\psfrag{s-1}{\footnotesize ${\mathcal O}(q_{{}_{\rm max}}\!\!-\!1)$}
\psfrag{s}{\footnotesize ${\mathcal O}(q_{{}_{\rm max}})$}
\psfrag{t}{\footnotesize $\!{\mathcal O}(q_{{}_{\rm min}})$}
\psfrag{t-1}{\footnotesize $\!\!\!\!\!{\mathcal O}(q_{{}_{\rm min}}\!\!+\!1)$}
\psfrag{T}{${\mathcal T}_{{\bf w}_{-,\theta}}$}
\centerline{\includegraphics{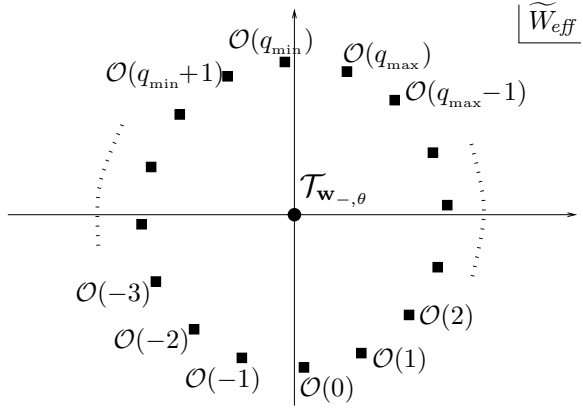}}
\caption{Low energy images of the branes}
\label{fig:FtoL}
\end{figure}

The above observations are enough to conclude the following, assuming
$\theta\not\equiv \pi d, \pi N$ (mod $2\pi \Z$).
It is enough to consider branes which are grade restricted with respect to
a zone of length $N$. A natural zone is
\beq
{\bf w}_{+,\theta}=
\left\{\,q\in\Z\,\,\Bigl|\,\,
-{N\over 2}~<~{\theta\over 2\pi}+q~<~{N\over 2}\,\,\right\}.
\label{GRR2UV}
\eeq
For $|\theta+\pi d+2\pi q|<\pi (N-d)$,
the brane $\mathfrak{B}_1(j,q)$ descends to a brane
of the massive vacuum at $\boldsigma_q$.
On the other hand,
branes which are grade restricted with respect to the zone
\beq
{\bf w}_{-,\theta}=
\left\{\,q\in\Z\,\,\Bigl|\,\,
-{d\over 2}~<~{\theta\over 2\pi}+q~<~{d\over 2}\,\,\right\}
\label{GRR2LGO}
\eeq
descend purely to the superconformal field theory.
A picture of the descent is shown in
Fig.~\ref{fig:FtoL} where the branes are plotted on
the $\wt{W}_{\!{\it eff}}$-plane,
for the value
$\theta=-\pi d+\delta$ with a small positive $\delta$.
The square dots are the values of the massive vacua (\ref{massivevacua}),
 and the origin is the value for the Landau-Ginzburg orbifold.
We plot the large volume image ${\mathcal O}(q)={\mathcal O}_{X_f}(q)$
of the brane $\mathfrak{B}_1(0,q)$ in the place of the critical value
of $\boldsigma_q$
The maximum and the minimum values of $q$ are
$q_{{}_{\rm max}}:=[{N-d-1\over 2}]$ and $q_{{}_{\rm min}}:=-[{N-d\over 2}]$.

At the special values $\theta\equiv \pi d, \pi N$, one
or two of the critical points (\ref{massivevacua}) are on
the imaginary axis. When a critical point crosses the negative imaginary
axis as we vary $\theta$, the zone ${\bf w}_{-,\theta}$ changes. For example,
if we move $\theta=-\pi d+\delta$ from a positive $\delta$ to a
negative $\delta$,
the zone changes from ${\bf w}_-=\{0,1,\ldots,d-1\}$ to
${\bf w}_-'=\{1,\ldots, d\}$.
If a brane $\mathfrak{B}'$ is grade restricted with
respect to the latter it is not grade rerstricted with respect to the former,
at the components of charge $d$. That can be cancelled by binding
the branes $\mathfrak{B}_1(j,0)$ there, and we obtain a brane
$\mathfrak{B}$ which is grade restricted with respect to ${\bf w}_-$.
But $\mathfrak{B}$ and $\mathfrak{B}'$ are not the same brane.
They differ by the attached branes $\mathfrak{B}_1(j,0)$ which are
not empty at the massive vacuum $\boldsigma_0$.
This is the ``brane creation'' in the sense of \cite{HIV}.
When the critical point crosses the positive imaginary axis,
there is again the change of zones ${\bf w}_{+,\theta}$.
The change is accompanied with a brane replacement using
$\mathfrak{B}_2(j,q_*)$ for a particular $q_*$.
Since $\mathfrak{B}_2(j,q_*)$ are genuinely empty, it is simply a change of
linear sigma model representatives of the same brane.

To summarize, let us redraw the diagram of the sets and maps of the branes.
\\[-0.2cm]
\begin{figure}[h]
\psfrag{a2}{${\mathcal T}_{{\bf w}_-}\!\!\subset {\mathcal T}_{{\bf w}_+}$}
\psfrag{b}{$\LSM$}
\psfrag{c}{$D_-$}
\psfrag{equiv}{$\cong$}
\psfrag{cup}{$\cup$}
\psfrag{cp}{$D_+$}
\psfrag{p}{$\pi_-$}
\psfrag{pp}{$\pi_+$}
\centerline{\includegraphics{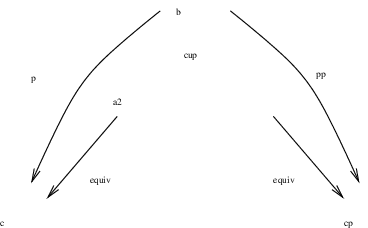}}
\end{figure}
\\
We emphasize that $D_{\pm}$ are the set of branes in the {\it classical}
low energy theory. $D_+$ is the set of branes
in the non-linear sigma model on the Fano-manifold $X_f$ and
$D_-$ is the set of branes in the Landau-Ginzburg orbifold.
For each $\theta$, we have a pair of grade restricted subsets,
${\mathcal T}_{{\bf w}_-}\subset {\mathcal T}_{{\bf w}_+}$, of $\LSM$.
This gives rize to an embedding $D_-\subset D_+$,
and the complement is given by the collection of $(N-d)$ branes,
${\mathcal O}(q_{{}_{\rm min}}),\ldots, {\mathcal O}(q_{{}_{\rm max}})$,
which descend to the $(N-d)$ massive vacua. As we vary $\theta$,
one or both of the pair
${\mathcal T}_{{\bf w}_-}\subset {\mathcal T}_{{\bf w}_+}$ can jump.
When ${\mathcal T}_{{\bf w}_-}$ jumps,
the embedding $D_-\subset D_+$ will also jump.

\subsection{\underline{$d>N$}: Flow from the Landau-Ginzburg orbifold}

When $d<N$, the FI parameter is smaller at higher energies and
the theory describes a relevant deformation of the
superconformal field theory with $c/3=N(1-2/d)$
associated to the Landau-Ginzburg orbifold $W=f(X_1,\ldots, X_N)/\Z_d$.
At low energies, the theory reduces to
the non-linear sigma model on the manifold of general type $X_f$
or one of $(d-N)$ massive vacua.
The nonilinear sigma model is free in the infra-red limit with
central charge
\beq
{c\over 3}=N-2.
\label{NLSMc}
\eeq
The massive vacua are at $\boldsigma_k$ in
(\ref{massivevacua}) which may be rewritten as
\beq
\boldsigma_k=\im\wt{\Lambda}\exp\left(\im{\theta+\pi N+2\pi k\over N-d}\right),
\quad\,k\in \Z/(d-N)\Z,
\label{massivevacuap}
\eeq
with the value $2\pi\wt{W}_{\!{\it eff}}=(N-d)\boldsigma_k$
for the twisted superpotential.

The analysis of the contour and the integral
goes in the same way as in the $d<N$ case, and the description can be brief.
Roughly speaking, we only need to exchange the r\^oles of $d$ and $N$
and flip the sign of the FI parameter and $\usigma'_2$.
The contour $\gamma$ can be taken as in Fig.~\ref{fig:contournonCY}-Right,
coming in from and going out to th eregion where ${\rm Im}(\usigma')$
is negative infinity.
The sum of residues at the poles on the positive imaginary axis
is dominated by the one at $\usigma'=\im\epsilon/2$ at $r\to\infty$
which behaves as
\beq
(r\Lambda)^{\wh{c}/2}\e^{\im t_R\im\epsilon/2}
~\sim~(r\Lambda)^{(N-2)/2}.
\label{pospol}
\eeq
This is the expected behaviour for the conformal field theory of
central charge (\ref{NLSMc}).
The integral for a fixed charge $q$ is dominated by (\ref{pospol})
when $|\theta+2\pi q|<{\pi\over 2}(N+d)$.
Assumimg $\theta\not\equiv \pi d, \pi N$,
the integrand has a unique critical point at (\ref{massivevacuap})
with 
\beq
k=\left\{\begin{array}{cl}
q-N&~~~(\pi N<\theta+2\pi q<\pi d)\\[0.2cm]
q&~~~(-\pi d<\theta+2\pi q<-\pi N).
\end{array}\right.
\eeq 
For these $q$'s, a part of the integral can be evaluated by th saddle
point approximation and has the exponential behaviour as
$r\to\infty$. For other values of $q$'s, the integrand
has no critical points. In particular, when $|\theta+2\pi q|<\pi N$,
the integral minus the sum of residues at the
poles on the positive imaginary exis decays more rapidly than
any of the above exponentials as $r\to\infty$.
The branes $\mathfrak{B}_1$ and $\mathfrak{B}_2$ and their shifts have
the following partition functions,
\beq
Z_{D^2}(\mathfrak{B}_2(j,q))~=~0,\quad\,\,\forall r,
\eeq
for any $(j,q)$ and
\beq
Z_{D^2}(\mathfrak{B}_1(j,q))~\sim~\exp\Bigl(-(N-d)\im r
\boldsigma_q\Bigr),
\quad\,\,\, r\to\infty.
\label{expogene}
\eeq
for any $j$ and for $|\theta+\pi N+2\pi q|<\pi(d-N)$.

The conclusion for $\theta\not\equiv \pi d,\pi N$ (mod $2\pi\Z$) is as follows:
It is enough to consider branes which are grade restricted with respect to
a zone of length $d$ and a natural choice is ${\bf w}_{-,\theta}$
as in (\ref{GRR2LGO}).
For $|\theta+\pi N+2\pi q|<\pi (d-N)$,
the brane $\mathfrak{B}_2(j,q)$ descends to a brane
of the massive vacuum at $\boldsigma_q$.
On the other hand,
branes which are grade restricted with respect to the zone
${\bf w}_{+,\theta}$ as in (\ref{GRR2UV})
descend purely to the non-linear sigma model.
\begin{figure}[htb]
\psfrag{W}{\footnotesize $\wt{W}_{\!{\it eff}}$}
\psfrag{2}{\footnotesize ${\mathcal B}(-2)$}
\psfrag{1}{\footnotesize ${\mathcal B}(-1)$}
\psfrag{0}{\footnotesize ${\mathcal B}(0)$}
\psfrag{-1}{\footnotesize ${\mathcal B}(1)$}
\psfrag{-2}{\footnotesize ${\mathcal B}(2)$}
\psfrag{-3}{\footnotesize ${\mathcal B}(-3)$}
\psfrag{s-1}{\footnotesize ${\mathcal B}(q_{{}_{\rm min}}\!\!\!+\!1)$}
\psfrag{s}{\footnotesize ${\mathcal B}(q_{{}_{\rm min}})$}
\psfrag{t}{\footnotesize $\!{\mathcal B}(q_{{}_{\rm max}})$}
\psfrag{t-1}{\footnotesize $\!\!\!\!\!{\mathcal B}(q_{{}_{\rm max}}\!\!\!-\!1)$}
\psfrag{T}{${\mathcal T}_{{\bf w}_{+,\theta}}$}
\centerline{\includegraphics{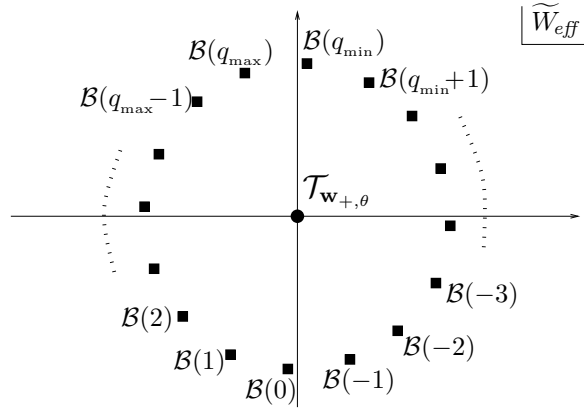}}
\caption{Low energy images of the branes}
\label{fig:LtoF}
\end{figure}
A picture of the descent is shown in
Fig.~\ref{fig:LtoF} where the branes are plotted on
the $\wt{W}_{\!{\it eff}}$-plane,
for a small positive $\theta$.
The square dots are the values of the massive vacua (\ref{massivevacuap}),
 and the origin is the value for the non-linear sigma model.
We plot the ${\bf L}=0^N$
Recknagel-Schomerus brane ${\mathcal B}(q)={\mathcal B}_{0^N,q,0}$
which is the Landau-Ginzburg orbifold image of the
brane $\mathfrak{B}_2(-N,q-N)$.
The maximum and the minimum values of $q$ are
$q_{{}_{\rm max}}:=[{d-N-1\over 2}]$ and $q_{{}_{\rm min}}:=-[{d-N\over 2}]$.

When $\theta$ varies across the special values $\theta\equiv\pi d$
and $\pi N$, the critical points crosses the imaginary axis,
and the zone change occurs. 
When a critical point crosses the positive imaginary axis, 
a non-linear sigma model brane creates branes at the massive vacua.

To summarize, let us draw the diagram of sets and maps of the branes.
\\[-0.2cm]
\begin{figure}[h]
\psfrag{a2}{${\mathcal T}_{{\bf w}_-}\!\!\supset {\mathcal T}_{{\bf w}_+}$}
\psfrag{b}{$\LSM$}
\psfrag{c}{$D_-$}
\psfrag{equiv}{$\cong$}
\psfrag{cup}{$\cup$}
\psfrag{cp}{$D_+$}
\psfrag{p}{$\pi_-$}
\psfrag{pp}{$\pi_+$}
\centerline{\includegraphics{diagram4.eps}}
\end{figure}
\\
For each $\theta$, we have a pair of grade restricted subsets,
${\mathcal T}_{{\bf w}_-}\supset {\mathcal T}_{{\bf w}_+}$, of $\LSM$.
This gives rize to an embedding $D_-\supset D_+$,
and the complement is given by the collection of $(d-N)$ branes,
${\mathcal B}(q_{{}_{\rm min}}),\ldots, {\mathcal B}(q_{{}_{\rm max}})$,
which descend to the $(d-N)$ massive vacua. As we vary $\theta$,
one or both of the pair
${\mathcal T}_{{\bf w}_-}\supset {\mathcal T}_{{\bf w}_+}$ can jump.
When ${\mathcal T}_{{\bf w}_+}$ jumps,
the embedding $D_-\supset D_+$ will also jump.

\section{Expressions In Phases}

In this section, we compute the partition function
for the theory deep inside various phases.
In particular, we find expressions at the Landau-Ginzburg orbifold points
and in the geometric phases.
The expression at a Landau-Ginzburg orbifold point agrees with
the result of the purely Landau-Ginzburg orbifold found in
Section~\ref{subsec:thecaseof} which in turn agrees with
the formula for the central charge.
The expression at the large volume limit mathces with the
expected formula for the central charge, except that the
class $\sqrt{\wh{A}}$ should be replaced by the Gamma-class,
a correction well-known among mathematicians.

\subsection{Landau-Ginzburg Orbifold Phase}

Let us first look at the Landau-Ginzburg orbifold phase.
We start with the $U(1)$ theories introduced in
Section~\ref{subsec:U1} as a warm up, and then consider 
more general theories.

\subsubsection{The $U(1)$ Theories}

The Landau-Ginzburg orbifold appears in the regime 
$\zeta\ll 0$ if $d=N$,
as a part of the theory in the long distance
regime $r\gg \Lambda$ if $d<N$,
and as the theory in the short distance regime $r\ll \Lambda$ if $d>N$. 
In either case, we look at the parameter region with $\zeta_R\ll 0$.

Before looking at the partition function, we describe the descent
rule of branes \cite{HHP}. The Landau-Ginzburg orbifold is
obtained by freezing the field $p$ at some value, say $1$, which breaks the
gauge group $G=U(1)$ to $G_L=\Z_d$. Therefore, it is natural to
go to the $\epsilon={2\over d}$ frame
where the R-charge of $p$ vanishes.
In this frame, the element $J\in G$ defined in (\ref{defJ})
becomes
\beq
J=\e^{\pi \im\epsilon}\,\,\stackrel{\epsilon\to{2\over d}}{\longrightarrow}
\, \e^{2\pi\im\over d}=:J_L
\eeq
The brane $\mathfrak{B}=(M,Q,\rho,{\bf r}_*)$ descends to the brane
$\mathfrak{B}_{\rm LG}=(M_L,Q_L,\rho_L,{\bf r}_{*,L})$ where
\beqa
&&M_L\,=\,M,\nn\\
&&Q_L(x)\,=\,Q(1,x),\nn\\
&&\rho_L(\omega)\,=\,\rho(\omega),\quad\omega^d=1,\nn\\
&&{\bf r}_{*,L}\,=\,{\bf r}_*\Bigr|_{\epsilon={2\over d}}.
\label{descLGO}
\eeqa
We recall that (\ref{intr*}) is satisfied, so that
$\e^{\pi\im {\bf r}_*}\rho(J)=\e^{\pi\im {\bf r}_{*,L}}\rho_L(J_L)$ is the $\Z_2$
grading, or equivalently,
$\e^{\pi\im r_j}\e^{q_j\pi\im \epsilon}=\e^{\pi\im r_{j,L}}\e^{q_j2\pi\im/d}=
(-1)^{r^o_j}$.

The formula for the partition function in the $\epsilon= {2\over d}$ frame is
\beq
Z_{D^2}(\mathfrak{B})=
(r\Lambda)^{\wh{c}_{\rm LG}\over 2}
\int_{\gamma}\dd\usigma' \,\Gamma(-d\,\im \usigma')\,
\Gamma\left({\textstyle \im\usigma'+{1\over d}}\right)^N
\e^{\im t_R\usigma'}f_{\mathfrak{B}}(\usigma'),
\eeq
with $\wh{c}_{\rm LG}=N(1-{2\over d})$.
Note that the contour $\gamma$ should be poked at $\usigma'=0$
so that it goes above $0$.
From a glance at the contours
(Figs.~\ref{fig:contour1}, \ref{fig:contournonCY}, \ref{fig:contour2}),
and from the discussion in the previous section,
we see that we only need to take the poles on the negative imaginary axis,
$\usigma'=-\im n/d$ for $n=0,1,2,\ldots$ .
Using (\ref{GammaPoles}), we obtain
\beq
Z_{D^2}^{{}^{\rm LG}}(\mathfrak{B})
={2\pi\over d}
(r\Lambda)^{\wh{c}_{\rm LG}\over 2}
\sum_{n=0}^{\infty}{(-1)^n\over n!}
\Gamma\left({\textstyle {n+1\over d}}\right)^N\e^{t_Rn/d}
f_{\mathfrak{B}}\left({\textstyle -\im {n\over d}}\right).
\eeq
The brane factor can be written as
\beqa
f_{\mathfrak{B}}\left({\textstyle -\im {n\over d}}\right)
&=&\sum_j\e^{\pi\im r_j}|_{\epsilon={2\over d}}
\e^{2\pi q_j(-\im {n\over d})}\nn\\
&=&\sum_j(-1)^{r^o_j}\e^{-2\pi\im q_j\left({1+n\over d}\right)}
={\rm Str}^{}_M\rho(J_L^{-1-n}).
\eeqa
Thus, we obtain
\beq
Z_{D^2}^{{}^{\rm LG}}(\mathfrak{B})
={2\pi\over d}
(r\Lambda)^{\wh{c}_{\rm LG}\over 2}
\sum_{n=0}^{\infty}{(-1)^n\over n!}
\Gamma\left({\textstyle {n+1\over d}}\right)^N\e^{t_Rn/d}\,
{\rm Str}^{}_M\rho(J_L^{-1-n}).
\eeq
This is the $\e^{t_R/d}$ expansion of the full partition function
for $d\geq N$ and of a part of it for $d<N$.
In the limit $\zeta_R\to-\infty$, that is,
$\zeta\to-\infty$, the infra-red and the ultra-violet limits
respectively for $d=N$, $d<N$ and $d>N$,
only the leading term remains, 
\beq
Z_{D^2}^{{}^{\rm LG}}(\mathfrak{B})
~\longrightarrow~
{2\pi\over d}\Gamma\left({\textstyle {1\over d}}\right)^N
(r\Lambda)^{\wh{c}_{\rm LG}\over 2}\,
{\rm Str}^{}_{M}\rho(J_L^{-1}).
\eeq
Up to the numerical factor, 
this agrees with the formula (\ref{LGOres})
for the brane (\ref{descLGO}) in the Landau-Ginzburg orbifold, which 
in turn is the same as the formula of \cite{WalcherLG}
for the central charge of the same brane.

To get back a general $\epsilon$, say for comparison with the 
other phase, we use  (\ref{effectgshift}) and find that
 the result must be multiplied by
$\exp\left(-t \left({\epsilon\over 2}-{1\over d}\right)\right)$.

\subsubsection{More General Theories}

Let us consider a theory with an Abelian and connected gauge group,
$G=T$, with charge integrality. 
We assume the situation as discussed in 
\cite{HHP},\footnote{Although this include
a wide class of examples, this is not the most general situation.
Some examples in \cite{Wphases} are not of this type.}
where the fields are grouped into two, $Y_1,\ldots, Y_k$ and
$X_1,\ldots,X_l$, such that $Q_{Y_1},\ldots, Q_{Y_k}$ span $\im \ttt^*$
and that $Q_{X_j}$ are non-positive spans of $Q_{Y_i}$'s,
\beq
Q_{X_j}=-\sum_{i=1}^k a_j^{\,\,i}Q_{Y_i};\qquad\,\, a_j^{\,\,i}\geq 0\,\quad 
\forall (i,j).
\label{posaij}
\eeq
If $\zeta$ is a positive linear span of $Q_{Y_i}$'s,
the D-term equation
$\sum_iQ_{Y_i}|y_i|^2=\sum_{i,j}a_j^{\,\,i}Q_{Y_i}|x_j|^2+\zeta$
has a solution with $y_i$'s all non-zero,
for any value of $x_j$'s. Therefore the gauge group is broken to
a finite subgroup $G_L$, consisting of elements that fix all $y_i$'s,
 and the classical low energy theory is the
$G_L$-orbifold of the Landau-Ginzburg model with the superpotential
\beq
W_L(X_1,\ldots, X_l)=W(1,\ldots,1,X_1,\ldots,X_l),
\eeq
where $W(Y,X)$ is the original superpotential.
By the charge integrality, we have the R$^o$ frame in which all
the R-charges of the bulk fields, $R^o_{Y_i}, R^o_{X_j}$, are $0$ or $2$.
Since $Q_{Y_i}$'s span $\im \ttt^*$,
there is a unique gauge shift $\sDelta$
that annihilates the R-charges of $Y_i$'s,
\beq
R_{Y_i,L}=R^o_{Y_i}+Q_{Y_i}(\sDelta)=0.
\eeq
The new R-charges for $X_j$'s
\beq
R_{X_j,L}=R^o_{X_j}+Q_{X_j}(\sDelta)
=R^o_{X_j}+\sum_{i=1}^ka_j^{\,\,i}R^o_{Y_i},
\eeq
are the R-charges of the low energy Landau-Ginzburg orbifold.
The element $\e^{\pi\im \sDelta}\in G$ acts trivially on $Y_i$'s 
and acts on $X_j$ by the phase $\e^{\pi\im R_{X_j,L}}$.
That is, it is the element $J_L\in G_L$ of the Landau-Ginzburg orbfiold,
\beq
J_L=\e^{\pi\im \sDelta}.
\eeq

The brane descent is as in (\ref{descLGO}):
$\mathfrak{B}=(M,Q,\rho,{\bf r}_*)\mapsto
\mathfrak{B}_{\rm LG}=(M_L,Q_L,\rho_L,{\bf r}_{*,L})$, where
\beqa
&&M_L\,=\,M,\nn\\
&&Q_L(x_1,\ldots,x_l)\,=\,Q(1,\ldots,1,x_1,\ldots,x_l),\nn\\
&&\rho_L\,=\,\rho\bigl|_{G_L},\nn\\
&&{\bf r}_{*,L}\,=\,{\bf r}^o_*-\rho(\sDelta).
\label{descLGOgen}
\eeqa

The partition function in the $R_L$-frame is 
\beqa
Z_{D^2}(\mathfrak{B})&=&(r\Lambda)^{\wh{c}_{\rm LG}/2}
\int_{\gamma}\dd^k\usigma\,\prod_{i=1}^k
\Gamma\left({\textstyle \im Q_{Y_i}(\usigma)}\right)
\prod_{j=1}^l
\Gamma\left({\textstyle 
-\im \sum_ia_j^{\,\,i}Q_{Y_i}(\usigma)+{R_{X_j,L}\over 2}}\right)\nn\\
&&~~~~~~~~~~~~~~~~~~~~~~~~~~
\times\e^{\im t(\usigma)}
\,{\rm tr}^{}_M\left(\e^{\pi\im ({\bf r}^o_*-\sDelta)}
\e^{2\pi\rho(\usigma)}\right),
\eeqa
where the contour $\gamma$ should be poked near 
$(Q_{Y_i}(\usigma)=0)$'s to avoid the poles that came down in the
$R_L$-frame limit. 
If the theory satisfies the Calabi-Yau condition,
the contour $\gamma$ in this phase 
can be taken as in (\ref{daijoubu}). For example, we can take
(the poked version of)
\beq
Q_{Y_i}(\Is)=(Q_{Y_i}(\Rs))^2,\quad\,\, i=1,\ldots, k.
\label{contourLGOgen}
\eeq
The growth condition is satisfied since $\zeta$ is a positive span
of $Q_{Y_i}$'s. The wedge condition to avoid poles, which is
trivially satisfied for $Y_i$'s, is also satisifed for $X_j$'s,
\beq
Q_{X_j}(\Is)=-\sum_{i=1}^ka_j^{\,\,i}Q_{Y_i}(\Is)=-\sum_{i=1}^k
a_j^{\,\,i}(Q_{Y_i}(\Rs))^2\leq 0,
\eeq
where (\ref{posaij}) is used.
If the theory is not Calabi-Yau, as in the $U(1)$
theory, the above contour may still be admissible,
or appears as a part of the admissible contour in the regime where
$\zeta_R$ is deep inside the positive span of $Q_{Y_i}$'s.
In either case, we decide to take (the poked version of)
(\ref{contourLGOgen}) as the contour.
In the Calabi-Yau case and some other cases, it is the full partition 
function but in some other cases it is only a part of it.

Taking the poles at
$Q_{Y_i}(\usigma)=\im n_i$, $n_i=0,1,2,\ldots$, for
$i=1,\ldots, k$, we obtain
\beqa
Z_{D^2}^{{}^{\rm LG}}(\mathfrak{B})&=&
{(2\pi)^k\over\det{Q_Y}}(r\Lambda)^{\wh{c}_{\rm LG}/2}
\sum_{n}{(-1)^{n_1+\cdots+n_k}\over
n_1!\cdots n_k!}\prod_{j=1}^l\Gamma\left({\textstyle
a_j(n)+{R_{X_j,L}\over 2}}\right)~~~~~~~~~~~\nn\\
&&~~~~~~~~~~~~~~~~~~~~~~~~~~~~~
\times\e^{- t (Q_Y^{-1}(n))}\,{\rm Str}^{}_M\,
\rho\!\left(J_L^{-1}\e^{2\pi\im Q_Y^{-1}(n)}\right)
\eeqa
In the limit $(t Q_Y^{-1})^i\to \infty$, only the $n=0$ term remains,
\beq
Z_{D^2}^{{}^{\rm LG}}(\mathfrak{B})\,\,\longrightarrow\,\,
{(2\pi)^k\over\det{Q_Y}}
\prod_{j=1}^l\Gamma\left({\textstyle {R_{X_j,L}\over 2}}\right)
(r\Lambda)^{\wh{c}_{\rm LG}/2}\,
\,{\rm Str}_M^{}\rho(J_L^{-1}).
\eeq
Up to the numerical factor, 
this agrees with the formula (\ref{LGOres})
for the brane (\ref{descLGOgen}) in the Landau-Ginzburg orbifold.

\subsection{Geometric Phase}

We next consider the geometric phase.

\subsubsection{The Gamma Classes}

Before starting, we describe some characteristic classes which will
enter into the formulae.
Let us introduce some functions of one variable $x$
with Taylor series at $x=0$ starting with $1$:
\beqa
&\displaystyle \wh{\rm A}(x)={x/2\over\sinh(x/2)},\\
&\displaystyle {\rm td}(x)={x\over 1-\e^{-x}},\\
&\displaystyle \wh{\Gamma}(x)=\Gamma\left(1-{x\over 2\pi\im}\right),\quad\,
\wh{\Gamma}^*(x)=\Gamma\left(1+{x\over 2\pi\im}\right).
\eeqa
They define characteristic classes 
$\,\wh{\rm A}_X,\,$ ${\rm td}_X,\,$ $\wh{\Gamma}_X\,$ and $\,\wh{\Gamma}^*_X\,$
of the tangent bundle of a complex manifold $X$
via the total Chern class $c(X)$ \cite{Hirzebruch}, in such a way as
\beq
c(X)={\prod_i(1+x_i)\over\prod_j(1+y_j)}\,\,\Longrightarrow\,\,
{\rm td}_X={\prod_i{\rm td}(x_i)\over\prod_j{\rm td}(y_j)}.
\eeq
These are called the 
{\it A-roof class}, {\it Todd class}, and {\it Gamma classes}.
To be more precise, the A-roof class can be defined for any real manifold
and can be expressed in terms of the Pontrjagin classes. Here we are
considering the specialization to complex manifolds, assuming the
usual relation between the Pontrjagin classes of the real tangent bundle
and the Chern classes of the complex tangent bundle.
Explicit expressions in terms of the Chern classes are well known for
$\wh{A}$ and ${\rm td}$. We write down first few terms for
the Gamma class:
\beqa
\wh{\Gamma}&=&1-\im{\gamma\over 2\pi}c_1
+{1\over 24}c_2
+\left(-{1\over 48}-\half\left({\gamma\over 2\pi}\right)^2\right)c_1^2
+\im{\zeta(3)\over (2\pi)^3}c_3\nn\\
&&
-\im\left({\gamma\over 24\cdot 2\pi}+{\zeta(3)\over (2\pi)^3}\right)c_2c_1
+\im\left({\gamma\over 48\cdot 2\pi}
+{1\over 6}\left({\gamma\over 2\pi}\right)^3
+{\zeta(3)\over 3(2\pi)^3}\right)c_1^3+\cdots,~~~
\eeqa
where $\gamma$ is Euler's consant.
There are some relations among the above
functions,
$\wh{\rm A}(x)=\e^{-x/2}{\rm td}(x)=\wh{\Gamma}(x)\wh{\Gamma}^*(x)$,
which are copied to the relations
among the associated classes,
\beq
\wh{\rm A}_X\,=\,\e^{-c_1(X)/2}{\rm td}_X\,=\,\wh{\Gamma}_X\wh{\Gamma}^*_X.
\label{relnch}
\eeq
Let us also recall that
the A-roof and Todd classes appears in some index formula.
If $X$ is an even dimensional smooth manifold with a spin structure,
and $E$ is a smooth vector bundle on $X$, we can consider
the Dirac operator acting on the spinors with values in $E$.
Then, the index of the Diract operator is given by
the Atiyah-Singer formula:
\beq
{\rm ind}\Dirac_E\,\,=\,\int_X\wh{\rm A}_X\,\ch(E).
\label{Diracindex}
\eeq
If $X$ is a complex manifold and ${\mathcal E}$ is a holomorphic
vector bundle on $X$, we can consider the Dolbeault operator acting on
anti-holomorphic differential forms with values in ${\mathcal E}$.
Then, the Euler characteristic of the Dolbeault complex
is given by the Riemann-Roch
formula
\beq
\chi({\mathcal E},\overline{\partial})
\,\,=\,\int_X{\rm td}_X\,\ch({\mathcal E}).
\label{RiemannRoch}
\eeq

\subsubsection{The $U(1)$ Theories}

As a warm up,
we start with the $U(1)$ theories.
The geometric phase is in the regime $\zeta\gg 0$ if $d=N$,
in the short distance regime $r\ll \Lambda$ if $d<N$,
 and in a part of the long distance regime $r\gg \Lambda$ if $d>N$,
In either case, we look at the parameter region with $\zeta_R\gg 0$.

Before starting, let us describe how branes in the linear sigma model
descend to branes in the non-linear sigma model \cite{HHP}.
What we have after imposing the Higgs mechanism (step (i) in the language of
Section~\ref{subsub:classical}) is
the non-linear sigma model on the total space of the line bundle
${\mathcal O}(-d)$ over $\CP^{N-1}$, with the superpotential $W=pf(x)$.
This $W$ is a Bott-Morse function and the critical set is
the locus $p=f(x)=0$, that is, the hypersurface $X_f$.
We obtain the non-linear sigma model on $X_f$ by
integrating out the massive modes, $p$ and $f(x)$.
(This is the step (ii).) The brane descent for
integrating out a pair of massive variables is known
as Kn\"orrer periodicity \cite{Knorrer} and we only have to apply it in the
current situation. How to do it is described in \cite{HHP}
and we simply record the procedure.

Branes in the non-linear sigma model with the target $X_f$
are represented by complexes of holomorphic
vector bundles on $X_f$, possibly of infinite
lengths but with truncation to finite lengths complexes of coherent sheaves.
A brane is therefore specified by a pair $({\mathcal E}, d)$:
${\mathcal E}$ is a $\Z$-graded vector bundle on $X_f$ which is 
of finite rank in each degree. $d$ is a local
endomorphism of ${\mathcal E}$ (holomorphic bundle map of ${\mathcal E}$)
of degree $1$ such that $d^2=0$ and that
$\{d,d^{\dag}\}$ has a finite rank kernel for some choice of
fibre metric on ${\mathcal E}$.

We shall simply write $M$ for $(M,\rho,{\bf r}_*)$ so that
the information of gauge group and R-symmetry group action
on the Cahn-Paton vector space is included into the notation $M$.
The brane $\mathfrak{B}=(M,Q)$ descends to the brane
$\mathfrak{B}_{\rm LV}=(M_L,Q_L)$ where
\beqa
&&M_L~=~\bigoplus_{i=0}^{\infty}M(2i,di),\\
&&Q_L~=~Q(p_L,x),
\eeqa
where $p_L$ is the shift of charges by $(2,d)$.
Here we regard $\C(j,q)$ as the line bundle ${\mathcal O}_{X_f}(q)$
at degree $j$.
The Chern character of this brane is given by
\beqa
\ch(\mathfrak{B}_{\rm LV})&=&\sum_{i=0}^{\infty}
\sum_j(-1)^{r^o_j+2i}\e^{(q_j+di) H}
\nn\\
&=&
{1\over 1-\exp (dH)}\,
f_{\mathfrak{B}}\!\left({\textstyle {1\over 2\pi}H}\right).
\label{chBLV}
\eeqa
It is also noticed that the Kn\"orrer procedure involves
the shift of the Chan-Paton charge, which can be absorbed into
the shift of the theta angle or a B-field:
\beq
2\pi B=(\theta+\pi d)H
\label{Bshift}
\eeq
In this paper, we normalize the B-field as
$[B]\in H^2(X_f,\Z)$ on the closed string sector
so that the instanton factor for the degree $\beta$ maps is
\beq
\exp\left(-\int_{\beta}(\omega-2\pi \im B)\right),
\eeq
where $\omega$ is the K\"ahler form.

Now we look at the partition function.
The formula in the R$^o$ frame is
\beq
Z_{D^2}(\mathfrak{B})=
(r\Lambda)^{\wh{c}_{\rm LV}\over 2}
\int_{\gamma}\dd\usigma' \,\Gamma(-d\,\im \usigma'+1)\,
\Gamma\left({\textstyle \im\usigma'}\right)^N
\e^{\im t_R\usigma'}f_{\mathfrak{B}}(\usigma'),
\eeq
with $\wh{c}_{\rm LV}=N-2$.
Note that the contour $\gamma$ should be poked at $\usigma'=0$
so that it goes below $0$.
Looking at the contours
(Figs.~\ref{fig:contour1}, \ref{fig:contournonCY}, \ref{fig:contour2}),
and from the discussion in the previous section,
we see that we only need to take the poles on the positive imaginary axis,
$\usigma'=\im n$ for $n=0,1,2,\ldots$ . At each $n$, we shift the 
integration variable as
$\usigma'=\im n+{z\over 2\pi}$.
This yields
\beq
Z_{D^2}^{{}^{\rm LV}}(\mathfrak{B})=
(r\Lambda)^{\wh{c}_{\rm LV}\over 2}
\sum_{n=1}^{\infty}
\oint_0{\dd z\over 2\pi} \,
\Gamma\left({\textstyle dn+{dz\over 2\pi \im}+1}\right)
\Gamma\left({\textstyle -n-{z\over 2\pi\im}}\right)^N\!
\e^{-t_Rn+{\im\over 2\pi}t_R z}
f_{\mathfrak{B}}\!\left({\textstyle \im n+{z\over 2\pi}}\right).
\eeq
Note that $f_{\mathfrak{B}}(\im n+{z\over 2\pi})
=f_{\mathfrak{B}}({z\over 2\pi})$ since $\e^{2\pi q_j(\im n)}=1$.
We also use the relation
\beq
\Gamma(x)\Gamma(1-x)={\pi\over \sin(\pi x)},
\label{Gammaid4}
\eeq
to rewrite a part of the gamma function factors. This yields
\beq
Z_{D^2}^{{}^{\rm LV}}(\mathfrak{B})=
-C(r\Lambda)^{\wh{c}_{\rm LV}\over 2}\sum_{n=0}^{\infty}
\oint_0{\dd z\over 2\pi\im}
\left(\!{(-1)^n\over
2\sinh\left({z\over 2}\right)}\!\right)^{\!\!N}\!
{\Gamma\left(1+{dz\over 2\pi \im}+dn\right)
\over
\Gamma\left(1+{z\over 2\pi\im}+n\right)^N}
\e^{-nt_R+{\im\over 2\pi}t_R z}
f_{\mathfrak{B}}\!\left({\textstyle {z\over 2\pi}}\right),
\eeq
with $C=-\im (-2\pi\im)^N$.
We further rewrite it as follows,
\beqa
Z_{D^2}^{{}^{\rm LV}}(\mathfrak{B})\!&=&\!
C(r\Lambda)^{\wh{c}_{\rm LV}\over 2}
\sum_{n=0}^{\infty}
\oint_0{\dd z\over 2\pi\im} \,{1\over z^N}\cdot dz\cdot
{z^{N-1}(1-\e^{-dz})\over
d(1-\e^{-z})^N}{\Gamma\left(1+{dz\over 2\pi \im}+dn\right)
\over
\Gamma\left(1+{z\over 2\pi\im}+n\right)^N}~~~~~~~~~~~~~~~~\nn\\
&&~~~~~~~~~~~~~~~~~~~~~~~~~~~~~~~~~~~~~
\times\exp\left(-nt'_R+{\im\over 2\pi}t'_R z\right)
{f_{\mathfrak{B}}\!\left({z\over 2\pi}\right)\over 1-\e^{dz}},
\label{hasai}
\eeqa
where
\beq
t'_R=t_R-d\pi \im +(N-d)\pi\im.
\label{tRp}
\eeq
This is in order to express each term
as an integral over $X_f$, using
\beq
\int_{X_f}g(H)=\int_{\CP^{N-1}} d\,H g(H)=\oint_0
{\dd z\over 2\pi\im} \,{1\over z^N}\cdot dz\cdot g(z),
\label{intXf}
\eeq
which holds for a power series $g(z)$ in $z$ where
 $H$ is the hyperplane class on $\CP^{N-1}$ or its restriction on $X_f$.

At this point, let us write down the expressions of
some characteristic classes of $X_f$.
By the exact sequences
\beqa
&&
0\to {\mathcal O}\longto{\mathcal O}(1)^{\oplus N}\longto
T_{\CP^{N-1}}\to 0,\nn\\
&&
0\to T_{X_f}\longto T_{\CP^{N-1}}|_{X_f}\longto N_{X_f/\CP^{N-1}}\to 0,
\nn
\eeqa
we have
\beq
c(X_f)={(1+H)^N\over (1+dH)},
\eeq
which implies $c_1(X_f)=(N-d)H$ and
\beq
{\rm td}_{X_f}={H^{N-1}(1-\e^{-dH})\over d (1-\e^{-H})^N}.
\eeq
Let us also introduce a cohomology class 
\beqa
\wh{\Gamma}_{X_f}(n)&:=&\wh{\rm A}_{X_f}\cdot
{\Gamma\left(1+d\left({H\over 2\pi\im}+n\right)\right)
\over \Gamma\left(1+{H\over 2\pi\im}+n\right)^N}\nn\\
&=&\e^{-{N-d\over 2}H}{H^{N-1}(1-\e^{-dH})\over d (1-\e^{-H})^N}
\cdot{\Gamma\left(1+d\left({H\over 2\pi\im}+n\right)\right)
\over \Gamma\left(1+{H\over 2\pi\im}+n\right)^N}.
\label{lookat}
\eeqa
At $n=0$, it reduces to the Gamma class,
\beq
\wh{\Gamma}_{X_f}(0)\,=\,
\wh{\rm A}_{X_f}\cdot{\Gamma(1+{d\over 2\pi\im}H)
\over \Gamma(1+{1\over 2\pi\im}H)^N}
\,=\,\wh{\rm A}_{X_f}\cdot{1\over\,\wh{\Gamma}^*_{X_f}\!\!\!}
\,\,=\,\wh{\Gamma}_{X_f},
\eeq
where we used (\ref{relnch}).

Using (\ref{intXf}) and looking at (\ref{chBLV}) and (\ref{lookat}),
we can write (\ref{hasai}) as
\beq
Z_{D^2}^{{}^{\rm LV}}(\mathfrak{B})=
C(r\Lambda)^{\wh{c}_{\rm LV}\over 2}
\sum_{n=0}^{\infty}
\int_{X_f}\e^{{N-d\over 2}H}\wh{\Gamma}_{X_f}(n)
\exp\left(-nt'_R+{\im\over 2\pi}t'_R H\right)
\ch(\mathfrak{B}_{\rm LV})
\label{LVexp}
\eeq
Let us denote the renormalized K\"ahler form as $\omega_R=\zeta_RH$.
In view of (\ref{Bshift}), we have $t'_RH=\omega_R-2\pi\im B+\pi\im (N-d)H$.
Then, we may also write the result as
\beq
Z_{D^2}^{{}^{\rm LV}}(\mathfrak{B})=
C(r\Lambda)^{\wh{c}_{\rm LV}\over 2}
\sum_{n=0}^{\infty}\e^{-nt_R'}
\int_{X_f}\wh{\Gamma}_{\!X_f}\!(n)
\exp\left(B+{\im\over 2\pi}\omega_R\right)
\ch(\mathfrak{B}_{\rm LV})
\label{LVexp2}
\eeq
This is the $\e^{-t_R}$ expansion of the full partition function
for $d\leq N$ and of a part of it for $d>N$.
In the limit $\zeta_R\to +\infty$, that is,
$\zeta\to +\infty$, the ultra-violet and the infra-red limits
respectively for $d=N$, $d<N$ and $d>N$,
only the leading term remains,
\beq
Z_{D^2}^{{}^{\rm LV}}(\mathfrak{B})
~\longrightarrow~
C(r\Lambda)^{\wh{c}_{\rm LV}\over 2}
\int_{X_f}\wh{\Gamma}_{X_f}
\exp\left(B+{\im\over 2\pi}\omega_R\right)
\ch(\mathfrak{B}_{\rm LV}).
\label{LVlimit}
\eeq
This is agrees with the expected formula for the central charge
of the brane $\mathfrak{B}_{\rm LV}$,
except that we have the Gama class in the place of
$\sqrt{\wh{A}_{X_f}}$.
That the Gamma class rather than $\sqrt{\wh{A}_{X_f}}$
should enter into
the asymptotic formula for the central charge had been well-known
to mathematicians.
In fact, formula of the type (\ref{LVexp}), (\ref{LVexp2})
were first presented by Hosono \cite{Hosono} for the quintic,
$N=d=5$, and that was a part of the motivation to
define the Gamma class \cite{Iritani09,Katzarkov,Iritani11}.

\subsubsection*{Gravitational Descendants and Loop Operators}

In one of such development \cite{Iritani09,Iritani11},
Iritani studied the D-brane central charge from the view point
of Gromov-Witten theory, i.e.,
topological A-model, on Fano manifolds. 
If we compare our results with his formula, it looks like 
that the central charge can be expressed in terms of the genus zero
topological string three point amplitudes as
\beq
Z_{D^2\!{}_{(-)}}(\mathfrak{B})
\,=\,\sum_{n=0}^{\infty}(-1)^n(r\Lambda)^n
\langle \tau_n(F(\mathfrak{B}))PP\rangle_{0},
\label{loop}
\eeq
where 
$\langle\cdots\rangle_0$ stands for the genus zero topological string amplitude
with sum over all worldsheet instantons,
$F(\mathfrak{B})$ is a certain cohomology class of $X_f$ 
constrcuted out of $\ch(\mathfrak{B})$, $\wh{\Gamma}_{X_f}$ and
$(r\Lambda)^{c_1(X_f)}$. 
$\tau_nF(\mathfrak{B})$ is the $n$-th gravitational descendant
of $F(\mathfrak{B})$ and $P$ is the puncture operator.
The series of the form
$w(\ell)=\sum_n(-1)^n\ell^n\tau_n$ is known as the ``loop operator''
in the study of 2d quantum gravity which creates a hole on the worldsheet.
It is interesting to observe that our formula came from the hemisphere, i.e.
a genus zero Riemann surface with a big hole, and that
the right hand side  of (\ref{loop}) is also associated to
a sphere amplitude with one hole. 
It would be interesting to understand the meaning of this 
observation.

\newcommand{\boldeps}{\mbox{\boldmath$\epsilon$}}

\subsubsection{More General Theories}\label{subsub:genLV}

Let us move on to a more general linear sigma model with a gauge group $G$
and the matter fields grouped into two, an $E$-valued field $X$ and 
an $F^*$-valued field $P$,
for some representations $E$ and $F$ of $G$ of dimensions
$d_E$ and $d_F$.
We assume the superpotential of the form 
\beq
W=\langle P,f(X)\rangle
\eeq
where $f:E\to F$ is a $G$-equivariant polynomial map, and
$\langle-,-\rangle$ is the pairing between $F^*$ and $F$.
 We assign
the R-charge $0$ to $X$ and $2$ to $P$ in the R$^o$-frame.
We assume that there is a phase in which the D-term equation requires
$X$ to have non-zero values which break the gauge group $G$
completely. We also assume that $f$ is generic enough so that
the D- and F-term equations force $P=0$ and that the vacuum manifold
is a smooth submanifold $X_f$, defined by $f=0$, of a smooth
compact symplectic quotient $\PP$ of $E$ by $G$.
We may also regard $\PP$ as the geometric invariant theory quotient
\beq
\PP=E/\!/G_{\C}
\eeq
with respect to the stability condition defined by the FI
parameter in the phase.
We write the weights of $E$ and $F$ with respect to a maximal torus $T$
by $Q_i$'s and $d_{\beta}$'s,
\beq
E|_T=\bigoplus_i\C(Q_i),\,\,\quad
F|_T=\bigoplus_{\beta}\C(d_{\beta}).
\eeq
Let us write down some characteristic classes of $X_f$.
By the exact sequences,
\beqa
&&
0\to {\mathcal O}(\mathfrak{g}_{\C})\longto{\mathcal O}(E)\longto
T_{\PP}\to 0,\nn\\
&&
0\to T_{X_f}\longto T_{\PP}|_{X_f}\longto N_{X_f/\PP}\to 0,
\eeqa
we have
\beq
c(X_f)={c(\PP)\over c(N_{X_f/\PP})}
={\prod_i(1+Q_i(H))\over
\prod_{\alpha>0}(1-\alpha(H)^2)\prod_{\beta}(1+d_{\beta}(H))},
\eeq
which implies $c_1(X_f)=\sum_iQ_i(H)-\sum_{\beta}d_{\beta}(H)$ and
\beq
{\rm td}(X_f)={\prod_iQ_i(H)\prod_{\alpha>0}
\left(2\sinh({\alpha(H)\over 2})\right)^2
\prod_{\beta}(1-\e^{-d_{\beta}(H)})\over
\prod_{\alpha>0}\alpha(H)^2\prod_{\beta}d_{\beta}(H)\prod_i(1-\e^{-Q_i(H)})}
\eeq
For a coroot $n\in{\rm Q}^{\vee}\subset\im \ttt$, we put
\beq
\wh{\Gamma}_{X_f}(n):=\wh{\rm A}_{X_f}\cdot
{\prod_{\beta}\Gamma\left(1+d_{\beta}\left({H\over 2\pi\im}+n\right)\right)
\over \prod_i\Gamma\left(1+Q_i\left({H\over 2\pi\im}+n\right)\right) }
\prod_{\alpha>0}\Gamma\!\left({\textstyle
1+\alpha\left({H\over 2\pi\im}+n\right)}
\right)
\Gamma\!\left({\textstyle
1-\alpha\left({H\over 2\pi\im}+n\right)}
\right).
\label{lookattt}
\eeq
It reduces to the Gamma class $\wh{\Gamma}_{X_f}$ at $n=0$.

The rule of brane descent is just as in the $U(1)$ case.
We shall denote the R-charge shift by $j$ by $M\mapsto M[j]$ ---
if the gauge group were Abelian
we could use the notation $M\mapsto M(j,0)$, but
that would not be appropriate for non-Abelian gauge group.
The brane $\mathfrak{B}=(M,Q)$ descends to the brane
$\mathfrak{B}_{\rm LV}=(M_L,Q_L)$ in the non-linear sigma model on $X_f$
where
\beqa
&&M_L~=~M\otimes {\rm Sym}\,F[2],\\
&&Q_L~=~Q(p_L,x)
\eeqa
where $p_L$ is the co-evaluation combined with the degree $2$ shift:
The component $p_L(v)$ for $v\in F$ is the multiplication by $v$ and the 
shift of the R-charge by $2$.
The Chern character of the image brane is
\beq
\ch(\mathfrak{B}_{\rm LV})={1\over
\prod_\beta\left(1-\exp\left(d_{\beta}(H)\right)\right)}
f_{\mathfrak{B}}\!\left({\textstyle {1\over 2\pi}H}\right)
\eeq
The theta angle shift is
\beq
2\pi B=\left(\theta+\pi\sum_{\beta}d_{\beta}\right)(H).
\label{Bgen}
\eeq

Now let us compute the partition function, which is given
in the R$^o$-frame by
\beqa
Z_{D^2}(\mathfrak{B})&=&(r\Lambda)^{\wh{c}_{\rm LV}/2}
\int_{\gamma}\dd^{l_G}\!\usigma'\,
\prod_{\alpha>0}\alpha(\usigma')\sinh(\pi\alpha(\usigma'))
\nn\\
&&~~~~~~~~~~~~~~
\times\prod_{\beta}\Gamma(-\im d_{\beta}(\usigma')+1)
\prod_i\Gamma(\im Q_i(\usigma'))
\e^{\im t_R(\usigma')}
f_{\mathfrak{B}}(\usigma').~~~~
\eeqa
with $\wh{c}=d_E-d_F-d_G$. The contour $\gamma$ should be poked
near $(Q_i(\usigma')=0)$'s to avoid poles that came down in the R$^o$
limit. Alternatively, we can uniformly shift $\gamma$
by $-\im \boldeps$ for some small $\boldeps\in \im \ttt$.
As in the $U(1)$ case, we would like to deform, or close, the contour
$\gamma$ so that we have a sum over residues. We may try to do it for
one coordinate after another, but that is not practical for high rank cases.
Fortunately, a machinery is developed for the situation like this. It is
called the {\it multivariable Jordan lemma} \cite{Tsikh,Passare}.

Let $C\subset \im\ttt$ be a cone with $l_G$ faces with $-\boldeps$
as its vertex,
and suppose $\mathcal{C}=\{{\rm Im}(\usigma')\in C\}$ is deep inside
the admissible region, i.e., the integrand decays exponentially
fast at infinity of $\mathcal{C}$.
We name the faces of ${\mathcal C}$ by $\{{\mathcal C}_a\}_{a=1}^{l_G}$.
We assume that the charges $\{Q_i\}$ are decomposed into $l_G$ groups,
$\{Q_i\}_{i\in I_a}$, for $a=1,\ldots, l_G$, so that the following condition
is satisfied. Let us define holomorphic functions
$f_1,\ldots, f_{l_G}$ of $\usigma'$ by 
${1\over f_a(\usigma')}:=\prod_{i\in I_a}\Gamma(\im Q_i(\usigma'))$.
Then the condition is that the divisor
$D_a:=(f_a=0)$ do not meet the face ${\mathcal C}_a$, for each $a$, and that
the intersection of $D_1,\ldots, D_{l_G}$
is a discrete point set $\im S\subset \ttt$ on the imaginary plane.
Under this situation,
the multivariable Jordan lemma says that
the integral is the sum of residues at $\usigma'=\im n$ for
$n\in C\cap S$,
\beqa
Z_{D^2}^{{}^{\rm LV}}(\mathfrak{B})&=&
(r\Lambda)^{\wh{c}_{\rm LV}\over 2}
\sum_{n\in C\cap S}\,
\oint\limits_{\,\,\,\,\,\,\gamma^{}_{\bf G}\!\!}
{\dd^{l_G} z\over (2\pi)^{l_G}} \,
\prod_{\alpha>0}\alpha\left({\textstyle \im n+{z\over 2\pi}}\right)
\sinh\left(\pi\alpha\left({\textstyle \im n+{z\over 2\pi}}\right)\right)
\nn\\
&&\!\!\!\!\!\!\!\!\!\!\!\!\!\!\!\!\!\!\!\!\!\!\!\!\!\!\!\!
\times\prod_{\beta}\Gamma\left({\textstyle 
-\im d_{\beta}\left(in+{z\over 2\pi}\right)
+1}\right)
\prod_i\Gamma\left({\textstyle \im Q_i\left(\im n+{z\over 2\pi}\right)}\right)
\exp\left({\textstyle \im t_R\left(\im n+{z\over 2\pi}\right)}\right)
f_{\mathfrak{B}}\!\left({\textstyle \im n+{z\over 2\pi}}\right).
\nn
\eeqa
The cycle $\gamma^{}_{\bf G}$, called the Grothendieck cycle, is
a small cycle of $z\in \ttt_{\C}$ defined
by the equation $|f_a(\im n+{z\over 2\pi})|=\varepsilon_a$ 
for all $a$, for some $0<\varepsilon_a\ll 1$.
One important property is that the integral does not depend on the
choice of $\varepsilon_a$'s.

At this point, we assume that the set $C\cap S$ is a subset of the coroot
lattice ${\rm Q}^{\vee}$, and denote it by ${\rm Q}^{\vee}_+$.
Then, we have $f_{\mathfrak{B}}\left(\im n+{z\over 2\pi}\right)=
f_{\mathfrak{B}}\left({z\over 2\pi}\right)$.
After some computation, we find
\beqa
Z_{D^2}^{{}^{\rm LV}}(\mathfrak{B})&=&
C(r\Lambda)^{\wh{c}_{\rm LV}\over 2}
\sum_{n\in {\rm Q}^{\vee}_+}\oint\limits_{\gamma_{\bf G}^{}\!\!\!}
{\dd^{l_G}z\over (2\pi\im)^{l_G}}\,
{\prod_{\alpha>0}\alpha(z)^2 \prod_{\beta}d_{\beta}(z)\over
\prod_i Q_i(z)}\,\wh{\Gamma}_{X_f}(n,z)
\nn\\
&&~~~~~~~~~~~~~~~~~~~~~~
\times
\e^{\half(\sum Q_i-\sum d_{\beta})(z)}
\e^{-t'_R(n)+{\im\over 2\pi}t'_R(z)}
{f_{\mathfrak{B}}\left({z\over 2\pi}\right)\over
\prod_{\beta}(1-\e^{d_{\beta}(z)})},~~~~
\eeqa
in which
$\wh{\Gamma}_{X_f}(n,z)=\wh{\Gamma}_{X_f}(n)|_{H\to z}$,
and 
\beq
t_R'=t_R-\pi\im\left(\sum_iQ_i-2\sum_{\beta}d_{\beta}\right).
\eeq
$C$ is a constant
$(-1)^{d_F}(-2\pi\im )^{d_E}(2\pi)^{-|\Delta_+|}\im^{d_G}$
($|\Delta_+|$ is the number of positive roots).

We would now like to convert each term into an integral over $X_f$
using an identity like (\ref{intXf}).
A generalization of (\ref{intXf}) to a possibly non-Abelian quotient
exists and is known as the
Jeffrey-Kirwan localization formula \cite{JK}:
\beq
\int_{X_f}g(H)=\int_{\PP}
\prod_{\beta}d_{\beta}(H) g(H)
=\oint\limits_{\gamma^{}_{\bf JK}\!\!\!}{\dd^{l_G}z\over (2\pi\im)^{l_G}}\,
{\prod_{\alpha>0}\alpha(z)^2\over
\prod_i Q_i(z)}\cdot \prod_{\beta}d_{\beta}(z)\cdot g(z),
\label{JK}
\eeq
where $\gamma_{\bf JK}^{}$ is a middle dimensional
homology class of the complement of $\prod_iQ_i(z)=0$, called the JK cycle.
The question is whether the integration over $\gamma_{\bf G}^{}$
and the one over $\gamma_{\bf JK}^{}$ are the same.
That is indeed the case in two examples which we will present below,
but we do not have a proof at the moment.
We simply assume this and proceed. 
Then we immediately see that the partition function can be written
as
\beq
Z_{D^2}^{{}^{\rm LV}}(\mathfrak{B})=
C(r\Lambda)^{\wh{c}_{\rm LV}\over 2}
\sum_{n\in {\rm Q}^{\vee}_+}
\int_{X_f}\e^{\half c_1(X_f)}\wh{\Gamma}_{X_f}(n)
\exp\left(-t'_R(n)+{\im\over 2\pi}t'_R(H)\right)
\ch(\mathfrak{B}_{\rm LV}).
\label{LVexpgen}
\eeq
In view of (\ref{Bgen}), we have
 $t_R'(H)=\omega_R-2\pi\im B+\pi\im c_1(X_f)$.
Using this, we may also rewrite the result as
\beq
Z_{D^2}^{{}^{\rm LV}}(\mathfrak{B})=
C(r\Lambda)^{\wh{c}_{\rm LV}\over 2}
\sum_{n\in {\rm Q}^{\vee}_+}\e^{-t_R'(n)}
\int_{X_f}\wh{\Gamma}_{\!X_f}\!(n)
\exp\left(B+{\im\over 2\pi}\omega_R\right)
\ch(\mathfrak{B}_{\rm LV}).
\label{LVexp2gen}
\eeq
In the large volume limit in this phase, only the $n=0$ term remains,
\beq
Z_{D^2}^{{}^{\rm LV}}(\mathfrak{B})
~\longrightarrow~
C(r\Lambda)^{\wh{c}_{\rm LV}\over 2}
\int_{X_f}\wh{\Gamma}_{X_f}
\exp\left(B+{\im\over 2\pi}\omega_R\right)
\ch(\mathfrak{B}_{\rm LV}).
\label{LVlimitgen}
\eeq
Again, this matches with the expected formula for the central charge
of the brane $\mathfrak{B}_{\rm LV}$.

Let us present two examples. These are simple enough so that 
one by one contour deformation can be done by hand. It is
instructive to do so and check that the multivariable Jordan lemma
gives the correct answer.

The first is the two parameter model considered in
Section~\ref{subsec:higher}. We are in Phase I.
$\{Q_i\}$ is the set of charges $\{(0,1),(1,0),(1,-2)\}$
 for $X_1,\ldots, X_6$.
As the shift, we can take $-\boldeps=(-\epsilon_1,-\epsilon_2)$ with
$\epsilon_1>0$, $\epsilon_2>0$ and $\epsilon_1>2\epsilon_2$.
The last condition comes from the requirement that $R_{X_6}>0$
before taking the R$^o$-frame limit.
\begin{figure}[htb]
\psfrag{u}{$\Is$}
\centerline{\includegraphics{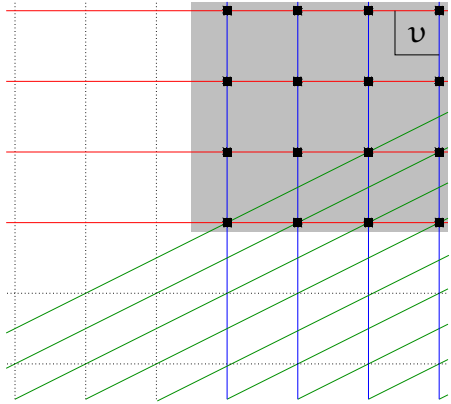}}
\caption{The cone and the poles: {\footnotesize 
The cone $C$ is the shaded region.
The poles for the charges $(0,1)$, $(1,0)$ and $(1,-2)$ are
shown as the red, blue and green lines respectively.
The red itself forms a group
while the blue and the green form the other group. The intersection of the two
groups are shown as the black dots. Note that the intersection only between
the blue and the green are not taken.}}
\label{fig:lines}
\end{figure}
As the cone $C$, we can take $C_{\rm I}-\boldeps$,
where $C_{\rm I}$ is the image cone of the map
$\Rs\mapsto\Is(\Rs)$ given in (\ref{ph1}). It is the first quadrant
shifted by $-\boldeps$. 
See Fig.~\ref{fig:lines}.
Let us regard the horizontal and vertical faces by
the first and the second respectively.
Let us group the charges of $X_i$'s so that
$\{(0,1)\}$ and $\{(1,0),(1,-2)\}$ are the first and the second groups
respectively.
Then, the grouping satisfies the condition for the multivariable Jordan
lemma. And $C\cap S$ is the first quadrant of the integral lattice
$\Z^{\oplus 2}$, that is, it is a subset of the coroot lattice
${\rm Q}^{\vee}=\Z^{\oplus 2}$.
The Grothendieck cycle is therefore
\beq
|z_2^2|=\varepsilon_1,\quad\,\,|z_1^3(z_1-2z_2)|=\varepsilon_2.
\eeq
On the other hand, the JK cycle is
\beq
|z_1|=\wt{\varepsilon}_1,\quad\,\, |z_2|=\wt{\varepsilon}_2,
\quad\,\,\wt{\varepsilon}_1\ll \wt{\varepsilon}_2.
\eeq
See for example, \cite{BEHT2}. If we choose $\varepsilon_1\gg\varepsilon_2$,
then, the two cycles are homotopic to each other.

The second example is the R\o dland model in the Grassmannian phase.
The set $\{Q_i\}$ is $\{(1,0),(0,1)\}$. We can take
$-\boldeps=(-\epsilon_1,-\epsilon_2)$ with arbitrary
positive $\epsilon_1$ and $\epsilon_2$ as the shift.
The cone $C$ is the first quadrant shifted by $-\boldeps$.
There is a unique grouping and the assumption of the lemma is trivially
satisfied. $C\cap S$ is again the first quadrant of the integral lattice
and hence is a subset of the coroot lattice. The Grothendieck cycle
and the JK cycle are the same, $|z_1|=|z_2|=\varepsilon$.

%\section{Annulus Partition Function}\label{sec:annulus}

\section{Factorization Of Two-Sphere Partition Function}
\label{sec:factorize}

We have collected a lot of evidence that
the parition function of the hemisphere
is equal to the central charge of the brane placed
at the boundary.
They agree whenever both can be computed, and the expressions in
some limits also match. This motivates us to conjecture
that this is the case in general:
\beq
Z_{D^2\!{}_{(+)}\!}(\mathfrak{B})
=\braGS|\mathfrak{B}\rangle_{{}_{\rm RR}},
\qquad
Z_{D^2\!{}_{(-)}\!}(\mathfrak{B})
={}_{{}_{\rm RR}}\!\langle \mathfrak{B}|\GSket.
\label{relrel}
\eeq
On the other hand, there is a conjecture
\cite{Romoetal} that the partition function on the whole sphere
is equal to the $00$ component of the tt$^*$ metric:
\beq
Z_{S^2}=\braGS|\GSket.
\label{Romoconj}
\eeq
If we admit these,
there is a certain relation between the partition functions on the
whole sphere and the hemisphere.
For any basis $\{|a\rangle\}_{a=1}^{\mu}$ 
of the space of supersymmetric ground states,
we have
\beq
\braGS|\GSket=\sum_{a,b=1}^{\mu}\braGS|a\rangle g^{ab}\langle b|\GSket,
\label{abf}
\eeq
where $(g^{ab})$ is inverse to the matrix $(\langle a|b\rangle)$.
Suppose there are $\mu$ D-branes
$\{\mathfrak{B}_i\}_{i=1}^{\mu}$ whose boundary states have components
which span the space of supersymmetric ground states. That is, the square
matrix $({}_{{}_{\rm RR}}\!\langle \mathfrak{B}_i|a\rangle)$ is invertible.
Then, we may use the ground state components of
$|\mathfrak{B}_i\rangle_{{}_{\rm RR}}$'s as a new basis
and obtain the formula like (\ref{abf}).
In the place of $g^{ab}$ we have the inverse to
\beq
{}_{{}_{\rm RR}}\!\langle\mathfrak{B}_i|P_G
|\mathfrak{B}_j\rangle_{{}_{\rm RR}},
\label{Lij}
\eeq
where $P_G$ is the orthogonal projection to the
space of supersymmetric ground states.
The matrix element (\ref{Lij}) can be represented by the partition function
on the infinitely long cylinder in which the fields including fermions 
are all periodic along the circle direction. In fact, by the supersymmetry,
the length and the thickness of the cylinder does not matter.
So, it is just a cylinder partition function of any size.
Viewed from the open string channel, (\ref{Lij})
is the open string Witten index,
\beq
\chi(\mathfrak{B}_i,\mathfrak{B}_j)
\,:=\,{\rm Tr}^{}_{{\mathcal H}_{\mathfrak{B}_i,\mathfrak{B}_j}}
(-1)^F\e^{-\beta H},
\eeq
where ${\mathcal H}_{\mathfrak{B}_i,\mathfrak{B}_j}$ is the space of states
of the open string with the boundary conditions $\mathfrak{B}_i$
and $\mathfrak{B}_j$ on the left and the right
ends of the string. $H$ and $F$ are the Hamiltonian
and a fermion number operator. Given (\ref{relrel}) and (\ref{Romoconj}),
we must have
\beq
Z_{S^2}=\sum_{i,j}
Z_{D^2\!\!{}_{(+)}\!}(\mathfrak{B}_i)\,\chi^{ij}\,
Z_{D^2\!\!{}_{(-)}\!}(\mathfrak{B}_j),
\label{factorization}
\eeq
where $\chi^{ij}$ is the inverse to $\chi(\mathfrak{B}_i,\mathfrak{B}_j)$.
In this section, we shall examine whether
this factorization equation holds.

\subsection{The Sphere}

First let us write down the formula for the two-sphere
partition function.
The result of \cite{Beninietal,Doroudetal} is essentially as follows:
\beqa
Z_{S^2}&=&(r\Lambda)^{\wh{c}}\sum_{m\in {\rm Q}^{\vee}}
\int\limits_{\mbox{$\im\ttt$}}\dd^{l_G}\usigma'\,
\exp\Bigl(\,2\,\im\, \zeta_R(\usigma')+\im\,(\theta+2\pi\rho)(m)\Bigr)\\
&&~~~~~~~~~~~~~~~~~~
\times\prod_{\alpha>0}\left({\alpha(m)^2\over 4}+\alpha(\usigma')^2\right)
\prod_i
{\Gamma\left(\im Q_i(\usigma')-{Q_i(m)\over 2}+{R_i\over 2}\right)\over
\Gamma\left(1-\im Q_i(\usigma')-{Q_i(m)\over 2}-{R_i\over 2}\right)}.\nn
\label{ZS2}
\eeqa
We say ``essentially'' because we have done one modification:
a shift of the theta angle,
\beq
\theta~\longrightarrow~\theta+\pi\sum_{\alpha>0}\pm\alpha\,\,\equiv\,\,
\theta+2\pi\rho\quad\,\,\mbox{mod $2\pi{\rm P}$}.
\label{thetamodi}
\eeq
Note that the choice of sign assignment $\pm\alpha$
does not matter since a root is always a weight $\alpha\in {\rm P}$
(so that it takes integer values on coroots $m\in {\rm Q}^{\vee}$).
$\rho$ is half the sum of positive roots,
$\rho:=\half\sum_{\alpha>0}\alpha$, which
may fail to land on the weight lattice ${\rm P}$
depending on the group $G$.
For example, for a $U(k)$ gauge theory this matters if and only if $k$ is even.
As we will see, this is needed for the factorization.
Necessity of the same modification is also noticed in \cite{HPT}
from a different point of view.
The factor $(r\Lambda)^{\wh{c}}$ is not in \cite{Beninietal,Doroudetal}
but is noticed by the authors of these papers, \cite{LeeTalks,privatecomm}.

\subsection{The Annulus}

Next, we dicuss the open string Witten index
$\chi(\mathfrak{B}_i,\mathfrak{B}_j)$, or equivalently,
the cylinder, or annulus, partition function.
At this moment, we do not have a complete results concerning the computation,
but let us make some preliminary remarks.

We may try to apply the localization, sending
the gauge coupling to zero and the K\"ahler metric of the matter to infinity.
However, that is plagued  by the presence of bosonic as well as fermionic
zero modes. It is similar to
the situation of the elliptic genus \cite{BEHT1,BEHT2} but it is worse
than that. In the case of elliptic genus, we have, by definition,
 the twist by R-symmetry, which usualy separates
the singular loci for ``positively charged'' and 
the ``negatively charged'' matter fields. That separation made it possible to
justify a certain manipulation of the path integral.
For the case of open string Witten index, we do not have that, so that
singular loci may collide and cannot be separated.
That makes the justification of computation based on
the free approximation difficult. 
But we may hope that there is a way to justify it some way,
and try to see if we obtain a reasonable answer.

The annulus partition function of each multiplet
in the free approximation is straightforward.
We choose the real boundary condition (\ref{realL}) for the vector multiplet,
so that we may need to consider only grade restricted branes.
For the matter sector, the computation is almost done in \cite{HHP}.
It immediately gives the result in the operator formalism
but the mode expansion presented there can also be used for the path integral.
The result is
\beq
Z_{\rm chiral}={1\over \prod_i2\sinh\left({Q_i(u)\over 2}\right)}.
\eeq
where $u$ parametrizes the bosonic zero mode of the vector multiplet,
\beq
u\,=\,\beta\sigma_1-\im a\,\in \ttt_{\C}/2\pi\im {\rm Q}^{\vee}.
\eeq
In the last expression,
$\beta$ is the circumference of the annulus,
$\sigma_1$ is the scalar zero mode and $a\in \im\ttt/2\pi {\rm Q}^{\vee}$
parametrizes the gauge holonomy along the circle.
The one for the vector multiplet can also be computed.
The W-boson pair with the roots $\pm\alpha$ yields
\beq
Z_{{\rm vector},\alpha}=
\left({\textstyle 2\sinh\left({\alpha(u)\over 2}\right)}\right)^2.
\eeq
The path integral is presented as the integration over the whole moduli space
\beq
(\ttt_{\C}/2\pi\im {\rm Q}^{\vee})/W_G
\label{modannulus}
\eeq
of the vector multiplet
bosonic zero modes.
A proper treatment of the bosonic zero modes from the matter and
the fermionic zero modes from the vector may results in an expression
of the integrand as a total derivative, which by Stokes theorem
leads to the integration over a
lower dimensional subspace, just as in \cite{BEHT1,BEHT2}.
This and some evidences which we will describe below
motivates us to make the following conjecture:
The annulus partition function is given by a contour integral
\beq
\chi(\mathfrak{B}_1,\mathfrak{B}_2)
={1\over |W_G|}\int\limits_{\Gamma}
{\dd^{l_G} u\over (2\pi\im )^{l_G}}
{ \prod_{\alpha>0}\left(2\sinh\left(
{\alpha(u)\over 2}\right)\right)^2\over
\prod_i2\sinh\left(
 {Q_i(u)\over 2}\right)}
f_{\mathfrak{B}_1}\left({\textstyle -{u\over 2\pi}}\right)
f_{\mathfrak{B_2}}\left({\textstyle {u\over 2\pi}}\right)
\label{annulus}
\eeq
where $f_{\mathfrak{B}}({u\over 2\pi})$ is the brane factor in the R$^o$-frame,
\beq
f_{\mathfrak{B}}\left({\textstyle {u\over 2\pi}}\right)
={\rm Str}_M^{}\rho\left(\e^{u}\right),
\eeq
and
$\Gamma\subset \ttt_{\C}/2\pi\im {\rm Q}^{\vee}$ is {\it some} middle
dimensional cycle which represents a homology class of the complement
of the divisor $\prod_i\sinh(Q_i(u)/2)=0$.

Let us comment on some anomaly, which was already noticed in \cite{HHP}.
The integrand of (\ref{annulus}) is not always single valued on the moduli
space (\ref{modannulus}). If one shifts $u$ by $2\pi\im n$ with
$n\in {\rm Q}^{\vee}$, then, the integrand changes by a sign,
$(-1)^{\sum_iQ_i(n)}$. The integrand is single valued if and only if
the sum of weights is even,
\beq
\sum_iQ_i\in 2{\rm P}.
\eeq
If the theory has a usual geometric phase, with a target K\"ahler manifold
$X$, this is equivalent to the condition that $c_1(X)$ is even, 
in other words, $X$ admits a spin structure.

Let us describe some evidences for the conjecture.
We first consider the $U(1)$ theories. We assume $N-d$ is even.
The formula is written as
\beq
I_{\Gamma}=\int\limits_{\Gamma}{\dd u\over 2\pi\im}
{f_{\mathfrak{B}_1}(-{u\over 2\pi})f_{\mathfrak{B}_2}({u\over 2\pi})\over 
\left(\e^{u\over 2}-\e^{-{u\over 2}}\right)^N
\left(\e^{-d{u\over 2}}-\e^{d{u\over 2}}\right)},
\eeq
where $\Gamma$ is some cycle in $\C/2\pi\im\Z$
minus the pole location which
is $\{\e^{2\pi\im n/d}\}_{n=0}^{d-1}$. It can be rewitten as
\beq
I_{\Gamma}=\int\limits_{\Gamma}{\dd u\over 2\pi\im}{1\over u^N}\cdot du\cdot
{u^{N-1}\left(\e^{du\over 2}-\e^{-{du\over 2}}\right)\over
d\left(\e^{u\over 2}-\e^{-{u\over 2}}\right)^N}
{f_{\mathfrak{B}_1}(-{u\over 2\pi})
\over \left(1-\e^{-du}\right)}
{f_{\mathfrak{B}_2}({u\over 2\pi})\over 
\left(1-\e^{du}\right)}
\eeq
Suppose the cycle is the small contour $\gamma_0$ around $u=0$. Then,
we can use the identity (\ref{intXf}) to write it as an
integration over $X_f$ and in fact it is nothing but
\beq
I_{\gamma_0}\,=\,\int_{X_f}\wh{\rm A}_{X_f}\,
\ch(\mathfrak{B}_{1\,{\rm LV}})^{\vee}\,
\ch(\mathfrak{B}_{2\,{\rm LV}}).
\label{here21}
\eeq
(For a $2i$ form $\omega$ we define $\omega^{\vee}:=(-1)^i\omega$.)
This is indeed an expected answer in the geometric phase.
For the Witten index, we may employ the zero mode approximation.
In the zero mode sector, open string states are spinors valued 
in $Hom(E_1,E_2)$, where $E_i$ is the vector bundle
for $\mathfrak{B}_{i\,{\rm LV}}$
 and a linear combination of the
supercharges is essentially the Dirac operator. Therefore, the Witten index
is the Dirac index gievn by the Atiyah-Singer formula (\ref{Diracindex}),
which is (\ref{here21}) in the present context.
To be more precise, the geometric phase can represent the full theory 
only for $d\leq N$. So, we obtain the expected correct answer 
if we choose $\Gamma=\gamma_0$ in the case $d\leq N$.

This can be generalized to any theory with the usual
geometric phase. In the set up of Section~\ref{subsub:genLV},
assuming $\sum_iQ_i-\sum_{\beta}d_{\beta}$ is even, if we take
the JK cycle near $u=0$,
$\Gamma=\gamma^{}_{\bf JK}$, then the same computation
yields the Dirac-type index,
\beq
I_{\gamma_{\bf JK}^{}}\,=\,\int_{X_f}\wh{\rm A}_{X_f}\,
\ch(\mathfrak{B}_{1\,{\rm LV}})^{\vee}\,
\ch(\mathfrak{B}_{2\,{\rm LV}}).
\label{here22}
\eeq

Let us come back to the $U(1)$ theory. Recall that
there are also $(d-1)$ poles at $u=\e^{2\pi\im n/d}$.
If we start from $\Gamma=\gamma_0$ and deform it, provided the behaviour
${\rm Re}(u)\to \pm\infty$ is good enough, we can arrive at
the $(d-1)$ small cycles around these poles, with the clockwise orientation.
Since each is a simple pole, it is easy to evaluate the residues.
The result is
\beq
\sum_{n=1}^{d-1}I_{-\gamma_n}~
~
={1\over d}\,\,\sum_{n=1}^{d-1}\e^{\pi\im (d-N)n\over d}
{f_{\mathfrak{B}_1}\left(-{\im n\over d}\right)
f_{\mathfrak{B}_2}\left({\im n\over d}\right)\over 
\left(1-\e^{-{2\pi\im n\over d}}\right)^N}
\eeq
When $d=N$ this is precisely the open string Witten index 
in the Landau-Ginzburg orbifold \cite{WalcherLG}.
When $d<N$ it does not agree with that. Indeed, we do not expect an agreement
since the Landau-Ginzburg orbifold is only a part of the whole theory.
However, some of the branes descends purely to the Landau-Ginzburg orbifold.
It would be interesting to see if the above gives the correct answer
for a pair of such branes.
When $d>N$ it {\it is} the whole theory, but the starting choice
$\Gamma=\gamma_0$ would not be the right choice in general since it
gives the formula in the non-linear sigma model,
which is only a part of the theory.

We would like to make a final comment on the behaviour at
${\rm Re}(u)\to\pm\infty$ in this $U(1)$ theory.
The charge $(q^{(1)},q^{(2)})$ term of the integrand behaves as
\beq
{\rm integrand}_{q^{(1)},q^{(2)}}\longrightarrow 
\exp\left(-q^{(1)}u+q^{(2)} u-{N+d\over 2}|u|\right)\quad
\mbox{as ${\rm Re}(u)\to \pm\infty$}.
\eeq
A good behaviour is guaranteed only if
\beq
\left|\,q^{(1)}-q^{(2)}\,\right|~<~{N+d\over 2}
\eeq
for any pair of Chan-Paton charges of $\mathfrak{B}_1$ and 
$\mathfrak{B}_2$. In the Calabi-Yau case, $d=N$, if the two
branes are grade  restricted with respect to a common window,
$-{N\over 2}<{\theta\over 2\pi}+q^{(a)}_{j_a}<{N\over 2}$, $a=1,2$,
then the above condition is indeed satisfied.

\subsection{Factorization}

Let us now come back to the question of factorization.
Since we do not yet know the general formula for the annulus,
we cannot make the most general check at this moment.
However, we do know the formula for the theory with a usual
geometric phase --- it is given by the Dirac index (\ref{here21})
and (\ref{here22}).
So, we shall test the factorization in such theories.

\newcommand{\bn}{\overline{n}}
\newcommand{\bt}{\overline{t}}

Let us first examine the $U(1)$ theory introduced in Section~\ref{subsec:U1}.
We shall only consider the case $d\leq N$ where the large volume
expression (\ref{LVexp}) is an expansion of the full partition function.
In this case, the formula (\ref{ZS2}) is
\beqa
Z_{S^2}&=&\sum_{m\in\Z}\,\,\,\int\limits_{\R-\im 0}
\dd\usigma'\e^{2\im \zeta_R\usigma'+\im\theta m}
{\Gamma\left(1-\im d\usigma'+{dm\over 2}\right)\over
\Gamma\left(\im d\usigma'+{dm\over 2}\right)}
{\Gamma\left(\im\usigma'-{m\over 2}\right)^N\over
\Gamma\left(1-\im\usigma'-{m\over 2}\right)^N}
\eeqa
We look at the geometric regime $\zeta_R\gg 0$ in which 
the integrand decays exponentially fast
 in the positive imaginary direction.
 due to the factor $\e^{2\in\zeta_R\usigma'}$.
Then, we can bend both ends of the contour upwards
and we only have to take the poles on the upper half plane.
The poles are at
\beq
~~~~~~~~~~~
\im\usigma'-{m\over 2}=-l\,;\qquad l\geq 0,\,\,\,l\geq m.
\eeq
They come from the factor $\Gamma(\im\usigma'-{m\over 2})^N$.
The condition $l\geq m$ is to omit the poles
which are cancelled by the zeroes from the gamma functions in the denominator.
The other gamma function in the numerator may have poles on
the upper half plane but they are all
cancelled from the other gamma function in the denominator.
With the reparametrization $l=n$, $m=n-\bn$, the condition $l\geq 0,m$
becomes $n, \bn\geq 0$.
If we shift the integration variable as
 $\usigma'=\im\left(l-{m\over 2}\right)+{z\over 2\pi}$ at each pole, we have
\beq
Z_{S^2}=\sum_{n,\bn\geq 0}\oint\limits_0{\dd z\over 2\pi}\,
\e^{-t_Rn-\bt_R\bn+\im(t_R+\bt_R){z\over 2\pi}}
{\Gamma\left(1+dn+{dz\over 2\pi\im}\right)\over
\Gamma\left(-d\bn-{dz\over 2\pi\im}\right)}
{\Gamma\left(-n-{z\over 2\pi\im}\right)^N\over
\Gamma\left(1+\bn+{z\over 2\pi\im}\right)^N}.
\label{exrrrr}
\eeq
On the other hand, we use the large volume formula
(\ref{LVexp}) for the hemisphere partition function,
\beqa
Z_{D^2\!{}_{(+)}\!}(\mathfrak{B}_i)&=&
\sum_{n=0}^{\infty}
\int_{X_f}\e^{{N-d\over 2}H}\wh{\Gamma}_{X_f}(n)
\exp\left(-nt'_R+{\im\over 2\pi}t'_R H\right)
\ch(\mathfrak{B}_{i{\rm LV}}),\\
Z_{D^2\!{}_{(-)}\!}(\mathfrak{B}_j)&=&
\sum_{\bn=0}^{\infty}
\int_{X_f}\e^{-{N-d\over 2}H}\wh{\Gamma}_{X_f}(\bn)
\exp\left(-\bn\bt'_R+{\im\over 2\pi}\bt'_R H\right)
\ch(\mathfrak{B}_{j{\rm LV}})^{\vee}.\label{formmm}
\eeqa
$\bt_R'$ is the complex conjugate of $t_R'$.
The latter expression (\ref{formmm}) is obtained from the former by using
(\ref{ccreln})
and the sign change of $H$. Note that $\wh{A}(-x)=\wh{A}(x)$.
We ignore overall nemerical factors.
To evaluate the right hand siade of (\ref{factorization}),
we employ the identity
\beq
\sum_{i,j}\int_{X_f}\omega\, \ch(\mathfrak{B}_{i{\rm LV}})\,\,\chi^{ij}
\int_{X_f}\eta\,\ch(\mathfrak{B}_{j{\rm LV}})^{\vee}
=\int_{X_f}\omega\,{1\over\,\wh{\rm A}_{X_f}\!\!}\,\eta
\label{idR}
\eeq
Then the right hand side is
\beq
{\rm RHS}=\sum_{n,\bn\geq 0}\,\,\int_{X_f}
{\wh{\Gamma}_{X_f}\!(n)\,\wh{\Gamma}_{X_f}\!(\bn)\over \,\wh{\rm A}_{X_f}\!\!}
\exp\left(-nt_R'-\bn\bt_R'+{\im\over 2\pi}
(t_R'+\bt_R')H\right)
\label{intermP}
\eeq
Recalling the definition (\ref{lookat}) and (\ref{tRp}),
after some computation using the gamma function identity (\ref{Gammaid4}),
we find
\beq
{\rm RHS}=\sum_{n,\bn\geq 0}\oint\limits_0{\dd z\over 2\pi}\,
(-1)^{(N-d)\bn}
\e^{-t_Rn-\bt_R\bn+\im(t_R+\bt_R){z\over 2\pi}}
{\Gamma\left(1+dn+{dz\over 2\pi\im}\right)\over
\Gamma\left(-d\bn-{dz\over 2\pi\im}\right)}
{\Gamma\left(-n-{z\over 2\pi\im}\right)^N\over
\Gamma\left(1+\bn+{z\over 2\pi\im}\right)^N}.
\eeq
This agrees with the expression (\ref{exrrrr}) when
$(N-d)$ is even, i.e., when $X_f$ is a spin manifold, which is the case 
where the Dirac index makes sense.

We next consider the more general theory with a geometric phase
from Section~\ref{subsub:genLV}. We take over the assumptions made
in that section (which are confirmed in the examples).
The two sphere partition function is
\beqa
Z_{S^2}&=&\sum_{m\in {\rm Q}^{\vee}}
\int\limits_{\mbox{$\im\ttt$}-\im 0}\dd^{l_G}\usigma'\,
\exp\Bigl(\,2\,\im\, \zeta_R(\usigma')+\im\,(\theta+2\pi\rho)(m)\Bigr)\\
&&
\times\prod_{\alpha>0}\left({\alpha(m)^2\over 4}+\alpha(\usigma')^2\right)
\prod_{\beta}
{\Gamma\left(1+d_{\beta}(-\im\usigma'+{m\over 2})\right)\over
\Gamma\left(d_{\beta}(\im\usigma'+{m\over 2})\right)}
\prod_i
{\Gamma\left(Q_i(\im \usigma'-{m\over 2})\right)\over
\Gamma\left(1+Q_i(-\im\usigma'-{m\over 2})\right)}.\nn
\eeqa
We deform the contour in the direction of the cone ${\mathcal C}$.
By the multi-dimensional Jordan lemma, we only have to take
the poles at
$\im\usigma'-{m\over 2}=-l$ with $l\in {\rm Q}^{\vee}_+$,
but we also need to omit the poles that are cancelled by the zeroes
from the gamma function on the denominator.
We assume that it can be done by requiring $l-m\in {\rm Q}^{\vee}_+$.
We also assume that the gamma function factors from the $P$-fields
do not have poles that contribute to this integral.
We do not have a proof of these claims,
although these indeed hold in the examples. To summarize, we take poles at
\beq
~~~~~~~~~~~
\im\usigma'-{m\over 2}=-l\,;\quad,\,\, l\in {\rm Q}^{\vee}_+,\,\,\,
l-m\in {\rm Q}^{\vee}_+.
\eeq
With the same reparametrization of $l$ and $m$ and
the shift of integration variables as in the $U(1)$ theory,
we find that the $S^2$ partition function can be written as
\beqa
Z_{S^2}&=&(-1)^{|\Delta_+|}\sum_{n,\bn\in {\rm Q}^{\vee}_+}\,
\oint\limits_{\gamma_{\bf G}^{}\!\!\!\!\!}{\dd^{l_G}z\over (2\pi)^{l_G}}
\e^{-(t_R-2\pi\im \rho)(n)-(\bt_R+2\pi\im\rho)(\bn)
+{\im\over 2\pi}(t_R+\bt_R)(z)}\label{ZS2nonAb}\\
&&\times
\prod_{\alpha>0}
\alpha\left({\textstyle n+{z\over 2\pi\im}}\right)
\alpha\left({\textstyle \bn+{z\over 2\pi\im}}\right)
\prod_{\beta}{\Gamma\left(1+d_{\beta}(n+{z\over 2\pi\im})\right)
\over
\Gamma\left(d_{\beta}(-\bn-{z\over 2\pi\im})\right)}
\prod_i{\Gamma\left(Q_i(-n-{z\over 2\pi\im})\right)\over
\Gamma\left(1+Q_i(\bn+{z\over 2\pi\im})\right)}\nn.
\eeqa
On the other hand, we use the expression (\ref{LVexpgen}) 
for the hemisphere partition function in the geometric phase. 
Using the identity (\ref{idR}), we see that the right hand side of 
(\ref{factorization}) can be written in the same way as
(\ref{intermP}) where the sum is over $n,\bn\in {\rm Q}_+^{\vee}$
and the exponent is $-t'_R(n)-\bt'_R(\bn)+{\im\over 2\pi}(t'_R+\bt'_R)(H)$.
Applying the Jeffrey-Kirwan fomula (\ref{JK}) and
after some computation using the identity (\ref{Gammaid4}),
we find
\beqa
{\rm RHS}\!\!\!&=&\!\!\!{\rm const}\!\!\!\sum_{n,\bn\in{\rm Q}^{\vee}_+}\,
\oint\limits_{\gamma^{}_{\bf JK}\!\!\!\!\!}
{\dd^{l_G}z\over (2\pi)^{l_G}}
\e^{-t_R(n)-\bt_R(\bn)
+{\im\over 2\pi}(t_R+\bt_R)(z)}(-1)^{2\rho(n+\bn)+
(\sum_iQ_i-\sum_{\beta}d_{\beta})(\bn)}\\
&&\times\prod_{\alpha>0}
\alpha\left({\textstyle n+{z\over 2\pi\im}}\right)
\alpha\left({\textstyle \bn+{z\over 2\pi\im}}\right)
\prod_{\beta}{\Gamma(1+d_{\beta}\left(n+{z\over 2\pi\im})\right)
\over
\Gamma\left(d_{\beta}(-\bn-{z\over 2\pi\im})\right)}
\prod_i{\Gamma(Q_i\left(-n-{z\over 2\pi\im})\right)\over
\Gamma\left(1+Q_i(\bn+{z\over 2\pi\im})\right)}.\nn
\eeqa
The sign $(-1)^{2\rho(n+\bn)}$ comes out during the
process of the following type,
$$
{\textstyle
\sin\left(\pi\alpha\left(n+{z\over 2\pi\im}\right)\right)
=(-1)^{\alpha(n)}
\sin\left(\pi\alpha\left({z\over 2\pi\im}\right)\right)}.
$$
We see that it agrees with (\ref{ZS2nonAb}) up to constant,
provided  $X_f$ is a spin manifold,
$c_1(X_f)\equiv 0$ mod 2, 
so that $\sum_iQ_i-\sum_{\beta}d_{\beta}$ takes even numbers
on the coroot lattice.
And we see why the shift (\ref{thetamodi}) is needed in order for the
factorization to work out.

\section{Mirror Symmetry}

In this final section, we use one more property of
the gamma function. That is,
the Euler integral of the second kind,
\beq
~~~~~~~~~~
\Gamma(z)=\int_0^{\infty}\e^{-t}\,t^{z-1}\,\dd t,\qquad {\rm Re}(z)>0.
\label{gamma4}
\eeq
which is usually used as the {\it definition} of the gamma function.

For convenience, let us write once again the formula for the
hemisphere partition function,
\beqa
Z_{D^2}(\mathfrak{B})\!&=&\!(r\Lambda)^{\wh{c}/2}
\int\limits_{\gamma}
\dd^{l_G}\usigma'
\prod_{\alpha>0}\alpha(\usigma')\sinh(\pi \alpha(\usigma'))
\prod_{i}\Gamma\left(\im Q_i(\usigma')+{R_i\over 2}\right)
\nn\\
&&~~~~~~~~~~~~~~~~~~~~\times
\exp\left(\im t_{\rm R}(\usigma')\right)
\sum_j\e^{\pi\im r_j}\e^{2\pi q_j(\usigma')}.
\label{again}
\eeqa
Let us apply (\ref{gamma4}) to the gamma function factor in 
(\ref{again}). Using the variable $\e^{-y'_i}$ instead of $t$,
we have
\beq
\Gamma\!\left(\im Q_i(\usigma')+{R_i\over 2}\right)
\,\,=\,\,\int_{-\infty}^{\infty}\dd y'_i\,
\exp\left(-y_i'\left(\im Q_i(\usigma')
+{R_i\over 2}\right)-\e^{-y_i}\right)
\eeq
which is valid when ${\rm Im}(Q_i(\usigma'))<{R_i\over 2}$.
Using this, we can write (\ref{again}) as
\beq
Z_{D^2}(\mathfrak{B})=
(r\Lambda)^{\wh{c}/2}\sum_{\varepsilon,j}
\left(\prod_{\alpha>0}{\varepsilon_{\alpha}\over 2}\right)
\e^{\pi\im r_j}
\int\limits_{\mbox{$\gamma\times \R^{d_V}$}\!\!\!\!\!\!\!\!\!}
%\limits_{\gamma\times \R^{d_V}}
\dd^{l_G}\usigma'\,\dd^{d_V}\!y'\,
\prod_{\alpha>0}\alpha(\usigma')\,\cdot\,\delta'\,
\cdot\,
\e^{F_{\varepsilon, q_j}(\usigma',y')}
\eeq
where the sum is over $j$ and the choice of
$\varepsilon_{\alpha}=\pm 1$ for each $\alpha>0$,
$\delta':=\prod_i\e^{-y_i'R_i/2}$ and
\beq
F_{\varepsilon,q_j}:=\im t_R(\usigma')-\im\sum_iy_i'Q_i(\usigma')
-\sum_i\e^{-y_i'}+\sum_{\alpha>0}\varepsilon_{\alpha}\pi\alpha(\usigma')
+2\pi q_j(\usigma').
\eeq 
Recalling $t_R=t-\sum_iQ_i\log(r\Lambda)$ and shifting the variables as 
$y_i'=y_i-\log(r\Lambda)$, we find that the partition function can be
rewritten as
\beqa
Z_{D^2}(\mathfrak{B})&=&
{(r\Lambda)^{d_V+l_G\over 2}\over
\Lambda^{d_G+l_G\over 2}}
\sum_{\varepsilon,j}
\left(\prod_{\alpha>0}{\varepsilon_{\alpha}\over 2}\right)
\e^{\pi\im r_j}\nn\\
&&
\times\int\limits_{\mbox{$\wt{\gamma}\times \R^{d_V}$}\!\!\!\!\!\!\!\!\!}
\dd^{l_G}\usigma\,\,\dd^{d_V}\!y\,
\prod_{\alpha>0}\alpha(\usigma)\,\cdot\,\delta\,
\cdot\,
\exp\left(-2\pi r\im \wt{W}_{\!\varepsilon, q_j}(\usigma,y)\right)
\eeqa
where
\beq
\delta=\prod_i\exp\left(-{R_i\over 2}y_i\right)
\eeq
and
\beq
2\pi\wt{W}_{\!\varepsilon, q_j}
=\left(\sum_iQ_iy_i-t_{\varepsilon, q_j}\right)(\usigma)
+(-\im \Lambda)\sum_i\exp\Bigl(\,-y_i\,\Bigr),
\eeq
in which $t_{\varepsilon,q_j}=\zeta-\im \theta_{\varepsilon,q_j}$
with 
\beq
\theta_{\varepsilon, q_j}=\theta+2\pi q_j+\sum_{\alpha>0}\varepsilon_{\alpha}
\pi\alpha
\eeq
This is valid when the contour $\wt{\gamma}$ lies in the region with
${\rm Im}(\usigma)<{R_i\over 2r}$. For convergence of the integral,
we may need to consider only the grade restricted branes.

This is the same as, or more precisely, similar to the expression for
the D-brane central charge found in \cite{HV} during
the derivation of mirror symmetry.
$2\pi\wt{W}_{\varepsilon,q}$ is essentially the mirror superpotential
found in \cite{HV}.
The factor $\delta$ is the factor found in \cite{HV} following
\cite{MP}, also denoted
by $\delta$, which is required
if there is a tree level superpotential in the original side.
The factor $\prod_{\alpha>0}\alpha(\usigma)$ is also in \cite{HV}.
We say ``essentially'', because they are not the same, even modulo
$2\pi \im {\rm P}(\usigma)$, because of the shift
$\sum_{\alpha>0}\pm\pi\alpha(\usigma)$ of the theta angle.
This is a simple mistake in \cite{HV}.
More importantly, even the integral part from $2\pi\im {\rm P}$
matters. Our formula shows precisely
how to fix this integral part
and then how to sum over the integrals with appropriate signs/phases,
depending on the choice of D-brane.

Our formula may be used as a string point to find explicit correspondence
between B-branes in the linear sigma model and A-branes in the mirror theory,
at least at the level of Ramond-Ramond charge.
We leave this problem for future works.

\section*{Acknowledgement}

We would like to thank
Matthew Ballard,
Francesco Benini,
Nima Doroud,
Richard Eager,
Jaume Gomis,
David Favero,
Bruno Le Floch,
Daniel Halpern-Leistner,
Simeon Hellerman,
Shinobu Hosono,
Daniel Jafferis,
Ludmil Katzarkov,
Johanna Knapp,
Maxim Kontsevich,
Sungjay Lee,
Todor Milanov,
Dave Morrison,
Hirosi Ooguri,
Chan Y. Park,
Daniel Pomerleano,
Yongbin Ruan,
Kyoji Saito,
Ed Segal,
Yuji Tachikawa,
Yukinobu Toda
and
Masahito Yamazaki for discussions, conversations, instructions,
and encouragement.

This work is supported by JSPS Grant-in-Aid for Scientific Research
No. 21340109 and WPI Initiative, MEXT, Japan at Kavli IPMU,
the University of Tokyo.

\appendix{Conventions}\label{app:spinors}

\subsection{Spinors On A Two-Manifold}

A two-dimensionsal oriented Riemannian manifold $(\Sigma, g)$ has a 
natural complex structure. 
The holomorphic and antiholomorphic cotangent bundles
are isomorphic as unitary bundles to the anti-holomorphic and
holomorphic tangent bundles,
$K_{\Sigma}\cong\overline{T}_{\Sigma}$,
$\overline{K}_{\Sigma}\cong T_{\Sigma}$.
A spin structure defines square roots of these bundles,
$S_-=\sqrt{K}_{\Sigma}\cong\sqrt{\overline{T}}_{\Sigma}$ and
$S_+=\sqrt{\overline{K}}_{\Sigma}\cong\sqrt{T}_{\Sigma}$. We
assume them be dual to each other.
The total spin bundle is the direct sum,
$S=S_-\oplus S_+$.
A local complex coordinate $z$ of $\Sigma$ yields a local frame
$(\sqrt{\dd z},\sqrt{\dd \bz})$ of $S$, 
with respect to which the gamma matrices are expressed as
\beq
\gamma^z\stackrel{\cdot}{=}\left(\begin{array}{cc}
0&(2g^{z\bz})^{\half}\\
0&0
\end{array}\right),\quad
\gamma^{\bz}\stackrel{\cdot}{=}\left(\begin{array}{cc}
0&0\\
(2g^{z\bz})^{\half}&0
\end{array}\right).
\eeq
The chirality operator $\gamma_3$ is defined to have the expression
\beq
\gamma_3\stackrel{\cdot}{=}\left(\begin{array}{cc}
1&0\\
0&-1
\end{array}\right),
\eeq
that is, $\gamma_3=+1$ on $S_-$ and $-1$ on $S_+$.
(We hope that this is not too confusing.)
The natural projections $P_{\mp}:S\to S_{\mp}$ have expressions
\beq
P_-={1+\gamma_3\over 2}\stackrel{\cdot}{=}
\left(\begin{array}{cc}
1&0\\
0&0
\end{array}\right),\qquad
P_+={1-\gamma_3\over 2}\stackrel{\cdot}{=}
\left(\begin{array}{cc}
0&0\\
0&1
\end{array}\right).
\eeq
Spinors are expressed as
\beq
\epsilon=\epsilon_-^{\{z\}}\sqrt{\dd z}
+\epsilon_+^{\{z\}}\sqrt{\dd \bz}.
\label{spinorexpression}
\eeq
We shall often suppress the superscript ``$\{z\}$'' when it is obvious.
The pairing between $S_-$ and $S_+$ is extended to an antisymmetric
bilinear form on $S$,
\beq
\langle\epsilon,\eta\rangle
=(2g_{z\bz})^{-\half}(\epsilon_+\eta_-
-\epsilon_-\eta_+).
\eeq
If the spinors are anticommuting, then it is {\it symmetric},
$\langle\epsilon,\eta\rangle=\langle\eta,\epsilon\rangle$.
It obeys other relations including Fierz identities,
\beqa
&&\langle\epsilon,\gamma^{\mu}\eta\rangle=-\langle\gamma^{\mu}\epsilon,
\eta\rangle,\qquad
\langle\epsilon,\gamma_3\eta\rangle=-\langle\gamma_3\epsilon,\eta\rangle,\nn\\
&&\epsilon\langle\eta,\lambda\rangle+\eta\langle\lambda,\epsilon\rangle
+\lambda\langle\epsilon,\eta\rangle=0,\nn\\
&&\gamma^{\mu}\epsilon\langle\eta,\gamma_{\mu}\lambda\rangle
+\gamma_3\epsilon\langle\eta,\gamma_3\lambda\rangle
+\epsilon\langle\eta,\lambda\rangle+2\lambda\langle\epsilon,\eta\rangle=0,
\nn\\
&&\epsilon\langle\eta,\lambda\rangle
-\gamma_3\epsilon\langle\gamma_3\eta,\lambda\rangle
+2\langle P_-\epsilon,\eta\rangle P_-\lambda
+2\langle P_+\epsilon,\eta\rangle P_+\lambda=0.
\label{Fierzetc}
\eeqa
We may also write spinors as
\beq
\epsilon=\epsilon_{\{z\}}^+\sqrt{\partial\over\partial\bz}
+\epsilon_{\{z\}}^-\sqrt{\partial\over\partial z},
\eeq
where 
\beq
\epsilon^{\{z\}}_{\pm}=\pm (2g_{z\bz})^{\half}\epsilon_{\{z\}}^{\mp},
\eeq
so that the bilinear form has the expression
\beq
\langle\epsilon,\eta\rangle
=(2g_{z\bz})^{\half}(-\epsilon^-\eta^+
+\epsilon^+\eta^-)
=\epsilon^-\eta_-+\epsilon^+\eta_+.
\eeq
These mean $\langle\sqrt{\partial\over\partial z},\sqrt{\dd z}\rangle
=\langle\sqrt{\partial\over\partial \bz},\sqrt{\dd \bz}\rangle=1$,
$\langle\sqrt{\dd\bz},\sqrt{\dd z}\rangle=(2g_{z\bz})^{-\half}$,
$\langle\sqrt{\partial\over\partial \bz},\sqrt{\partial\over\partial z}
\rangle=(2g_{z\bz})^{\half}$, and
\beq
\sqrt{\dd z}=-(2g_{z\bz})^{-\half}\sqrt{\partial\over\partial \bz},\qquad
\sqrt{\dd \bz}=(2g_{z\bz})^{-\half}\sqrt{\partial\over\partial z}.
\eeq
 On the flat space with metric 
$\dd^2s=|\dd z|^2$ we have $2g_{z\bz}=1$. The above spinor 
convention matches with
the dimensionally reduced and Wick rotated version of the
standard one in four dimensions \cite{WessBagger}.

A conformal Killing spinor is a section $\epsilon$ of $S$ obeying
\beq
\nabla_{\mu}\epsilon=\gamma_{\mu}\epsilon'
\eeq
for some other section $\epsilon'$. Obviously,
$\epsilon'=\half\nbirac\epsilon$, and the condition
is equivalent to
\beq
\partial_z\epsilon^+=0,\qquad
\partial_{\bz}\epsilon^-=0.
\eeq
That is, the $S_-$ and $S_+$ components of $\epsilon$
are antiholomorphic and holomorphic
sections of $\sqrt{\overline{T}}_{\Sigma}$ and $\sqrt{T}_{\Sigma}$
respectively. If $\Sigma$ is closed, such a spinor exists only when 
$\Sigma$ is a sphere or a torus.

When the manifold $\Sigma$ has a boundary $\partial\Sigma$,
a spin structure includes, as a part of the information, an identification
\beq
\varsigma: S_{\mp}|_{\partial\Sigma}\longrightarrow 
S_{\pm}|_{\partial\Sigma},\qquad
\varsigma^2={\rm id},
\eeq
whose second tensor power equals
a canonical isomorphism between $K_{\Sigma}|_{\Sigma}$ and
$\overline{K}_{\Sigma}|_{\Sigma}$.
As the canonical isomorphism, we may take the one
that sends $\dd\zeta$ to $-\dd\overline{\zeta}$ for
 a complex coordinate $\zeta$ near the
boundary that maps the chart of $\Sigma$ to the upper half plane.
A conformal Killing spinor is assumed to be anti-invariant
under $\varsigma$ at the boundary. Then it defines a conformal Killing spinor
of the double, $\Sigma\,\sharp\overline{\Sigma}$, which exists only when
the latter is a sphere or a torus. That is, a conformal Killing spinor
exists only when $\Sigma$ is a hemisphere or an annulus.

\subsection{Sphere And Hemisphere}

A two-sphere is $\CP^1$ as a complex manifold and is
covered by two charts. One with coordinate $z$
and the other with $w$ which are related by $zw=1$.
The round sphere metric of radius $r$ is
\beq
\dd s^2={4r^2|\dd z|^2\over (1+|z|^2)^2}\quad\mbox{or}
\quad
g_{z\bz}={2r^2\over (1+|z|^2)^2},
\eeq
with the Christoffel symbols given by
$\Gamma^z_{zz}=-{2\bz\over 1+|z|^2}$, $\Gamma^{\bz}_{\bz\bz}=-{2z\over 1+|z|^2}$.
The expressions in terms of the $w$ coordinate are the same.
It is useful to note
\beq
(2g_{z\bz})^{\half}={2r\over 1+|z|^2}\stackrel{|z|\to 1}{\longrightarrow} r.
\eeq
There is a unique spin structure on $\CP^1$. 
$\sqrt{T}_{\CP^1}$ as a holomorphic bundle is isomorphic to ${\mathcal O}(1)$ 
 and has two holomorphic sections. Thus, there are
four conformal Killing spinors,
\beq
{\bf s}_{-\half}=\sqrt{{\partial\over\partial z}},\quad\,
{\bf s}_{\half}=z\sqrt{{\partial\over\partial z}},\quad\,
\wt{\bf s}_{-\half}=\sqrt{{\partial\over\partial \bz}},\quad\,
\wt{\bf s}_{\half}=\bz\sqrt{{\partial\over\partial \bz}}.
\eeq
It is useful to note that
\beq
\nbirac\,{\bf s}_{\pm\half}=\mp\,{1\over r}\,\wt{\bf s}_{\mp\half},\qquad
\nbirac\,\wt{\bf s}_{\pm\half}=\mp\,{1\over r}\,{\bf s}_{\mp\half}.
\label{diracons}
\eeq

Let us consider the southern hemisphere $D^2_0=\{|z|\leq 1\}$.
There are two spin structures, $(+)_0$ and $(-)_0$, given by
\beq
\varsigma_{(\pm)_0}:\sqrt{\dd z\over z}\longleftrightarrow
\pm\sqrt{\dd \bz\over\bz},\quad
\sqrt{z{\partial\over\partial z}}
\longleftrightarrow
\mp \sqrt{\bz{\partial\over\partial \bz}},\qquad
\mbox{at $|z|=1$.}
\label{defspst}
\eeq
There are two conformal Killing spinors for each,
\beqa
(+)_0:&&
{\bf s}_{(+)+}={\bf s}_{\half}+\wt{\bf s}_{-\half},\quad
{\bf s}_{(+)-}={\bf s}_{-\half}+\wt{\bf s}_{\half},\\
(-)_0:&&
{\bf s}_{(-)+}={\bf s}_{\half}-\wt{\bf s}_{-\half},\quad
{\bf s}_{(-)-}={\bf s}_{-\half}-\wt{\bf s}_{\half}
\eeqa
For the outward unit normal vector at the boundary
\beq
\wh{n}={1\over r}
\left(z{\partial\over\partial z}+\bz{\partial\over\partial\bz}\right),
\eeq
$\gamma^{\whn}=g_{\mu\nu}\whn^{\mu}\gamma^{\nu}$ acts on the above conformal
Killing spinors as
\beq
\gamma^{\whn} \,\,{\bf s}_{(\pm )\nu}=\mp\,{\bf s}_{(\pm)\nu}\,\,\quad
\mbox{at $|z|=1$.}
\label{ngammaons}
\eeq 

Finally, let us consider the northern hemisphere $D^2_{\infty}=\{|w|\leq 1\}$.
We define two spin structures $(\pm)_{\infty}$
in the same way as (\ref{defspst}) but with the replacement $z,\bz
\to w,\overline{w}$.
Conformal Killing spinors are
${\bf s}_{(-)\pm}$ for $(+)_{\infty}$
and ${\bf s}_{(+)\pm}$ for $(-)_{\infty}$.

\appendix{Graded Chan-Paton Factor}\label{app:gradedCP}

Chan-Paton factors which appear in this paper takes the following form 
\beq
{\rm tr}_M^{}\left[P\exp\left(\oint_{S^1}\left(\upsi^a T_a
+V\right)\dd \tau\right)\right]\quad\mbox{or}\quad
{\rm Str}_M^{}\left[P\exp\left(\oint_{S^1}\left(\upsi^a T_a
+V\right)\dd \tau\right)\right],
\label{genCP}
\eeq
where $M$ is a $\Z_2$-graded vector space,
$\tau\equiv \tau+\beta$ is a periodic coordinate of a circle $S^1$,
$T_a$ and $V$ are functions on $S^1$ with values in
${\rm End}^{\it od}(M)$ and ${\rm End}^{\it ev}(M)$
respectively, $\upsi^a$ are fermionic fields
(i.e. anticommuting functions) on $S^1$.
${\rm tr}_M$ is the usual trace over $M$
and ${\rm Str}_M$ is the supertrace defined by
${\rm Str}_M(U)={\rm tr}_{M^{\it ev}}(U)-{\rm tr}_{M^{\it od}}(U)$.
We take the usual trace when the fermions are anit-periodic
$\upsi^a(\tau+\beta)=-\upsi^a(\tau)$ and the supertrance
when they are periodic $\upsi^a(\tau+\beta)=\upsi^a(\tau)$.
In this appendix, we give a definition to the expression
like (\ref{genCP}), and explain why we take the trace or the supertrace
depending on the periodicity of $\upsi^a(\tau)$.\footnote{{\bf Warning:}
We are not reviewing the well understood rule of
quantum mechanics that the trace and the supertrace correspond
respectively
to path integrals over fermions with the antiperiodic and periodic
boundary conditions along a (time) circle. The present problem can be
related to that, as we will mention below, but only in a special case.}

We start with defining
\beq
U(\tau_f,\tau_i)=P\exp\left(\int_{\tau_i}^{\tau_f}
\left(\upsi^a T_a+V\right)\dd \tau\right),
\label{formal1}
\eeq
for an interval $[\tau_i,\tau_f]$.
First, we formally apply the usual rule of
path ordered exponential. 
If we set $V=0$ for simplicity just for now,
the $n$-th order term is of the form
\beq
\int^{\tau_f}_{\tau_i}\dd \tau_n\cdots\int^{\tau_3}_{\tau_i}\dd \tau_2
\int^{\tau_2}_{\tau_i}\dd \tau_1\,
(\upsi^{a_n}T_{a_n})(\tau_n)
\cdots
(\upsi^{a_2}T_{a_2})(\tau_2)
(\upsi^{a_1}T_{a_1})(\tau_1)
\label{formal2}
\eeq
We {\it now define} this expression by
\beqa
\lefteqn{:=
\int^{\tau_f}_{\tau_i}\dd \tau_n\cdots\int^{\tau_3}_{\tau_i}\dd \tau_2
\int^{\tau_2}_{\tau_i}\dd \tau_1\,(-1)^{1+2+\cdots+(n-1)}\,
\upsi^{a_n}(\tau_n)\cdots\upsi^{a_2}(\tau_2)\upsi^{a_1}(\tau_1)}\nn\\
&&\qquad\qquad\qquad\qquad\qquad\qquad\qquad\times\,
T_{a_n}(\tau_n)
\cdots T_{a_2}(\tau_2)T_{a_1}(\tau_1).
\label{actual}
\eeqa
The last line is the usual matrix multiplcation of $T_{a_j}(\tau_j)$'s.
We can recover $V\ne 0$ by inserting
 $U_0(\tau_{j+1},\tau_j):=P\exp\left(\int^{\tau_{j+1}}_{\tau_j}V(\tau)
\dd \tau\right)$ between $T_{a_{j+1}}(\tau_{j+1})$ and $T_{a_j}(\tau_j)$,
as well as $U_0(\tau_f,\tau_n)$ to the left of
$T_{a_n}(\tau_n)$ and $U_0(\tau_1,\tau_i)$ to the right of $T_{a_1}(\tau_1)$.
By the sign $(-1)^{1+\cdots +(n-1)}$, we may treat  
$T_a(\tau)$'s as fermionic quantities inside the formal expressions
like (\ref{formal1}) and (\ref{formal2}). But in the actual
definition (\ref{actual}), they are genuine
(``bosonic'') functions with values in the space
${\rm End}^{\it od}(M)$ of usual matrices.
Let us express $U=U(\tau_f,\tau_i)$ with respect to a basis of $M$,
where the first entries are even and the last entries are odd,
\beq
U\stackrel{\cdot}{=}\left(\begin{array}{cc}
A&B\\
C&D
\end{array}\right).
\label{Uexpr}
\eeq
In view of the above definition of $U$,
we see that $A$ and $D$ have even powers of
$\upsi^a(\tau)$'s and hence are bosonic
while $B$ and $C$ have odd powers of
$\upsi^a(\tau)$'s and hence are fermionic.

We next consider the case where $\tau$ is a coordinate of
 a circle with periodocity $\tau\equiv \tau+\beta$.
In the usual case, say the case $T^a=0$,
we can simply take the trace
of $U(\tau_0+\beta,\tau_0)$ to define an invariant.
This does not depend on the choice of the initial time
 $\tau_0$, because
\beqa
{\rm tr}\,U(\tau_0+\beta,\tau_0)&=&
{\rm tr}\,\left[U(\tau_0+\beta,\tau_1)U(\tau_1,\tau_0)\right]
={\rm tr}\,\left[U(\tau_1,\tau_0)U(\tau_0+\beta,\tau_1)\right]\nn\\
&=&{\rm tr}\,\left[U(\tau_1+\beta,\tau_0+\beta)U(\tau_0+\beta,\tau_1)\right]
={\rm tr}\,U(\tau_1+\beta,\tau_1).
\eeqa
In this proof, we used the following properties
\beqa
\mbox{composition rule}&&
U(\tau_2,\tau_1)=U(\tau_2,\tau_*)U(\tau_*,\tau_1),\nn\\
\mbox{cyclicity of the trace}&&{\rm tr}\,[U_1U_2]={\rm tr}\,[U_2U_1],\nn\\
\mbox{periodicity}&& U(\tau_2+\beta,\tau_1+\beta)=U(\tau_2,\tau_1).\nn
\eeqa
In the graded case, the composition rule holds for (\ref{formal1}).
However, the cyclicity of the trace or supertrace may fail since
some of the matrix entries are fermionic.
To examine how it may fail or hold, let us write
\beq
U_i=\left(\begin{array}{cc}
A_i&B_i\\
C_i&D_i
\end{array}\right),\qquad i=1,2,
\eeq
with respect to the basis where the first entries are even and last entries
are odd. We have
\beqa
{\rm tr}\,[U_1U_2]&=&{\rm tr}\,[A_1A_2+B_1C_2+C_1B_2+D_1D_2],\nn\\
{\rm Str}\,[U_1U_2]&=&{\rm tr}\,[A_1A_2+B_1C_2-C_1B_2-D_1D_2],\nn
\eeqa
When $A$, $D$ are bosonic and $B$, $C$ are fermionic as in (\ref{Uexpr}),
then we
see that the supertrace has the right cyclicity
\beq
{\rm Str}\,[U_1U_2]={\rm Str}\,[U_2U_1]
\eeq
but the usual trace violates it in the middle two terms.
However, we can say
\beq
{\rm tr}\,[U_1U_2]={\rm tr}\,[U_2U_1]|_{B_2\to -B_2, \atop C_2\to -C_2}
\eeq
For $U$ in (\ref{Uexpr}), the sign flip of the $B$ and $C$ components
can be realized by $\tau\to \tau+\beta$ provided $\upsi^a(\tau)$
are antiperiodic.
This proves that the following is independent of the choice of
the initial time $\tau_0$:
\beqa
{\rm tr}_M^{}\,U(\tau_0+\beta,\tau_0)&&
\mbox{if $\upsi^a(\tau)$ are antiperiodic,}\nn\\
{\rm Str}_M^{}\,U(\tau_0+\beta,\tau_0)&&
\mbox{if $\upsi^a(\tau)$ are periodic.}\nn
\eeqa

In the special case where the rank of $M$ is a power of $2$,
there is a very familar way to understand the above
construction. Let us consider the simplest case where ${\rm rank}(M)=2$,
${\rm rank}(M^{\it ev})={\rm ranl}(M^{\it od})=1$.
Let us write
\beq
T_a=\left(\begin{array}{cc}
0&f_a\\
g_a&0
\end{array}\right),\qquad
V=\left(\begin{array}{cc}
V_0&0\\
0&V_0
\end{array}\right).
\eeq
(We take this special form for $V$ for simplicity.)
Using
\beq
\eta=\left(\begin{array}{cc}
0&1\\
0&0
\end{array}\right),\qquad
\overline{\eta}=\left(\begin{array}{cc}
0&0\\
1&0
\end{array}\right),
\eeq
the matrix $\upsi^aT_a+V$ may be written as
$\upsi^a(f_a\eta+g_a\overline{\eta})+V_0=:-H$. We may regard $H$ as
a time dependent Hamiltonian of a quantum mechanical system
whose space of states is $M$.
In the path-integral formulation, such a system can be realized by
a pair of anticommuting variables $\eta(t)$, 
$\overline{\eta}(t)$, with the Lagrangian
$L=\im\overline{\eta}{\dd\over \dd t}\eta-H$.
The matrix (\ref{formal1}), which can be regarded as the evolution
in the imaginary time, $\tau=\im t$, is represented by the path-integral
with an appropriate boundary condition ${\bf B}^{\tau_f}_{\,\,\tau_i}$
\beq
U(\tau_f,\tau_i)=\int_{{\bf B}^{\tau_f}_{\,\,\tau_i}}
{\mathcal D}\overline{\eta}{\mathcal D}\eta
\exp\left(\int_{\tau_i}^{\tau_f}\left(-\overline{\eta}{\dd\over \dd\tau}\eta
+\upsi^a(f_a\eta+g_a\overline{\eta})
+V_0
\right)
\dd\tau\right).
\eeq
Let us now discuss the case where $\tau$ is a periodic coordinate,
$\tau\equiv\tau+\beta$. 
If $\upsi^a(\tau)$ is anti-periodic ({\it resp}. periodic),
we need $\eta(\tau)$ and $\overline{\eta}(\tau)$ to be also
anti-periodic ({\it resp}. periodic), in order for the Lagrangian to be
periodic. By the standard quantization rule, we have
\beqa
{\rm tr}_M^{}\,U(\beta,0)\!\!&=&\!\!
\int_{\bf A}{\mathcal D}\overline{\eta}{\mathcal D}\eta
\exp\left(\int_{\tau_i}^{\tau_f}\left(-\overline{\eta}{\dd\over \dd\tau}\eta
+\upsi^a(f_a\eta+g_a\overline{\eta})
+V_0
\right)
\dd\tau\right)\!,\\
{\rm Str}_M^{}\,U(\beta,0)\!\!&=&\!\!
\int_{\bf P}{\mathcal D}\overline{\eta}{\mathcal D}\eta
\exp\left(\int_{\tau_i}^{\tau_f}\left(-\overline{\eta}{\dd\over \dd\tau}\eta
+\upsi^a(f_a\eta+g_a\overline{\eta})
+V_0
\right)
\dd\tau\right)\!,
\eeqa
where ${\bf A}$ and ${\bf P}$ stand for the anti-periodic and 
the periodic boundary
conditions for both $\eta(\tau),\overline{\eta}(\tau)$ and $\upsi^a(\tau)$.
In this presentation,
we explicitly see that $T_a=f_a\eta+g_a\overline{\eta}$
is a fermionic opeartor which is anti-periodic ({\it resp}. periodic)
in the former ({\it resp}. latter) case.

\appendix{Explict Expressions For Supersymmetry Transformations}
\label{app:SUSY}

For the study of supersymmetry of the boundary conditions,
we explicitly write down the A$_{(\pm)}$-type supersymmetry transformation
of the chiral multiplet and vector multiplet fields.
Spinors are written in components (\ref{fcomp}) 
with respect to the natural frames
near the boundary $\partial D^2$. 
We also use the variational parameter
$\varepsilon(\tau)$ and $\bvarepsilon(\tau)$
defined in (\ref{defvarep}).
Expressions are simplified a little by partially using
$\upsi_-'=|z|^{\mp\half}\upsi_-$,
$\upsi_+'=|z|^{\pm\half}\upsi_+$,
$\bupsi_-'=|z|^{\pm\half}\bupsi_-$,
$\bupsi_+'=|z|^{\mp\half}\bupsi_+$, 
$\ulambda_-'=|z|^{\pm\half}\ulambda_-$,
$\ulambda_+'=|z|^{\mp\half}\ulambda_+$,
$\bulambda_-'=|z|^{\mp\half}\bulambda_-$,
$\bulambda_+'=|z|^{\pm\half}\bulambda_+$.
(Here and elsewhere the multiple signs $\pm$ or $\mp$
are always correlated with the spin structure $(\pm)$ or equivalently
the type A$_{(\pm)}$ of supersymmetry.)
We also use
$$
|x_{1,2}|={2|z|\over 1+|z|^2}\stackrel{\partial D^2}{=}1,\qquad
x_3={|z|^2-1\over 1+|z|^2}\stackrel{\partial D^2}{=}0.
$$
The transformation of the chiral multiplet fields is
\beqa
&&\delta\phi=\varepsilon(\upsi_-'+\upsi_+'),\quad\,\,
\delta\bphi=-\bvarepsilon(\bupsi_-'+\bupsi_+'),\label{SAchiral}\\
&&\delta(\upsi'_-+\upsi'_+)=2\bvarepsilon\left[
D_{\tau}\phi+\left(\pm\left({\im\over 2r}R-\sigma_1\right)
+\im x_3\sigma_2\right)\phi\right],
\nn\\
&&\delta(\bupsi'_-+\bupsi'_+)=2\varepsilon\left[
-D_{\tau}\bphi+\bphi\left(\pm\left({\im\over 2r}R-\sigma_1\right)
+\im x_3\sigma_2\right)
\right],
\nn\\
&&\delta(\upsi'_--\upsi'_+)=2\bvarepsilon\left[\im D_{\rho}\phi
-\left(x_3\left({\im\over 2r}R-\sigma_1\right)\pm\im \sigma_2\right)\phi\right]
\mp 2\im\varepsilon |x_{1,2}|f,
\nn\\
&&\delta(\bupsi'_--\bupsi'_+)=2\varepsilon\left[-\im D_{\rho}\bphi
+\bphi\left(x_3\left({\im\over 2r}R-\sigma_1\right)\pm\im \sigma_2\right)\right]
\mp 2\im\bvarepsilon |x_{1,2}|\overline{f},
\nn\\
&&\delta f=\bvarepsilon\Biggl[\pm{2\over r|x_{1,2}|}
\left(|z|^{\pm \half}zD_z\upsi_+
+|z|^{\mp\half}\bz D_{\bz}\upsi_-\right)\nn\\
&&~~~~~~~~~
-|z|^{\pm\half}\left({R\over 2r}+\im\bsigma\right)\upsi_-
+|z|^{\mp\half}\left({R\over 2r}+\im\sigma\right)\upsi_+
-\left(|z|^{\pm\half}\bulambda_-+|z|^{\mp\half}\bulambda_+\right)\phi
\Biggr],\nn\\
&&\delta \overline{f}=\varepsilon\Biggl[\pm{2\over r|x_{1,2}|}
\left(|z|^{\mp \half}zD_z\bupsi_+
+|z|^{\pm\half}\bz D_{\bz}\bupsi_-\right)\nn\\
&&~~~~~~~~~
+|z|^{\mp\half}\bupsi_-\left({R\over 2r}+\im\sigma\right)
-|z|^{\pm\half}\bupsi_+\left({R\over 2r}+\im\bsigma\right)
-\bphi\left(|z|^{\mp\half}\ulambda_-+|z|^{\pm\half}\ulambda_+\right)
\Biggr].\nn
\eeqa
The transformation of the vector multiplet fields is
\beqa
&&\delta v_{\tau}=
\mp{|x_{1,2}|\over 2}\varepsilon\left(
|z|^{\pm \half}\bulambda_--|z|^{\mp \half}\bulambda_+\right)
\mp{|x_{1,2}|\over 2}\bvarepsilon\left(
|z|^{\mp \half}\ulambda_--|z|^{\pm \half}\ulambda_+\right),
\label{SAvector}\\
&&\delta v_{\rho}=
\pm\im{|x_{1,2}|\over 2}\varepsilon\left(
|z|^{\pm \half}\bulambda_-+|z|^{\mp \half}\bulambda_+\right)
\pm\im{|x_{1,2}|\over 2}\bvarepsilon\left(
|z|^{\mp \half}\ulambda_-+|z|^{\pm \half}\ulambda_+\right),\nn\\
&&\delta\sigma_1=-{\im\over 2}\varepsilon(\bulambda_-'-\bulambda_+')
-{\im\over 2}\bvarepsilon\left(\ulambda_-'-\ulambda_+'\right),\nn\\
&&\delta\sigma_2={1\over 2}\varepsilon(\bulambda_-'+\bulambda_+')
-{1\over 2}\bvarepsilon\left(\ulambda_-'+\ulambda_+'\right),\nn\\
&&\delta(\ulambda_-'-\ulambda_+')=2\varepsilon\left[
\,\im D_{\tau}\sigma_1-\im D_{\rho}\sigma_2
\pm\left(D_E+{\sigma_1\over r}\right)
+\im x_3\left({v_{12}\over\sqrt{g}}+{\sigma_2\over r}+\half[\sigma,\bsigma]
\right)\right],
\nn\\
&&\delta(\bulambda_-'-\bulambda_+')=2\bvarepsilon\left[
\,\im D_{\tau}\sigma_1+\im D_{\rho}\sigma_2
\mp\left(D_E+{\sigma_1\over r}\right)
-\im x_3\left({v_{12}\over\sqrt{g}}+{\sigma_2\over r}-\half[\sigma,\bsigma]
\right)\right],
\nn\\
&&\delta(\ulambda_-'+\ulambda_+')=2\varepsilon\left[
-D_{\rho}\sigma_1-D_{\tau}\sigma_2
+x_3\left(D_E+{\sigma_1\over r}\right)
\pm\im \left({v_{12}\over\sqrt{g}}+{\sigma_2\over r}+\half[\sigma,\bsigma]
\right)\right],
\nn\\
&&\delta(\bulambda_-'+\bulambda_+')=2\bvarepsilon\left[
-D_{\rho}\sigma_1+D_{\tau}\sigma_2
+x_3\left(D_E+{\sigma_1\over r}\right)
\pm\im \left({v_{12}\over\sqrt{g}}+{\sigma_2\over r}-\half[\sigma,\bsigma]
\right)\right],
\nn\\
&&\delta D_E={\im\over r|x_{1,2}|}\left\{
\mp\bvarepsilon\left(
|z|^{\pm\half}zD_z\ulambda_++|z|^{\mp\half}\bz D_{\bz}\ulambda_-\right)
\pm\varepsilon\left(
|z|^{\mp\half}zD_z\bulambda_++|z|^{\pm\half}\bz D_{\bz}\bulambda_-\right)
\right\}
\nn\\
&&~~~~~~~~~
+{\im\over 2r}\left\{\varepsilon(\bulambda_-'-\bulambda_+')
+\bvarepsilon(\ulambda_-'-\ulambda_+')\right\}\nn\\
&&~~~~~~~~~
+{1\over 2}\left[\sigma_1,
\varepsilon(\bulambda'_--\bulambda'_+)
-\bvarepsilon(\ulambda'_--\ulambda'_+)
\right]
+{\im\over 2}\left[\sigma_2,
\varepsilon(\bulambda'_-+\bulambda'_+)
+\bvarepsilon(\ulambda'_-+\ulambda'_+)
\right],\nn\\
&&\delta{v_{12}\over \sqrt{g}}=
{1\over r|x_{1,2}|}\left\{
\pm\bvarepsilon\left(
|z|^{\pm\half}zD_z\ulambda_+-|z|^{\mp\half}\bz D_{\bz}\ulambda_-\right)
\pm\varepsilon\left(
|z|^{\mp\half}zD_z\bulambda_+-|z|^{\pm\half}\bz D_{\bz}\bulambda_-\right)
\right\}
\nn\\
&&~~~~~~~~~
+{1\over 2r}\left\{\varepsilon(-\bulambda_-'-\bulambda_+')
+\bvarepsilon(\ulambda_-'+\ulambda_+')\right\}.\nn
\eeqa

If we set $|z|=1$, the above transformation rule simplifies.
The rule (\ref{SAchiral}) for the chiral multiplet just becomes
(\ref{chSUSYb}).
For the vector multiplet, we write $\sigma^a=\sigma_a$ for $a=1,2$
and introduce
\beqa
&&\ulambda^1={\im\over 2}(\ulambda_--\ulambda_+)
-{\im\over 2}(\bulambda_--\bulambda_+),\quad
\ulambda^2={1\over 2}(\ulambda_-+\ulambda_+)
+{1\over 2}(\bulambda_-+\bulambda_+),\nn\\
&&\tulambda^1=-{\im\over 2}(\ulambda_-+\ulambda_+)
+{\im\over 2}(\bulambda_-+\bulambda_+),\quad
\tulambda^2=-{1\over 2}(\ulambda_--\ulambda_+)
-{1\over 2}(\bulambda_--\bulambda_+),\nn\\
&&D^1_0=\mp\left({v_{12}\over\sqrt{g}}+{\sigma_2\over r}\right),\quad
D^2_0=\pm\left(D_E+{\sigma_1\over r}\right),
\eeqa
and $N^a=D_{\rho}\sigma^a+\im D^a_0$. We also use
$\varepsilon_1$ and $\varepsilon_2$ given by
$\varepsilon=\im\varepsilon_1-\varepsilon_2$ and
$\bvarepsilon=-\im\varepsilon_1-\varepsilon_2$.
Then a part of (\ref{SAvector}) at $|z|=1$ is written as (\ref{vctSUSYb}).


\begin{thebibliography}{99}

\small
\parskip=0pt plus 2pt


\bibitem{Windex}
E.~Witten,
  ``Constraints on Supersymmetry Breaking,''
  Nucl.\ Phys.\ B {\bf 202} (1982) 253.

\bibitem{Pestun}
V.~Pestun,
  ``Localization of gauge theory on a four-sphere
and supersymmetric Wilson loops,''
  Commun.\ Math.\ Phys.\  {\bf 313} (2012) 71
  [arXiv:0712.2824 [hep-th]].


\bibitem{Beninietal}
  F.~Benini and S.~Cremonesi,
  ``Partition functions of N=(2,2) gauge theories on $S^2$ and vortices,''
  arXiv:1206.2356 [hep-th].


\bibitem{Doroudetal}
  N.~Doroud, J.~Gomis, B.~Le Floch and S.~Lee,
  ``Exact Results in D=2 Supersymmetric Gauge Theories,''
  JHEP {\bf 1305} (2013) 093
  [arXiv:1206.2606 [hep-th]].

\bibitem{Romoetal}
H.~Jockers, V.~Kumar, J.~M.~Lapan, D.~R.~Morrison and M.~Romo,
  ``Two-Sphere Partition Functions and Gromov-Witten Invariants,''
  arXiv:1208.6244 [hep-th].


\bibitem{WalcherLG}
J.~Walcher,
``Stability of Landau-Ginzburg branes,''
  J.\ Math.\ Phys.\  {\bf 46} (2005) 082305
  [hep-th/0412274].


\bibitem{Hosono}
  S.~Hosono,
  ``Central charges, symplectic forms, and hypergeometric series
in local mirro\
r symmetry,''
  hep-th/0404043.
  %%CITATION = HEP-TH/0404043;%%                                                


\bibitem{Iritani09}
  H.~Iritani,
  ``An integral structure in quantum cohomology and mirror symmetry
for toric o\
rbifolds,''
  Adv. Math. {\bf 222} (2009), no.3, 1016-1079
  arXiv:0903.1463 [math.AG].

%\cite{Katzarkov:2008hs}                                                        
\bibitem{Katzarkov}
  L.~Katzarkov, M.~Kontsevich and T.~Pantev,
  ``Hodge theoretic aspects of mirror symmetry,''
  arXiv:0806.0107 [math.AG].
  %%CITATION = ARXIV:0806.0107;%%                                               

\bibitem{Iritani11}
  H.~Iritani,
  ``Quantum Cohomology and Periods,''
   arXiv:1101.4512 [math.AG].

\bibitem{HHP}
M.~Herbst, K.~Hori and D.~Page,
  ``Phases Of N=2 Theories In 1+1 Dimensions With Boundary,''
  arXiv:0803.2045 [hep-th].

\bibitem{HV}
K.~Hori and C.~Vafa,
  ``Mirror symmetry,''
  hep-th/0002222.

\bibitem{HIV}
  K.~Hori, A.~Iqbal and C.~Vafa,
  ``D-branes and mirror symmetry,''
  hep-th/0005247.


\bibitem{WessBagger}
J.~Wess and J.~Bagger,
{\it Supersymmetry and Supergravity},
(Princeton Univ. Press, 1992).

\bibitem{Wphases}
E.~Witten,
  ``Phases of N=2 theories in two-dimensions,''
  Nucl.\ Phys.\ B {\bf 403} (1993) 159
  [hep-th/9301042].

\bibitem{cmibook}
K.~Hori, S.~Katz, A.~Klemm, R.~Pandharipande, R.~Thomas, C.~Vafa,
R.~Vakil and E.~Zaslow, {\it Mirror Symmetry},
(AMS/Clay Math. Inst., 2003).


\bibitem{LVW}
  W.~Lerche, C.~Vafa and N.~P.~Warner,
  ``Chiral Rings in N=2 Superconformal Theories,''
  Nucl.\ Phys.\ B {\bf 324} (1989) 427.

\bibitem{OOY}
  H.~Ooguri, Y.~Oz and Z.~Yin,
  ``D-branes on Calabi-Yau spaces and their mirrors,''
  Nucl.\ Phys.\ B {\bf 477} (1996) 407
  [hep-th/9606112].


\bibitem{BPZ}
A.~A.~Belavin, A.~M.~Polyakov and A.~B.~Zamolodchikov,
  ``Infinite Conformal Symmetry in Two-Dimensional Quantum Field Theory,''
  Nucl.\ Phys.\ B {\bf 241} (1984) 333.


\bibitem{CV}
S.~Cecotti and C.~Vafa,
  ``Topological antitopological fusion,''
  Nucl.\ Phys.\ B {\bf 367} (1991) 359.

\bibitem{GomisLee}
J.~Gomis and S.~Lee,
  ``Exact Kahler Potential from Gauge Theory and Mirror Symmetry,''
  JHEP {\bf 1304} (2013) 019
  [arXiv:1210.6022 [hep-th]].


\bibitem{Warner}
N.~P.~Warner,
  ``Supersymmetry in boundary integrable models,''
  Nucl.\ Phys.\ B {\bf 450} (1995) 663
  [hep-th/9506064].

\bibitem{KapLi}
  A.~Kapustin and Y.~Li,
  ``D branes in Landau-Ginzburg models and algebraic geometry,''
  JHEP {\bf 0312} (2003) 005
  [hep-th/0210296].

\bibitem{BHLS}
  I.~Brunner, M.~Herbst, W.~Lerche and B.~Scheuner,
  ``Landau-Ginzburg realization of open string TFT,''
  JHEP {\bf 0611} (2006) 043
  [hep-th/0305133].

\bibitem{HW}
  K.~Hori and J.~Walcher,
  ``D-branes from matrix factorizations,''
  Comptes Rendus Physique {\bf 5} (2004) 1061
  [hep-th/0409204].

\bibitem{Pontrjagin}
L.~Pontrjagin,
{\it Topological Groups},
(Princeton Univ. Press, 1939).


\bibitem{Rose}
M.~E.~Rose, {\it Elementary Theory of Angular Momentum},
(John Wiley \& Sons, 1957).


\bibitem{LeeTalks}
S.~Lee, talks at {\it Geometry and Physics of the Gauged Linear Sigma Model},
Univ. Michigan, March 4-8, 2013 and at
{\it Strings 2013}, Seoul, June 24-28, 2013.

\bibitem{privatecomm}
Private communication with J.~Gomis, March 1-3, 2013.



\bibitem{Polyakov}
A.~M.~Polyakov,
  ``Quantum Geometry of Bosonic Strings,''
  Phys.\ Lett.\ B {\bf 103} (1981) 207.


\bibitem{Martinec}
E.~J.~Martinec,
  ``Algebraic Geometry and Effective Lagrangians,''
  Phys.\ Lett.\ B {\bf 217} (1989) 431.

\bibitem{VafaWarner}
C.~Vafa and N.~P.~Warner,
  ``Catastrophes and the Classification of Conformal Theories,''
  Phys.\ Lett.\ B {\bf 218} (1989) 51.
  


\bibitem{SilWi}
E.~Silverstein and E.~Witten,
  ``Global U(1) R symmetry and conformal invariance of (0,2) models,''
  Phys.\ Lett.\ B {\bf 328} (1994) 307
  [hep-th/9403054].

\bibitem{HoTo}
K.~Hori and D.~Tong,
 ``Aspects of Non-Abelian Gauge Dynamics in Two-Dimensional N=(2,2) Theories,''
  JHEP {\bf 0705} (2007) 079
  [hep-th/0609032].



\bibitem{HPT}K.~Hori, C.Y.~Park and Y.~Tachikawa,
``2d SCFT from M2-branes'' to appear.


\bibitem{2dduality}
K.~Hori,
 ``Duality In Two-Dimensional (2,2)
Supersymmetric Non-Abelian Gauge Theories,''
  arXiv:1104.2853 [hep-th].

\bibitem{WVerlinde}
E.~Witten,
  ``The Verlinde algebra and the cohomology of the Grassmannian,''
  In *Cambridge 1993, Geometry, topology, and physics* 357-422
  [hep-th/9312104].


\bibitem{Candelasetal}
P.~Candelas, X.~C.~De La Ossa, P.~S.~Green and L.~Parkes,
 ``A Pair of Calabi-Yau manifolds as an exactly soluble
superconformal theory,''
  Nucl.\ Phys.\ B {\bf 359} (1991) 21.


\bibitem{MP}
D.~R.~Morrison and M.~R.~Plesser,
  ``Summing the instantons: Quantum cohomology
and mirror symmetry in toric varieties,''
  Nucl.\ Phys.\ B {\bf 440} (1995) 279
  [hep-th/9412236].


\bibitem{Rodland}
E.A. R\o dland,
``The Pfaffian Calabi-Yau, its mirror, and their link to the Grassmannian
$G(2,7)$,''
Composito Math. {bf 122} (2000) 135-149;
arXiv:math/9801092.

\bibitem{DoSe}
W.~Donovan and E.~Segal,
``Window shifts, flop equivalences and Grassmannian twists,''
 arXiv:1206.0219 [math.AG].


\bibitem{ReSc}
A.~Recknagel and V.~Schomerus,
  ``D-branes in Gepner models,''
  Nucl.\ Phys.\ B {\bf 531} (1998) 185
  [hep-th/9712186].


\bibitem{Segal}
E.~Segal,
``Equivalences between GIT quotients of Landau-Ginzburg B-models,''
Commun. Math. Phys.{\bf 304} (2011) 411-432.


\bibitem{DHL}
D.~Halpern-Leistner,
``The derived category of a GIT quotient,''\\
 arXiv:1203.0276 [math.AG].

\bibitem{BFK}
M.~Ballard, D.~Favero and L.~Katzarkov,
``Variation of geometric invariant theory quotients and derived categories,''
arXiv:1203.6643 [math.AG]

\bibitem{Orlov}
D.~Orlov,
``Derived categories of coherent sheaves and triangulated
categories of singularities,''
Progress. Math. {\bf 270} (2009) 503-531;
[arXiv:math/0503632].


\bibitem{Strominger}
  A.~Strominger,
  ``Massless black holes and conifolds in string theory,''
  Nucl.\ Phys.\ B {\bf 451}, 96 (1995)
  [arXiv:hep-th/9504090].


\bibitem{SeTh}
P.~Seidel and R.~Thomas,
``Braid group actions on derived categories of coherent sheaves,''
arXiv:math/0001043;
Duke Math.\ Jour.\ {\bf 108} (2001) 37--108.

\bibitem{Horja}
  R.~P.~Horja,
  ``Hypergeometric functions and Mirror Symmetry in Toric Varieties,''
  [arXiv:math.AG/9912109].

\bibitem{DHLmono}
D.~Halpern-Leistner and I.~Shipman,
``Autoequivalences of derived categories via geometric invariant theory,''
arXiv:1303.5531 [math.AG].


\bibitem{Hirzebruch}
F.~Hirzebruch,
{\it Topological methods in algebraic geometry}, (Springer 1966).


\bibitem{Knorrer}
H.~Kn\"orrer,
``Cohen-Macaulay modules on hypersurface singularities. I''
Invent.~Math. {\bf 88} (1987) 153-164.




\bibitem{Tsikh}
Zhdanov, O. N., and A. K. Tsikh,
``Studying the multiple Mellin-Barnes integrals by means of
multidimensional re\
sidues,''
Siberian Mathematical Journal {\bf 39}, 2 (1998): 245-260.

%\cite{Passare:1996db}                                                          
\bibitem{Passare}
  M.~Passare, A.~K.~Tsikh and A.~A.~Cheshel,
  ``Multiple Mellin-Barnes integrals as periods of Calabi-Yau
manifolds with se\
veral moduli,''
  Theor.\ Math.\ Phys.\  {\bf 109}, 1544 (1997)
  [Teor.\ Mat.\ Fiz.\  {\bf 109N3}, 381 (1996)]
  [hep-th/9609215].
  %%CITATION = HEP-TH/9609215;%%    

\bibitem{JK}
L.C.~Jeffrey and F.C.~Kirwan,
``Localization for nonabelian group actions,''
Topology {\bf 34} (1995) 291-327;
 [arXiv:alg-geom/9307001].

\bibitem{BEHT2}
F.~Benini, R.~Eager, K.~Hori and Y.~Tachikawa,
``Elliptic genera of 2d ${\mathcal N}=2$ gauge theories,''
to appear.

\bibitem{BEHT1}
F.~Benini, R.~Eager, K.~Hori and Y.~Tachikawa,
 ``Elliptic genera of two-dimensional N=2 gauge
theories with rank-one gauge groups,''
  arXiv:1305.0533 [hep-th].




\end{thebibliography}
\end{document}